\documentclass[aps,prd,a4paper,twoside,onecolumn,english,superscriptaddress,showkeys,nofootinbib,amssymb,amsmath,amsfonts]{revtex4}

\ifx\pdfoutput\undefined
  \usepackage[dvips]{graphicx}
  \DeclareGraphicsExtensions{.eps}
\else
  \usepackage[pdftex]{graphicx}
  \DeclareGraphicsExtensions{.pdf}
\fi

\usepackage{babel}
\usepackage{dcolumn}

\newcolumntype{e}[1]{D{.}{.}{#1}}

\begin{document}

\title{New perspectives in physics and astrophysics from the theoretical understanding of Gamma-Ray Bursts}

\author{\firstname{Remo} \surname{Ruffini}}\email{ruffini@icra.it}
\affiliation{ICRA --- International Center for Relativistic Astrophysics.}
\affiliation{Dipartimento di Fisica, Universit\`a di Roma ``La Sapienza'', Piazzale Aldo Moro 5, I-00185 Roma, Italy.}

\author{\firstname{Carlo Luciano} \surname{Bianco}}\email{bianco@icra.it}
\affiliation{ICRA --- International Center for Relativistic Astrophysics.}
\affiliation{Dipartimento di Fisica, Universit\`a di Roma ``La Sapienza'', Piazzale Aldo Moro 5, I-00185 Roma, Italy.}

\author{\firstname{Pascal} \surname{Chardonnet}}\email{chardonnet@icra.it}
\affiliation{ICRA --- International Center for Relativistic Astrophysics.}
\affiliation{Universit\'e de Savoie, LAPTH - LAPP, BP 110, F­74941 Annecy-le-Vieux Cedex, France.}

\author{\firstname{Federico} \surname{Fraschetti}}\email{fraschetti@icra.it}
\affiliation{ICRA --- International Center for Relativistic Astrophysics.}
\affiliation{Universit\`a di Trento, Via Sommarive 14, I-38050 Povo (Trento), Italy.}
 
\author{\firstname{Luca} \surname{Vitagliano}}\email{vitagliano@icra.it}
\affiliation{ICRA --- International Center for Relativistic Astrophysics.}
\affiliation{Dipartimento di Fisica, Universit\`a di Roma ``La Sapienza'', Piazzale Aldo Moro 5, I-00185 Roma, Italy.}

\author{\firstname{She-Sheng} \surname{Xue}}\email{xue@icra.it}
\affiliation{ICRA --- International Center for Relativistic Astrophysics.}
\affiliation{Dipartimento di Fisica, Universit\`a di Roma ``La Sapienza'', Piazzale Aldo Moro 5, I-00185 Roma, Italy.}

\begin{abstract}
If due attention is given in formulating the basic equations for the Gamma-Ray Burst (GRB) phenomenon and in performing the corresponding quantitative analysis, GRBs open a main avenue of inquiring on totally new physical and astrophysical regimes. This program is very likely one of the greatest computational efforts in physics and astrophysics and cannot be actuated using shortcuts. A systematic approach is needed which has been highlighted in three basic new paradigms: the relative space-time transformation (RSTT) paradigm (\textcite{lett1}), the interpretation of the burst structure (IBS) paradigm (\textcite{lett2}), the GRB-supernova time sequence (GSTS) paradigm (\textcite{lett3}). From the point of view of fundamental physics new regimes are explored: (1) the process of energy extraction from black holes; (2) the quantum and general relativistic effects of matter-antimatter creation near the black hole horizon; (3) the physics of ultrarelativisitc shock waves with Lorentz gamma factor $\gamma > 100$. From the point of view of astronomy and astrophysics also new regimes are explored: (i) the occurrence of gravitational collapse to a black hole from a critical mass core of mass $M\agt 10M_\odot$, which clearly differs from the values of the critical mass encountered in the study of stars ``catalyzed at the endpoint of thermonuclear evolution" (white dwarfs and neutron stars); (ii) the extremely high efficiency of the spherical collapse to a black hole, where almost $99.99\%$ of the core mass collapses leaving negligible remnant; (iii) the necessity of developing a fine tuning in the final phases of thermonuclear evolution of the stars, both for the star collapsing to the black hole and the surrounding ones, in order to explain the possible occurrence of the ``induced gravitational collapse". New regimes are as well encountered from the point of view of nature of GRBs: (I) the basic structure of GRBs is uniquely composed by a proper-GRB (P-GRB) and the afterglow; (II) the long bursts are then simply explained as the peak of the afterglow (the E-APE) and their observed time variability is explained in terms of inhomogeneities in the interstellar medium (ISM); (III) the short bursts are identified with the P-GRBs and the crucial information on general relativistic and vacuum polarization effects are encoded in their spectra and intensity time variability. A new class of space missions to acquire information on such extreme new regimes are urgently needed.
\end{abstract}


\maketitle

\tableofcontents

\section{Introduction}\label{newint}

In understanding new astrophysical phenomena, the solution has been found as soon as the energy source of the phenomena has been identified. This has been the case for pulsars (see \textcite{hbpsc68}) where the rotational energy of the neutron star was identified as the energy source (see e.g. \textcite{g68,g69}). Similarly, in binary X-ray sources the accretion process from a normal companion star in the deep potential well of a neutron star or a black hole has clearly pointed to the gravitational energy of the accreting matter as the basic energy source and all the main features of the light curves of the sources have been clearly understood (\textcite{gr78}). In this spirit, our work in the field of Gamma-Ray Bursts (GRBs) has focused to identify the energy extraction process from the black hole (\textcite{cr71}) as the basic energy sources for the GRB phenomenon: a distinguishing feature of this process is a theoretically predicted energetics of the source all the way up to $1.8\times 10^{54}\left(M_{BH}/M_{\odot}\right) {\rm ergs}$ for $3.2 M_{\odot} \le M_{BH} \le 7.2 \times 10^6 M_{\odot}$ (\textcite{dr75}). In particular, the very specific process of the formation of a ``dyadosphere'', during the process of gravitational collapse leading to a black hole endowed with electromagnetic structure (EMBH), has been indicated as originating and giving the initial boundary conditions of the onset of the GRB process (\textcite{rukyoto,prx98}). Our model has been referred as ``the EMBH model for GRBs'', although the EMBH physics only determines the initial boundary conditions of the GRB process by specifying the physical parameters and spatial extension of the neutral electron positron plasma originating the phenomenon.

Traditionally, following the observations of the {\em Vela} (\textcite{s75}) and {\em CGRO}\footnote{see http://cossc.gsfc.nasa.gov/batse/} satellites, GRBs have been characterized by few parameters such as the fluence, the characteristic duration ($T_{90}$ or $T_{50}$) and the global time averaged spectral distribution (\textcite{b93}). With the observations of {\em BeppoSAX}\footnote{see http://www.asdc.asi.it/bepposax/} and the discovery of the afterglow, and the consequent optical identification, the distance of the GRB source has been determined and consequently the total energetics of the source has been added as a crucial parameter.

The observed energetics of GRBs, coinciding for spherically symmetric explosions with the ones theoretically predicted in (\textcite{dr75}), has convinced us to develop in full details the EMBH model. For simplicity, we have considered the vacuum polarization process occurring in an already formed Riessner-Nordstr\"om black hole (\textcite{rukyoto,prx98}), whose dyadosphere has an energy $E_{dya}$. It is clear, however, that this is only an approximation to the real dynamical description of the process of gravitational collapse to an EMBH. In order to prepare the background for attacking this extremely complex dynamical process, we have clarified some basic theoretical issues, necessary to be implemented prior to the description of the fully dynamical process of gravitational collapse to an EMBH (\textcite{rv02a,rv02b,crv02}, see section~\ref{luca}). We have then described the following five eras in our model. {\em Era I}: the $e^+e^-$ pairs plasma, initially at $\gamma=1$, expands away from the dyadosphere as a sharp pulse (the PEM pulse), reaching Lorentz gamma factor of the order of 100 (\textcite{rswx99}). {\em Era II}: the PEM pulse, still optically thick, engulfs the remnant left over in the process of gravitational collapse of the progenitor star with a drastic reduction of the gamma factor; the mass $M_B$ of this engulfed baryonic material is expressed by the dimensionless parameter $B=M_Bc^2/E_{dya}$ (\textcite{rswx00}). {\em Era III}: the newly formed pair-electromagnetic-baryonic (PEMB) pulse, composed of $e^+e^-$ pair and of the electrons and baryons of the engulfed material, self-propels itself outward reaching in some sources Lorentz gamma factors of $10^3$--$10^4$; this era stops when the transparency condition is reached and the emission of the proper-GRB (P-GRB) occurs (\textcite{brx00}). {\em Era IV}: the resulting accelerated baryonic matter (ABM) pulse, ballistically expanding after the transparency condition has been reached, collides at ultrarelativistic velocities with the baryons and electrons of the interstellar matter (ISM) which is assumed to have a average constant number density, giving origin to the afterglow. {\em Era V}: this era represents the transition from the ultrarelativistic regime to the relativistic and then to the non relativistic ones (\textcite{rbcfx02e_paperI}).

Our approach differs in many respect from the ones in the current literature. The major difference consists in the appropriate theoretical description of all the above five eras, as well as in the evaluation of the process of vacuum polarization originating the dyadosphere. The dynamical equations as well as the description of the phenomenon in the laboratory time and the time sequence carried by light signals recorded at the detector have been explicitly integrated (see e.g. Tab.~\ref{tab1} and \textcite{rbcfx02e_paperI,rbcfx02e_paperII}). In doing so we have also corrected a basic conceptual mistake, common to all the current works on GRBs, which led to the wrong spacetime parametrization of the GRB phenomenon, preempting all these theoretical works from their predictive power. The description of the inner engine originating the GRBs has never been addressed in the necessary details in the literature. In this sense neither the specific boundary conditions originating in the dyadosphere nor the needed solutions of the relativistic hydrodynamic and pair equations for the first three eras described above have been considered. Only the treatment of the afterglow has been widely considered in the literature by the so-called ``fireball model'' (see e.g. \textcite{mr92mnras,mr93,rm94,p99} and references therein).

However, also in the description of the afterglow, which is represented by the two conceptually and technically simplest eras in our model, there are major differences between the works in the literature and our approach:\\
{\bf a)} Processes of synchrotron radiation and inverse Compton as well as an adiabatic expansion in the source generating the afterglow are usually adopted in the current literature. On the contrary, in our approach a ``fully radiative'' condition is systematically adopted in the description of the X-ray and $\gamma$-ray emission of the afterglow. The basic microphysical emission process is traced back to the physics of shock waves as considered by \textcite{zr66}. A special attention is given to identify such processes in the comoving frame of the shock front generating the observed spectra of the afterglow (see \textcite{rbcfx02c_spectrum}).\\
{\bf b)} In the literature the variation of the gamma Lorentz factor during the afterglow is expressed by a unique power-law of the radial co-ordinate of the source and a similar power-law relation is assumed also between the radial coordinate of the source and the asymptotic observer frame time. Such simple approximations appear to be quite inadequate and do contrast with the almost hundred pages summarizing the needed computations which we recall in the rest of this article. In our approach the dynamical equations of the source are integrated self-consistently with the constitutive equations relating the observer frame time to the laboratory time and the boundary conditions are adopted and uniquely determined by each previous era of the GRB source (see e.g. \textcite{rbcfx02a_sub,rbcfx02e_paperI,rbcfx02e_paperII,rbcfx02c_spectrum}).\\
{\bf c)} At variance with the many power-laws for the observed afterglow flux found in the literature, our treatment naturally leads to a ``golden value'' for the power-law index $n=-1.6$. The fit of the EMBH model to the observed afterglow data fixes the only two free parameters of our theory: the $E_{dya}$ and the $B$ parameter, measuring the remnant mass left over by the gravitational collapse of the progenitor star (\textcite{rbcfx02a_sub,rbcfx02e_paperI,rbcfx02e_paperII,rbcfx02c_spectrum}).

It is not surprising that such large differences in the theoretical treatment have led to a different interpretation of the GRB phenomenon as well as to the identification of new fundamental physical regimes. The introduction of new interpretative paradigms has been necessary and the theory has been confirmed by the observation to extremely high accuracy. 

In particular from the definition of the complete space-time coordinates of the GRB phenomenon as a function of the radial coordinate, the comoving time, the laboratory time, the arrival time and the arrival time at the detector, expressed in Tab.~\ref{tab1}, it has been concluded that in no way a description of a given era is possible in the GRB phenomena without the knowledge of the previous ones. Therefore the afterglow as such cannot be interpreted unless all the previous eras have been correctly computed and estimated. It has also become clear that a great accuracy in the analysis of each era is necessary in order to identify the theoretically predicted features with the observed ones. If this is done, the GRB phenomena presents an extraordinary and extremely precise correspondence between the theoretically predicted features and the observations leading to the exploration of totally new physical and astrophysical process with unprecedented accuracy.
This has been expressed in the relative space-time transformation (RSTT) paradigm: ``the necessary condition in order to interpret the GRB data, given in terms of the arrival time at the detector, is the knowledge of the {\em entire} worldline of the source from the gravitational collapse. In order to meet this condition, given a proper theoretical description and the correct constitutive equations, it is sufficient to know the energy of the dyadosphere and the mass of the remnant of the progenitor star'' (\textcite{lett1}).

Having determined the two independent parameters of the EMBH model, namely $E_{dya}$ and $B$, by the fit of the afterglow we have introduced a new interpretative paradigm for the burst structure: the IBS paradigm (\textcite{lett2}). In it we reconsider the relative roles of the afterglow and the burst in the GRBs by defining in this complex phenomenon two new phases:\\
{\bf 1)} the {\em injector phase} starting with the process of gravitational collapse, encompassing the above Eras I, II, III and ending with the emission of the Proper-GRB (P-GRB);\\
{\bf 2)} the {\em beam-target phase} encompassing the above Eras IV and V giving rise to the afterglow. In particular in the afterglow three different regimes are present for the average bolometric intensity : one increasing with arrival time, a second one with an Extended Afterglow Peak Emission (E-APE) and finally one decreasing as a function of the arrival time. Only this last one appears to have been considered in the current literature (\textcite{lett2}).

The EMBH model allows, in the case of GRB~991216, to compute the intensity ratio of the afterglow to the P-GRB ($1.45 \cdot 10^{-2}$), and the arrival time of the P-GRB ($8.413 \cdot 10^{-2}$s) as well as the arrival time of the peak of the afterglow ($19.87$s) (see Figs.~\ref{991216},\ref{fintkin},\ref{dtab}). The fact that the theoretically predicted intensities coincide within a few percent with the observed ones and that the arrival time of the P-GRB and the peak of the afterglow also do coincide within a tenth of millisecond with the observed one can be certainly considered a clear success of the predictive power of the EMBH model.

As a by-product of this successful analysis, we have reached the following conclusions:\\
{\bf a)} The most general GRB is composed by a P-GRB, an E-APE and the rest of the afterglow. The ratio between the P-GRB and the E-APE intensities is a function of the B parameter.\\
{\bf b)} In the limit B=0 all the energy is emitted in the P-GRB. These events represent the ``short burst'' class, for which no afterglows has been observed.\\
{\bf c)} The ``long bursts'' do not exist, they are just part of the afterglow, the E-APEs.

We are currently verifying these theoretical predictions on the following GRBs: GRB~991216, GRB~980425, GRB~970228, GRB~980519. It is very remarkable that, although the energetics of GRB~980425 (see Fig.~\ref{991216}) differs from the one of GRB~991216 by roughly five orders of magnitude, the model applies also to this case with success. Furthermore from these analysis we can claim that both in the case of GRB~991216 and in the case of GRB~980425 there is not significant departure from spherical symmetry.

While this analysis of the average bolometric intensity of GRB was going on in the radial approximation, we have proceeded to the full non-radial approximation, taking into account all the relativistic corrections for the off-axis emission from the spherically symmetric expansion of the ABM pulse (\textcite{rbcfx02a_sub,rbcfx02e_paperII}). We have so defined the temporal evolution of the ABM pulse visible area (see Fig.~\ref{opening_p}), as well as the equitemporal surfaces (see Fig.~\ref{ETSNCF_p}) (\textcite{rbcfx02a_sub,rbcfx02e_paperII}).

We have then addressed the issue whether the fast temporal variations observed in the so-called long bursts, on time scales as short as fraction of a second (\textcite{rbcfx02a_sub}), can indeed be explained as an effect of inhomogeneities in the interstellar medium.

We are making further progress in identifying the basic mechanisms of energy release in the afterglow by presenting a new theoretical formalism which as a function of only one parameter fits the entire spectral distribution of the X-ray and $\gamma$-ray radiation in GRB~991216 (\textcite{rbcfx02c_spectrum}).

Finally the GRB-supernova time sequence (GSTS) paradigm introduces the concept of {\em induced supernova explosion} in the supernovae-GRB association (\textcite{lett3}) leading to the very novel possibility of a process of gravitational collapse induced on a companion star in a very special evolution phase by the GRB explosion.

Before concluding, we also present some theoretical developments which have been motivated by preparing the analysis of the general relativistic effects during the process of gravitational collapse itself and we also show how such results motivated by GRB studies have already generated new results in the fundamental understanding of black hole physics.

In the next section we briefly summarize the main results and we will then give the summary of the treatment in the following sections. For the complete details we refer to the quoted papers.

\section{Summary of the main results}\label{int}

\subsection{The physical and astrophysical background}

Gamma-ray bursts (GRBs) are rapidly fueling one of the broadest scientific pursuit in the entire field of science, both in the observational and theoretical domains. Following the discovery of GRBs by the Vela satellites (\textcite{s75}), the observations from the Compton satellite and BATSE had shown the isotropic distribution of the GRBs strongly suggesting a cosmological nature for their origin. It was still through the data of BATSE that the existence of two families of bursts, the ``short bursts'' and the ``long bursts'' was presented, opening an intense scientific dialogue on their origin still active today, see e.g. \textcite{s01} and section \ref{new}.

An enormous momentum was gained in this field by the discovery of the afterglow phenomena by the BeppoSAX satellite and the optical identification of GRBs which have allowed the unequivocal identification of their sources at cosmological distances (see e.g. \textcite{c00}). It has become apparent that fluxes of $10^{54}$ erg/s are reached: during the peak emission the energy of a single GRB equals the energy emitted by all the stars of the Universe (see e.g. \textcite{rk01}).

From an observational point of view, an unprecedented campaign of observations is at work using the largest deployment of observational techniques from space with the satellites CGRO-BATSE, Beppo-SAX, Chandra\footnote{see http://chandra.harvard.edu/}, R-XTE\footnote{see http://heasarc.gsfc.nasa.gov/docs/xte/}, XMM-Newton\footnote{see http://xmm.vilspa.esa.es/}, HETE-2\footnote{see http://space.mit.edu/HETE/}, as well as the HST\footnote{see http://www.stsci.edu/}, and from the ground with optical (KECK\footnote{see http://www2.keck.hawaii.edu:3636/}, VLT\footnote{see http://www.eso.org/projects/vlt/}) and radio (VLA\footnote{see http://www.aoc.nrao.edu/vla/html/VLAhome.shtml}) observatories. The further possibility of examining correlations with the detection of ultra high energy cosmic rays, UHECR for short, and in coincidence neutrinos should be reachable in the near future thanks to developments of AUGER\footnote{see http://www.auger.org/} and AMANDA\footnote{see http://amanda.berkeley.edu/amanda/amanda.html} (see also \textcite{h00}).

\begin{figure}
\includegraphics[width=10cm,clip]{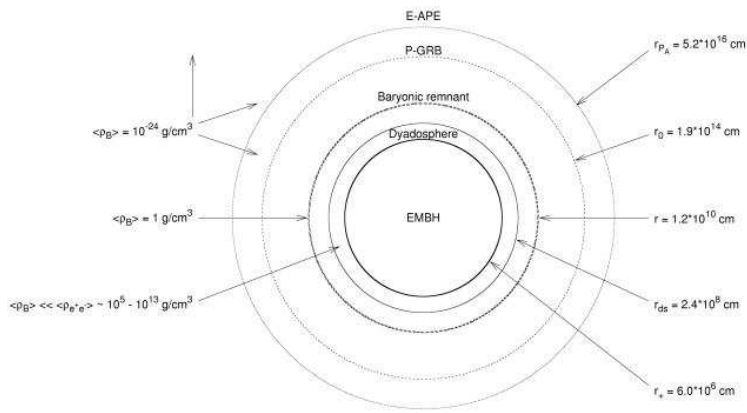}
\caption{Selected events in the EMBH theory are represented. For each one the values of the energy density of the medium and the distances from the EMBH, in the laboratory frame and in logarithmic scale, are given.}
\label{raggi2}
\end{figure}

From a theoretical point of view, GRBs offer comparable opportunities to develop entire new domains in yet untested directions of fundamental science. For the first time within the theory based on the vacuum polarization process occurring in an electromagnetic black hole, the EMBH theory, see Fig.~\ref{raggi2}, the opportunity exists to 
theoretically approach the following fundamental issues:
\begin{enumerate}
\item The extremely relativistic hydrodynamic phenomena of an electron-positron plasma expanding with sharply varying gamma factors in the range $10^2$ to $10^4$ and the analysis of the very high energy collision of such an expanding plasma with baryonic matter reaching intensities $10^{38}$ larger than the ones usually obtained in Earth-based accelerators.
\item The bulk process of vacuum polarization created by overcritical electromagnetic fields, in the sense of Heisenberg, Euler (\textcite{he35}) and Schwinger (\textcite{s51}). This longly sought quantum ultrarelativistic effect has not been yet unequivocally observed in heavy ion collision on the Earth (see e.g. \textcite{ga96,la97,la98,ha98}). The difficulty of the heavy ion collision experiments appears to be that the overcritical field is reached only for time scales of the order $\hbar/m_pc^2$, which is much shorter than the characteristic time for the $e^+e^-$ pair creation process which is of the order of $\hbar/m_ec^2$, where $m_p$ and $m_e$ are respectively the proton and the electron mass. It is therefore very possible that the first appearance of such an effect occurs in the present general relativistic context: in the strong electromagnetic fields developed in astrophysical conditions during the process of gravitational collapse to an EMBH, where no problem of confinement exists.
\item A novel form of energy source: the extractable energy of a black hole. The enormous energies released almost instantly in the observed GRBs, points to the possibility that for the first time we are witnessing the release of the extractable energy of an EMBH, during the process of gravitational collapse itself. This problem presents still some outstanding theoretical issues in black hole physics. Having progressed in some of these issues (see \textcite{crv02,rv02a,rv02b,rvx02}) we can now compute and have the opportunity to study all general relativistic as well as the associated ultrahigh energy quantum phenomena as the horizon of the EMBH is approached and is being formed (see section~\ref{luca}).
\end{enumerate}

It is clear that in approaching such a vast new field of research, implying previously unobserved relativistic and quantum regimes, it is not possible to proceed {\itshape as usual} with an uncritical comparison of observational data to theoretical models within the classical schemes of astronomy and astrophysics. Some insight to the new approach needed can be gained from past experience in the interpretation of relativistic effects in high energy particle physics as well as from the explanation of some observed relativistic effects in the astrophysical domain. Those relativistic regimes, both in physics and astrophysics, are however much less extreme than those encountered now in GRBs.

There are three major new features in relativistic systems which have to be properly taken into account:

\begin{figure}
\includegraphics[width=10cm,clip]{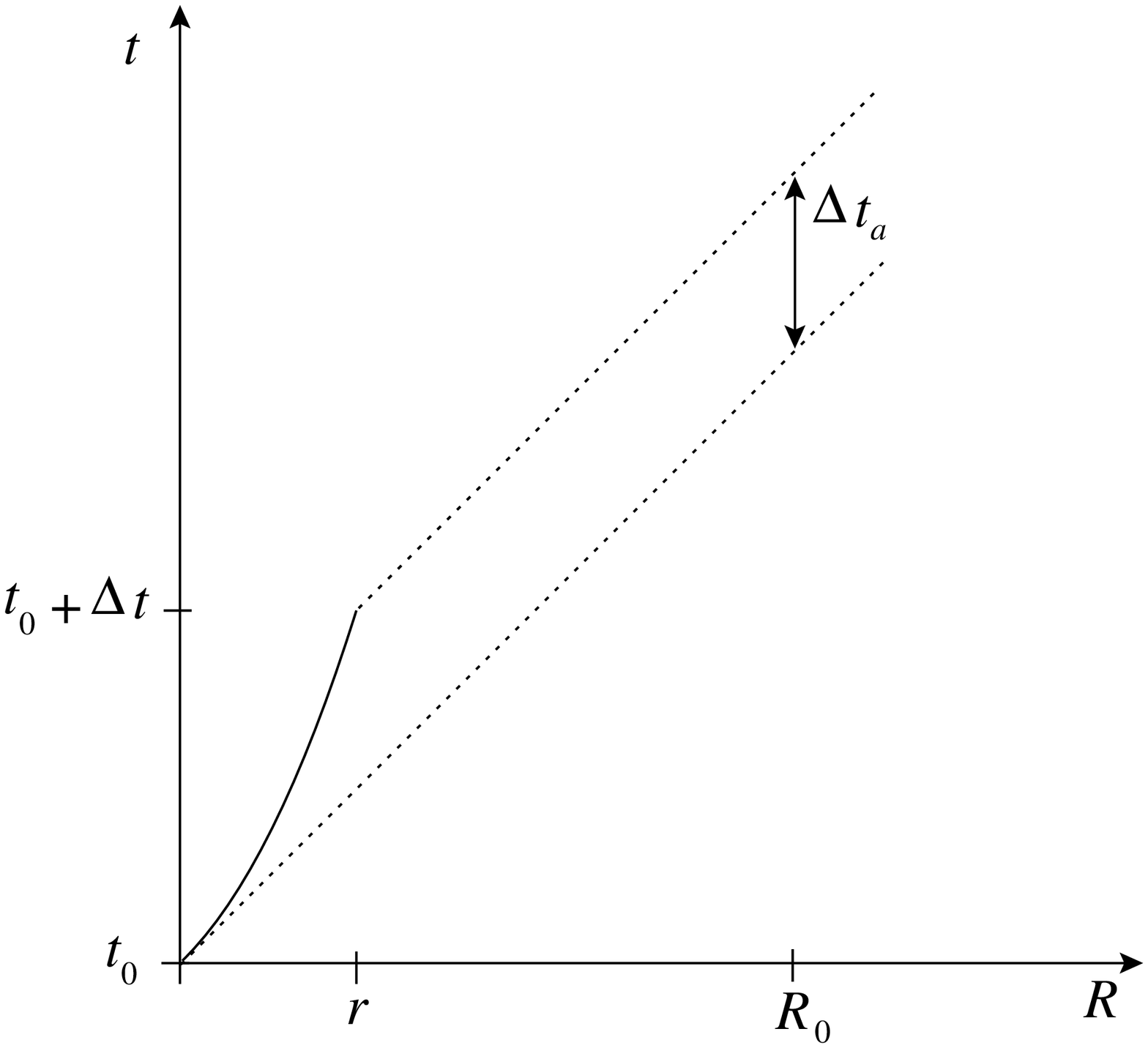}
\caption{This qualitative diagram illustrates the relation between the laboratory time interval $\Delta t$ and the arrival time interval $\Delta t_a$ for a pulse moving with velocity $v$ in the laboratory time (solid line). We have indicated here the case where the motion of the source has a nonzero acceleration. The arrival time is measured using light signals emitted by the pulse (dotted lines). $R_0$ is the distance of the observer from the EMBH, $t_0$ is the laboratory time corresponding to the onset of the gravitational collapse, and $r$ is the radius of the expanding pulse at a time $t=t_0 + \Delta t$. See also \textcite{lett1}.}
\label{ttasch_p}
\end{figure}

\begin{enumerate}
\item Practically all data on astronomical and astrophysical systems is acquired by using photon arrival times. It was \textcite{e05} at the very initial steps of special relativity who cautioned about the use of such an arrival time analysis and stated that when dealing with objects in motion proper care should be taken in defining the time synchronization procedure in order to construct the correct space-time coordinate grid (see Fig.~\ref{ttasch_p}). It is not surprising that as soon as the first relativistic bulk motion effects were observed their interpretations within the classical framework of astrophysics led to the concept of ``superluminal'' motion. These were observations of extragalactic radio sources, with gamma factors $\sim 10$ (\textcite{bsm99}) and of microquasars in our own galaxy with gamma factor $\sim 5$ (\textcite{mr99}). It has been recognized (\textcite{r66}) that no ``superluminal'' motion exists if the prescriptions indicated by Einstein are used in order to establish the correct space-time grid for the astrophysical systems. In the present context of GRBs, where the gamma factor can easily surpass $10^2$ and is very highly varying, this approximation breaks down (\textcite{brx00,lett1,rbcfx02e_paperII}). The direct application of classical concepts in this context would lead to enormous ``superluminal'' behaviors (see e.g. Tab.~\ref{tab1}). An approach based on classical arrival time considerations as sometimes done in the current literature completely subverts the causal relation in the observed astrophysical phenomenon.
\item One of the clear successes of relativistic field theories has been the understanding of the role of four-momentum conservation laws in multiparticle collisions and decays such as in the reaction: $n\rightarrow p+e^-+\bar\nu_e$. From the works of Pauli and Fermi it became clear how in such a process, contrary to the case of classical mechanics, it is impossible to analyze a single term of the decay, the electron or the proton or the neutrino or the neutron, out of the context of the global point of view of the relativistic conservation of the total four momentum of the system. This in turn involves the knowledge of the system during the entire decay process. These rules are routinely used by workers in high energy particle physics and have become part of their cultural background. If we apply these same rules to the case of the relativistic system of a GRB it is clear that it is just impossible to consider a part of the system, e.g. the afterglow, without taking into account the general conservation laws and whole relativistic history of the entire system. Especially since in astrophysics the ``somewhat pathological" arrival time coordinate is basically used (see Fig.~\ref{ttasch_p}). The description of the afterglow alone, as has been given at times in the literature, indeed possible within the framework of classical astronomy and astrophysics, is not viable in a relativistic astrophysics context where the space-time grid necessary for the description of the afterglow depends on the entire previous relativistic part of the worldline of the system (see also section \ref{bf}).
\item The lifetime of a process has not an absolute meaning as special and general relativity have shown. It depends both on the inertial reference frame of the laboratory and of the observer and on their relative motion. Such a phenomenon, generally expressed in the ``twin paradox'', has been extensively checked and confirmed to extremely high accuracy as a byproduct of the elementary particle physics (g-2) experiment (see e.g. \textcite{vd77}). This situation is much more extreme in GRBs due to the very large (in the range $10^2$--$10^4$) and time varying (on time scales ranging from fractions of seconds to months) gamma factors between the comoving frame and the far away observer (see Fig.~\ref{gamma}). Moreover in the GRB context such an observer is also affected by the cosmological recession velocities of its local Lorentz frame.
\end{enumerate}

\subsection{The Relative Space-Time Transformations: the RSTT paradigm and current scientific literature}

Here are some of the reasons why we have presented a basic relative space-time transformation (RSTT) paradigm (\textcite{lett1}) to be applied prior to the interpretation of GRB data.

The first step is the establishment of the governing equations relating:\\
a) The comoving time of the pulse ($\tau$)\\
b) The laboratory time ($t$)\\
c) The arrival time at the detector ($t_a$)\\
d) The arrival time at the detector corrected for cosmological expansion ($t_a^d$)\\
The book-keeping of the four different times and corresponding space variables must be done carefully in order to keep the correct causal relation in the time sequence of the events involved. 

As formulated the RSTT paradigm contains two parts: the first one is a necessary condition, the second one a sufficient condition. The first part reads: ``the necessary condition in order to interpret the GRB data, given in terms of the arrival time at the detector, is the knowledge of the {\em entire} worldline of the source from the gravitational collapse''.

Clearly such an approach is in contrast with articles in the current literature which emphasize either some too qualitative description of the sources and the quantitative description of the sole afterglow era. In this quantitative description they oversimplify the relations between the radial coordinate of the source and its gamma Lorentz factor as well as the relation between the radial coordinate and the arrival time using power-law relations which do not correctly take into account the complexity of the problem.

In the current literature several attempts have addressed the issue of the sources of GRBs. They include scenarios of binary neutron stars mergers (see e.g. \textcite{elps89,npp92,mr92mnras,mr92apj}), black hole~/~white dwarf (\textcite{fwhd99}) and black hole~/~neutron star binaries (\textcite{p91,mr97b}), hypernovae (see \textcite{p98}), failed supernovae or collapsars (see \textcite{w93,mw99}), supranovae (see \textcite{vs98,vs99}). Only those based on binary neutron stars have reached the stage of a definite model and detailed quantitative estimates have been made. In this case, however, various problems have surfaced: in the general energetics which cannot be greater than $\sim 3\times 10^{52}$ erg, in the explanation of ``long bursts'' (see \textcite{swm00,wmm96}), and in the observed location of the GRB sources in star forming regions (see \textcite{bkd00}). In the remaining cases attention was directed to a qualitative analysis of the sources without addressing the overall problem from the source to the observations. Also generally missing are the necessary details to formulate the equations of the dynamical evolution of the system and to develop a complete theory to be compared with the observations.

Other models in the literature have addressed the problem of only fitting the data of the afterglow observations by simple power-laws. They are separated into two major classes:
 
The ``internal shock model'', introduced by \textcite{rm94}, by far the most popular one, has been developed in many different aspects, e.g. by \textcite{px94,sp97,f99,fcrsyn99}. The underlying assumption is that all the variabilities of GRBs in the range $\Delta t\sim 1\, {\rm ms}$ up to the overall duration $T$ of the order of $50\, {\rm s}$ are determined by a yet undetermined ``inner engine''. The difficulties of explaining the long time scale bursts by a single explosive model has evolved into a subclass of approaches assuming an ``inner engine'' with extended activity (see e.g. \textcite{p01} and references therein).

The ``external shock model'', see e.g. \textcite{cr78,sp90,mr93}, is less popular today. Paradoxically, some of the authors who have qualitatively highlighted distinctive features of this model have later disclaimed its validity (see e.g. \textcite{rm94,mr00,p99} and references therein). Possibly they were carried to this extreme conclusion by an impressive sequence of mistakes they made in implementing the basic physical processes of the model. This model relates the GRB light curves and time variabilities to interactions of a single thin blast wave with clouds in the external medium. The interesting possibility has been also recognized within this model, that GRB light curves ``are tomographic images of the density distribution of the medium surrounding the sources of GRBs'' (\textcite{dm99}), see also \textcite{dcb99,d00} and references therein. In this case, the structure of the burst is assumed not to depend directly on the ``inner engine'' (see e.g. \textcite{p01} and references therein).

All these works encounter the above mentioned difficulty: they present either a purely qualitative or phenomenological or a piecewise description of the GRB phenomenon. By neglecting the earlier phases, the relation of the space-time grid to the photon arrival time is not properly estimated. To tell more explicitly, their clocks are out of the proper synchronization and the theory is emptied of any predictive power!

We will explicitly show in the following how an unified description naturally leads to the identification of new characteristic features both in the burst and afterglow of GRBs. Our theory, in respect to the afterglow description, can be generally considered an ``external shock model'' and fits most satisfactorily all the observations.

\subsection{The EMBH Theory}

In a series of papers, we have developed the EMBH theory (\textcite{rukyoto}) which has the advantage, despite its simplicity, that all eras following the process of gravitational collapse are described by precise field equations which can then be numerically integrated.

Starting from the vacuum polarization process {\it \`a la} Heisenberg-Euler-Schwinger (\textcite{he35,s51}) in the overcritical field of an EMBH first computed in \textcite{dr75}, we have developed  the dyadosphere concept (\textcite{prx98}). 

The dynamics of the $e^+e^-$-pairs and electromagnetic radiation of the plasma generated in the dyadosphere propagating away from the EMBH in a sharp pulse (PEM pulse) has been studied by the Rome group and validated by the numerical codes developed at Livermore Lab (\textcite{rswx99}).

The collision of the still optically thick $e^+e^-$-pairs and electromagnetic radiation plasma with the baryonic matter of the remnant of the progenitor star has been again studied by the Rome group and validated by the Livermore Lab codes (\textcite{rswx00}). The further evolution of the sharp pulse of pairs, electromagnetic radiation and baryons (PEMB pulse) has been followed for increasing values of the gamma factor until the condition of transparency is reached (\textcite{brx00}). 

As this PEMB pulse reaches transparency the proper GRB (P-GRB) is emitted (\textcite{lett2}) and a pulse of accelerated baryonic matter (the ABM pulse) is injected into the interstellar medium (ISM) giving rise to the afterglow.

\subsection{The GRB~991216 as a prototypical source}

In the early phases of development of our model, the EMBH theory was developed from first principles by the EMBH uniqueness theorem (\textcite{rw71}), the energetics of black hole (\textcite{cr71}) as well as the quantum description of the vacuum polarization process in overcritical electromagnetic fields (\textcite{dr75}). Turning now to the afterglow, the variety of physical situations that can possibly be encountered are very large and far from unique: the description from first principles is just impossible. We have therefore proceeded to properly identify what we consider a prototypical GRB source and to develop a theoretical framework in close correspondence with the observational data.

The criteria which have guided us in the selection of the GRB source to be used as a prototype before proceeding to an uncritical comparison with the theory are expressed in the following. It is now clear, since the observations of GRB~980425, GRB~991216, GRB~970514 and GRB~980326 that the afterglow phenomena can present, especially in the optical and radio wavelengths, features originating from phenomena spatially and causally distinct from the GRB phenomena. There is also the distinct possibility that phenomena related to a supernova can be erroneously attributed to a GRB. This problem has been clearly addressed by the GRB supernova time sequence (GSTS) paradigm in which the time sequence of the events in the GRB supernova phenomena has been outlined (\textcite{lett3}). This has led to the novel concept of an induced supernova (\textcite{lett3}). This problem will be addressed in a forthcoming paper (\textcite{rbcfx02d_supernova}).

\begin{figure}
\includegraphics[width=\hsize,clip]{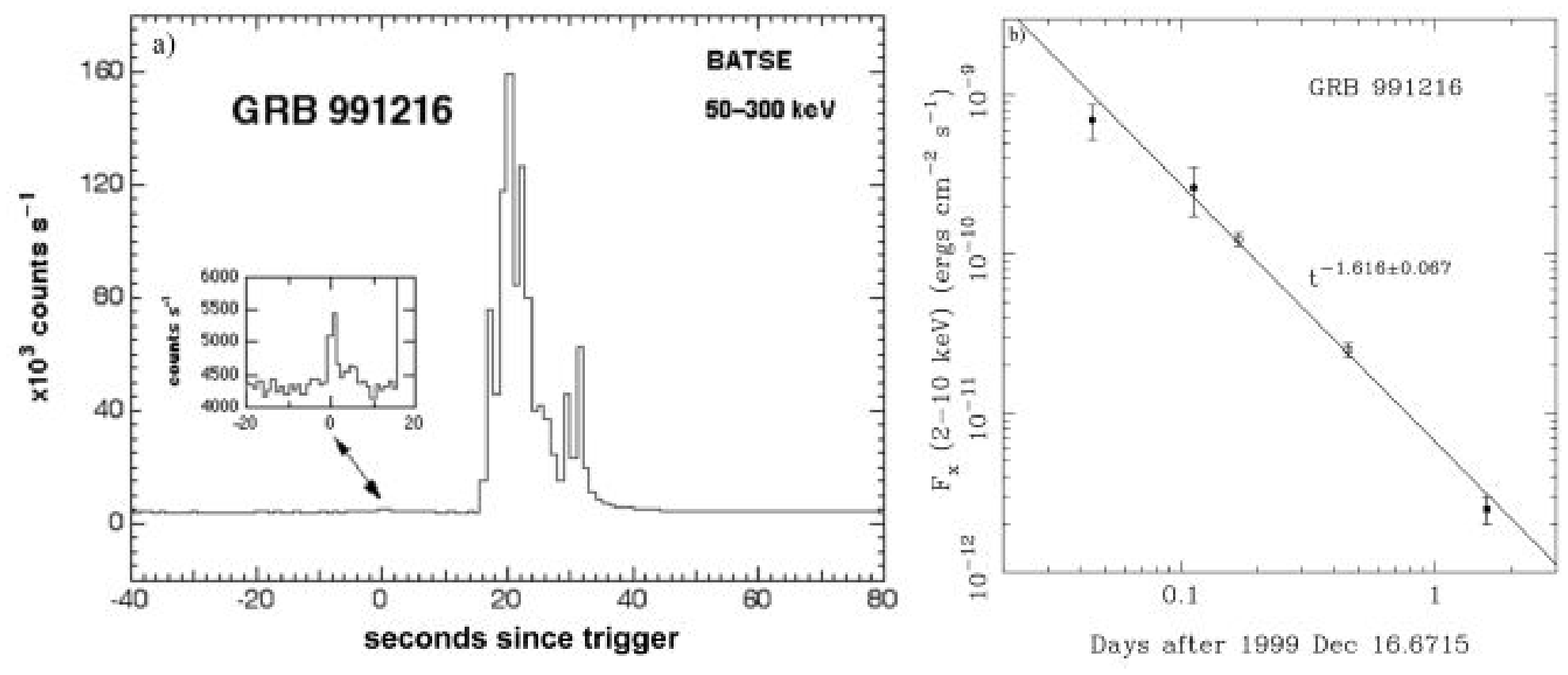}
\caption{{\bf a)} The peak emission of GRB~991216 as seen by BATSE (reproduced from \textcite{brbr99}); {\bf b)} The afterglow emission of GRB~991216 as seen by XTE and Chandra (reproduced from \textcite{ha00}).}
\label{grb991216}
\end{figure}

In view of these considerations we have selected GRB~991216 as a prototypical case (see Fig.~\ref{grb991216}) for the following reasons:
\begin{enumerate}
\item GRB~991216 is one of the strongest GRBs in X-rays and is also quite general in the sense that it shows relevant cosmological effects. It radiates mainly in X-rays and in $\gamma$-rays and less than 3\% is emitted in the optical and radio bands (see \textcite{ha00}).
\item The excellent data obtained by BATSE on the burst (\textcite{brbr99}) is complemented by the data on the afterglow acquired by Chandra (\textcite{p00}) and RXTE (\textcite{cs00}). Also superb data have been obtained from spectroscopy of the iron lines (\textcite{p00}).
\item A value for the slope of the energy emission during the afterglow as a function of time has been obtained: $n=-1.64$ (\textcite{tmmgk99}) and $n=-1.616\pm 0.067$ (\textcite{ha00}).
\end{enumerate}

\subsection{The interpretation of the burst structure: the IBS paradigm and the different eras of the EMBH theory}

The comparison of the EMBH theory with the data of the GRB~991216 and its afterglow has naturally led to a new paradigm for the interpretation of the burst structures (IBS paradigm)) of GRBs (\textcite{lett2}). The IBS paradigm reads: {\itshape ``In GRBs we can distinguish an injector phase and a beam-target phase. The injector phase includes the process of gravitational collapse, the formation of the dyadosphere, as well as Era I (the PEM pulse), Era II (the engulfment of the baryonic matter of the remnant) and Era III (the PEMB pulse). The injector phase terminates with the P-GRB emission. The beam-target phase addresses the interaction of the ABM pulse, namely the beam generated during the injection phase, with the ISM as the target. It gives rise to the E-APE and the decaying part of the afterglow''}. The detailed presentations of these results are a major topic in this article.

\begin{figure}
\includegraphics[width=10cm,clip]{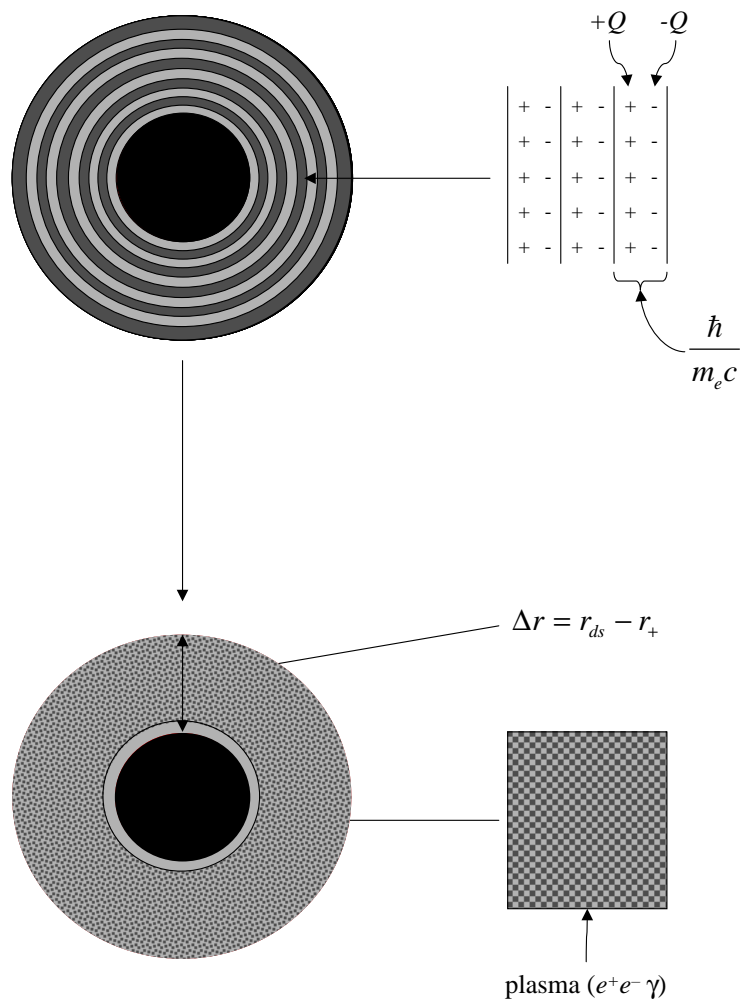}
\caption{The dyadosphere of a Reissner-Nordstr\"{o}m black hole can be represented as constituted by a concentric set of shells of capacitors, each one of thickness $\hbar/m_ec$ and producing a number of $e^+e^-$ pairs of the order of $\sim Q/e$ on a time scale of $10^{-21}$ s, where $Q$ is the EMBH charge. The shells extend in a region $\Delta r$, from the horizon $r_{+}$ to the dyadosphere outer radius $r_{\rm ds}$ (see text). The system evolves to a thermalised plasma configuration.}
\label{dyaon}
\end{figure}

We recall that the {\bf injector phase} starts from the moment of gravitational collapse and encompasses the following eras:

{\itshape The zeroth Era: the formation of the dyadosphere}. In section \ref{dyadosphere} we review the basic scientific results which lie at the basis of the EMBH theory: the black hole uniqueness theorem, the mass formula of an EMBH, the process of vacuum polarization in the field of an EMBH. We also point out how after the discovery of the GRB afterglow the reexamination of these results has led to the novel concept of the dyadosphere of an EMBH. We have investigated this concept in the simplest possible case of an EMBH depending only on two parameters: the mass and charge, corresponding to the Reissner-Nordstr\"{o}m spacetime. We recall the definition of the energy $E_{dya}$ of the dyadosphere as well as the spatial distribution and energetics of the $e^+e^-$ pairs. See Fig.~\ref{dyaon}. We return in section~\ref{luca} to the theoretical development of the time varying process lasting less than a second in the process of a realistic gravitational collapse. In reality the vacuum polarization process will lead to a final uncharged black hole, but the analysis based on a Reissner-Nordstr\"{o}m black hole is an excellent approximation to the description of this phenomenon (\textcite{rvx03}).

In order to analyse the time evolution of the dyadosphere we give in the three following sections the theoretical background for the needed equations.

In section \ref{hydro_pem} we give the general relativistic equations governing the hydrodynamics and the rate equations for the plasma of $e^+e^-$-pairs. 

In section \ref{arrival_time} we give the governing equations relating the comoving time $\tau$ to the laboratory time $t$ corresponding to an inertial reference frame in which the EMBH is at rest and finally to the time measured at the detector $t_a$ which, to finally get $t_a^d$, must be corrected to take into account the cosmological expansion. 

In section \ref{num_int} we describe the numerical integration of the hydrodynamical equations and the rate equation developed by the Rome and Livermore groups. This entire research program could never have materialized without the fortunate interaction between the complementary computational techniques developed by these two groups. The validation of the results of the Rome group by the fully general relativistic Livermore codes has been essential both from the point of view of the validity of the numerical results and the interpretation of the scientific content of the results.  

{\itshape The Era I: the PEM pulse}. In section \ref{hydro_pem} by the direct comparison of the integrations performed with the Rome and Livermore codes we show that among all possible geometries the $e^+e^-$ plasma moves outward from the EMBH reaching a very unique relativistic configuration: the plasma self-organizes in a sharp pulse which expands in the comoving frame exactly by the amount which compensates for the Lorentz contraction in the laboratory frame. The sharp pulse remains of constant thickness in the laboratory frame and self-propels outwards reaching ultrarelativistic regimes, with gamma factors larger than $10^2$, in a few dyadosphere crossing times. We recall that, in analogy with the electromagnetic (EM) pulse observed in a thermonuclear explosion on the Earth, we have defined this more energetic pulse formed of electron-positron pairs and electromagnetic radiation a pair-electromagnetic-pulse or PEM pulse.

{\itshape The Era II}: We describe the interaction of the PEM pulse with the baryonic remnant of mass $M_B$ left over from the gravitational collapse of the progenitor star. We give the details of the decrease of the gamma factor and the corresponding increase in the internal energy during the collision. The dimensionless parameter $B={M_Bc^2}/{E_{dya}}$ which measures the baryonic mass of the remnant in units of the $E_{dya}$ is introduced. This is the second fundamental free parameter of the EMBH theory.

\begin{figure}
\includegraphics[width=10cm,clip]{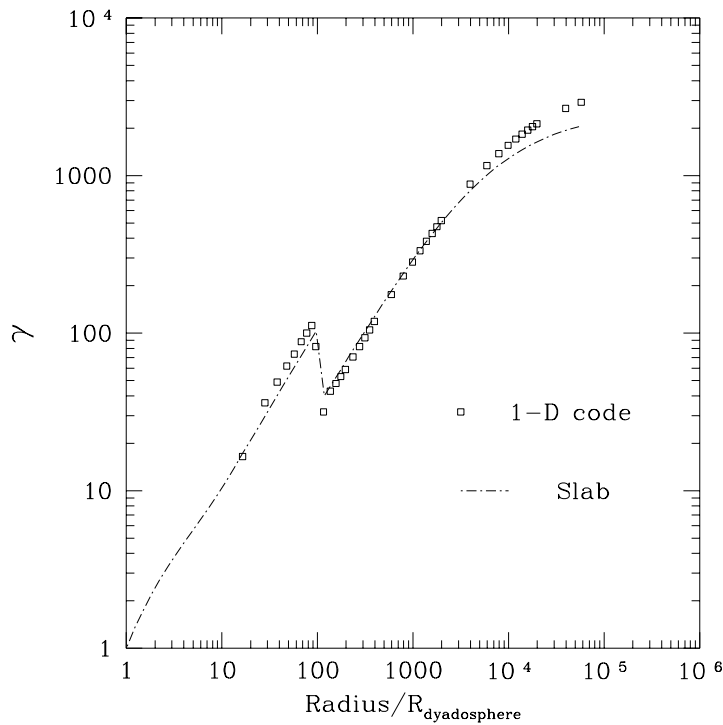}
\caption{Comparison of gamma factor for the one-dimensional (1-D) hydrodynamic calculations (Livermore code) and slab calculations (Rome code) as a function of the radial coordinate (in units of dyadosphere radius) in the laboratory frame. The calculations show an excellent agreement.}
\label{twocodecompare}
\end{figure}

{\itshape The Era III}: We describe in section~\ref{era3} the further expansion of the $e^+e^-$ plasma, after the engulfment of the baryonic remnant of the progenitor star. By direct comparison of the results of integration obtained with the Rome and the Livermore codes it is shown how the pair-electromagnetic-baryon (PEMB) plasma further expands and self organizes in a sharp pulse of constant length in the laboratory frame (see Fig.~\ref{twocodecompare}).
We have examined the formation of this PEMB pulse in a wide range of values $10^{-8}<B<10^{-2}$ of the parameter $B$, the upper limit corresponding to the limit of validity of the theoretical framework developed.

In section~\ref{fp} it is shown how the effect of baryonic matter of the remnant, expressed by the parameter $B$, is to smear out all the detailed information on the EMBH parameters. The evolution of the PEMB pulse is shown to depend only on $E_{dya}$ and $B$: the PEMB pulse is degenerate in the mass and charge parameters of the EMBH and rather independent of the exact location of the baryonic matter of the remnant.

In section~\ref{at} the relevant thermodynamical quantities of the PEMB pulse, the temperature in the different frames and the $e^+e^-$ pair densities, are given and the approach to the transparency condition is examined. Particular attention is given to the gradual transfer of the energy of the dyadosphere $E_{dya}$ to the kinetic energy of the baryons $E_{Baryons}$ during the optically thick part of the PEMB pulse. 

\begin{figure}
\includegraphics[width=8.5cm,clip]{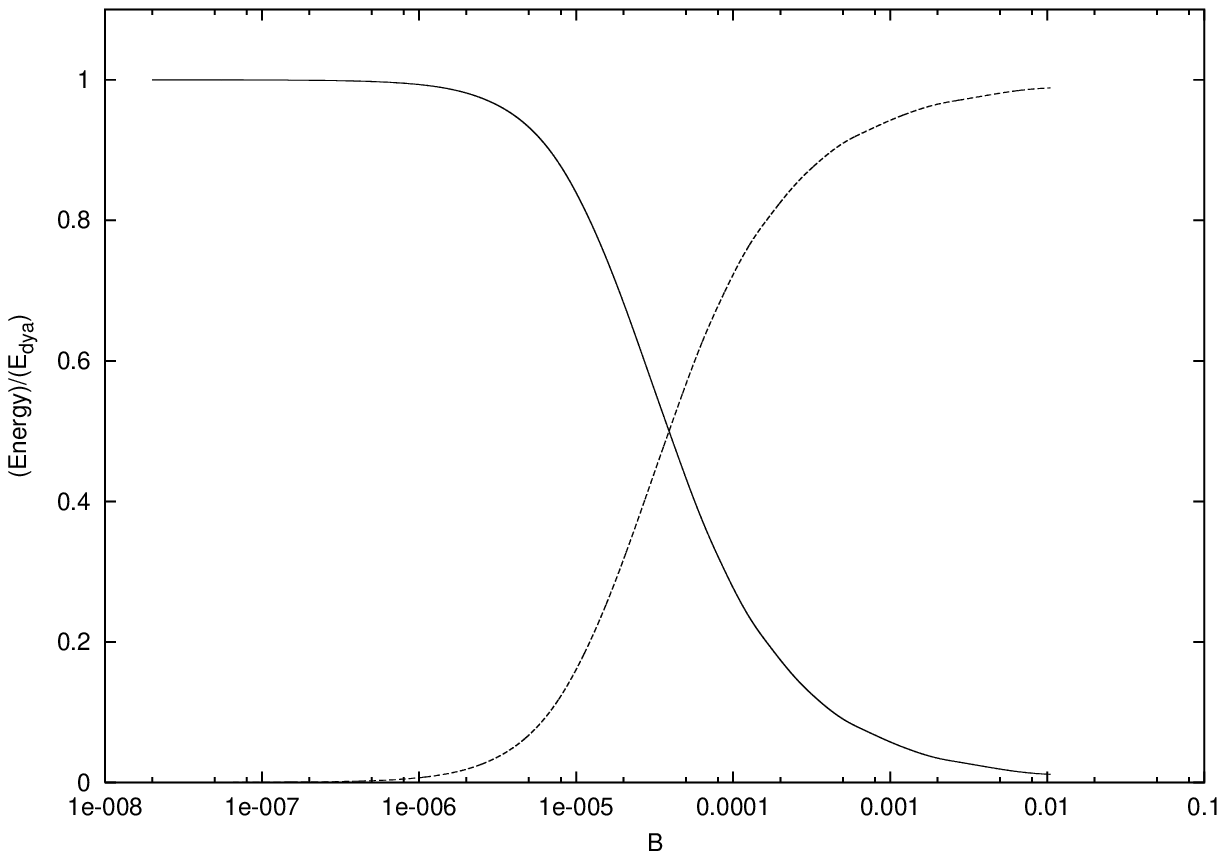}
\includegraphics[width=8.5cm,clip]{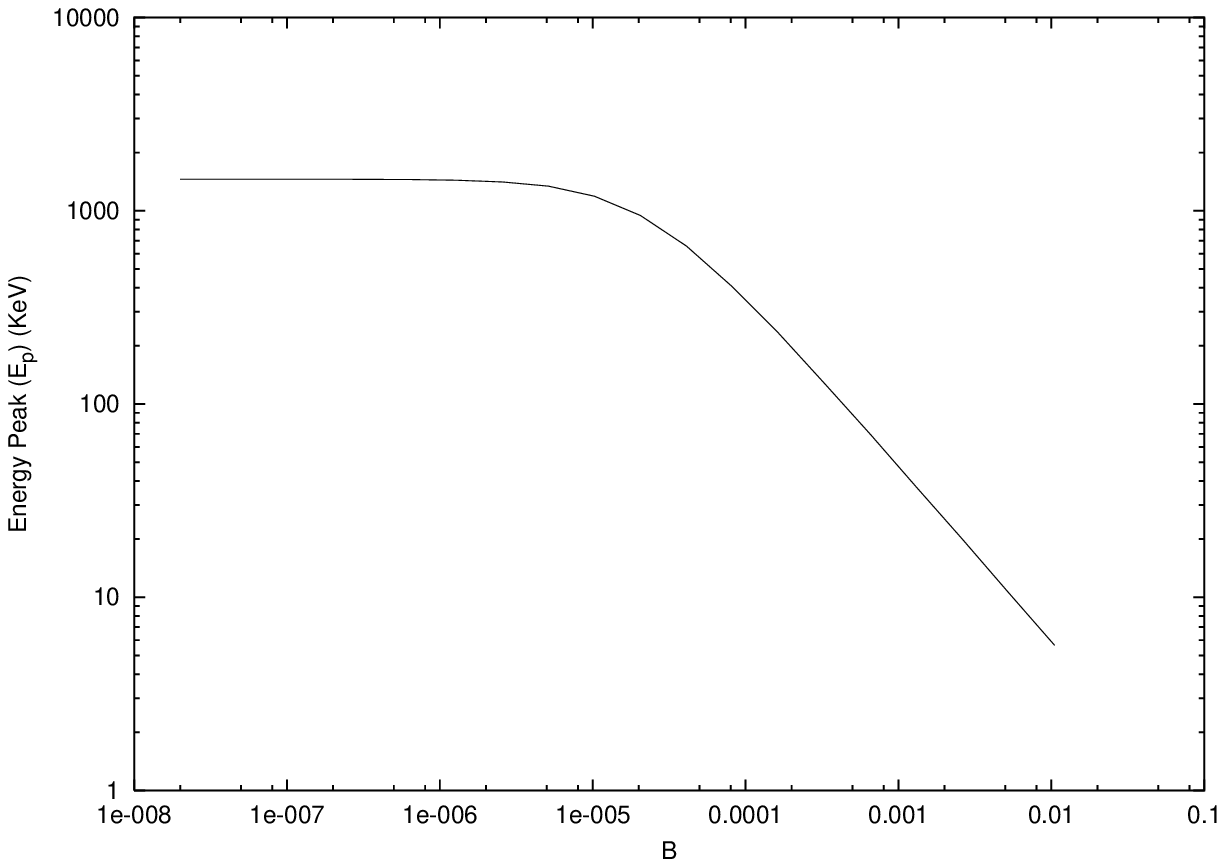}
\caption{{\bf Left)} At the transparent point, the energy radiated in the P-GRB (the solid line) and (the dashed line) the final kinetic energy of baryonic matter, $E_{Baryons}$, in units of the total energy of the dyadosphere ($E_{dya}$), are plotted as functions of the $B$ parameter. {\bf Right)} The energy corresponding to the peak of the photon number spectrum in the P-GRB as measured in the laboratory frame is plotted as function of the $B$ parameter.}
\label{fintkin}
\label{energypeak}
\end{figure}

In section~\ref{new}, as the condition of transparency is reached, the injector phase is concluded with the emission of a sharp burst of electromagnetic radiation and an accelerated beam of highly relativistic baryons. We recall that we have respectively defined the radiation burst (the proper GRB or for short P-GRB) and the accelerated-baryonic-matter (ABM) pulse. By computing for a fixed value of the EMBH different PEMB pulses corresponding to selected values of $B$ in the range $\left[10^{-8}\right.$--$\left.10^{-2}\right]$, it has been possible to obtain a crucial universal diagram which is reproduced in Fig.\ref{fintkin}. In the limit of $B\rightarrow 10^{-8}$ or smaller almost all $E_{dya}$ is emitted in the P-GRB and a negligible fraction is emitted in the kinetic energy $E_{Baryons}$ of the baryonic matter and therefore in the afterglow. On the other hand in the limit $B\rightarrow 10^{-2}$ which is also the limit of validity of our theoretical framework, almost all $E_{dya}$ is transferred to $E_{Baryons}$ and gives origin to the afterglow and the intensity of the P-GRB correspondingly decreases. We have identified the limiting case of negligible values of $B$ with the process of emission of the so called ``short bursts''. A complementary result reinforcing such an identification comes from the thermodynamical properties of the P-GRB: the hardness of the spectrum decreases for increasing values of $B$, see Fig.~\ref{energypeak}.

The injector phase is concluded by the emission of the P-GRB and the ABM pulse, as the condition of transparency is reached. 

The {\bf beam-target phase}, in which the accelerated baryonic matter (ABM) generated in the injector phase collides with the ISM, gives origin to the afterglow. Again for simplicity we have adopted a minimum set of assumptions:

\begin{enumerate}
\item The ABM pulse is assumed to collide with a constant homogeneous interstellar medium of number density $n_{\rm ism} \sim 1 {\mathrm cm}^{-3}$. The energy emitted in the collision is assumed to be instantaneously radiated away (fully radiative condition). The description of the collision and emission process is done using spherical symmetry, taking only the radial approximation neglecting all the delayed emission due to off-axis scattered radiation.
\item Special attention is given to numerically compute the power of the afterglow as a function of the arrival time using the correct governing equations for the space-time transformations in line with the RSTT paradigm.
\item Finally some approximate solutions are adopted in order to obtain the determination of the power law exponents of the afterglow flux and compare and contrast them with the observational results as well as with the alternative results in the literature. 
\end{enumerate}
We first consider the above mentioned radial approximation and a spherically symmetric distribution in order to concentrate on the role of the correct space-time transformations in the RSTT paradigm and illustrate their impact on the determination of the power law index of the afterglow. This topic has been seriously neglected in the literature. We then turn to the fully relativistic analysis of the off-axis emission and of the temporal structure in the long bursts (see also \textcite{rbcfx02a_sub} and sections~\ref{angle}--\ref{off-axis}) and of their spectral distribution (see also \textcite{rbcfx02c_spectrum} and section~\ref{spectrum}). Details of the role of beaming are going to be discussed elsewhere (\textcite{rbcfx02b_beam}).

We can now turn to the two eras of the beam-target phase:

{\itshape The Era IV}: the ultrarelativistic and relativistic regimes in the afterglow. In section~\ref{era4} the hydrodynamic relativistic equations governing the collision of the ABM pulse with the interstellar matter are given in the form of a set of finite difference equations to be numerically integrated. Expressions for the internal energy developed in the collision as well as for the gamma factor are given as a function of the mass of the swept up interstellar material and of the initial conditions. In section~\ref{approximation} the infinitesimal limit of these equations is given as well as analytic power-law expansions in selected regimes.

{\itshape The Era V}: the approach to the nonrelativistic regimes in the afterglow. In section~\ref{era5} it is stressed that this last era often discussed in the current literature can be described by the same equations used for era IV. 

Having established all the governing equations for all the eras of the EMBH theory, we can proceed to compare and contrast the predictions of this theory with the observational data.

\subsection{The Best fit of the EMBH theory to the GRB~991216: the global features of the solution}

\begin{figure}
\includegraphics[width=\hsize,clip]{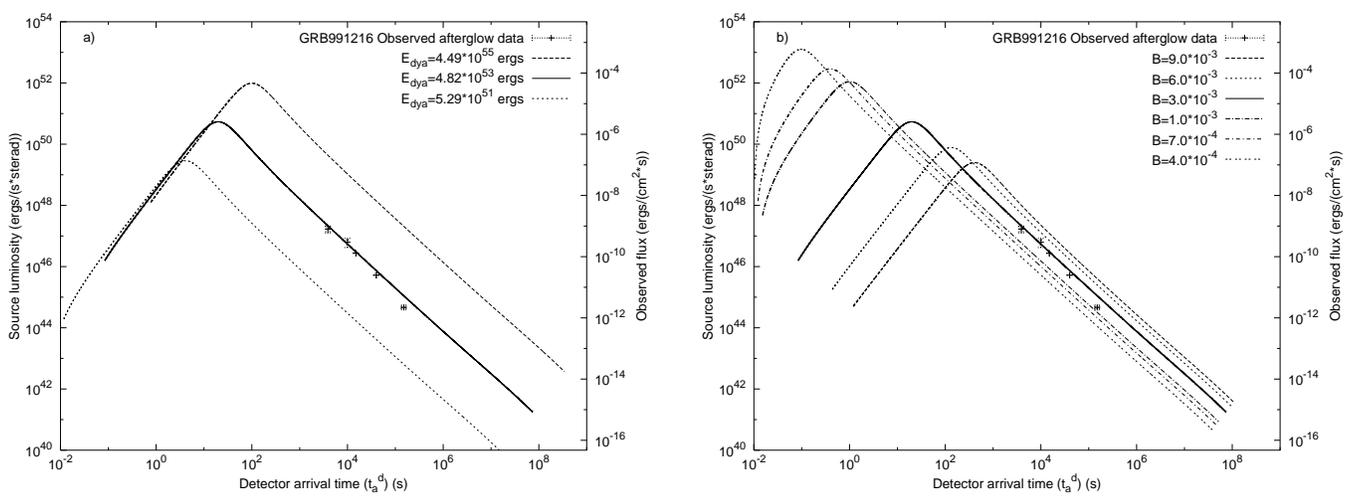}
\caption{a) Afterglow luminosity computed for an EMBH of $E_{dya}= 5.29\times 10^{51}$ erg, $E_{dya}= 4.83\times 10^{53}$ erg, $E_{dya}= 4.49 \times 10^{55}$ erg and $B=3\times 10^{-3}$. b) for the $E_{dya}= 4.83\times 10^{53}$, we give the afterglow luminosities corresponding respectively to $B=9\times 10^{-3}$, $6\times 10^{-3}$, $3\times 10^{-3}$, $1\times 10^{-3}$, $7\times 10^{-4}$, $4\times 10^{-4}$.}
\label{ii-fig2}
\end{figure}

As expressed in section~\ref{bf}, we have proceeded to the identification of the only two free parameters of the EMBH theory, $E_{dya}$ and $B$, by fitting the observational data from R-XTE and Chandra on the decaying part of the GRB~991216 afterglow. The afterglow appears to have three different parts: in the first part the luminosity increases as a function of the arrival time, it then reaches a maximum and finally monotonically decreases. In Fig.~\ref{ii-fig2}, we show how such a fit is actually made and how changing the two free parameters affects the intensity and the location in time of the peak of the afterglow. The best fit is obtained for $E_{dya}=4.83\times 10^{53}\,erg$ and $B=3\times 10^{-3}$. 

\begin{figure}
\includegraphics[width=\hsize,clip]{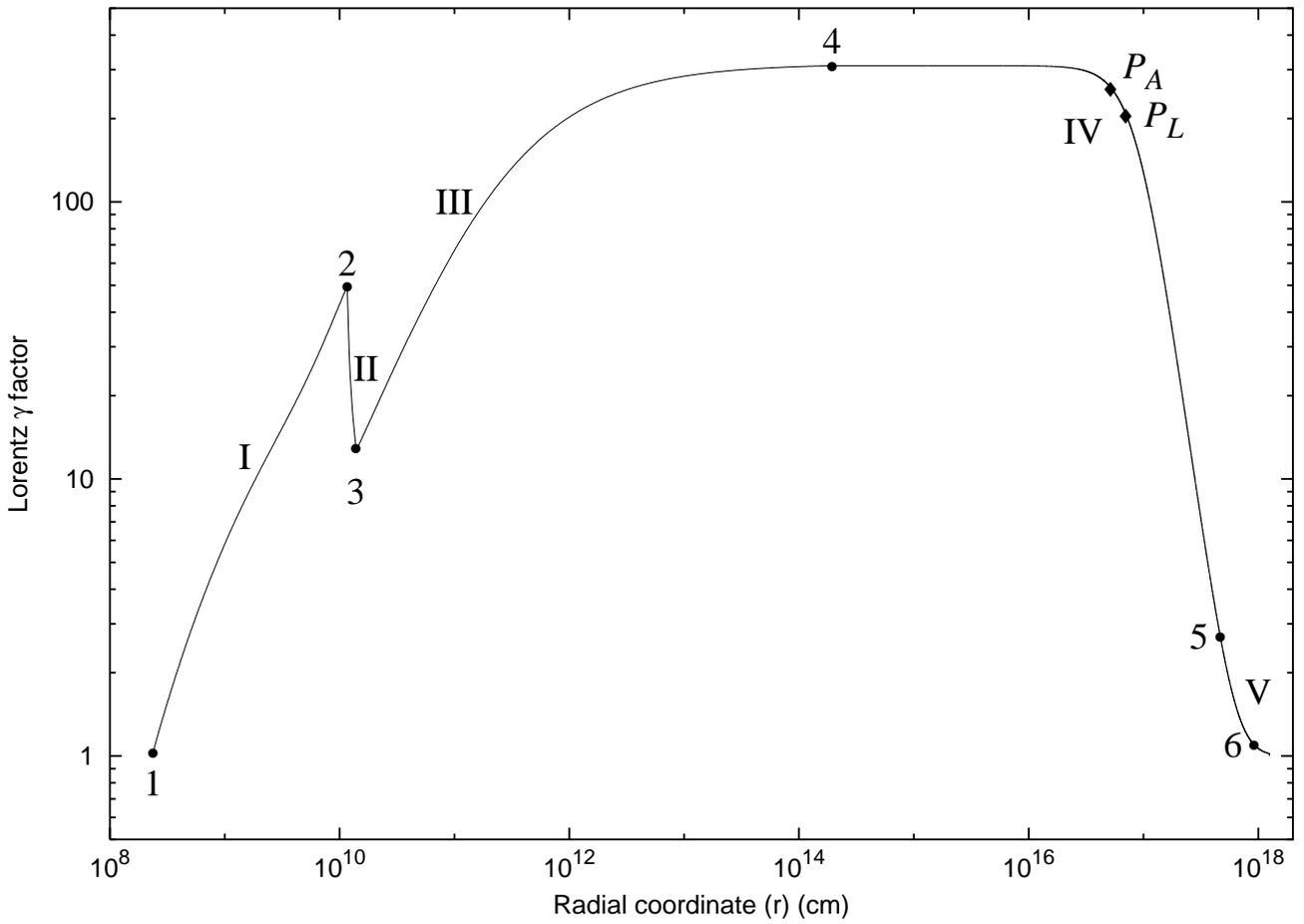}
\caption{The theoretically computed gamma factor for the parameter values $E_{dya}=4.83\times 10^{53}$ erg, $B=3\times 10^{-3}$ is given as a function of the radial coordinate in the laboratory frame. The corresponding values in the comoving time, laboratory time and arrival time are given in Tab.~\ref{tab1}. The different eras indicated by roman numerals are illustrated in the text (see sections~\ref{era1},\ref{era2},\ref{era3},\ref{era4},\ref{era5}), while the points 1,2,3,4,5 mark the beginning and end of each of these eras. The points $P_L$ and $P_A$ mark the maximum of the afterglow flux, respectively in emission time and in arrival time (see \textcite{lett2} and sections~\ref{era4},\ref{approximation}). The point 6 is the beginning of Phase D in Era V (see sections~\ref{era5},\ref{approximation}). At point 4 the transparency condition is reached and the P-GRB is emitted. This diagram clearly shows the inadequacy of considering a simple power-law relation $\gamma\propto r^{-3/2}$ for the relation between the radius of the source and its Lorentz gamma factor as assumed in the large majority of current papers on GRBs (see e.g. \textcite{s97,w97,s98,spn98,pm98b,p99} and references therein). Actually, such a power-law behaviour is never found to exist.}
\label{gamma}
\end{figure}

\begin{figure}
\includegraphics[width=\hsize,clip]{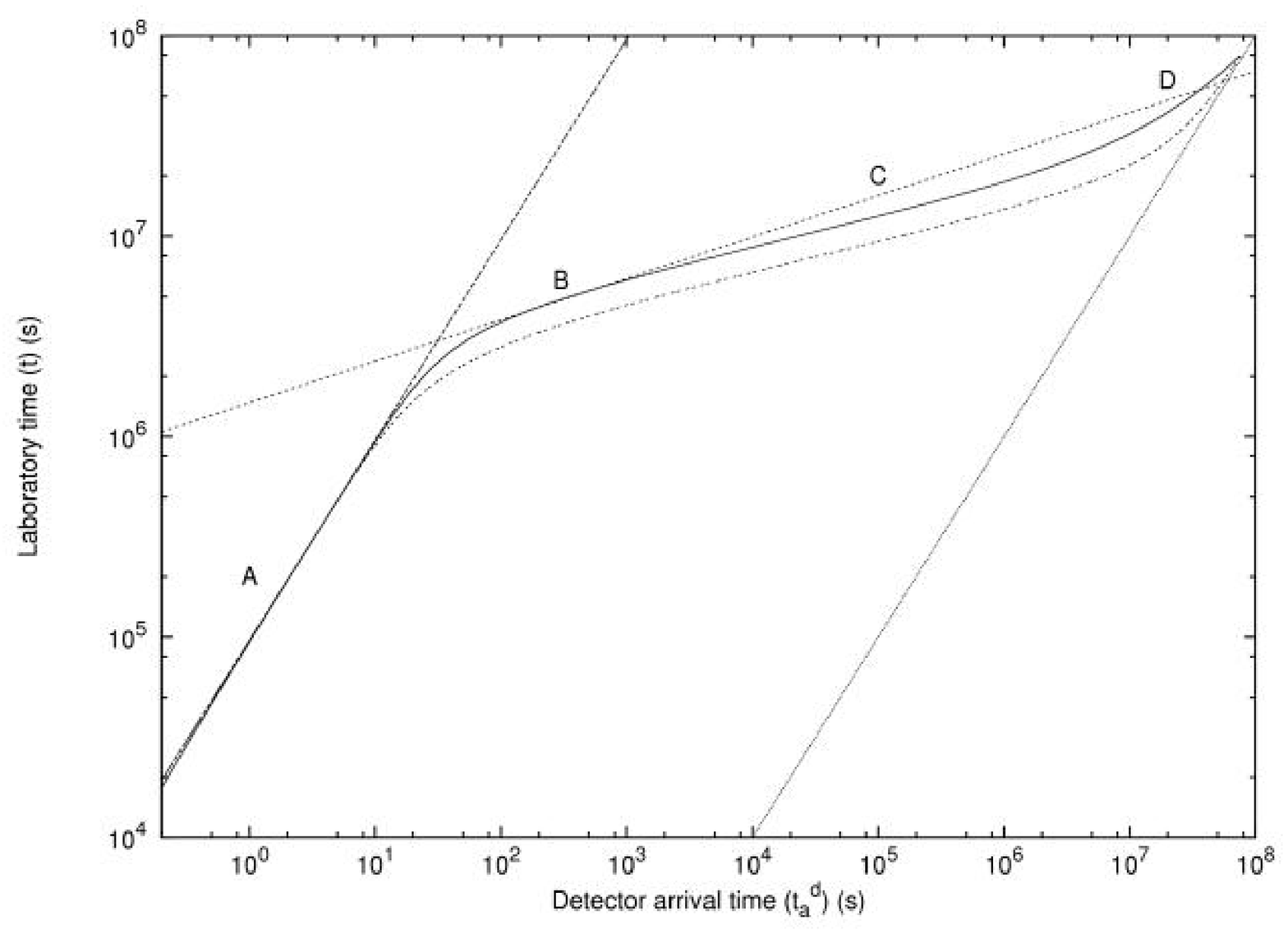}
\caption{Relation between the arrival time ($t_a^d$) measured at the detector and the laboratory time ($t$) measured at the GRB source. The solid curve is computed using the exact formula prescribed by the RSTT paradigm $t_a^d= \left(1+z\right)\left(t - \int_0^t{\frac{\sqrt{\gamma^2\left(t'\right)-1}}{\gamma\left(t'\right)} dt' } - \frac{r_{ds}}{c}\right)$ (see Eq.(\ref{tadef}) in section~\ref{arrival_time}). The dashed-dotted curve is computed using the approximate formula $t_a^d=\left(1+z\right)\left(t/2\gamma^2\left(t\right)\right)$ (see Eq.(\ref{taexp3})) often used in the current literature (see e.g. \textcite{fmn96,s97,w97,s98,p99} and references therein). The difference between the solid line and the dashed-dotted line clearly shows the inadequacy of using such an approximate relation. We like to stress that the difference between the above two curves is especially marked in the afterglow region. Note that this difference as been estimated assuming in both curves the correct relation between the Lorentz gamma factor and the radial coordinated of the source given in Fig.~\ref{gamma}. In the case that the wrong relation $\gamma\propto r^{-3/2}$ is adopted as done in the literature (see e.g. \textcite{s97,w97,s98,spn98,pm98b,p99} and references therein) the discrepancy between the two curves will be much larger. It is anyway clear that, even knowing quantitatively the exact Lorentz gamma factor curve reported in Fig.~\ref{gamma}, the use of the approximate relation given in Eq.(\ref{taexp3}) is enough to miss the correct clock synchronization and to obtain a wrong value for the power-law index $n$ in the decaying phases of the afterglow (see sections~\ref{approximation}--\ref{power-law} and Tab.~\ref{tab2_p}). We distinguish four different phases. {\bf Phase A}: There is a linear relation between $t$ and $t_a^d$, given by Eq.(\ref{appA}) in the text (dashed line). {\bf Phase B}: There is an ``effective'' power-law relation between $t$ and $t_a^d$, given by Eq.(\ref{appC}) (dotted line). {\bf Phase C}: No analytic formula holds and the relation between $t$ and $t_a^d$ has to be directly computed by the integration of the complete equations of energy and momentum conservation (Eqs.(\ref{heat},\ref{dgamma})). {\bf Phase D}: As the gamma factor approaches $\gamma=1$, the relation between $t$ and $t_a^d$ asymptotically goes to $t=t_a^d$ (light gray line). See also \textcite{lett1}.}
\label{tvsta}
\end{figure}

Having determined the two free parameters of the theory, we have integrated the governing equations corresponding to these values and then obtained for the first time the complete history of the gamma factor from the moment of gravitational collapse to the latest phases of the afterglow observations (see Fig.~\ref{gamma}). This diagram clearly shows the inadequacy of considering a simple power-law relation $\gamma\propto r^{-3/2}$ for the relation between the radius of the source and its Lorentz gamma factor as assumed in the large majority of current papers on GRBs (see e.g. \textcite{s97,w97,s98,spn98,pm98b,p99} and references therein). Actually, such a power-law behaviour is never found to exist.

We have also determined the different regimes encountered in the relation between the laboratory time and the detector arrival time within the RSTT paradigm compared and contrasted with the ones in the current literature (see Fig.~\ref{tvsta}). The solid curve is computed using the exact formula prescribed by the RSTT paradigm (see Eq.(\ref{tadef}) in section~\ref{arrival_time})
\begin{displaymath}
t_a^d= \left(1+z\right)\left(t - \int_0^t{\frac{\sqrt{\gamma^2\left(t'\right)-1}}{\gamma\left(t'\right)} dt' } - \frac{r_{ds}}{c}\right)\, .
\end{displaymath}
The dashed-dotted curve is computed using the approximate formula (see Eq.(\ref{taexp3}))
\begin{displaymath}
t_a^d=\left(1+z\right)\frac{t}{2\gamma^2\left(t\right)}\, ,
\end{displaymath}
often used in the current literature (see e.g. \textcite{fmn96,s97,w97,s98,p99} and references therein). The difference between the solid line and the dashed-dotted line clearly shows the inadequacy of using such an approximate relation. We like to stress that the difference between the above two curves is especially marked in the afterglow region. Note that this difference as been estimated assuming in both curves the correct relation between the Lorentz gamma factor and the radial coordinated of the source given in Fig.~\ref{gamma}. In the case that the wrong relation $\gamma\propto r^{-3/2}$ is adopted as done in the literature (see e.g. \textcite{s97,w97,s98,spn98,pm98b,p99} and references therein) the discrepancy between the two curves will be much larger. It is anyway clear that, even knowing quantitatively the exact Lorentz gamma factor curve reported in Fig.~\ref{gamma}, the use of the approximate relation given in Eq.(\ref{taexp3}) is enough to miss the correct clock synchronization and to obtain a wrong value for the power-law index $n$ in the decaying phases of the afterglow (see sections~\ref{approximation}--\ref{power-law} and Tab.~\ref{tab2_p}).

To be more explicit, from the result given in Figs.~\ref{gamma}--\ref{tvsta} follows that all existing GRB models, with the exception of ours, have the wrong spacetime coordinatization of the GRB phenomenon and they therefore lack the fundamental toola to compare the theoretical prediction in the laboratory time to the observations carried out in the asymptotic photon arrival time. This extreme situation affects all considerations on GRBs: as an example, all the considerations on the afterglow slopes, which drastically depend on the functional dependence between the laboratory time and the photon arrival time, are drastically affected (see subsection~\ref{subpl} below and Tab.~\ref{tab2_p}). In turn, all the considerations about the possible existence of beaming in GRBs inferred from the afterglow slopes are in this circumstance deprived of any meaning.

We have thus determined the entire space-time grid of the GRB~991216 by giving (see Tab.~\ref{tab1}) the radial coordinate of the GRB phenomenon as a function of the four coordinate time variables. A quick glance to Tab.~\ref{tab1} shows how the extreme relativistic regimes at work lead to enormous superluminal behaviour (up to $10^5 c$!) if the classical astrophysical concepts are adopted using the arrival time as the independent variable. In turn this implies that any causal relation based on classical astrophysics and the arrival time data, as at times found in the current GRB literature, is incorrect.

\begin{table}
\centering
\caption{Gamma factors for selected events and their space-time coordinates. The points marked 1,2,3,4,5,6,$P_L$,$P_A$ are the same reported in Fig.~\ref{gamma}, while the point $F$ is the endpoint of the simulation. It is particularly important to read the last column, where the apparent motion in the radial coordinate, evaluated in the arrival time at the detector, leads to an enormous ``superluminal'' behaviour, up to $9.55\times 10^4\,c$. This illustrates well the impossibility of using such a classical estimate in regimes with gamma factors up to $310.1$. \label{tab1}}
\begin{ruledtabular}
\begin{tabular}{c|e{9}|e{9}|e{9}|e{9}|e{9}|e{3}|e{8}}
 Point & r (cm)& \tau(s) & t(s) & t_a(s) & t_a^d (s)& \gamma & \begin{array}{c} {\rm ``Superluminal"} \\ v\equiv\frac{r}{t_a^d} \\ \\ \end{array}\\
\hline
\multicolumn{8}{c}{ }\\
\multicolumn{8}{c}{{\bf The Injector Phase}}\\
\hline
 & & & & & & & \\
1 & 2.354\times10^8    & 0.0                & 0.0                & 0.0                & 0.0                & 1.000 & 0\\
  & 1.871\times10^9    & 1.550\times10^{-2} & 5.886\times10^{-2} & 4.312\times10^{-3} & 8.625\times10^{-3} & 10.08 & 7.23c\\
  & 4.486\times10^9    & 2.141\times10^{-2} & 1.463\times10^{-1} & 4.523\times10^{-4} & 9.046\times10^{-3} & 20.26 & 16.5c\\
  & 7.080\times10^9    & 2.485\times10^{-2} & 2.329\times10^{-1} & 4.594\times10^{-3} & 9.187\times10^{-3} & 30.46 & 25.7c\\
  & 9.533\times10^9    & 2.715\times10^{-2} & 3.148\times10^{-1} & 4.627\times10^{-3} & 9.253\times10^{-3} & 40.74 & 34.4c\\
  & 1.162\times10^{10} & 2.868\times10^{-2} & 3.845\times10^{-1} & 4.644\times10^{-3} & 9.288\times10^{-3} & 49.70 & 41.7c\\
\hline
 & & & & & & & \\
2 & 1.162\times10^{10} & 2.868\times10^{-2} & 3.845\times10^{-1} & 4.644\times10^{-3} & 9.288\times10^{-3} & 49.70& 41.7c\\
  & 1.186\times10^{10} & 2.889\times10^{-2} & 3.923\times10^{-1} & 4.646\times10^{-3} & 9.292\times10^{-3} & 38.06& 42.6c\\
  & 1.234\times10^{10} & 2.949\times10^{-2} & 4.083\times10^{-1} & 4.655\times10^{-3} & 9.311\times10^{-3} & 24.21& 44.2c\\
  & 1.335\times10^{10} & 3.144\times10^{-2} & 4.423\times10^{-1} & 4.706\times10^{-3} & 9.413\times10^{-3} & 15.14& 47.3c\\
  & 1.389\times10^{10} & 3.279\times10^{-2} & 4.603\times10^{-1} & 4.753\times10^{-3} & 9.506\times10^{-3} & 12.94& 48.7c\\
\hline
 & & & & & & & \\
3 & 1.389\times10^{10} & 3.279\times10^{-2} & 4.603\times10^{-1} & 4.753\times10^{-3} & 9.506\times10^{-3} & 12.94& 48.7c\\
  & 2.326\times10^{10} & 5.208\times10^{-2} & 7.733\times10^{-1} & 5.369\times10^{-3} & 1.074\times10^{-2} & 20.09& 72.2c\\
  & 6.913\times10^{10} & 9.694\times10^{-2} & 2.304 &              6.086\times10^{-3} & 1.217\times10^{-2} & 50.66& 1.89\times10^2c\\ 
  & 1.861\times10^{11} & 1.486\times10^{-1} & 6.206 &              6.446\times10^{-3} & 1.289\times10^{-2} & 100.1& 4.82\times10^2c\\
  & 9.629\times10^{11} & 3.112\times10^{-1} & 32.12 &              6.978\times10^{-3} & 1.396\times10^{-2} & 200.3& 2.30\times10^3c\\
  & 3.205\times10^{13} & 3.958 &              1.069\times10^{ 3} & 1.343\times10^{-2} & 2.685\times10^{-2} & 300.1& 3.98\times10^4c\\
  & 1.943\times10^{14} & 21.57 &              6.481\times10^{ 3} & 4.206\times10^{-2} & 8.413\times10^{-2} & 310.1& 7.70\times10^4c\\
\hline
\multicolumn{8}{c}{ }\\
\multicolumn{8}{c}{{\bf The Beam-Target Phase}}\\
\hline
 & & & & & & & \\
4     & 1.943\times10^{14} & 21.57 &             6.481\times10^{3} & 4.206\times10^{-2}& 8.413\times10^{-2} & 310.1& 7.70\times10^4c\\
      & 6.663\times10^{15} & 7.982\times10^{2} & 6.481\times10^{3} & 1.164 &             2.328 &              310.0& 9.55\times10^4c\\
      & 2.863\times10^{16} & 3.114\times10^{3} & 9.549\times10^{5} & 5.057 &             10.11 &              300.0& 9.45\times10^4c\\
      & 4.692\times10^{16} & 5.241\times10^{3} & 1.565\times10^{6} & 8.775 &             17.55 &              270.0& 8.92\times10^4c\\
$P_A$ & 5.177\times10^{16} & 5.853\times10^{3} & 1.727\times10^{6} & 9.933 &             19.87 &              258.5& 8.69\times10^4c\\
      & 5.878\times10^{16} & 6.791\times10^{3} & 1.961\times10^{6} & 11.82 &             23.63 &              240.0& 8.30\times10^4c\\
      & 6.580\times10^{16} & 7.811\times10^{3} & 2.195\times10^{6} & 14.03 &             28.06 &              220.0& 7.82\times10^4c\\
$P_L$ & 7.025\times10^{16} & 8.506\times10^{3} & 2.343\times10^{6} & 15.66 &             31.32 &              207.0& 7.48\times10^4c\\
      & 7.262\times10^{16} & 8.895\times10^{3} & 2.422\times10^{6} & 16.61 &             33.23 &              200.0& 7.29\times10^4c\\
      & 9.058\times10^{16} & 1.236\times10^{4} & 3.021\times10^{6} & 26.66 &             53.32 &              150.0& 5.67\times10^4c\\
      & 1.136\times10^{17} & 1.866\times10^{4} & 3.788\times10^{6} & 52.84 &             1.057\times10^{2} &  100.0& 3.58\times10^4c\\
      & 1.539\times10^{17} & 3.819\times10^{4} & 5.134\times10^{6} & 2.000\times10^{2} & 4.000\times10^{2} &  50.02& 1.28\times10^4c\\
      & 2.801\times10^{17} & 2.622\times10^{5} & 9.351\times10^{6} & 7.278\times10^{3} & 1.455\times10^{4} &  10.00& 6.42\times10^2c\\
      & 3.624\times10^{17} & 6.702\times10^{5} & 1.213\times10^{7} & 3.860\times10^{4} & 7.719\times10^{4} &  5.001& 1.57\times10^2c\\
      & 4.454\times10^{17} & 1.433\times10^{6} & 1.500\times10^{7} & 1.439\times10^{5} & 2.877\times10^{5} &  2.998& 51.6c\\
\hline
 & & & & & & & \\
5 & 4.454\times10^{17} & 1.433\times10^{6} & 1.500\times10^{7} & 1.439\times10^{5} & 2.877\times10^{5} &  2.998& 51.6c\\
  & 4.830\times10^{17} & 1.928\times10^{6} & 1.635\times10^{7} & 2.381\times10^{5} & 4.762\times10^{5} &  2.500& 33.8c\\
  & 5.390\times10^{17} & 2.873\times10^{6} & 1.844\times10^{7} & 4.643\times10^{5} & 9.285\times10^{5} &  2.000& 19.4c\\
  & 6.422\times10^{17} & 5.387\times10^{6} & 2.271\times10^{7} & 1.291\times10^{6} & 2.581\times10^{6} &  1.500& 8.30c\\
  & 1.034\times10^{18} & 2.903\times10^{7} & 5.002\times10^{7} & 1.552\times10^{7} & 3.103\times10^{7} &  1.054& 1.11c\\
\hline
 & & & & & & & \\
6 & 1.034\times10^{18} & 2.903\times10^{7} & 5.002\times10^{7} & 1.552\times10^{7} & 3.103\times10^{7} &  1.054& 1.11c\\
  & 1.202\times10^{18} & 4.979\times10^{7} & 7.150\times10^{7} & 3.140\times10^{7} & 6.280\times10^{7} &  1.025& 6.38\times10^{-1}c\\
\hline
 & & & & & & & \\
$F$ & 1.248\times10^{18} & 5.706\times10^{7} & 7.894\times10^{7} & 3.731\times10^{7} & 7.461\times10^{7} & 1.000& 5.58\times10^{-1}c\\
\end{tabular}
\end{ruledtabular}
\end{table}

\subsection{The explanation of the ``long bursts'' and the identification of the proper gamma ray burst(P-GRB)}

\begin{figure}
\includegraphics[width=10cm,clip]{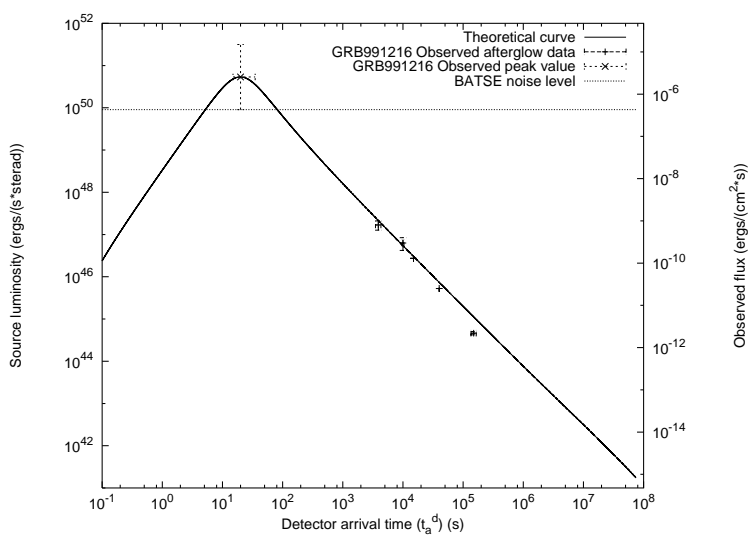}
\caption{Best fit of the afterglow data of Chandra, RXTE as well as of the range of variability of the BATSE data on the major burst, by a unique afterglow curve leading to the parameter values $E_{dya}=4.83\times 10^{53}erg, B=3\times 10^{-3}$. The horizontal dotted line indicates the BATSE noise threshold. On the left axis the luminosity is given in units of the energy emitted at the source, while the right axis gives the flux as received by the detectors.}
\label{fit_1}
\end{figure}

In section~\ref{shortlongburst}, having determined the two free parameters of the EMBH theory, we analyze the theoretical predictions of this theory for the general structure of GRBs. The first striking result, illustrated in Fig.~\ref{fit_1}, shows that the peak of the afterglow emission coincides both in intensity and in arrival time ($19.87\,s$) with the average emission of the long burst observed by BATSE. For this we have introduced the new concept of {\em extended afterglow peak emission (E-APE)}. Once the proper space-time grid is given (see Tab.~\ref{tab1}) it is immediately clear that the E-APE is generated at distances of $~5\times 10^{16}$ cm from the EMBH. The long bursts are then identified with the E-APEs and are not bursts at all: they have been interpreted as bursts only because of the high threshold of the BATSE detectors (see Fig.~\ref{fit_1}). Thus the long standing unsolved problem of explaining the long GRBs (see e.g. \textcite{wmm96,swm00,p01}) is radically resolved.

\begin{figure}
\includegraphics[width=8.5cm,clip]{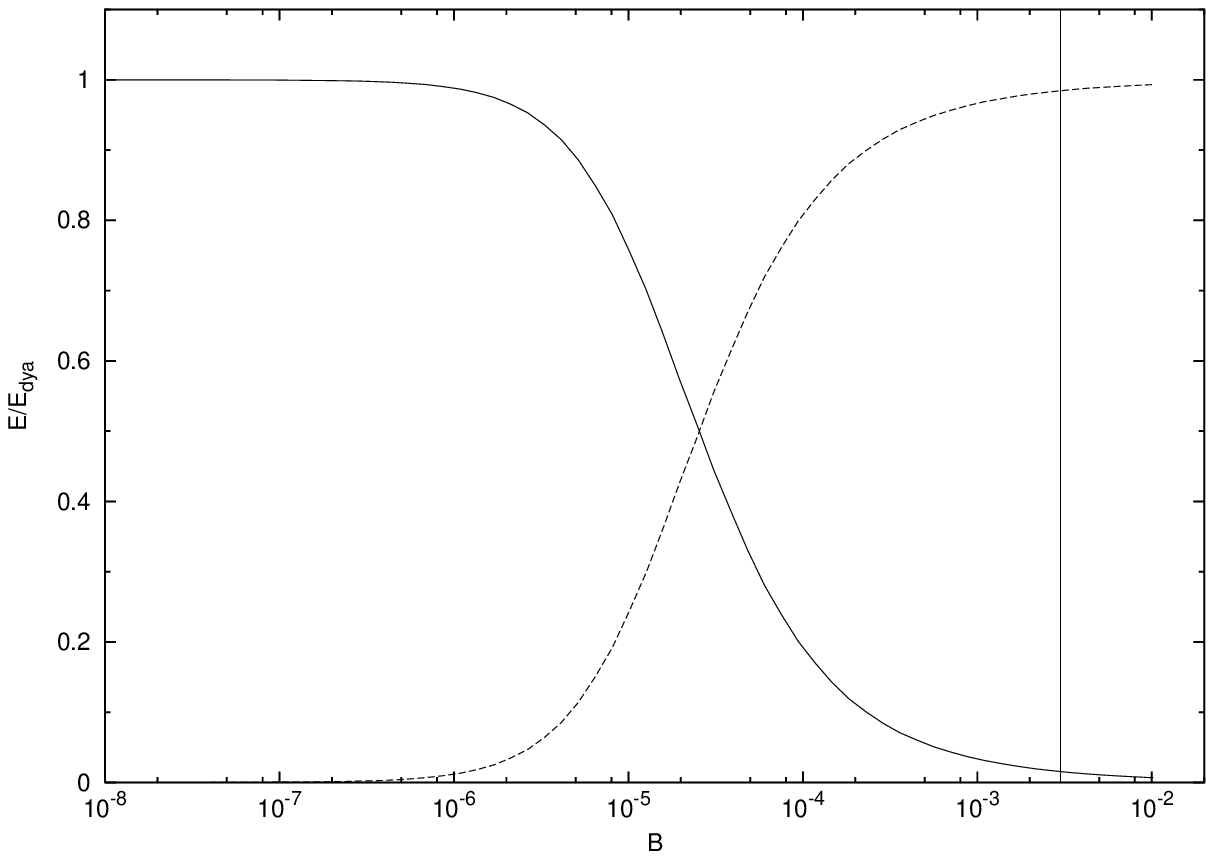}
\includegraphics[width=8.5cm,clip]{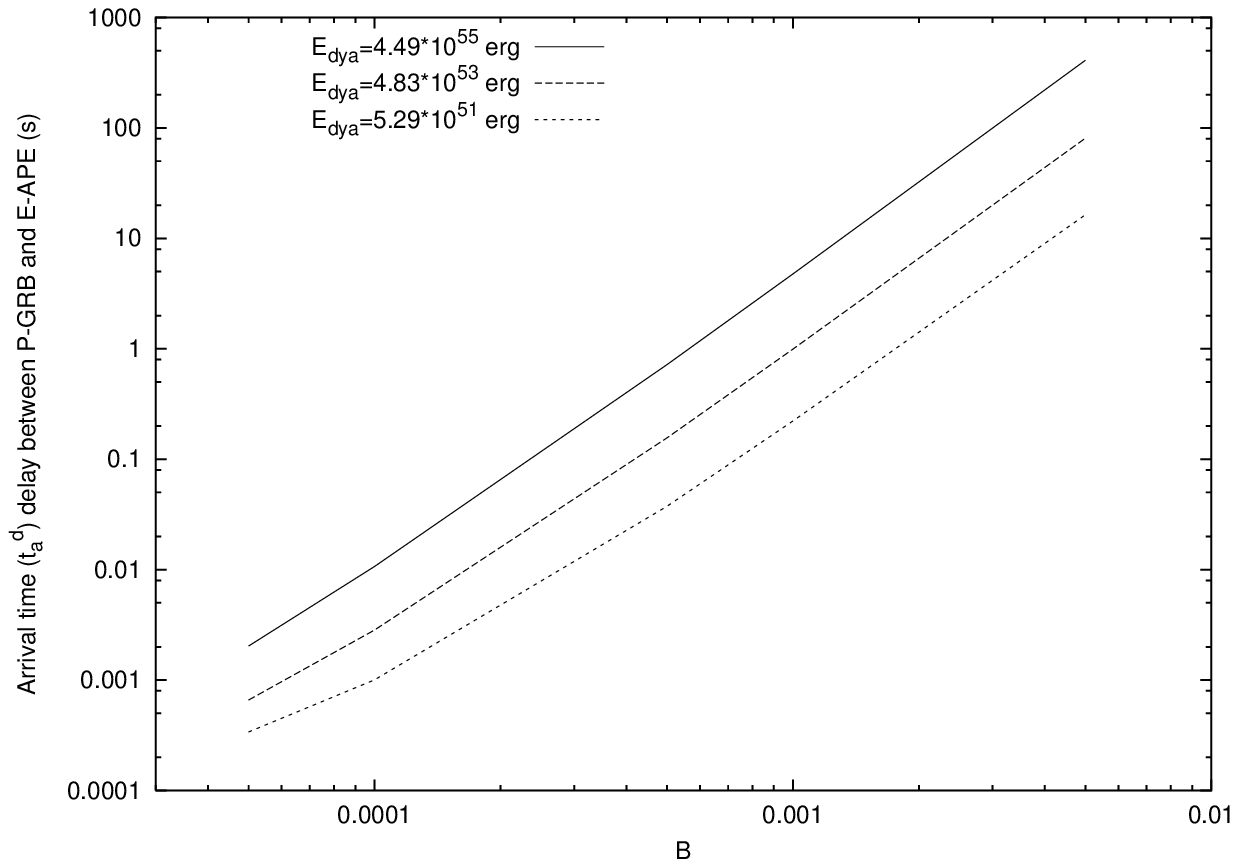}
\caption{{\bf Left)} Relative intensities of the E-APE (dashed line) and the P-GRB (solid line), as predicted by the EMBH theory corresponding to the values of the parameters determined in Fig.~\ref{fit_1}, as a function of $B$. Details are given in section~\ref{shortlongburst}. The vertical line corresponds to the value $B=3\times 10^{-3}$. {\bf Right)} The arrival time delay between the P-GRB and the peak of the E-APE is plotted as a function of the $B$ parameter for three selected values of $E_{dya}$.}
\label{crossen}
\label{dtab}
\end{figure}

Still in section~\ref{shortlongburst}, the search for the identification of the P-GRB in the BATSE data is described. This identification is made using the two fundamental diagrams shown in Fig.~\ref{crossen}. Having established the value of $E_{dya}=4.83\times 10^{53}\, erg$ and of $B=3\times 10^{-3}$, it is possible from the dashed line and the solid line in Fig.~\ref{crossen} to evaluate the ratio of the energy $E_{P\hbox{-}GRB}$ emitted in the P-GRB to the energy $E_{Baryons}$ emitted in the afterglow corresponding to the determined value of $B$, see the vertical line in Fig.~\ref{crossen}. We obtain $E_{P\hbox{-}GRB}/E_{Baryons}=1.58\times 10^{-2}$, which gives $E_{P\hbox{-}GRB}=7.54\times 10^{51}\, erg$. Having so determined the theoretically expected intensity of the P-GRB, a second fundamental observable parameter, which is also a function of $E_{dya}$ and $B$, is the arrival time delay between the P-GRB and the peak E-APE, determined in Fig.~\ref{dtab}. From Tab.~\ref{tab1}, we have that the detector arrival time of the P-GRB occurs at $8.41\times 10^{-2}\, s$, corresponding to a radial coordinate of $1.94\times 10^{14}\,cm$, a comoving time of $21.57\, s$, a laboratory time of $6.48\times 10^3\, s$ and an arrival time of $4.21\times 10^{-2}\, s$. At this point, the gamma factor is $310.1$. The peak of the E-APE occurs at a detector arrival time of $19.87\, s$, corresponding to a radial coordinate of $5.18\times 10^{16}\, cm$, a comoving time of $5.85\times 10^3\,s$, a laboratory time of $1.73\times 10^6\,s$ and an arrival time of $9.93\,s$ (see Tab.~\ref{tab1}). The delay between the P-GRB and the peak of the E-APE is therefore $19.78\,s$, see Fig.~\ref{dtab}. The theoretical prediction on the intensity and the arrival time uniquely identifies the P-GRB with the ``precursor'' in the GRB~991216 (see Fig.~\ref{grb991216}). Moreover, the hardness of the P-GRB spectra is also evaluated in this section. As pointed out in the conclusions, the fact that both the absolute and relative intensities of the P-GRB and E-APE have been predicted within a few percent accuracy as well as the fact that their arrival time has been computed with the precision of a few tenths of milliseconds, see Tab.~\ref{tab1} and Fig.~\ref{991216}, can be considered one of the major successes of the EMBH theory. 

\subsection{On the power-laws and beaming in the afterglow of GRB~991216.}\label{subpl}

In section~\ref{approximation} a piecewise description of the afterglow by the expansion of the fundamental hydrodynamical equations given by \textcite{taub} and \textcite{ll} have allowed the determination of a power-law index for the dependence of the afterglow luminosity on the photon arrival time at the detector. It is evident that the determination of the power-law index is very sensitive to the basic assumptions made for the description of the afterglow, as well as to the relations between the different temporal coordinates which have been clarified by the RSTT paradigm (see \textcite{lett1}). The different power-law indexes obtained are compared and contrasted with the ones in the current literature (see Tab.~\ref{tab2_p} and section.~\ref{power-law}). As a byproduct of this analysis, see also the conclusions, there is a perfect agreement between the observational data and the theoretical predictions, implying that the assumptions we have adopted for the description of the afterglow (see section~\ref{era4}) must be necessarily all valid and therefore, in particular, there is no evidence for a beamed emission in GRB~991216.

\begin{table}
\centering
\caption{We compare and contrast the results on the power-law index {\em n} of the afterglow in the EMBH theory with other treatments in the current literature, in the limit of high energy and fully radiative conditions. The differences between the values of $-10/7\sim -1.43$ (Dermer) and the results $-1.6$ in the EMBH theory can be retraced to the use of the two different approximation in the arrival time versus the laboratory time given in Fig.~\ref{tvsta}. See details in section~\ref{approximation}.\label{tab2_p}}
\begin{tabular}{c|c|c|c|c|c}
&& \textcite{cd99} & \textcite{p99} &&\\
& EMBH theory & \textcite{dcb99} & \textcite{sp99} & \textcite{v97} & \textcite{ha00}\\
&& \textcite{bd00} & \textcite{p01} &&\\
\hline \hline
&&&&&\\
Ultra-relativistic & $\displaystyle{\gamma=\gamma_\circ}$  & $\displaystyle{\gamma=\gamma_\circ}$ & $\displaystyle{\gamma=\gamma_\circ}$ & &\\
&&&&&\\
& $\gamma_\circ=310.1$ &&&& \\
&&&&&\\
 & $n=2$ & $n=2$ & $n\simeq2$ && \\
&&&&&\\
\hline
&&&&&\\
Relativistic & $\displaystyle{\gamma \simeq r^{-3}}$ & $\displaystyle{\gamma \sim r^{-3}}$ & $\gamma \sim r^{-3}$  && 
$n>-1.47$\\
&&&&&\\
& $3.0<\gamma<258.5$ &&&&  \\
&&&&&\\
 & $n=-1.6$ & $n=-\frac{10}{7}=-1.43$ & $n=-\frac{5.5}{4}=-1.375$ && \\
&&&&&\\
\hline
&&&&&\\
Non-relativistic & $n=-1.36$ &&& $n=-1.7$ & \\
&&&&&\\
& $1.05<\gamma<3.0$ &&&&\\
&&&&&\\
\hline
&&&&&\\
Newtonian & $n=-1.45$ &&&&\\
&&&&&\\
& $1<\gamma<1.05$ &&&&\\
&&&&&\\
\hline
\multicolumn{3}{c}{}\\
\end{tabular}
\end{table}

\begin{figure}
\begin{center}
\includegraphics[width=8.5cm,clip]{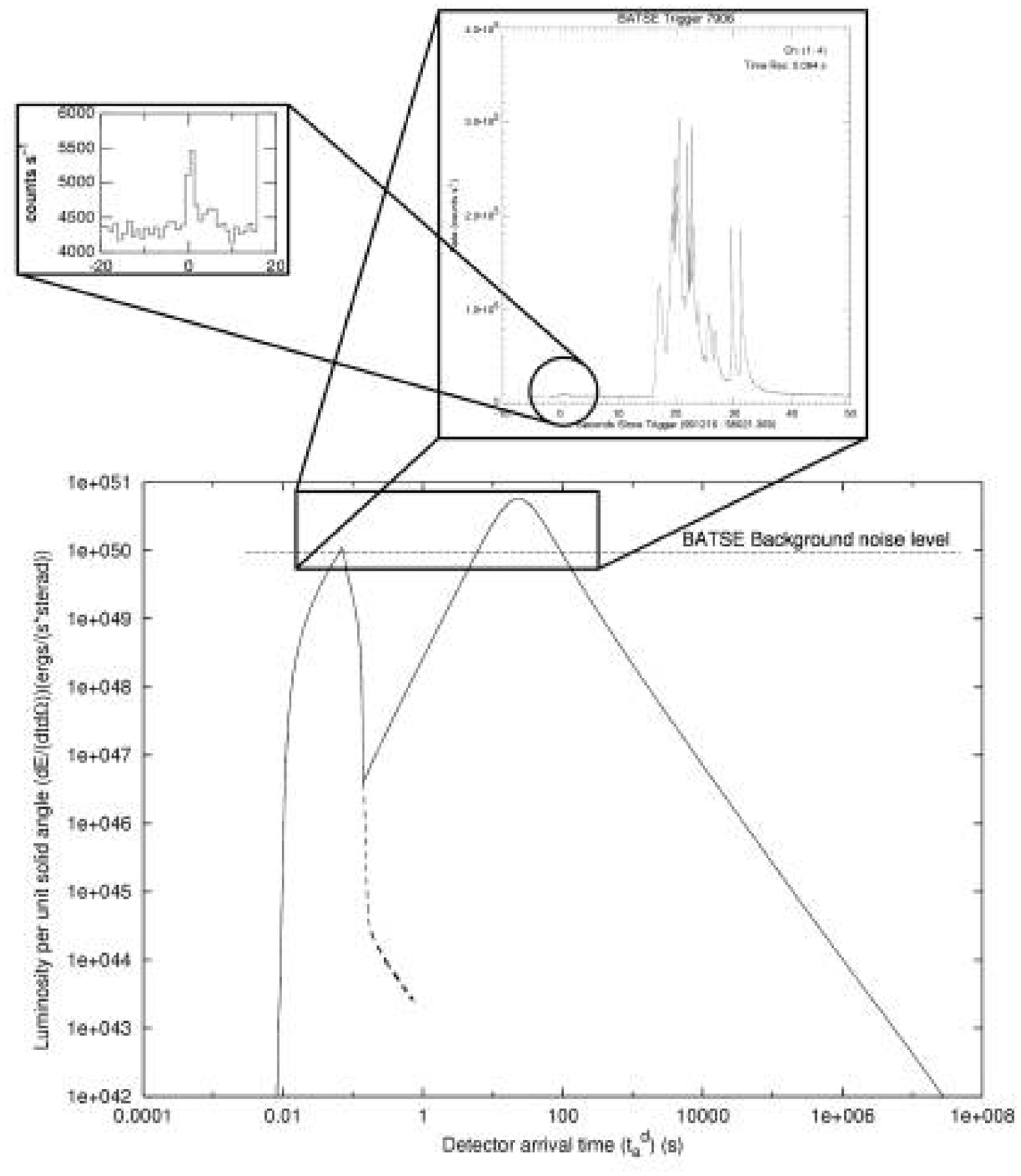}
\includegraphics[width=8.5cm,clip]{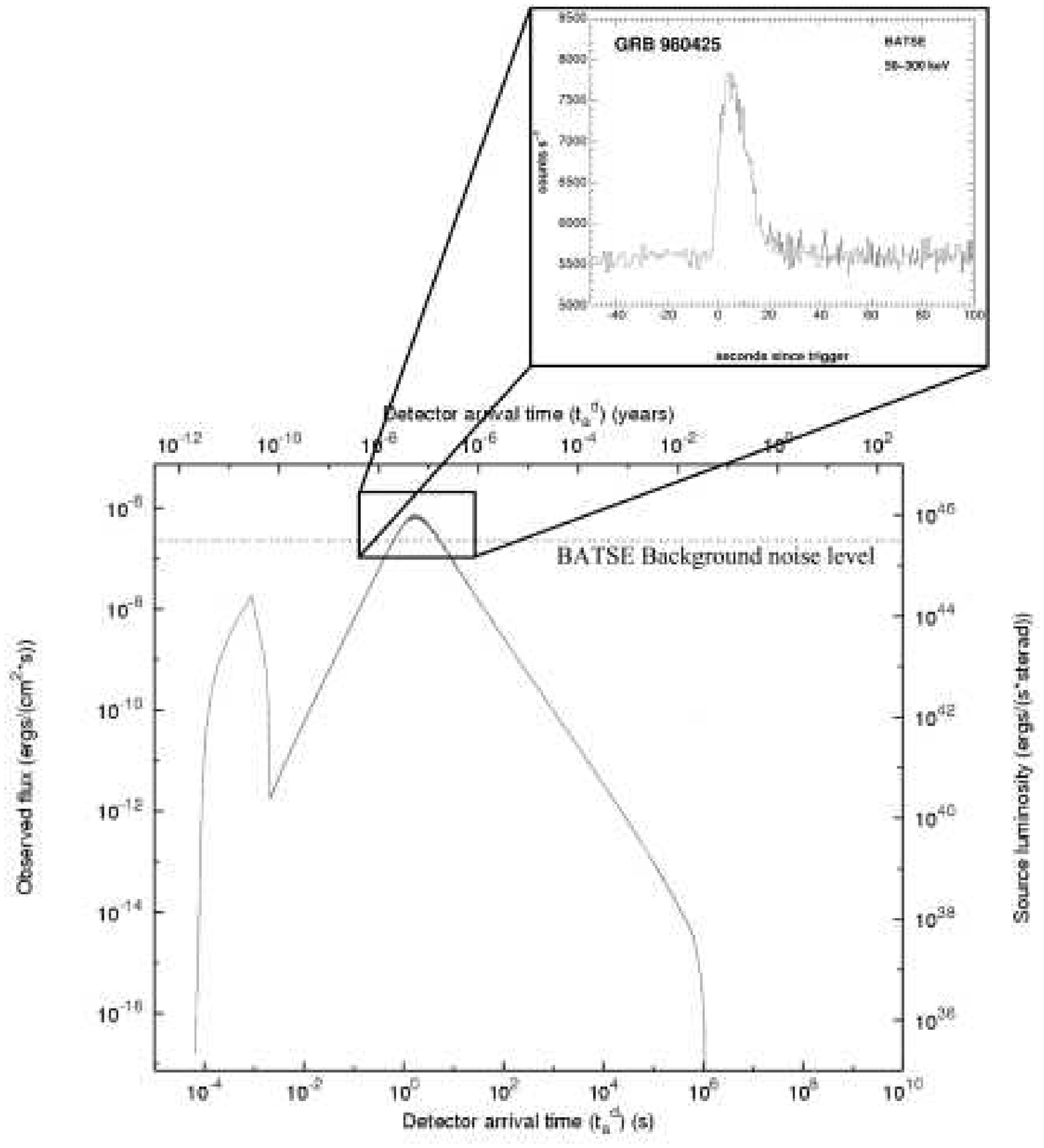}
\caption{{\bf Left)} The overall description of the EMBH theory applied to GRB~991216. The BATSE noise threshold is represented and the observations both of the P-GRB and of the E-APE are clearly shown in the subpanels. The continuos line in the picture represents the theoretical prediction of the EMBH model. {\bf Right)} The same diagrams are represented for GRB~980425. Two aspects are especially important to be mentioned: a) in this source the theoretical prediction of the P-GRB intensity is lower than the BATSE noise treshold and is therefore unobservable and unobserved; b) the E-APE is especially smooth as a consequence of the low value of the gamma Lorentz factor (see also section~\ref{structure_angle} and \textcite{r03cospar02}).}
\label{991216}
\end{center}
\end{figure}

We then summarize in Fig.~\ref{991216} the results for the average bolometric luminosity of GRB~991216 with particular attention to the striking agreement, both in arrival time and in intensity, for the theoretically predicted structure of the P-GRB and the E-APE with the observational data. To show the generality of application of the EMBH theory, we have applied it also to GRB~980425 (see \textcite{r03cospar02}) and the excellent results are also shown, for comparsion, in Fig.~\ref{991216}.

\subsection{Substructures in the E-APE due to inhomogeneities in the Interstellar medium}

In section~\ref{substructures} the role of the inhomogeneities in the interstellar matter has been analyzed in order to explain the observed temporal substructures in the BATSE data on GRB~991216. Having satisfactorily identified the average intensity distribution of the afterglow and the relative position of the P-GRB, in \textcite{lett5} we have addressed the issue whether the fast temporal variation observed in the so-called long bursts, on time scales as short as fraction of a second (see e.g. \textcite{fm95}), can indeed be explained as an effect of inhomogeneities in the interstellar medium. Such a possibility was pioneered in the work by \textcite{dm99}, purporting that such a time variability corresponds to a tomographic analysis of the ISM. In order to probe the validity of such an explanation, we have first considered the simplified case of the radial approximation (\textcite{lett5}). The aim has been to explore the possibility of explaining the observed fluctuation in intensity on a fraction of a second as originated from inhomogeneities in ISM, typically of the order of $10^{16}$ due to apparent superluminal behaviour of roughly $10^5$c. We have shown there that this approach is indeed viable: both the intensity variation and the time scale of the variability in the E-APE region can be explained by the interaction of the ABM pulse with inhomogeneities in the ISM, taking into due account the apparent superluminal effects. These effects, in turn, can be derived and computed self consistently from the dynamics of the source. We have then described the inhomogeneities of the ISM by an appropriate density profile (mask) of an ISM cloud. Of course at this stage,  for simplicity, only the case of spherically symmetric ``spikes'' with over-density separated by low-energy regions, has been considered. Each spike has been assumed to have the spatial extension of $10^{15}$cm. The cloud average density is $<n_{ism}> = 1\, {\rm particle}/{\rm cm}^3$. In conclusion, from the data of Tab.~\ref{tab1} and the highly ``superluminal'' behaviour of the source in the region of the E-APE, it is concluded that the observed time variability in the intensity of the emission $\left(\Delta I/ \overline{I} \right)\sim 5$ can be traced to inhomogeneities in the interstellar matter: $\left(\Delta n_{ism}/n_{ism}\right)\sim 5$. The typical size of the scattering region is estimated to be $5\times 10^{16}\, cm$, and these are the typical sizes and density contrasts found in interstellar clouds. Since the emission of the E-APE occurs at typical dimensions of the order of $5\times 10^{16}\, cm$, the observed inhomogeneities are probing the structure of the interstellar medium, and have nothing to do with the ``inner engine'' of the source.

The big issue was then open if all these results, obtained in the radial approximation, would still be valid in the more general case when off-axis emission in the description of the afterglow is taken into account. This is the reason why we have proceeded to the topic summarized in the next subsections (see \textcite{rbcfx02a_sub}).

\subsection{The definition of the equitemporal surfaces (EQTS) and the afterglow delayed intensity as a function of the viewing angle}

While the analysis of the average bolometric intensity of GRB was going on in the radial approximation, we have proceeded to develop the full non-radial approximation, taking into account all the relativistic corrections for the off-axis emission from the spherically symmetric expansion of the ABM pulse (see \textcite{rbcfx02a_sub,rbcfx02e_paperII} and sections~\ref{angle}--\ref{off-axis}). Photons emitted at the same time but at different angles of displacement from the line of sight reach the detector at very different arrival times. Correspondingly, photons detected at the same arrival time are emitted at very different times and angles. We have so defined the temporal evolution of the ABM pulse visible area as well as the equitemporal surfaces (EQTS), i.e. the locus of points on the ABM pulse emitting surface corresponding to a constant value of the photon arrival time at the detector.

\begin{figure}
\begin{center}
\includegraphics[width=6.2cm,clip]{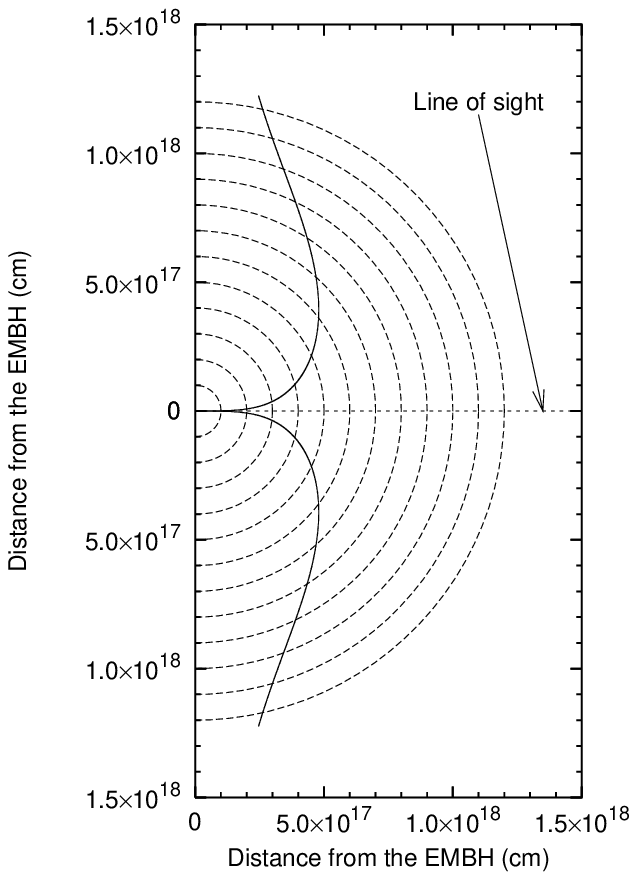}
\includegraphics[width=11cm,clip]{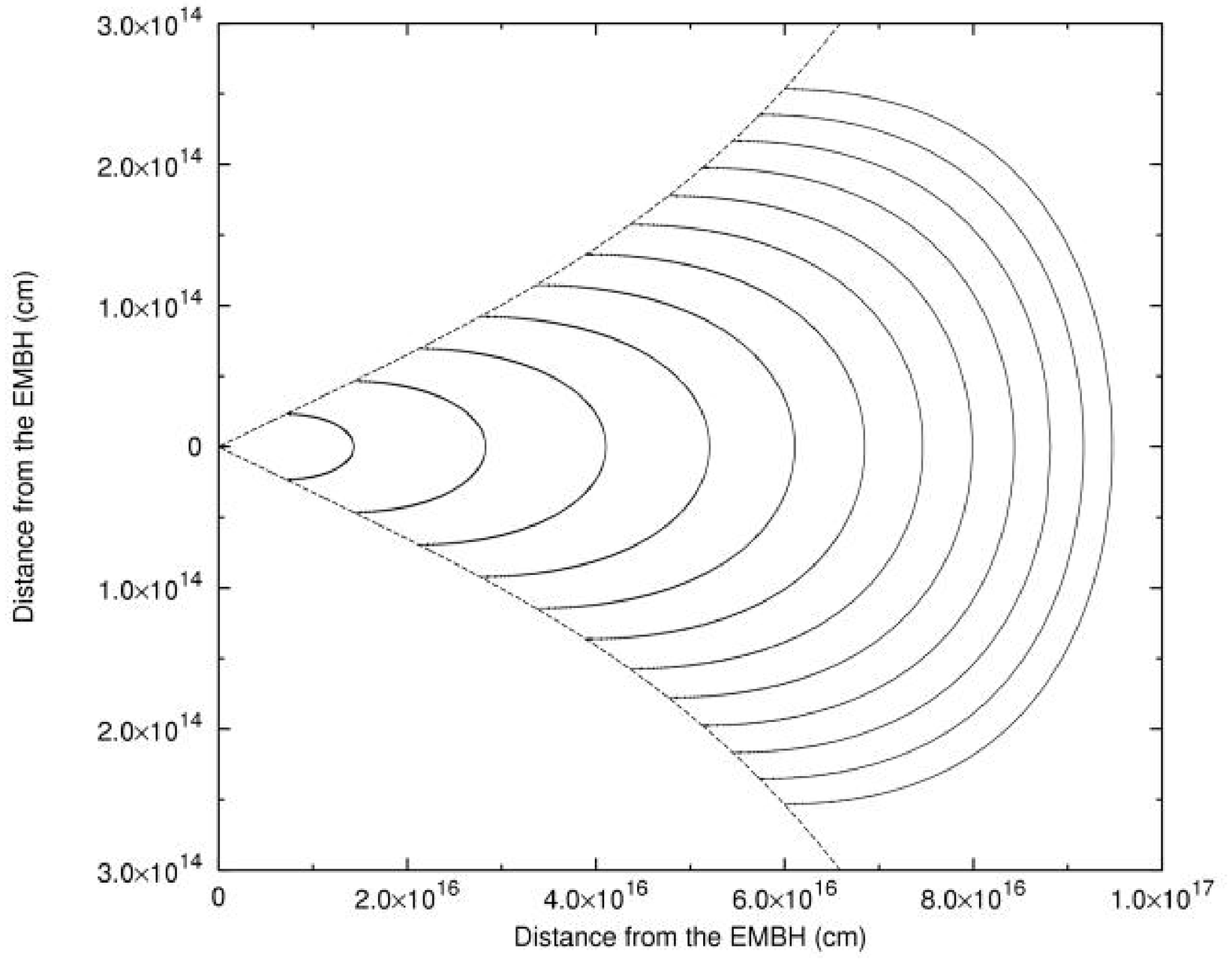}
\caption{{\bf Left)} This figure shows the temporal evolution of visible area of the ABM pulse. The dashed half-circles are the expanding ABM pulse at radii corresponding to different laboratory times. The black curve marks the boundary of the visible region. The EMBH is located at position (0,0) in this plot. Again, in the earliest GRB phases the visible region is squeezed along the line of sight, while in the final part of the afterglow phase almost all the emitted photons reach the observer. This time evolution of the visible area is crucial to the explanation of the GRB temporal structure. {\bf Right)} Due to the extremely high and extremely varying Lorentz gamma factor, photons reaching the detector on the Earth at the same arrival time are actually emitted at very different times and positions. We represent here the surfaces of photon emission corresponding to selected values of the photon arrival time at the detector: the {\em equitemporal surfaces} (EQTS). Such surfaces differ from the ellipsoids described by Rees in the context of the expanding radio sources with typical Lorentz factor $\gamma\sim 4$ and constant. In fact, in GRB~991216 the Lorentz gamma factor ranges from $310$ to $1$. The EQTSes represented here (solid lines) correspond respectively to values of the arrival time ranging from $5\, s$ (the smallest surface on the left of the plot) to $60\, s$ (the largest one on the right). Each surface differs from the previous one by $5\, s$. To each EQTS contributes emission processes occurring at different values of the Lorentz gamma factor. The dashed lines are the boundaries of the  visible area of the ABM pulse and the EMBH is located at position $(0,0)$ in this plot. Note the different scale on the two axes, indicating the very high EQTS ``effective eccentricity''. The time interval from $5\, {\rm s}$ to $60\, {\rm s}$ has been chosen to encompass the E-APE emission, ranging from $\gamma=308.8$ to $\gamma=56.84$.}
\label{opening_p}
\label{ETSNCF_p}
\end{center}
\end{figure}

The very same difficulties found in the current literature, relating the laboratory time to the photon arrival time at the detector (see Figs.~\ref{gamma}--\ref{tvsta}), still exists in the present context and are even magnified in the definition of the EQTS. In a classical article, \textcite{r66} expressed the relation between the laboratory time and the arrival time at the detector in order to explain observations in radio sources with a constant expansion velocity $v$ and Lorentz gamma factor $\gamma \sim 5$. He pointed out the EQTS are ellipsoids of constant eccentricity $v/c$. In the current literature, the Rees approach has been adapted to the analysis of GRBs (see e.g. \textcite{fmn96,s97,w97,s98,p99} and references therein). In addition to the very crucial relation between the laboratory time and the photon arrival time, which has not been properly treated, there have been a variety of other approximation and averaging processes on which we do not agree. Instead of specifically criticizing each assumption which we consider not correct, such comparison will be made in a forthcoming paper (\textcite{nl03b}), we just report here in the following the results of the EQTS surfaces (see Fig.~\ref{ETSNCF_p}) obtained in conformity with the RSTT paradigm. In the present case of GRBs, the gamma factor is not only much larger than the one observed in radio sources, but is also strongly time varying (see Fig.~\ref{gamma}). The Rees treatment has to be significantly improved to take into account the huge time variations in the Lorentz gamma factor: this is not just a technical point of modifying a formula by the introduction of a new integral. There is in the present context the crucial point expressed in the RSTT paradigm that the relation between the laboratory time and the arrival time at the detector is a function of all the the previous Lorentz gamma factors in the history of the source since $\gamma=1$ (see Fig.~\ref{tvsta}). In the definition of each EQTS, therefore, the entire previous past history of the source does concur and the EQTS surfaces become therefore a very refined and sensitive test of the correct description of the entire spacetime evolution of the source. In this case, we no longer have ellipsoids of constant eccentricity $\frac{v}{c}$. Since the velocity is strongly varying from point to point, we have more complicated surfaces like the profiles reported in Fig.~\ref{ETSNCF_p} where at every point there will be a tangent ellipsoid of a given eccentricity, but such an ellipsoid varies in eccentricity from point to point (see Fig.~\ref{ETSNCF_p} and section~\ref{angle}). Any departure from the correct equation of motion strongly alters the EQTS surfaces and accordingly modifies all the results of the integrations based on the EQTS surfaces, e.g. the spectral distribution or the afterglow (\textcite{nl03}).

\begin{figure}
\begin{center}
\includegraphics[width=8.5cm,clip]{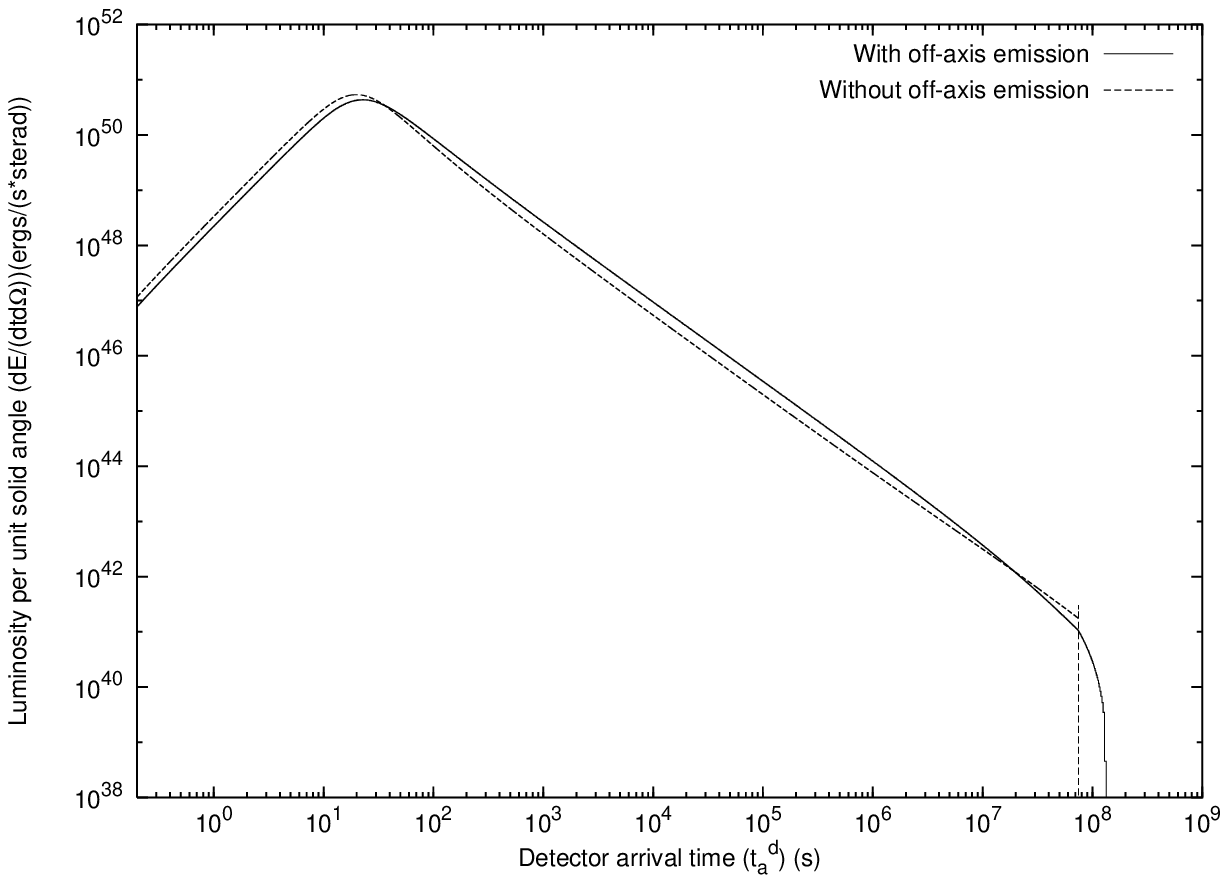}
\includegraphics[width=8.5cm,clip]{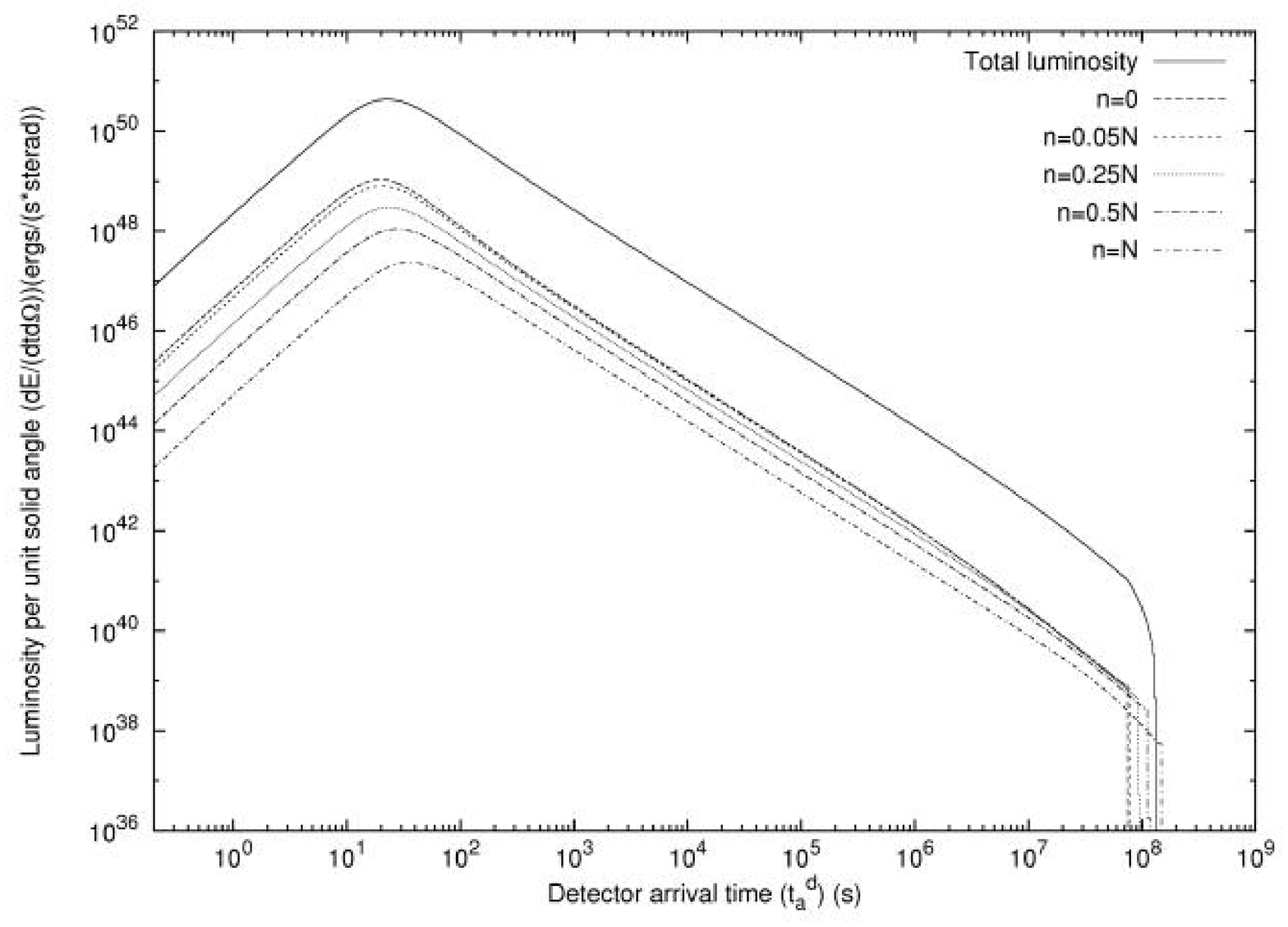}
\caption{{\bf Left)} The predicted afterglow curve for GRB~991216 assuming a constant ISM density equal to $1\, {\rm particle}/{\rm cm}^3$ and taking into account all the effects due to off-axis emission (solid line). For comparison we plot also  the corresponding curve obtained in the simple radial approximation (dashed line). We see that this last curve falls sharply to zero when the ABM pulse reaches $\gamma=1$, while the first one has a much smoother behavior due to the time delay in the arrival of the photons emitted at large $\vartheta$. Recall that when $\gamma$ tends to $1$, the maximum allowed values of $\vartheta$ tend to $90^\circ$. {\bf Right)} This figure shows how the radiation emitted from different angles contributes to the afterglow luminosity. The solid line on the top of the picture is the total luminosity as in the previous plots. The other dashed and dotted curves represent the radiation components corresponding to selected values of $n$ in Eq.(\ref{ct_def}). From the upper to the lower one they corresponds respectively to $n=0$, $n=0.05N$, $n=0.25N$, $n=0.5N$, $n=N$, where in this plot $N=200$. We can easily see that the radiation emitted at large angles ($n=N$) is time shifted with respect to that emitted near the line of sight ($n=0$).}
\label{afterang_p}
\label{angspre_p}
\end{center}
\end{figure}

Having determined the EQTS surfaces we have computed the observed GRB flux at selected values of the photon arrival time at the detector, taking into due account the delayed contributions at different angles and we have presented the results in section~\ref{off-axis} and Fig.~\ref{angspre_p}.

We have then recomputed the afterglow emission of GRB~991216 taking into account all the effects due to this temporal spreading in the arrival time as well as the ones due to the dependency of the photon Doppler shift on the angle of displacement from the line of sight of the emission location (see section~\ref{off-axis}). The result is reported in Fig.~\ref{angspre_p}.

From now on all the afterglow intensities are estimated using this very complex and extensive numerical program which is rooted in all previous history of the source: the general considerations on simple analytic expansion expressed in section~\ref{approximation} are kept only as an heuristic procedure as a guideline to comprehend these more complex results.

\subsection{The E-APE temporal substructures taking into account the off-axis emission}

\begin{figure}
\begin{center}
\includegraphics[width=\hsize,clip]{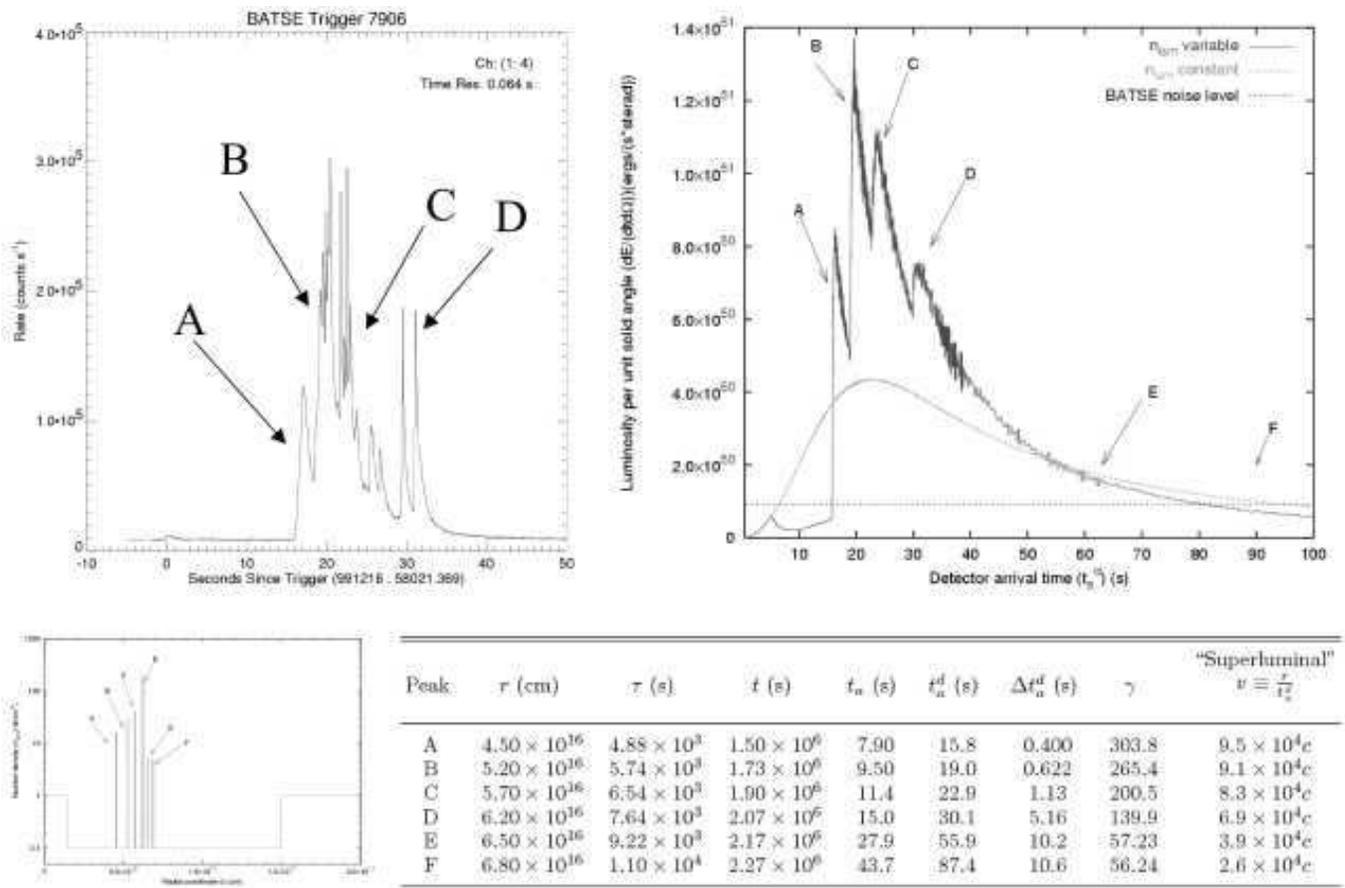}
\caption{In this figure we summarize the main results of the fit obtained by the EMBH model for the E-APE intensity in the case of GRB~991216 taking into account all off-axis contributions. The upper two diagrams represent respectively the observational data and the corresponding theoretically computed results. On the lower left the ``mask'' of the spherically symmetric density inhomogeneities with average $<n_{ism}>=1\, {\rm particle}/{\rm cm}^3$ is represented. The table summarizes all the parameters corresponding to the inhomogeneities including the vary large apparent superluminal effect up to $\sim 10^5c$. Details in section~23.}
\label{diap}
\end{center}
\end{figure}

Having determined the EQTS surfaces, we have reconsidered the E-APE temporal substructure taking into due account the off-axis emission contribution (see Fig.~\ref{diap} and section~\ref{structure_angle}).

We can distinguish two different regimes corresponding respectively to $\gamma > 150$ and to $\gamma < 150$. In the E-APE region ($\gamma > 150$) the GRB substructure intensities indeed correlate with the ISM inhomogeneities. In this limited region (see peaks A, B, C) the Lorentz gamma factor of the ABM pulse ranges from $\gamma\sim 304$ to $\gamma\sim 200$. The boundary of the visible region is smaller than the thickness $\Delta R$ of the inhomogeneities (see Figs.~\ref{diap},\ref{opening_p}, Tab.~\ref{tab3} and \textcite{rbcfx02a_sub,rbcfx02e_paperII}). Under these conditions the adopted spherical symmetry for the density spikes is not only mathematically simpler but also fully justified. The angular spreading is not strong enough to wipe out the signal from the inhomogeneity spike.

As we descend in the afterglow ($\gamma < 150$), a  border-line case occurs at peak D where $\gamma\sim 140$. There the visible region is comparable to the thickness $\Delta R$: to fit the observed data a three dimensional description would be necessary, breaking the spherical symmetry and making the computation more difficult, but we do not foresee any conceptual difficulty. For the peaks E and F we have $\gamma\sim 50$: under these circumstances the boundary of the visible region becomes much larger than the thickness $\Delta R$. The spherically symmetric description of the inhomogeneities is already enough to prove the overwhelming effect of the angular spreading and no three dimensional description is needed (\textcite{rbcfx02a_sub,rbcfx02e_paperII}).

From our analysis we can conclude that Dermer's expectations do indeed hold for $\gamma > 150$. However, as the gamma factor drops from $\gamma\sim 150$ to $\gamma\sim 1$ the intensity due to the inhomogeneities markedly decreases due to the angular spreading (events E and F). The initial Lorentz factor of the ABM pulse $\gamma\sim 310$ decreases very rapidly to $\gamma\sim 150$ as soon as a fraction of a typical ISM cloud is engulfed (see Figs.~\ref{diap},\ref{gamma}, Tab.~\ref{tab3} and \textcite{rbcfx02a_sub,rbcfx02e_paperII}). We conclude that the ``tomography" is indeed effective, but uniquely in the first ISM region close to the source and for GRBs with $\gamma > 150$.

It is then clear that no information on the nature of the GRB source can be inferred by the analysis of the $T_{90}$, nor by the intensity variability structure of the so-called ``long burts'': the only indirect information can be obtained from the value of Lorentz gamma factor, which has to be $\gamma > 150$ in presence of significant observed substructure. In this sense compare and contrast the two cases of GRB 991216 and GRB 980425 where the $\gamma$ value in the E-APE is found to be $\gamma\sim 120$ (see \textcite{r03cospar02}). The intensity substructures in the E-APE only carry information on the structure of the ISM clouds.

\subsection{The observation of the iron lines in GRB~991216: on a possible GRB-supernova time sequence}

In section~\ref{gsts} the program of using GRBs to further explore the region surrounding the newly formed EMBH is carried one step further by using the observations of the emitted iron lines (\textcite{p00}). This gives us the opportunity to introduce the GRB-supernova time sequence (GSTS) paradigm and to introduce as well the novel concept of an {\em induced supernova explosion}. The GSTS paradigm reads: {\em A massive GRB-progenitor star $P_1$ of mass $M_1$ undergoes gravitational collapse to an EMBH. During this process a dyadosphere is formed and subsequently the P-GRB and the E-APE are generated in sequence. They propagate and impact, with their photon and neutrino components, on a second supernova-progenitor star $P_2$ of mass $M_2$. Assuming that both stars were generated approximately at the same time, we expect to have $M_2 < M_1$. Under some special conditions of the thermonuclear evolution of the supernova-progenitor star $P_2$, the collision of the P-GRB and the E-APE with the star $P_2$ can induce its supernova explosion}.

Using the result presented in Tab.~\ref{tab1} and in all preceding sections, the GSTS paradigm is illustrated in the case of GRB~991216. Some general considerations on the nature of the supernova progenitor star are also advanced.

Some general considerations on the EMBH formation are presented in section~\ref{gc}. The general conclusions are presented in section~\ref{conclusions}.

We now proceed to a more detailed presentation of the results and we refer to the already published material for the complete details.

\section{The zeroth Era: the process of gravitational collapse and the formation of the dyadosphere}\label{dyadosphere}

We first recall the three theoretical results which lie at the basis of the EMBH theory.

In 1971 in the article {\itshape ``Introducing the Black Hole''} (\textcite{rw71}), the theorem was advanced that the most general black hole is characterized uniquely by three independent parameters: the mass-energy $M$, the angular momentum $L$ and the charge  $Q$ making it an EMBH. Such an ansatz, which came to be known as the ``uniqueness theorem'' has turned out to be one of the most difficult theorems to be proven in all of physics and mathematics. The progress in the proof has been authoritatively summarized by \textcite{c97}. The situation can be considered satisfactory from the point of view of the physical and astrophysical considerations. Nevertheless some fundamental mathematical and physical issues concerning the most general perturbation analysis of an EMBH are still the topic of active scientific discussion (\textcite{bcjr02}). 
 
In 1971 it was shown that the energy extractable from an EMBH is governed by the mass-energy formula (\textcite{cr71}),
\begin{equation}
E_{BH}^2=M^2c^4=\left(M_{\rm ir}c^2 + {Q^2\over2\rho_+}\right)^2+{L^2c^2\over \rho_+^2},\label{em}
\end{equation}
with
\begin{equation}
{1\over \rho_+^4}\left({G^2\over c^8}\right)\left( Q^4+4L^2c^2\right)\leq 1,
\label{s1}
\end{equation}
where
\begin{equation}
S=4\pi \rho_+^2=4\pi(r_+^2+{L^2\over c^2M^2})=16\pi\left({G^2\over c^4}\right) M^2_{\rm ir},
\label{sa}
\end{equation}
is the horizon surface area, $M_{\rm ir}$ is the irreducible mass, $r_{+}$ is the horizon radius and $\rho_+$ is the quasi-spheroidal cylindrical coordinate of the horizon evaluated at the equatorial plane. Extreme EMBHs satisfy the equality in Eq.(\ref{s1}). Up to 50\% of the mass-energy of an extreme EMBH can in principle be extracted by a special set of transformations: the reversible transformations (\textcite{cr71}).

In 1975, generalizing some previous results of \textcite{za75}, and \textcite{gb75}, \textcite{dr75} showed that the vacuum polarization process {\it \`a la} Heisenberg-Euler-Schwinger (\textcite{he35,s51}) created by an electric field of strength larger than 
\begin{equation}
{\cal E}_c=\frac{m_e^2c^3}{\hbar e}
\label{ecrit}
\end{equation}
can indeed occur in the field of a Kerr-Newmann EMBH. Here $m_e$ and $e$ are respectively the mass and charge of the electron.  There Damour and Ruffini considered an axially symmetric EMBH, due to the presence of rotation, and limited themselves to EMBH masses larger then the upper limit of a neutron star for astrophysical applications. They purposely avoided all complications of black holes with mass smaller then the dual electron mass of the electron $\left(m_e^\star=\frac{c\hbar}{G m_e}=\frac{m_{Planck}^2}{m_e}\right)$ which may lead to quantum evaporation processes (\textcite{h74}). They pointed out that:
\begin{enumerate}
\item The vacuum polarization process can occur for an EMBH mass larger than the maximum critical mass for neutron stars all the way up to $7.2\times 10^6 M_\odot$. 
\item The process of pair creation occurs on very short time scales, typically $\frac{\hbar}{m_e c^2}$, and is an almost perfect reversible process, in the sense defined by Christodoulou-Ruffini, leading to a very efficient mechanism of extracting energy from an EMBH.
\item The energy generated by the energy extraction process of an EMBH was found to be of the order of $10^{54}$ erg, released almost instantaneously. They concluded at the time {\itshape ``this work naturally leads to a most simple model for the explanation of the recently discovered $\gamma$-ray bursts''}.
\end{enumerate}

After the discovery of the afterglow of GRBs and the determination of the cosmological distance of their sources we noticed the coincidence between the theoretically predicted energetics and the observed ones in \textcite{dr75}: we returned to our theoretical results developing some new basic theoretical concepts (\textcite{rukyoto,prxprl,prx98,rswx99,rswx00}), which have led to the EMBH theory.

As a first simplifying assumption we have developed our considerations in the absence of rotation with spherically symmetric distributions. The space-time is then described by the Reissner-Nordstr\"{o}m geometry, whose spherically symmetric metric is given by
\begin{equation}
d^2s=g_{tt}(r)d^2t+g_{rr}(r)d^2r+r^2d^2\theta +r^2\sin^2\theta
d^2\phi ~,
\label{s}
\end{equation}
where $g_{tt}(r)= - \left[1-{2GM\over c^2r}+{Q^2G\over c^4r^2}\right] \equiv - \alpha^2(r)$ and $g_{rr}(r)= \alpha^{-2}(r)$.

The first new result we obtained is that the pair creation process does not occur at the horizon of the EMBH: it extends over the entire region outside the horizon in which the electric field exceeds the critical value given by Eq. \ref{ecrit}. Since the electric field in the Reissner-Nordstr\"{o}m geometry has only a radial component given by (see \textcite{r78})
\begin{equation}
{\cal E}\left(r\right)=\frac{Q}{r^2}\, ,
\label{edir}
\end{equation}
this region extends from the horizon radius
\begin{eqnarray}
r_{+}&=&1.47 \cdot 10^5\mu (1+\sqrt{1-\xi^2})\hskip0.1cm {\rm cm}
\label{r+}
\end{eqnarray}
out to an outer radius (\textcite{rukyoto})
\begin{equation}
r^\star=\left({\hbar\over mc}\right)^{1\over2}\left({GM\over
c^2}\right)^{1\over2} \left({m_{\rm p}\over m}\right)^{1\over2}\left({e\over
q_{\rm p}}\right)^{1\over2}\left({Q\over \sqrt{G}M}\right)^{1\over2}=1.12\cdot 10^8\sqrt{\mu\xi} \hskip0.1cm {\rm cm},
\label{rc}
\end{equation}
where we have introduced the dimensionless mass and charge parameters $\mu={M\over M_{\odot}}$, $\xi={Q\over (M\sqrt{G})}\le 1$, see Fig.~\ref{dyaon}.

The second new result has been to realize that the local number density of electron and positron pairs created in this region as a function of radius is given by
\begin{equation}
n_{e^+e^-}(r) = {Q\over 4\pi r^2\left({\hbar\over
mc}\right)e}\left[1-\left({r\over r^\star}\right)^2\right] ~,
\label{nd}
\end{equation}
and consequently the total number of electron and positron pairs in this region is
\begin{equation}
N^\circ_{e^+e^-}\simeq {Q-Q_c\over e}\left[1+{
(r^\star-r_+)\over {\hbar\over mc}}\right],
\label{tn}
\end{equation}
\begin{figure}
\includegraphics[width=10cm,clip]{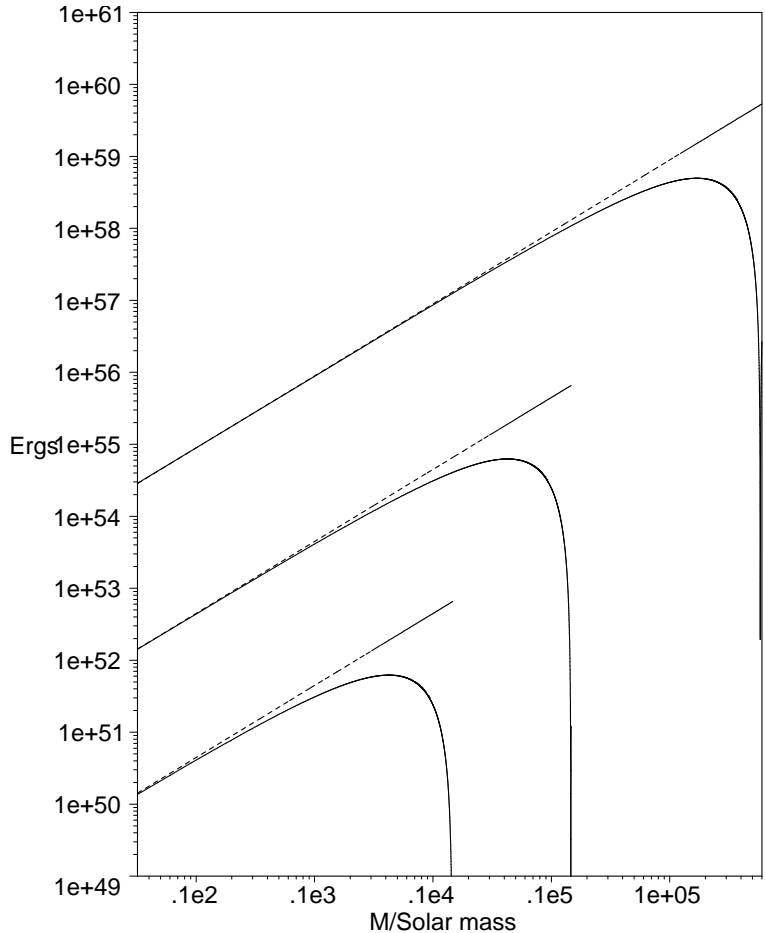}
\caption{The energy extracted by the process of vacuum polarization is plotted (solid lines) as a function of the mass $M$ in solar mass units for selected values of the charge parameter $\xi=1,0.1,0.01$ (from top to bottom) for an EMBH, the case $\xi=1$ reachable only as a limiting process. For comparison we have also plotted the maximum energy extractable from an EMBH (dotted lines) given by eq.~(\ref{em}). Details in \textcite{prx01}.}
\label{prep}
\end{figure}
where $Q_c={\cal E}_{\rm c}r_+^2$.

The total number of pairs is larger by an enormous factor $r^{\star}/\left(\hbar/mc\right) > 10^{18}$ than the value $Q/e$ which a naive estimate of the discharge of the EMBH would have predicted. Due to this enormous amplification factor in the number of pairs created, the region between the horizon and $r^{\star}$ is dominated by an essentially high density neutral plasma of electron-positron pairs. We have defined this region as the dyadosphere of the EMBH from the Greek duas, duadsos for pairs. Consequently we have called $r^\star$ the dyadosphere radius $r^\star \equiv r_{\rm ds}$ (\textcite{rukyoto,prxprl,prx98}). The vacuum polarization process occurs as if the entire dyadosphere are subdivided into a concentric set of shells of capacitors each of thickness $\hbar/m_ec$ and each producing a number of $e^+e^-$ pairs on the order of $\sim Q/e$ (see Fig.~\ref{dyaon}). The energy density of the electron-positron pairs is given by
\begin{equation}
\epsilon(r) = {Q^2 \over 8 \pi r^4} \biggl(1 - \biggl({r \over
r_{\rm ds}}\biggr)^4\biggr) ~, \label{jayet}
\end{equation}
(see Figs.~2--3 of \textcite{prxprl}). The total energy of pairs converted from the static electric energy and deposited within the dyadosphere is then
\begin{equation}
E_{\rm dya}={1\over2}{Q^2\over r_+}\left(1-{r_+\over r_{\rm ds}}\right)\left[1-\left({r_+\over r_{\rm ds}}\right)^4\right] ~.
\label{tee}
\end{equation}

\begin{figure}
\includegraphics[width=8.5cm,clip]{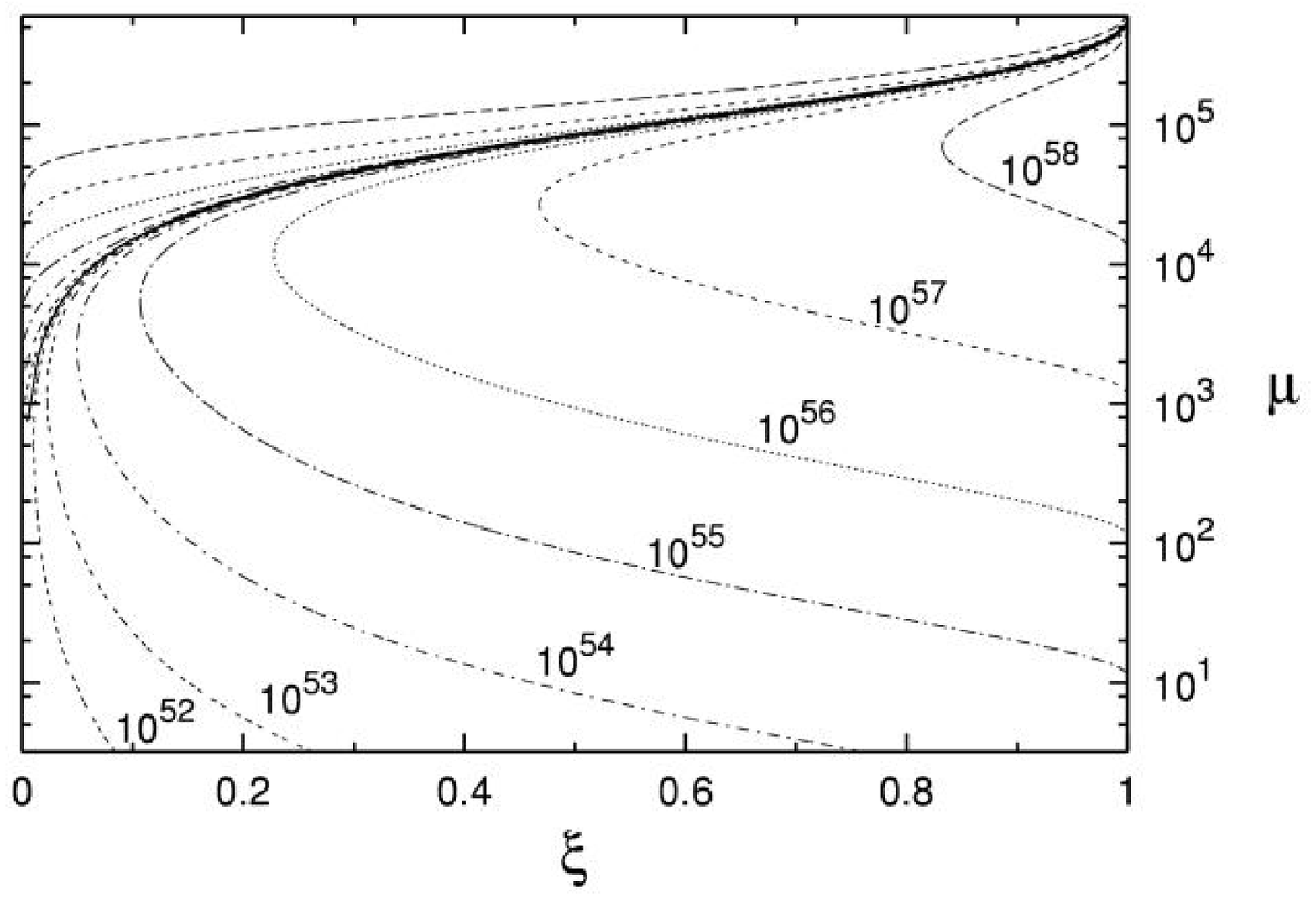}
\includegraphics[width=8.5cm,clip]{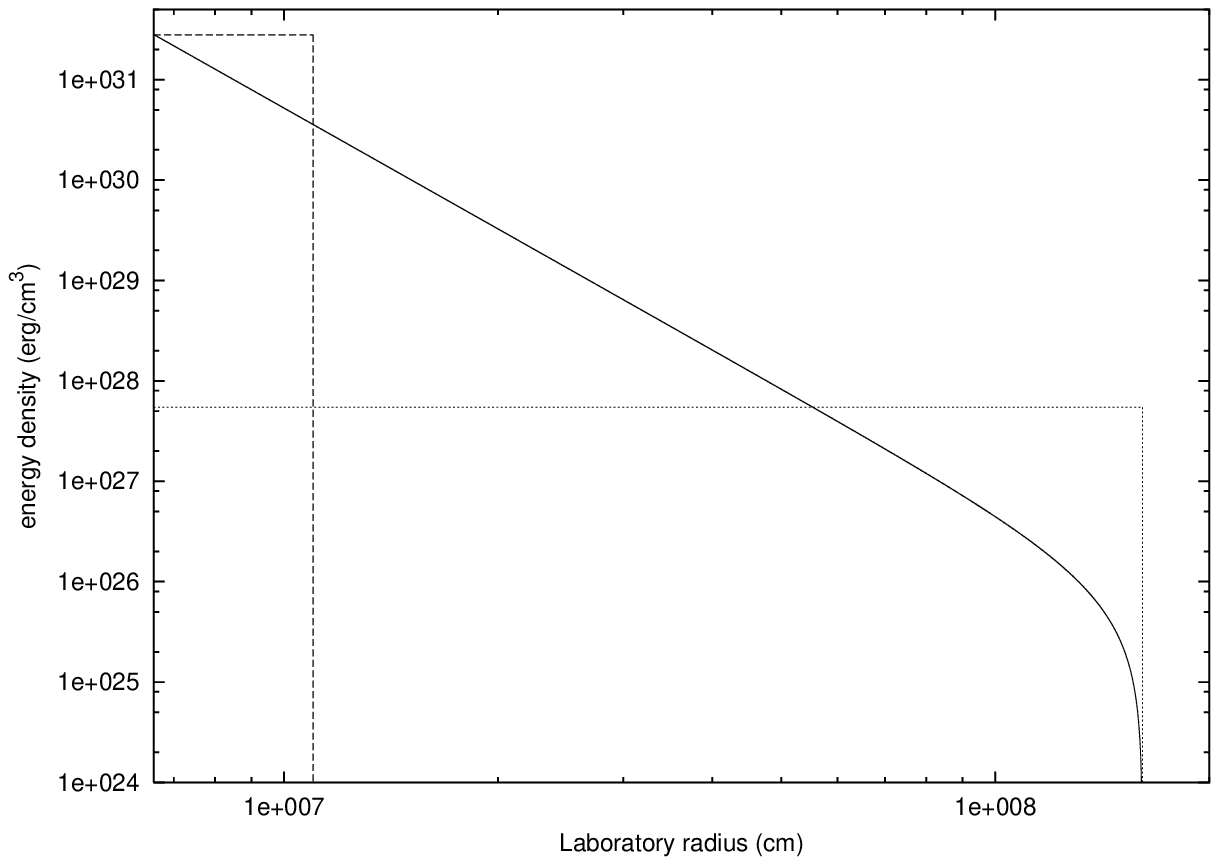}
\caption{{\bf Left)} Selected lines corresponding to fixed values of the $E_{dya}$ are given as a function of the two parameters $\mu$ $\xi$, only the solutions below the continuous heavy line are physically relevant.The configurations above the continuous heavy lines correspond to unphysical solutions with $r_{\rm ds} < r_+$. {\bf Right)} Two different approximations for the energy density profile inside the dyadosphere. The first one (dashed line) fixes the energy density equal to its peak value, and computes an ``effective'' dyadosphere radius accordingly. The second one (dotted line) fixes the dyadosphere radius to its correct value, and assumes an uniform energy density over the dyadosphere volume. The total energy in the dyadosphere is of course the same in both cases. The solid curve represents the real energy density profile.}
\label{muxi}
\label{dens}
\end{figure}

As we will see in the following this is one of the two fundamental parameters of the EMBH theory (see Fig.~\ref{muxi}). In the limit ${r_+\over r_{\rm ds}}\rightarrow 0$, Eq.(\ref{tee}) leads to $E_{\rm dya}\rightarrow {1\over2}{Q^2\over r_+}$, which coincides with the energy extractable from EMBHs by reversible processes ($M_{\rm ir}={\rm const.}$), namely $E_{BH}-M_{\rm ir}={1\over2}{Q^2\over r_+}$ (\textcite{cr71}), see Fig.~\ref{prep}. Due to the very large pair density given by Eq.(\ref{nd}) and to the sizes of the cross-sections for the process $e^+e^-\leftrightarrow \gamma+\gamma$, the system is expected to thermalize to a plasma configuration for which
\begin{equation}
n_{e^+}=n_{e^-} \sim n_{\gamma} \sim n^\circ_{e^+e^-},
\label{plasma}
\end{equation}
where $n^\circ_{e^+e^-}$ is the total number density of $e^+e^-$-pairs created in the dyadosphere (see \textcite{prxprl,prx98}). 

The third new result which we have introduced for simplicity is that for a given $E_{dya}$ we have assumed either a constant average energy density over the entire dyadosphere volume, or a more compact configuration with energy density equal to the peak value. These are the two possible initial conditions for the evolution of the dyadosphere (see Fig.~\ref{dens}).

These three old and three new theoretical results permit a good estimate of the general energetics processes originating in the dyadosphere, assuming an already formed EMBH. In reality, if the data become accurate enough, the full dynamical description of the dyadosphere formation mentioned above will be needed in order to follow all the general relativistic effects and characteristic time scales of the approach to the EMBH horizon (\textcite{crv02,rv02a,rv02b,rvx02} see also section~\ref{gc}).

Below we shall concentrate on the dynamical evolution of the electron-positron  plasma created in the dyadosphere. We shall first examine in the next three sections the governing equations necessary to approach such a dynamical description. 

\section{The hydrodynamics and the rate equations for the plasma of $e^+e^-$-pairs}\label{hydro_pem}

The evolution of the $e^+e^-$-pair plasma generated in the dyadosphere has been treated in two papers (\textcite{rswx99,rswx00}). We recall here the basic governing equations in the most general case in which the plasma fluid is composed of $e^+e^-$-pairs, photons and baryonic matter. The plasma is described by the stress-energy tensor
\begin{equation}
T^{\mu\nu}=pg^{\mu\nu}+(p+\rho)U^\mu U^\nu\, ,
\label{tensor}
\end{equation}
where $\rho$ and $p$ are respectively the total proper energy density and pressure in the comoving frame of the plasma fluid and $U^\mu$ is its four-velocity, satisfying
\begin{equation}
g_{tt}(U^t)^2+g_{rr}(U^r)^2=-1 ~,
\label{tt}
\end{equation}
where $U^r$ and $U^t$ are the radial and temporal contravariant components of the 4-velocity.

The conservation law for baryon number can be expressed in terms of the proper baryon number density $n_B$
\begin{eqnarray}
(n_B U^\mu)_{;\mu}&=& g^{-{1\over2}}(g^{1\over2}n_B
U^\nu)_{,\nu}\nonumber\\
&=&(n_BU^t)_{,t}+{1\over r^2}(r^2 n_BU^r)_{,r}=0 ~.
\label{contin}
\end{eqnarray}
The radial component of the energy-momentum conservation law of the plasma fluid reduces to 
\begin{equation}
{\partial p\over\partial r}+{\partial \over\partial t}\left((p+\rho)U^t U_r\right)+{1\over r^2} { \partial
\over \partial r}  \left(r^2(p+\rho)U^r U_r\right)
-{1\over2}(p+\rho)\left[{\partial g_{tt}
 \over\partial r}(U^t)^2+{\partial g_{rr}
 \over\partial r}(U^r)^2\right] =0 ~.
\label{cmom2}
\end{equation}
The component of the energy-momentum conservation law of the plasma fluid equation along a flow line is
\begin{eqnarray}
U_\mu(T^{\mu\nu})_{;\nu}&=&-(\rho U^\nu)_{;\nu}
-p(U^\nu)_{;\nu},\nonumber\\ &=&-g^{-{1\over2}}(g^{1\over2}\rho
U^\nu)_{,\nu} - pg^{-{1\over2}}(g^{1\over2} U^\nu)_{,\nu}\nonumber\\
&=&(\rho U^t)_{,t}+{1\over r^2}(r^2\rho
U^r)_{,r}\nonumber\\
&+&p\left[(U^t)_{,t}+{1\over r^2}(r^2U^r)_{,r}\right]=0 ~.
\label{conse1}
\end{eqnarray}

Defining the total proper internal energy density $\epsilon$ and the baryonic mass density $\rho_B$ in the comoving frame of the plasma fluid,
\begin{equation}
\epsilon \equiv \rho - \rho_B,\hskip0.5cm \rho_B\equiv n_Bmc^2 ~,
\label{cpp}
\end{equation} 
and using the law (\ref{contin}) of baryon-number conservation, from Eq. (\ref{conse1}) we have
\begin{equation}
(\epsilon U^\nu)_{;\nu} +p(U^\nu)_{;\nu}=0 ~.
\label{conse'}
\end{equation}
Recalling that ${dV\over d\tau}=V(U^\mu)_{;\mu}$, where $V$ is the comoving volume and $\tau$ is the proper time for the plasma fluid, we have along each flow line
\begin{equation}
{d(V\epsilon)\over d\tau}+p{dV\over d\tau}={dE\over d\tau}+p{dV\over
d\tau}=0 ~,
\label{f'}
\end{equation}
where $E=V\epsilon$ is the total proper internal energy of the plasma fluid. We express the equation of state by introducing a thermal index $\Gamma(\rho,T)$
\begin{equation}
\Gamma = 1 + { p\over \epsilon} ~.
\label{state}
\end{equation}

We now turn to the second set of governing equations describing the evolution of the $e^+e^-$ pairs. Letting $n_{e^-}$ and $n_{e^+}$  be the proper number densities of electrons and positrons associated with pairs and $n^b_{e^-}$ the proper number densities of ionized electrons, we clearly have
\begin{equation}
n_{e^-}=n_{e^+}=n_{\rm pair},\hskip0.5cm n^b_{e^-}=\bar Z n_B,
\label{eee}
\end{equation}
where $n_{\rm pair}$ is the number of $e^+e^-$ pairs and $\bar Z$ the average atomic number ${1\over2}<\bar Z< 1$ ($\bar Z=1$ for hydrogen atom and $\bar Z={1\over2}$ for general baryonic matter). The rate equation for electrons and positrons gives,
\begin{eqnarray}
(n_{e^+}U^\mu)_{;\mu}&=&(n_{e^+}U^t)_{,t}+{1\over r^2}(r^2 n_{e^+}U^r)_{,r}\nonumber\\
&=&\overline{\sigma v} \big[(n_{e^-}(T)+n^b_{e^-}(T))n_{e^+}(T)\nonumber\\
& - &(n_{e^-}+n^b_{e^-})n_{e^+}\big],
\label{e+contin}\\
(n_{e^-}U^\mu)_{;\mu}&=&(n_{e^-}U^t)_{,t}+{1\over r^2}(r^2 n_{e^-}U^r)_{,r}\nonumber\\
&=&\overline{\sigma v} \left[n_{e^-}(T)n_{e^+}(T) - n_{e^-}n_{e^+}\right],
\label{e-contin}\\
(n^b_{e^-}U^\mu)_{;\mu}&=&(n^b_{e^-}U^t)_{,t}+{1\over r^2}(r^2 n^b_{e^-}U^r)_{,r}\nonumber\\
&=&\overline{\sigma v} \left[n^b_{e^-}(T)n_{e^+}(T) - n^b_{e^-}n_{e^+}\right],
\label{tbe-contin}
\end{eqnarray}
where $\overline{\sigma v}$ is the mean of the product of the annihilation cross-section and the thermal velocity of the electrons and positrons, $n_{e^\pm}(T)$ are the proper number densities of electrons and positrons associated with the pairs, given by appropriate Fermi integrals with zero chemical potential, and $n^b_{e^-}(T)$ is the proper number density of ionized electrons, given by appropriate Fermi integrals with non-zero chemical potential $\mu_e$ at an appropriate equilibrium temperature $T$. These rate equations can be reduced to 
\begin{eqnarray}
(n_{e^\pm}U^\mu)_{;\mu}&=&(n_{e^\pm}U^t)_{,t}+{1\over r^2}(r^2 n_{e^\pm}U^r)_{,r}\nonumber\\
&=&\overline{\sigma v} \big[n_{e^-}(T)n_{e^+}(T)- n_{e^-}n_{e^+}\big],
\label{econtin}\\
(n^b_{e^-}U^\mu)_{;\mu}&=&(n^b_{e^-}U^t)_{,t}+{1\over r^2}(r^2 n^b_{e^-}U^r)_{,r}=0,
\label{becontin}\\
Frac&\equiv&{n_{e^\pm}\over n_{e^\pm}(T)}={n^b_{e^-}(T)\over n^b_{e^-}}.
\label{be-contin}
\end{eqnarray}
Equation (\ref{becontin}) is just the baryon-number conservation law (\ref{contin}) and (\ref{be-contin}) is a relationship satisfied by $n_{e^\pm}, n_{e^\pm}(T)$ and $n^b_{e^-}, n^b_{e^-}(T)$.

The equilibrium temperature $T$ is determined by the thermalization processes occurring in the expanding plasma fluid with a total proper energy density $\rho$ governed by the hydrodynamical equations (\ref{contin},\ref{cmom2},\ref{conse1}). We have
\begin{equation}
\rho = \rho_\gamma + \rho_{e^+}+\rho_{e^-}+\rho^b_{e^-}+\rho_B,
\label{eeq}
\end{equation}
where $\rho_\gamma$ is the photon energy density, $\rho_B\simeq m_Bc^2n_B$ is the baryonic mass density which is considered to be nonrelativistic in the range of temperature $T$ under consideration, and $\rho_{e^\pm}$ is the proper energy density of electrons and positrons pairs given by
\begin{equation}
\rho_{e^\pm}= {n_{e^\pm}\over n_{e^\pm}(T)}\rho_{e^\pm}(T),
\label{hat}
\end{equation}
where $n_{e^\pm}$ is obtained by integration of Eq.(\ref{econtin}) and $\rho_{e^\pm}(T)$ is the proper energy density of electrons(positrons) obtained from zero chemical potential Fermi integrals at the equilibrium temperature $T$. On the other hand $\rho^b_{e^-}$ is the energy density of the ionized electrons coming from the ionization of baryonic matter
\begin{equation}
\rho^b_{e^-}= {n^b_{e^-}\over n^b_{e^-}(T)}\rho^b_{e^-}(T),
\label{bhat}
\end{equation}
where $n^b_{e^-}$ is obtained by integration of Eq.(\ref{becontin}) and $\rho_{e^-}(T)$ is the proper energy density of ionized electrons obtained from an appropriate Fermi integral of non-zero chemical potential $\mu_e$ at the equilibrium temperature $T$.

Having intrinsically defined the equilibrium temperature $T$ in Eq.(\ref{eeq}), we can also analogously evaluate the total pressure 
\begin{equation}
p = p_\gamma + p_{e^+}+p_{e^-}+p^b_{e^-}+p_B,
\label{eep}
\end{equation}
where $p_\gamma$ is the photon pressure, $p_{e^\pm}$ and $p^b_{e^-}$ are given by
\begin{eqnarray}
p_{e^\pm}&=& {n_{e^\pm}\over n_{e^\pm}(T)}p_{e^\pm}(T),
\label{hat'}\\
p^b_{e^-}&=& {n^b_{e^-}\over n^b_{e^-}(T)}p^b_{e^-}(T),
\label{bhat'}
\end{eqnarray}
the pressures $p_{e^\pm}(T)$ are determined by zero chemical potential Fermi integrals, and $p^b_{e^-}(T)$ is the pressure of the ionized electrons, evaluated by an appropriate Fermi integral of non-zero chemical potential $\mu_e$ at the equilibrium temperature $T$. In Eq.(\ref{eep}), the ion pressure $p_B$ is negligible by comparison with the pressures $p_{\gamma, e^\pm, e^-}(T)$, since baryons and ions are expected to be nonrelativistic in the range of temperature $T$ under consideration. Finally using Eqs.(\ref{eeq},\ref{eep}) we compute the thermal factor $\Gamma$ of the equation of state (\ref{state}). 

It is clear that the entire set of equations considered above, namely Eqs.(\ref{contin},\ref{cmom2},\ref{conse1}) with equation of state given by Eq.(\ref{state}) and the rate equation (\ref{econtin}), have to be integrated satisfying the total energy conservation for the system. The boundary conditions adopted here are simply purely ingoing conditions at the horizon and purely outgoing conditions at radial infinity. The calculation is initiated by depositing a proper energy density (\ref{jayet}) between the Reissner-Nordstr\"{o}m horizon radius $r_+$ and the dyadosphere radius $r_{ds}$, following the approximation presented in Fig.\ref{prep}  The total energy deposited is given by Eq.(\ref{tee}).

\section{The equations leading to the relative space-time transformations}\label{arrival_time}

In order to relate the above hydrodynamic and pair equations with the observations we need the governing equations relating the comoving time to the laboratory time corresponding to an inertial reference frame in which the EMBH is at rest and finally to the time measured at the detector, which must also include the effect of the cosmological expansion. These transformations have been the object of the Relative space-time Transformations (RSTT) Paradigm, (\textcite{lett1}). 

\begin{figure}
\includegraphics[width=10cm,clip]{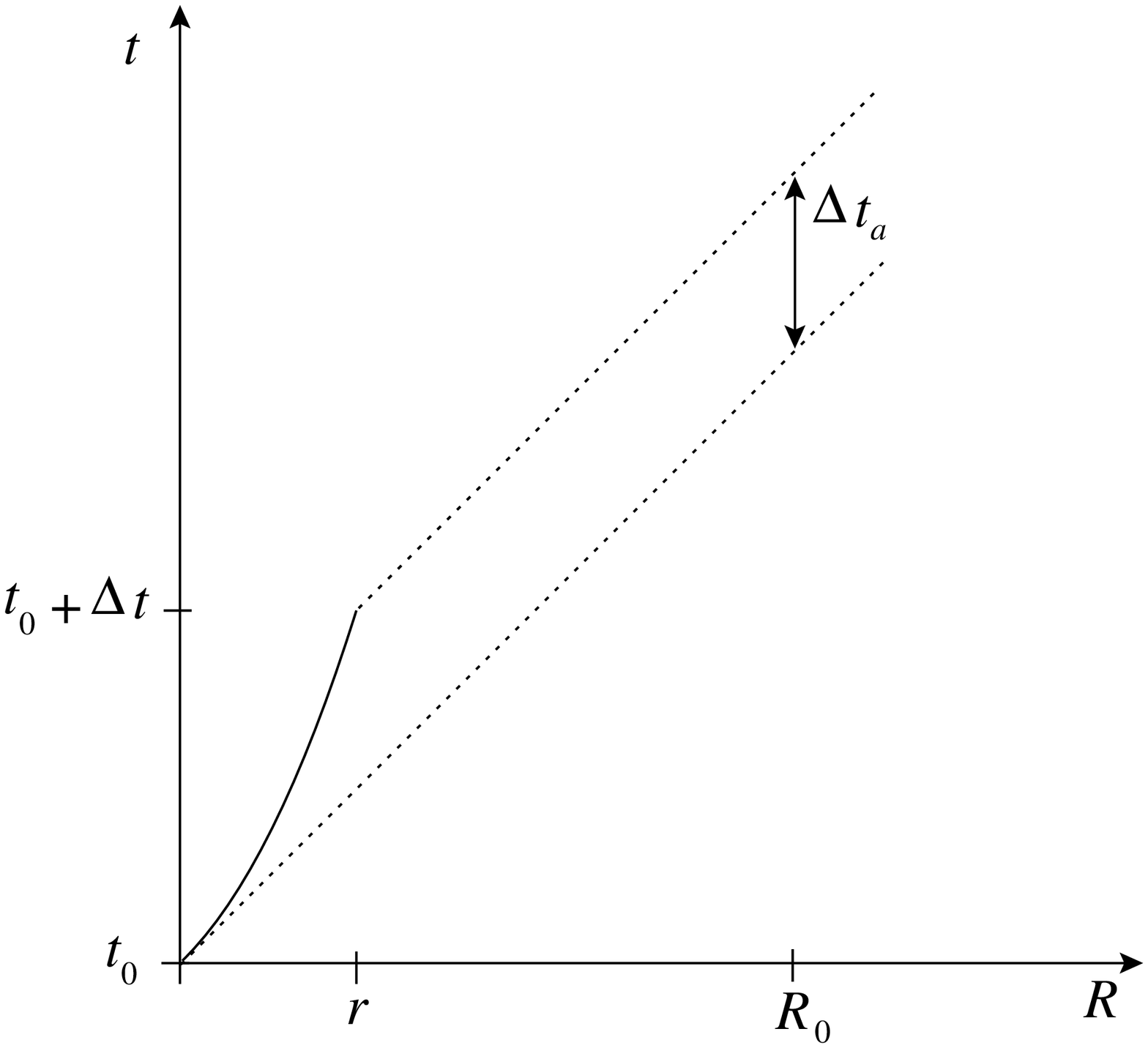}
\caption{This qualitative diagram illustrates the relation between the laboratory time interval $\Delta t$ and the arrival time interval $\Delta t_a$ for a pulse moving with velocity $v$ in the laboratory time (solid line). We have indicated here the case where the motion of the source has a nonzero acceleration. The arrival time is measured using light signals emitted by the pulse (dotted lines). $R_0$ is the distance of the observer from the EMBH, $t_0$ is the laboratory time corresponding to the onset of the gravitational collapse, and $r$ is the radius of the expanding pulse at a time $t=t_0 + \Delta t$. See also \textcite{lett1}.}
\label{ttasch}
\end{figure}

For signals emitted by a pulse moving with velocity $v$ in the laboratory frame (see also \textcite{lett1}), we have the following relation between the interval of arrival time $\Delta t_a$ and the corresponding interval of laboratory time $\Delta t$ (see Fig.~\ref{ttasch}):

\begin{equation}
\Delta t_a  = \left( {t_0  + \Delta t + \frac{{R_0  - r}}{c}} \right) - \left( {t_0  + \frac{{R_0 }}{c}} \right) = \Delta t - \frac{r}{c}\, .
\label{taintr}
\end{equation}

For simplicity in what follows we indicate by $t_a$ the interval of arrival time measured from the reception of a light signal emitted at the onset of the gravitational collapse. Analogously, $t$ indicates the laboratory time interval measured from the time of the gravitational collapse. In this case, Eq.(\ref{taintr}) can be written simply as:

\begin{equation}
t_a  = t - \frac{r}{c} = t - \frac{{\int_0^t {v\left( {t'} \right)dt'}  + r_{ds} }}{c}=t - \int_0^t{\frac{\sqrt{\gamma^2\left(t'\right)-1}}{\gamma\left(t'\right)} dt' } - \frac{r_{ds}}{c}\, ,
\label{tadef}
\end{equation}
where, as usual, $\gamma\left(t`\right)=1/\sqrt{1-v^2\left(t'\right)/c^2}$ and the dyadosphere radius $r_{ds}$ is the value of $r$ at $t=0$. It is important to stress that, although there is the presence of the Lorentz gamma factor, Eq.(\ref{tadef}) is not a Lorentz transformation, which by its own nature is linear and refers to a specific value of the Lorentz gamma factor at a given laboratory time. The transformation in Eq.(\ref{tadef}) is nonlinear in the Lorentz gamma factor and do depend on all the values of the gamma factor of the source from the time $t=0$ to the laboratory time $t$. This transformation is the price to pay to relate the laboratory time $t$, relativistically correct, to the ``highly pathological'' time usually considered by the astronomers, even in the case of object moving close to the speed of light, against the correct synchronization procedures established by Einstein in his classical paper of 1905 (\textcite{e05}). We consider here only the photons emitted along the line of sight from the external surface of the pulse. The arrival time spreading due to the angular dependence and that due to the thickness has also been given (see section~\ref{angle} and \textcite{rbcfx02a_sub,rbcfx02b_beam}). The solution of Eq.(\ref{tadef}) has the expansion:

\begin{equation}
t_a  = t - \frac{r_{ds}}{c} - \frac{{v\left(0\right)}}{c}t - \frac{1}{2}\frac{v'\left(0\right)}{c}t^2  -  \frac{1}{3}\frac{v''\left(0\right)}{c}t^3 - \ldots , 
\label{taex}
\end{equation}
so the relation between $t_a$ and $t$ in the specific case of GRBs is very highly nonlinear: it is sufficient to recall that in the early GRB phases we are witnessing the strongest acceleration ever recorded in the universe, since the PEM pulse goes from Lorentz factor $\gamma=1$ to Lorentz factor $\gamma=1000$ in $10^2$ seconds in the laboratory time (see section~\ref{era1}). The series in Eq.(\ref{taex}) will definitely converge, but the number of terms needed to reach a good approximation will strongly depend on the variability of the functions around the initial values $\gamma=1$. It is clear that the precise knowledge of $t_a$ as a function of the laboratory time, which is indeed essential for any physical interpretation of GRB data, depends on the definite integral given in Eq.(\ref{tadef}) whose limits in the laboratory time extend from the onset of the gravitational collapse to the time $t$ relevant for the observations. Such an integral depends on all previous values of the Lorentz gamma factor in the history of the source and is not generally expressible by a simple linear relation or even by any explicit analytic relation since we are dealing with processes with variable gamma factor unprecedented in the entire realm of physics (see Figs.~\ref{gamma} and Fig.~\ref{tvsta}). This is the crucial point of the RSTT paradigm (\textcite{lett1}) and this is the reason why we have spent a very large amount of work to develop the exact equations of motion of all different eras of the GRB phenomenon, starting from the onset of gravitational collapse and the creation of dyadosphere (see the following sections). It is clear then that, in order to express the arrival time $t_a$ and the radial coordinate of the source at the start of the afterglow phase, we need the explicit knowledge of all the previous eras of the GRB phenomenon, starting from $\gamma=1$ (\textcite{lett1}).

What has been currently done in the literature, is an extremely different approach. First they have assumed $\gamma$ constant. Therefore Eq.(\ref{tadef}) has been modified in:
\begin{equation}
t_a=t-\frac{\sqrt{\gamma^2-1}}{\gamma}\int_0^t{dt'}-\frac{r_{ds}}{c}\simeq t-\frac{\sqrt{\gamma^2-1}}{\gamma}t\, ,
\label{taexp1}
\end{equation}
where in the last approximation the contribution of the initial size of the source has been neglected. Even the validity of this last approximation has to be actually carefully verified since it is only valid in the late phases of the GRB expansion. They have further assumed $\gamma\gg 1$ and obtained:
\begin{equation}
t_a\simeq t-\left(1-\frac{1}{2\gamma^2}\right)t=\frac{t}{2\gamma^2}\, .
\label{taexp2}
\end{equation}
At this stage, they emphasize the existence of a linear relation between the arrival time $t_a$ and the laboratory time $t$. After this they proceed in two different directions. One to assume (see e.g. \textcite{fmn96,w97,sp97,f99,fcrsyn99})
\begin{equation}
t_a=t/\left(2\gamma^2\left(t\right)\right)\, ,
\label{taexp3}
\end{equation}
concurrently advancing the belief that the relation between the arrival time and the laboratory time does not depend from an integral on all the previous values of the gamma Lorentz factor of the source but from the instantaneous value of the gamma Lorentz factor at the time $t$, much like in a Lorentz transformation. This claim is clearly absurd from a physical point of view.

They further assume (see e.g. \textcite{pm98,s97,s98,p99} and references therein)
\begin{equation}
\delta t_a=\delta t/\left(2\gamma^2\left(t\right)\right)\, {\rm or, alternatively,}\, dt_a=dt/\left(2\gamma^2\left(t\right)\right)\, ,
\label{taexp4}
\end{equation}
and they proceed to develop all the observable quantities of the GRB phenomenon by integrating using the ``differential'' given in Eq.(\ref{taexp4}), reaching clearly meaningless results. As we show later, this also leads to the unfortunate attempt to obtain the gamma Lorentz factor and its time variability from the astrophysical data of the afterglow, neglecting all previous GRB source history what is clearly physically and astrophysically impossible.

Having established the correct relations between the laboratory time $t$ and the arrival time $t_a$ in Eq.(\ref{tadef}), we now proceed to relate the time in the laboratory frame $t$ to the time in the detector frame $t_a^d$. We have to do one additional step: the two frames are related by a transformation which is a function of the cosmological expansion. We recall that the geometry of the space-time of the universe is described by the Robertson-Walker metric:
\begin{equation}
ds^2 = dt^2 - {\cal R}^2(t) \left(\frac{dr^2}{1-kr^2} + r^2 d\vartheta^2 + r^2 sin\vartheta^2 d\varphi^2 \right),
\label{RW}
\end{equation}
where ${\cal R}\left(t\right)$ is the cosmic scale factor and $k$ is a constant related to the curvature of the three-dimensional space ($k=0, +1, -1$ corresponds to flat, close and open space respectively). 
The wavelength of an electromagnetic wave traveling from the point $P_1(t_1, r_1, \vartheta_1, \varphi_1)$ to the point $P_{\circ}(t_{\circ}, r_{\circ}, \vartheta_{\circ}, \varphi_{\circ})$ where the observer is located is related to the red-shift parameter $z$ by
\begin{equation}
z = \frac{\lambda_{\circ}-\lambda_1}{\lambda_1},
\label{z}
\end{equation}
where $\lambda_{\circ}$ is the wavelength of the radiation for the observer and $\lambda_1$ for the emitter.
We have the following general relation:
\begin{equation}
1 + z = ( 1 + z_u ) ( 1 + z_o ) ( 1 + z_s )\, ,
\end{equation}
where  $z$ is the total redshift due to the motion of the source $ z_s$, the motion of the observer $ z_o$ and the cosmological redshift $ z_u$. In the following we will assume $ z_o << 1 $ and  $ z_s << 1 $ so $z = z_u$. In terms of the scale factor ${\cal R}\left(t\right)$ the relation (\ref{z}) gives
\begin{equation}
\frac{\lambda_{\circ}}{{ \lambda_1}}  = \frac{ {\cal R}\left(t_o\right)}{ {\cal R}\left(t_1\right)} = 1 + z=\frac{\omega_1}{\omega_0}
\label{R}
\end{equation}
where $\omega_1$ and $\omega_0$ are the frequencies associated to $\lambda_1$ and $\lambda_0$ respectively. This frequency ratio then relates the time elapsing at the source with the time elapsing at the detector due to the cosmological expansion.

We can now define the corrected arrival time $t_a^d$ measured at the detector, which is related to $t_a$, clearly defined by Eq.(\ref{tadef}), by
\begin{equation}
t_a^d = t_a \left(1+z\right),
\label{taddef}
\end{equation}
where $z$ is the cosmological redshift of the GRB source. In the case of GRB~991216 we have $z\simeq 1.00$.
 
The observed flux is the flux which crosses the surface $ 4 \pi ( {\cal R}\left(t_o\right) r)^2 $ but this flux is lower by a factor $1 + z$ due to the redshift energy of the photons and by another factor $1 + z$ due to the fact that the number of photons at reception is less than the number at emission. Thus we can define a luminosity distance by:
\begin{equation}
d_L^2 =  {\cal R}_o^2 r^2 (1 + z )^2.
\end{equation}
Then the observed flux is related to the absolute luminosity of the GRB by the following relation:
\begin{equation}
l = \frac{L }{ 4 \pi d_L^2}\, ,
\end{equation}
where the luminosity distance $ d_L$ is simply related to the proper distance $ d_p=  {\cal R}_o r $ by  $ d_L = d_p ( 1 + z ) $. The observed total fluence $ f $ is related to the total energy  E of the GRB by the following relation:
\begin{equation}
f = \frac{E (1 + z) }{ 4 \pi d_L^2} 
\end{equation}

Then the cosmological effect is taken into account by the definition of the proper distance $ {\cal R}_o r$ which depends on the cosmological parameters: the Hubble constant $ H_\circ = \dot {\cal R}\left(t_\circ\right)/{\cal R}\left(t_\circ\right)$ at time $t_\circ$ and the matter density $\rho_\circ$ or $\Omega_M = \rho_\circ / \rho_{crit}$, where $\rho_{crit}= \frac{3 H_\circ^2}{8 \pi G}$.

The computation of the proper distance is then simply given by the relation :\\
\begin{equation}
d_p = \frac{c}{H_o}  \int_0^z  \frac{dz}{F(z)}\, ,
\end{equation}
where $ F(z)  = \sqrt{ \Omega_M (1+z)^3 } $.

In the case of the Friedman flat universe, $ \Omega_M = 1$ and we have:
\begin{equation}
d_p (z) = \frac{2 c}{H_o} \left[ 1 - \frac{1}{\sqrt{1 + z}}  \right]\, .
\end{equation}

So the measurement of the redshift gives us the luminosity distance via a cosmological scenario. With the measurement of the flux we can deduce the proper luminosity of the burst and from the measurement of the total fluence the total energy so we are then able to find the $E_{dya}$.

\section{The numerical integration of the hydrodynamics and the rate equations}\label{num_int}

\subsection{The Livermore code}

A computer code (\textcite{wsm97,wsm98}) has been used to
evolve the spherically symmetric general relativistic hydrodynamic equations starting from the dyadosphere (\textcite{rswx99}).

We define the generalized gamma factor $\gamma$ and the radial 3-velocity in the laboratory frame $V^r$
\begin{equation}
\gamma \equiv \sqrt{ 1 + U^r U_r},\hskip0.5cm V^r\equiv {U^r\over U^t}.
\label{asww}
\end{equation}
From Eqs.(\ref{s}, \ref{tt}), we then have
\begin{equation}
(U^t)^2=-{1\over g_{tt}}(1+g_{rr}(U^r)^2)={1\over\alpha^2}\gamma^2.
\label{rr}
\end{equation}
Following Eq.(\ref{cpp}), we also define 
\begin{equation}
E \equiv \epsilon \gamma,\hskip0.5cmD \equiv \rho_B \gamma,
\hskip0.3cm {\rm and}\hskip0.3cm\tilde\rho \equiv \rho\gamma
\label{cp}
\end{equation} 
so that the conservation law of baryon number (\ref{contin}) can then
be written as
\begin{equation}
{\partial D \over \partial t} = - {\alpha \over r^2} {
\partial \over \partial r} ({r^2 \over \alpha} D V^r).
\label{jay1}
\end{equation}
Eq.(\ref{conse1}) then takes the form,
\begin{equation}
{\partial E \over \partial t} = - {\alpha \over r^2} {
\partial \over \partial r} ({r^2 \over \alpha} E V^r) - p
\biggl[ {\partial \gamma \over \partial t} + {\alpha \over r^2}
{\partial \over \partial r} ({ r^2 \over \alpha} \gamma V^r)
\biggr].
\label{jay2}
\end{equation}
Defining the radial momentum density in the laboratory frame
\begin{equation}
S_r\equiv \alpha (p+\rho)U^tU_r = (D + \Gamma E) U_r,  
\label{mstate}
\end{equation}
we can express the radial component of the energy-momentum
conservation law given in Eq.(\ref{cmom2}) by
\begin{eqnarray}
{\partial S_r \over \partial t} &=& - {\alpha \over r^2} { \partial
\over \partial r} ({r^2 \over \alpha} S_r V^r) - \alpha {\partial p
\over \partial r}\nonumber\\ 
&-&{\alpha\over2}(p+\rho)\left[{\partial g_{tt}
\over\partial r}(U^t)^2+{\partial g_{rr} \over\partial
r}(U^r)^2\right]\nonumber\\ 
&=& - {\alpha \over r^2} { \partial \over
\partial r} ({r^2 \over \alpha} S_r V^r) - \alpha {\partial p \over
\partial r}\nonumber\\
&-& \alpha\left({M \over r^2}-{Q^2 \over r^3}\right)
\biggl({D + \Gamma E \over \gamma} \biggr) \biggl[ \left({\gamma \over
\alpha}\right)^2 + {(U^r)^2 \over \alpha^4 } \biggr] ~.
\label{jay3}
\end{eqnarray}

In order to determine the number-density of $e^+e^-$ pairs, we turn to Eq.(\ref{econtin}). 
Defining the $e^+e^-$-pair density in the laboratory frame 
$N_{e^\pm} \equiv\gamma n_{e^\pm}$ and $N_{e^\pm}(T) \equiv\gamma
n_{e^\pm}(T)$, where the equilibrium temperature $T$ has been obtained from
Eqs.(\ref{eeq}) and (\ref{hat}), and using Eq.(\ref{rr}), we rewrite the rate equation given by Eq.(\ref{econtin}) in the form
\begin{equation}
{\partial N_{e^\pm} \over \partial t} = - {\alpha \over r^2} {
\partial \over \partial r} ({r^2 \over \alpha} N_{e^\pm} V^r) +
\overline{\sigma v} (N^2_{e^\pm} (T) - N^2_{e^\pm})/\gamma^2~,
\label{jay:E:ndiff}
\end{equation}
These equations are integrated starting from the dyadosphere distributions given in Fig.~\ref{dens} and assuming as usual ingoing boundary conditions on the horizon of the EMBH.

\subsection{The Rome code}

In the following we recall a zeroth order approximation of the fully relativistic equations of the previous section (\textcite{rswx99}):\\
(i) Since we are mainly interested in the expansion of the $e^+e^-$ plasma away from the EMBH, we neglect the gravitational interaction.\\
(ii) We describe the expanding plasma by a
special relativistic set of equations.\\
(iii) In contrast with the previous treatment where the evolution of the density profiles given in Fig.~\ref{dens} are followed in their temporal evolution leading to a pulse-like structure, selected geometries of the pulse are a priori adopted and the correct one validated by the complete integration of the equations given by the Livermore codes.\\

Analogously to Eq.(\ref{f'}), from Eq.(\ref{contin}) we
have along each flow line in the general case in which baryonic matter is present 
\begin{equation}
{d(n_BV)\over d\tau}=0\, .
\label{f0}
\end{equation}
For the expansion
of a shell from its initial volume $\Delta V_\circ$ to the volume $\Delta  V$,
we obtain
\begin{equation}
{n_B^\circ\over n_B}= {\Delta V\over \Delta V_\circ}={\Delta {\cal V}\gamma(r)
\over \Delta {\cal V}_\circ\gamma_\circ(r)},
\label{be'_1}
\end{equation}
where $\Delta {\cal V}$ is the volume of the shell in the laboratory
frame, related to the proper volume $\Delta V$ in the comoving
frame by $\Delta V=\gamma(r) \Delta {\cal V}$, where $\gamma(r)$
defined in Eq.(\ref{asww}) is the gamma factor of the shell
at the radius $r$.

Similarly from Eq.(\ref{f'}), using the equation of state
(\ref{state}),  along the flow lines we obtain
\begin{equation}
d\ln\epsilon + \Gamma d\ln V=0.
\label{scale''}
\end{equation}
Correspondingly we obtain for the internal energy density $\epsilon$ along the flow lines 
\begin{equation}
{\epsilon_\circ\over \epsilon} = 
\left({\Delta V\over \Delta V_\circ}\right)^\Gamma=
\left({\Delta {\cal V}\over \Delta {\cal V}_\circ}\right)^\Gamma\left({\gamma(r)
\over \gamma_\circ(r)}\right)^\Gamma ~,
\label{scale'}
\end{equation}
where the thermal index $\Gamma$ given by (\ref{state}) is a
slowly-varying function with values around $4/3$. It can be computed for each value of $\epsilon,p$ as a function of $\Delta V$.

The overall energy conservation
requires that the change of the internal proper energy of a shell is compensated by a change in its  bulk kinetic energy. We then have (\textcite{rswx99})
\begin{equation}
dK=[\gamma(r)-1](dE+\rho_BdV).
\label{dk}
\end{equation}

In order to model the relativistic expansion of the plasma
fluid, we assume that $E$ and $D$ as defined by Eq.(\ref{cp}) are
constant in space over the volume $\Delta V$.  As a consequence the total energy conservation for the shell implies (\textcite{rswx99})
\begin{equation}
(\epsilon_\circ+\rho^\circ_B)\gamma_\circ^2(r) {\Delta \cal V}_\circ =(\epsilon+\rho_B)
\gamma^2 (r){\Delta \cal V},
\label{res'}
\end{equation}
which leads the solution
\begin{equation}
\gamma (r)=\gamma_\circ(r)\sqrt{{(\epsilon_\circ+\rho^\circ_B){\Delta \cal V}_\circ
\over(\epsilon+\rho_B) {\Delta \cal V}}}.
\label{result'}
\end{equation}
Corresponding to Eq.(\ref{jay:E:ndiff}) we obtain the equation for the evolution of the $e^\pm$
number-density as seen by an observer in the laboratory frame
\begin{equation}
{\partial \over \partial t}(N_{e^\pm}) = -N_{e^\pm}{1\over\Delta {\cal V}}{\partial \Delta {\cal V}\over \partial t}+\overline{\sigma v}{1\over\gamma^2(r)}  (N^2_{e^\pm} (T) - N^2_{e^\pm})~.
\label{paira'}
\end{equation}
Eqs.(\ref{be'_1}), (\ref{scale'}), (\ref{result'}) and
(\ref{paira'}) are a complete set of equations describing the
relativistic expansion of the shell.
If we now turn from a single shell to a finite distribution of shells, we can introduce the average values of the proper internal-energy, baryon-mass, baryon-number and pair-number densities ($\bar\epsilon, \bar\rho_B,\bar n_B,\bar n_{e^\pm}$) and $\bar E\equiv\bar\gamma\bar\epsilon$, $\bar D\equiv\bar\gamma\bar\rho_B$, $\bar N_{e^\pm}\equiv\bar\gamma(r) \bar n_{e^\pm}$ for the PEM-pulse, where the average $\bar\gamma$-factor is defined by
\begin{equation}
\bar\gamma={1\over{\cal V}}\int_{\cal V}\gamma(r) d{\cal V},
\label{ga}
\end{equation}
and ${\cal V}$ is the total volume of the shell in the laboratory frame. The corresponding equations are given in \textcite{rswx99}.
Having defined all its governing equations we can now return to the description of the different eras of the GRB phenomena.

\section{The Era I: the PEM pulse}\label{era1}
We have assumed that, following the gravitational collapse process, a region of very low baryonic contamination exists in the dyadosphere all the way to the remnant of the progenitor star.

Recalling Eq.(\ref{nd}) the limit on such baryonic contamination, where $\rho_{B_c}$ is the mass-energy density of baryons, is given by 
\begin{equation}
\rho_{B_c}\ll m_pn_{e^+e^-}(r) = 3.2\cdot 10^8\left({r_{ds}\over r}\right)^2\left[1-\left({r\over r_{ds}}\right)^2\right](g/cm^3).
\label{nb} 
\end{equation}
Near the horizon $r\simeq r_+$, this gives
\begin{equation}
\rho_{B_c}\ll m_pn_{e^+e^-}(r) =1.86 \cdot 10^{14}\left({\xi\over\mu}\right)(g/cm^3)\, ,
\label{nb1} 
\end{equation}
and near the radius of the dyadosphere $r_{ds}$:
\begin{equation}
\rho_{B_c}\ll m_pn_{e^+e^-}(r) = 3.2\cdot 10^8\left[1-\left({r\over r_{ds}}\right)^2\right]_{r\rightarrow r_{ds}}(g/cm^3)\, .
\label{nb2} 
\end{equation}
Such conditions can be easily satisfied in the collapse to an EMBH, but not necessarily in a collapse to a neutron star. 

Consequently we have solved the equations governing a plasma composed solely of $e^+e^-$-pairs and electromagnetic radiation, starting at time zero from the dyadosphere configurations corresponding to constant density in Fig.~\ref{dens}. The Livermore code (\textcite{rswx99}) has shown very clearly the self organization of the expanding plasma in a very sharp pulse which we have defined as the pair-electromagnetic pulse (PEM pulse), in analogy with the EM pulse observed in nuclear explosions. 
In order to further examine the structure of the PEM pulse with the simpler procedures of the Rome codes we have assumed (\textcite{rswx99}) three alternative patterns of expansion of the PEM pulse on which to try the simplified special relativistic treatment and then compared the results with the fully general relativistic hydrodynamical results:
\begin{itemize}
\item Spherical model: we assume the radial component of the four-velocity $U_r(r)=U{r\over {\cal R}}$, where $U$ is the radial component of the four-velocity at the moving outer surface ${r=\cal R}(t)$ of the PEM pulse and the $\bar\gamma$-factor and the velocity $V_r$ are
\begin{eqnarray} 
\bar\gamma &=& {3\over 8U^3}\Big[2U(1+U^2)^{3\over2}
- U(1+U^2)^{1\over2}\nonumber\\
&-& \ln(U+\sqrt{1+U^2})\Big],\hskip0.3cm V_r={U_r\over\bar\gamma}~;
\label{sp}
\end{eqnarray}
this distribution expands keeping an uniform density profile which decreases with time similar to a portion of a Friedmann Universe.
\item Slab 1: we assume $U(r)=U_r={\rm const.}$, the constant width of the
expanding slab ${\cal D}= R_\circ$ in the laboratory frame of the PEM pulse, while $\bar\gamma$  and $V_r$ are
\begin{equation}
\bar\gamma=\sqrt{1+U_r^2},\hskip0.3cm V_r={U_r\over\bar\gamma}~;
\label{sl1_1}
\end{equation}
this distribution does not need any averaging process.
\item Slab 2: we assume a constant width $R_2-R_1=R_\circ$ of the expanding slab in the
comoving frame of the PEM pulse, while $\bar\gamma$  and  $V_r$ are
\begin{equation}
\bar\gamma=\sqrt{1+U_r^2(\tilde r)},\hskip0.3cm V_r={U_r\over\bar\gamma},
\label{sp2}
\end{equation}
This distribution needs an averaging procedure and $R_1<\tilde r <R_2$, i.e.~$\tilde r$ is an intermediate radius in the slab.
\end{itemize}

\begin{figure}
\includegraphics[width=10cm,clip]{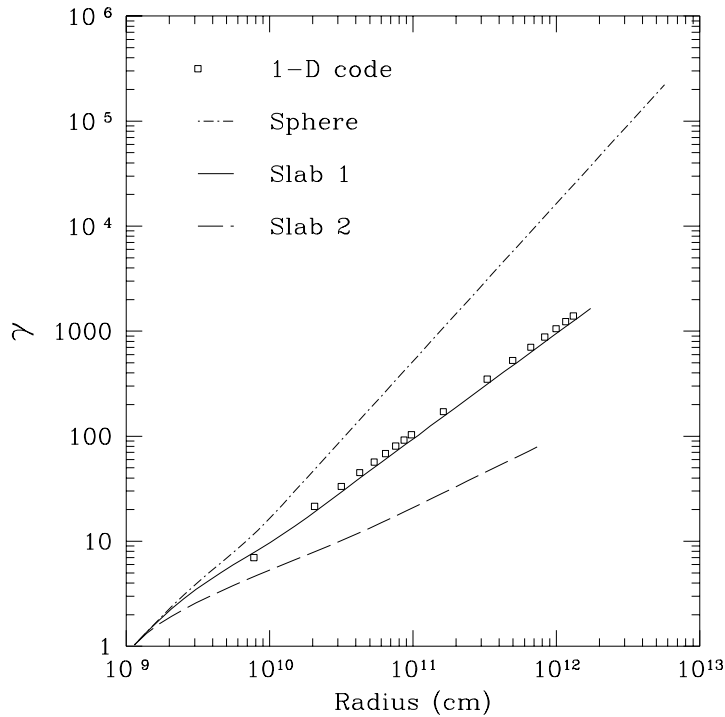}
\caption{gamma factor as a function of radius. Three models for the expansion pattern of the PEM-pulse are compared with the results of the one dimensional hydrodynamic code for an energy of dyadosphere $E_{dya}=3.1\times 10^{54}$ erg.  The 1-D code has an expansion pattern that strongly resembles that of a shell with constant thickness in the laboratory frame.
\label{figshells}}
\end{figure}

These different assumptions lead to three different distinct slopes for the monotonically increasing $\bar\gamma$-factor as a function of the radius (or time) in the laboratory frame, having assumed for the energy of dyadosphere $E_{dya}=3.1\times 10^{54}$ erg (see Fig.~\ref{figshells}). In principle, we could have an infinite number of models by defining arbitrarily the geometry of the expanding fluid in the special relativistic treatment given above. To find out which expanding pattern of PEM pulses is the physically realistic one, we need to compare and contrast the results of our simplified models (performed in Rome) with the numerical results based on the hydrodynamic Eqs.(\ref{jay1},\ref{jay2},\ref{jay3}) (obtained at Livermore) (\textcite{rswx99}). Details of the iterative method used to solve the special relativistic equation can be found in \textcite{rswx99}. 

It is manifest from the results (see Fig.~\ref{figshells}) that the slab 1 approximation (constant thickness in the laboratory frame) is in excellent agreement with the Livermore results (open squares).

The remarkable validation of the special relativistic treatment of the PEM pulse (\textcite{rswx99}), allows us to easily estimate the related quantities of physical and astrophysical interest in the model, like the $e^+e^-$-pair densities as a function of the laboratory time, the  temperature of the plasma in the comoving and laboratory frames, the reheating ratio as a function of the $e^+e^-$-pair annihilation for a variety of initial conditions (\textcite{rswx99}). 

\section{The Era II: the interaction of the PEM pulse with the remnant of the progenitor star}\label{era2}

The PEM pulse expands initially in a region of very low baryonic contamination created  by the process of gravitational collapse. As it moves further out the baryonic remnant (see Fig.~\ref{raggi2}) of the progenitor star is encountered.
As discussed in section \ref{gc} below, the existence of such a remnant is necessary in order to guarantee the overall charge neutrality of the system: the collapsing core has the opposite charge of the remnant and the system as a whole is clearly neutral. The number of extra charges in the baryonic remnant negligibly affects the overall charge neutrality of the PEM pulse (\textcite{r01mg9,rvx02}).

The baryonic matter remnant is assumed to be distributed well outside the dyadosphere in a shell of thickness $\Delta$ between an inner radius $r_{\rm in}$ and an outer radius $r_{\rm out}=r_{\rm in}+\Delta$ at a distance from the EMBH at which the original PEM pulse expanding in vacuum has not yet reached transparency. For the sake of an example we choose
\begin{equation}
r_{\rm in}=100r_{\rm ds},\hskip 0.5cm \Delta = 10r_{\rm ds}.
\label{bshell_1}
\end{equation}
The total baryonic mass $M_B=N_Bm_p$ is assumed to be a  fraction of the dyadosphere initial total
energy $(E_{\rm dya})$. The total baryon-number $N_B$ is then expressed as a function of the dimensionless parameter $B$ given by 
\begin{equation}
B=\frac{N_Bm_pc^2}{E_{\rm dya}}\, ,
\label{chimical1}
\end{equation}
where  $B$ is a parameter in the range $10^{-8}-10^{-2}$ and $m_p$ is the proton mass. We shall see below the paramount importance of $B$ in the determination of the features of the GRBs. We will see in section \ref{fp} the sense in which $B$ and $E_{dya}$ can be considered to be the only two free parameters of the EMBH theory for the entire GRB family, the so called ``long bursts''. We shall see in section~\ref{new} that for the so called ``short bursts'' the EMBH theory depends on the two other parameters $\mu$, $\xi$, since in that case $B=0$. 
The baryon number density $n^\circ_B$ is assumed to be a constant
\begin{equation}
\bar n^\circ_B={N_B\over V_B},\hskip0.5cm \bar\rho^\circ_B=m_p\bar n^\circ_B c^2.
\label{bnd}
\end{equation}
 
As the PEM pulse reaches the region $r_{\rm in}<r<r_{\rm out}$, it interacts with the baryonic matter which is assumed to be at rest. In our simplified quasi-analytic model we make the following assumptions to describe this interaction: 

\begin{itemize}
\item the PEM pulse does not change its geometry during the interaction;
\item the collision between the PEM pulse and the baryonic matter is assumed to be inelastic,
\item the baryonic matter reaches thermal equilibrium with the photons and pairs of the PEM pulse.
\end{itemize}

These assumptions are valid if: (i) the total energy of the PEM pulse is much larger than the total mass-energy of baryonic matter $M_B$, $10^{-8}<B<10^{-2}$, (ii) the ratio of the comoving number density  of pairs and baryons at the moment of collision $n_{e^+e^-}/n^\circ_B$ is very high (e.g., $10^6 <n_{e^+e^-}/ n^\circ_B <10^{12}$) and (iii)  the PEM pulse has a large value of the gamma factor ($100<\bar\gamma $).  
  
In the collision between the PEM pulse and the baryonic matter at $r_{\rm out}>r>r_{\rm in}$ , we impose total conservation of energy and momentum. We consider the collision process between two radii $r_2,r_1$ satisfying 
$r_{\rm out}>r_2>r_1>r_{\rm in}$ and $r_2-r_1\ll \Delta$. The amount of baryonic mass acquired by the PEM pulse is
\begin{equation}
\Delta M = {M_B\over V_B}{4\pi\over3}(r_2^3-r_1^3) ,
\label{mcc_2}
\end{equation}
where $M_B/ V_B$ is the mean-density of baryonic matter at rest.
The conservation of total energy leads to the estimate of the corresponding quantities before (with ``$\circ$'') and after such a collision  
\begin{equation}
(\Gamma\bar\epsilon_\circ + \bar\rho^\circ_B)\bar\gamma_\circ^2{\cal V}_\circ + \Delta M = (\Gamma\bar\epsilon + \bar\rho_B + {\Delta M\over V} + \Gamma\Delta\bar\epsilon)\bar\gamma^2{\cal V},
\label{ecc_2}
\end{equation}
where $\Delta\bar\epsilon$ is the corresponding increase of internal energy due to the collision. Similarly the momentum-conservation gives
\begin{equation}
(\Gamma\bar\epsilon_\circ + \bar\rho^\circ_B)\bar\gamma_\circ U^\circ_r{\cal V}_\circ = (\Gamma\bar\epsilon + \bar\rho_B + {\Delta M\over V} + 
\Gamma\Delta\bar\epsilon)\bar\gamma U_r{\cal V},
\label{pcc_2}
\end{equation}
where the radial component of the four-velocity of the PEM pulse is $U^\circ_r=\sqrt{\bar\gamma_\circ^2-1}$ and $\Gamma$ is the thermal index. 
We then find 
\begin{eqnarray}
\Delta\bar\epsilon & = & {1\over\Gamma}\left[(\Gamma\bar\epsilon_\circ + \bar\rho^\circ_B) {\bar\gamma_\circ U^\circ_r{\cal V}_\circ \over \bar\gamma U_r{\cal V}} - (\Gamma\bar\epsilon + \bar\rho_B + {\Delta M\over V})\right],\label{heat_2}\\
\bar\gamma & = & {a\over\sqrt{a^2-1}},\hskip0.5cm a\equiv {\bar\gamma_\circ  \over  
U^\circ_r}+ {\Delta M\over (\Gamma\bar\epsilon_\circ + \bar\rho^\circ_B)\bar\gamma_\circ U^\circ_r{\cal V}_\circ}.
\label{dgamma_2}
\end{eqnarray}
These equations determine the gamma factor $\bar\gamma$ and the internal energy density $\bar\epsilon=\bar\epsilon_\circ +\Delta\bar\epsilon$ in the capture process of baryonic matter by the PEM pulse.

The effect of the collision of the PEM pulse with the remnant leads to the following results (\textcite{rswx00}) as a function of the $B$ parameter defined in Eq.(\ref{chimical1}):\\
1) an abrupt decrease of the gamma factor given by
\begin{equation}
\gamma_{coll} = \gamma_\circ \frac{1+B}{\sqrt{ {\gamma_\circ}^2 \left(2B+B^2 \right) +1}}\, ,
\label{gamma_circ}
\end{equation}
where $\gamma_\circ$ is the gamma factor of the PEM pulse prior to the collision and $B$ is given by Eq.(\ref{chimical1}),\\
2) an increase of the internal energy in the comoving frame $E_{coll}$ developed in the collision given by
\begin{equation}
\frac{E_{coll}}{E_{dya}} =  \frac{\sqrt{ {\gamma_\circ}^2 \left(2B+B^2 \right) +1}}{\gamma_\circ} - \left(\frac{1}{\gamma_\circ} + B \right)\, ,
\label{E_int/E}
\end{equation}
3) a corresponding reheating of the plasma in the comoving frame but not in the laboratory frame, an increase of the number of $e^+e^-$ pairs and correspondingly an  overall increase of the opacity of the pulse. See details in section~\ref{at}.

\section{The Era III: the PEMB pulse}\label{era3}

After the engulfment of the baryonic matter of the remnant the plasma formed of $e^+e^-$-pairs, electromagnetic radiation and baryonic matter expands again as a sharp pulse, namely the PEMB pulse. The calculation is continued as the plasma fluid expands,
cools and the $e^+e^-$ pairs recombine until it becomes optically
thin:
\begin{equation} 
\int_R dr(n_{e^\pm}+\bar
Zn_B)\sigma_T\simeq O(1),
\label{thin_1}
\end{equation}
where $\sigma_T =0.665\cdot 10^{-24}
{\rm cm^2}$ is the Thomson cross-section and the integration is over the radial interval of the PEMB pulse in the
comoving frame. 
We have first explored the general problem of the PEMB pulse evolution by integrating the general relativistic hydrodynamical equations with the Livermore codes, for a total energy in the dyadosphere of $3.1\times 10^{54}$ erg and a baryonic shell  
of thickness $\Delta =10 r_{\rm ds}$ at rest at a radius
of $100 r_{\rm ds}$ and $B\simeq 1.3\cdot 10^{-4}$. 

In total analogy with the special relativistic treatment for the PEM pulse, presented in section~\ref{era1} (see also \textcite{rswx99}), we obtain for the adiabatic expansion of the PEMB pulse in the constant-slab approximation described by the Rome codes the following hydrodynamical equations with $\rho_B\not=0$
\begin{eqnarray}
{\bar n_B^\circ\over \bar n_B}&=& { V\over  V_\circ}={ {\cal V}\bar\gamma
\over {\cal V}_\circ\bar\gamma_\circ},
\label{be'}\\
{\bar\epsilon_\circ\over \bar\epsilon} &=& 
\left({V\over V_\circ}\right)^\Gamma=
\left({ {\cal V}\over  {\cal V}_\circ}\right)^\Gamma\left({\bar\gamma
\over \bar\gamma_\circ}\right)^\Gamma,
\label{scale1'}\\
\bar\gamma &=&\bar\gamma_\circ\sqrt{{(\Gamma\bar\epsilon_\circ+\bar\rho^\circ_B){\cal V}_\circ
\over(\Gamma\bar\epsilon+\bar\rho_B) {\cal V}}},
\label{result1'}\\
{\partial \over \partial t}(N_{e^\pm}) &=& -N_{e^\pm}{1\over{\cal V}}{\partial {\cal V}\over \partial t}+\overline{\sigma v}{1\over\bar\gamma^2}  (N^2_{e^\pm} (T) - N^2_{e^\pm}).
\label{paira'_2}
\end{eqnarray}
In these equations ($r>r_{\rm out}$) the comoving baryonic mass- and number densities are $\bar\rho_B=M_B/V$ and $\bar n_B=N_B/V$, where $V$ is the comoving volume of the PEMB pulse.

We compare and contrast (see Fig.~\ref{twocodecompare}) the bulk gamma factor as computed from the Rome and Livermore codes, where excellent agreement has been found. This validates the constant-thickness approximation in the case of the PEMB pulse as well. On this basis we easily estimate a variety of physical quantities for an entire range of values of $B$.

For the same EMBH we have considered five different cases: a shell of baryonic mass with (1) $B\simeq 1.3\cdot 10^{-4}$; (2) $B\simeq 3.8\cdot 10^{-4}$; (3) $B\simeq 1.3\cdot 10^{-3}$; (4) $B\simeq 3.8\cdot 10^{-3}$; (5) $B\simeq 1.3\cdot 10^{-3}$). The results of the integration given in detail in \textcite{rswx00} show that for the first parameter range the PEMB pulse propagates as a sharp pulse of constant thickness in the laboratory frame, but already for $B\simeq 1.3\cdot 10^{-2}$ the expansion of the PEMB pulse becomes much more complex and the constant-thickness approximation ceases to be valid; see \textcite{rswx00} for details.

\begin{figure}
\includegraphics[width=10cm,clip]{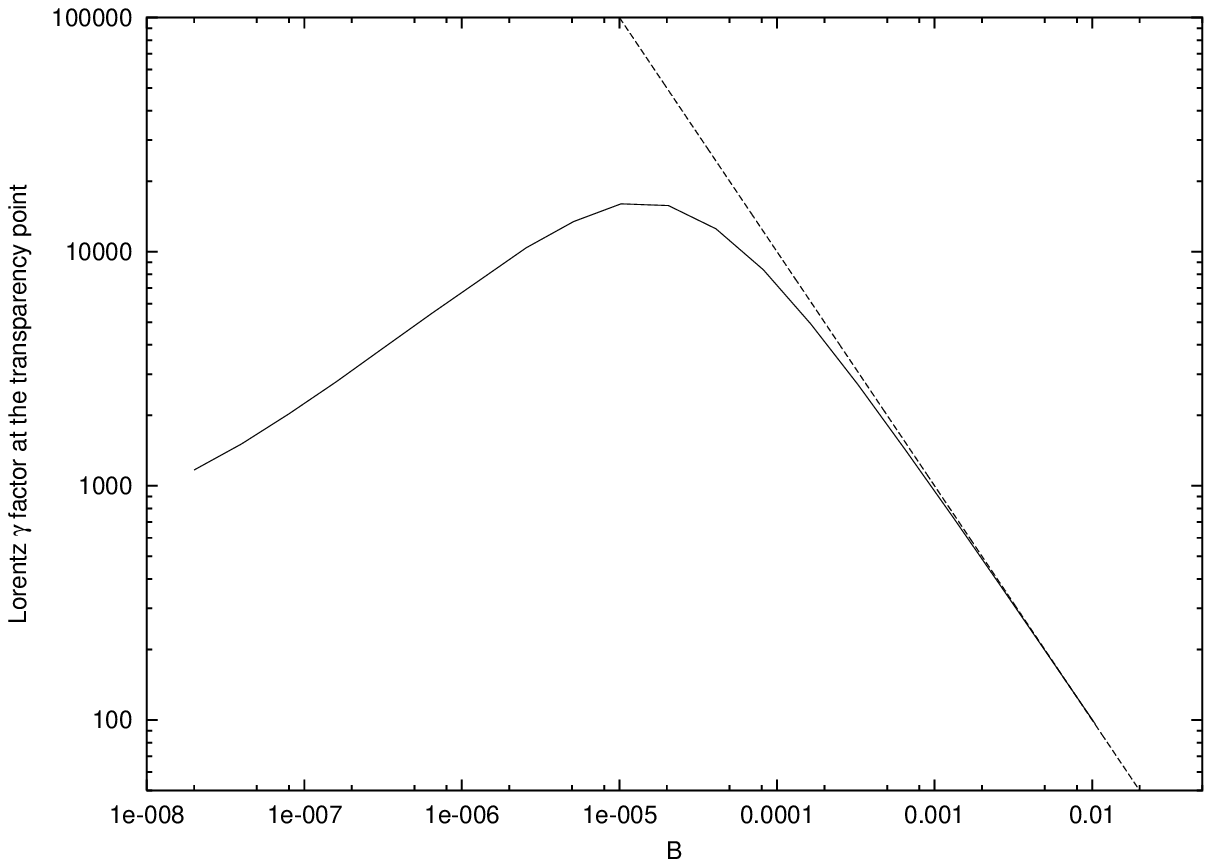}
\includegraphics[width=7.3cm,clip]{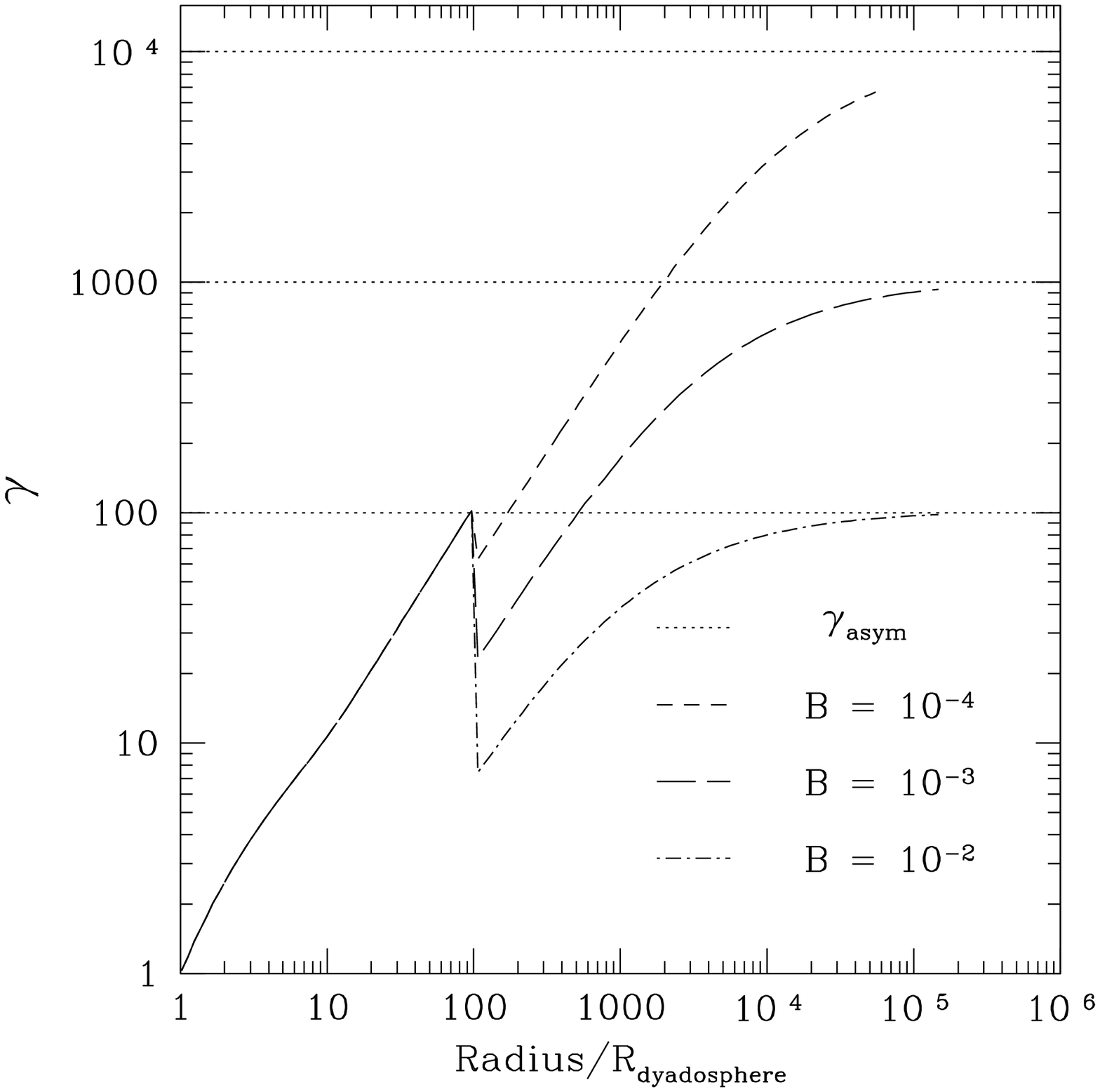}
\caption{{\bf Left)} The gamma factor (the solid line) at the transparent point is plotted as a function of the $B$ parameter. The asymptotic value (the dashed line) $E_{\rm dya}/ (M_Bc^2)$ is also plotted. {\bf Right)} The gamma factors are given as functions of the radius in units of the dyadosphere radius for selected values of $B$ for the typical case $E_{dya}=3.1\times 10^{54}$ erg. The asymptotic values $\gamma_{\rm asym} = E_{\rm dya}/(M_Bc^2)=10^4,10^3,10^2$ are also plotted. The collision of the PEM pulse with the baryonic remnant occurs at $r/r_{ds}=100$ where the jump occurs and the PEMB pulse starts.}
\label{fgamma}
\label{3gamma}
\end{figure}

It is particularly interesting to evaluate the final value of the gamma factor of the PEMB pulse when the transparency condition given by Eq.(\ref{thin_1}) is reached as a function of $B$, see Fig.~\ref{fgamma}. For a given EMBH, there is a {\em maximum} value of the gamma factor at transparency. By further increasing the value of $B$ the entire $E_{dya}$ is transferred into the kinetic energy of the baryons; see also section~\ref{new}.  
Details are given in \textcite{rswx00}.

In Fig.~\ref{3gamma} we plot the gamma factor of the PEMB pulse versus the radius for different amounts of baryonic matter. The diagram extends to values of the radial coordinate at which the transparency condition given by Eq.(\ref{thin_1}) is reached. The ``asymptotic'' gamma factor
\begin{equation}
\bar\gamma_{\rm asym}\equiv {E_{\rm dya}\over M_B c^2}
\label{asymp}
\end{equation}
is also shown for each curve. The closer the gamma value approaches the ``asymptotic'' value (\ref{asymp}) at transparency, the smaller the intensity of the radiation emitted in the burst and the larger the amount of kinetic energy left in the baryonic matter.

\section{The identification of the free parameters of the EMBH theory}\label{fp}

Within the approximation presented in section~\ref{dyadosphere} the EMBH is characterized by two parameters: $\mu$ and $\xi$. The energy of the dyadosphere is expressed in terms of these two parameters by Eq.(\ref{tee}).

\begin{figure}
\includegraphics[width=10cm,clip]{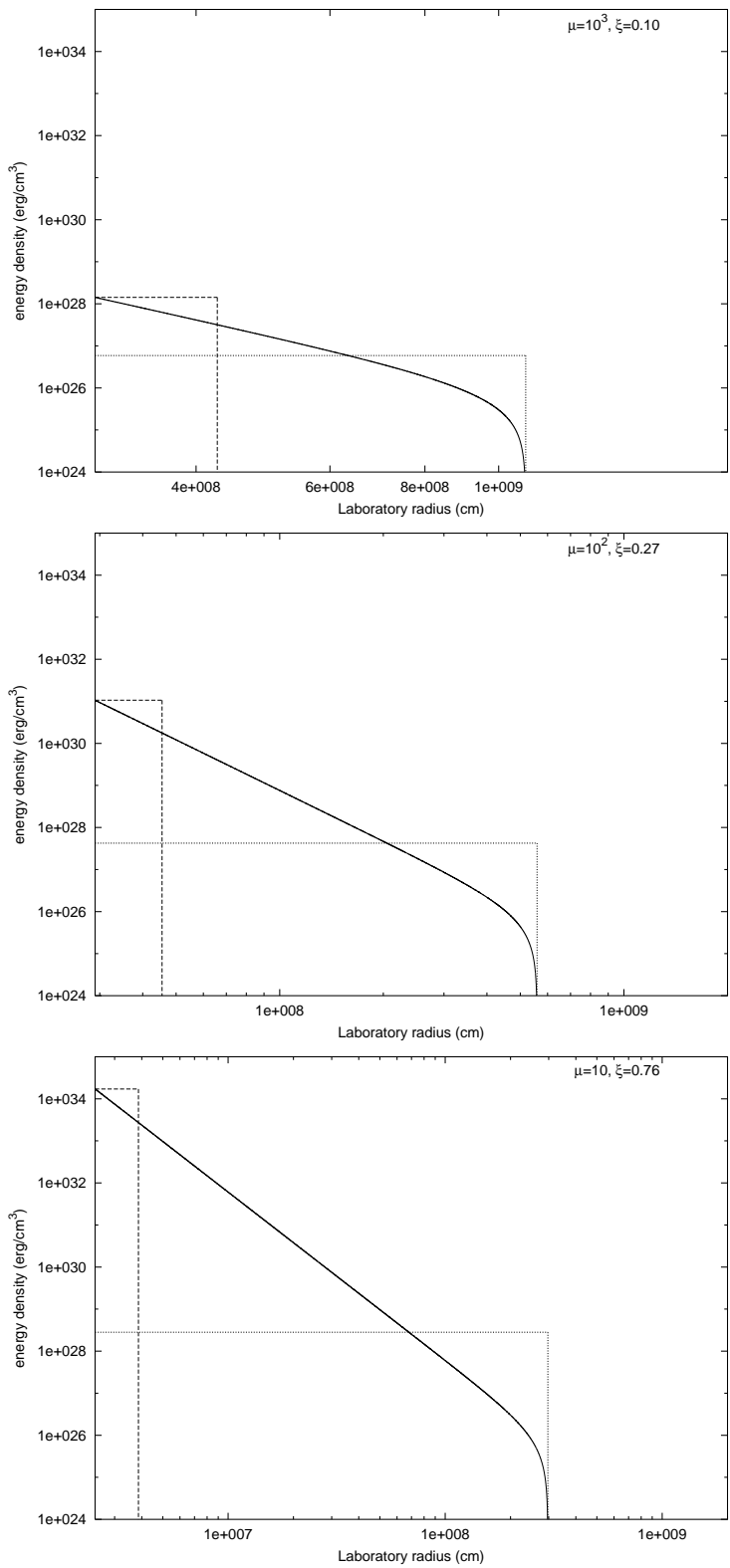}
\caption{Three different dyadospheres corresponding to the same value of $E_{dya}=3.1\times 10^{54}$ erg and with different values of the two parameters $\mu$ and $\xi$ are given. The three different configurations are markedly different in their spatial extent as well as in their energy-density distribution.}
\label{3dens}
\end{figure}

There is an entire family of EMBH solutions with different values of $\mu$ and $\xi$ corresponding to the same value of $E_{dya}$ (see Fig.~\ref{muxi}). These solutions are physically different with respect to the density of electron-positron pair distributions given by Eq.(\ref{nd}), as well as to their energy density given by Eq.(\ref{jayet}). A clear example of such a degeneracy is given in Fig.~\ref{3dens} where the two limiting energy density profiles approximating the dyadosphere as introduced in Fig.~\ref{dens} are given for three different EMBH configurations corresponding to the same value of $E_{dya}=3.1\times 10^{54}$ erg. The three configurations correspond respectively to the three different pairs $\left(\mu,\xi\right)$: $\left(10,0.76\right)$, $\left(10^2,0.27\right)$, $\left(10^3,0.10\right)$.

\begin{figure}
\includegraphics[width=10cm,clip]{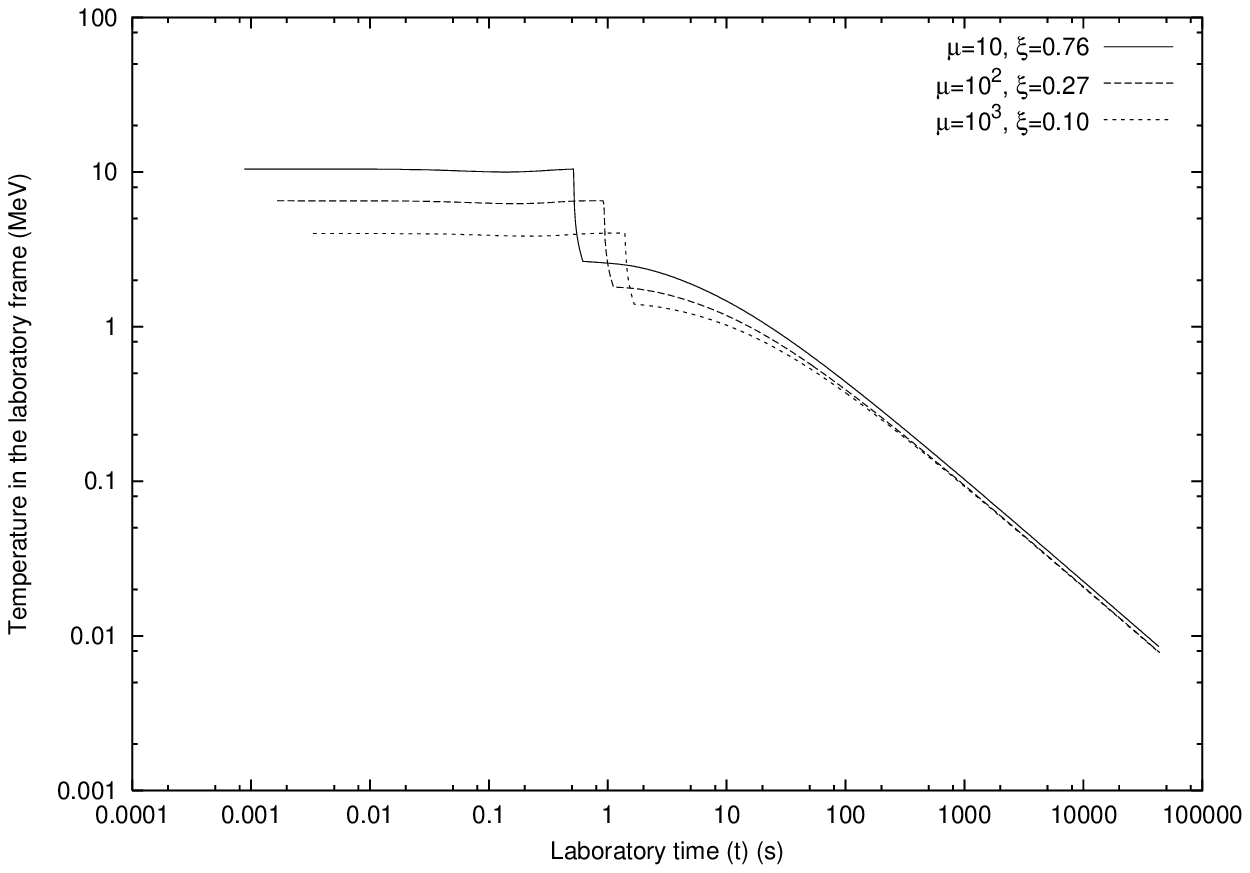}
\caption{The temperature of the plasma during the PEM pulse and PEMB pulse eras, measured in the laboratory frame, corresponding to the three configurations presented in Fig.~\ref{3dens} is given as a function of the laboratory time. The three different curves converge to a common one in the PEMB pulse era, which is therefore only a function of the $E_{dya}$ and $B$. The difference among the three curves in the early part of the PEMB pulse follows from having located the baryonic matter at a distance of $50(r_{\rm ds}-r_+)$, which is different in the three cases. Such difference  become negligible at  large distances in the later phases of the evolution. }
\label{3tem}
\end{figure}

The corresponding dynamical evolution of the PEM pulse introduced in section~\ref{era1} and \textcite{rswx99} is clearly different in the three cases. It is  remarkable that when the collision with the remnant of the progenitor star is considered all these differences disappear. As usual (see section~\ref{era2}) we describe the baryonic content of the remnant by the parameter $B$. The PEMB pulse generated after the collision with the baryonic matter depends uniquely on the two parameters $E_{dya}$ and $B$. In Fig.~\ref{3tem} the temperature in the laboratory frame is given for the PEM pulse and the PEMB pulse corresponding to the three configurations of Fig.\ref{3dens} and $B=4\times 10^{-3}$. It is clear that while for the PEM pulse era the three configurations are markedly different, they do converge to a common behaviour in the PEMB pulse era.

\begin{figure}
\includegraphics[width=\hsize,clip]{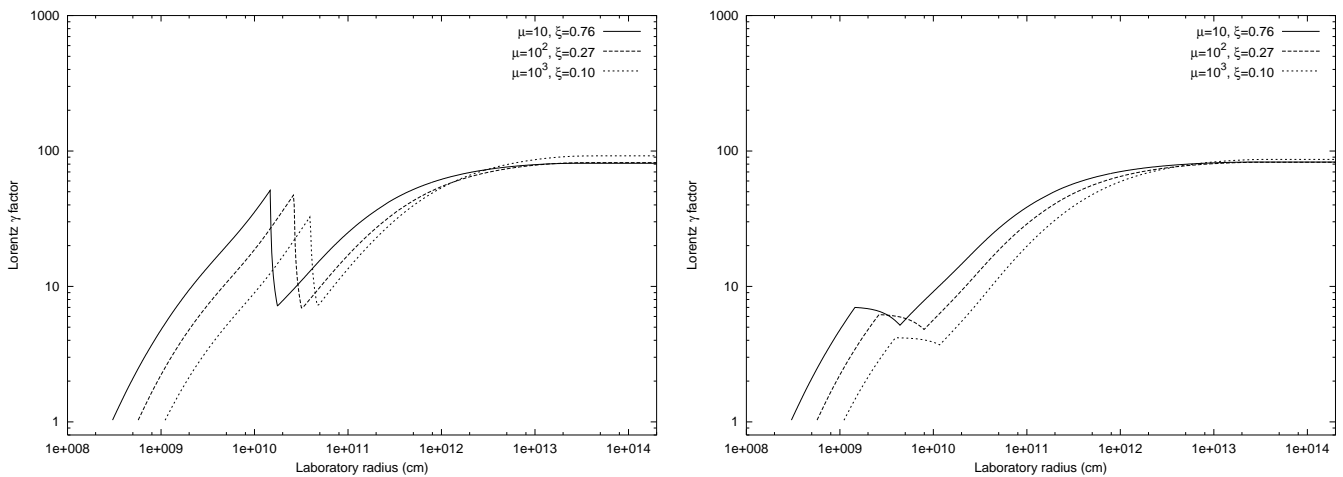}
\caption{The gamma factors for the three configurations considered in Fig.~\ref{3dens} are given as a function of the radial coordinate in the laboratory frame. The two figures correspond to a baryonic remnant positioned respectively at $r_{in}=50(r_{\rm ds} - r_+)$ (left) and at $r_{in}=5(r_{\rm ds} - r_+)$ (right). Again the convergence to a common behaviour, uniquely a function of $E_{dya}$ and $B$ for the late stages of the PEMB pulse, is manifest. }
\label{3gamma2b}
\end{figure}

If we turn now to the effect of the distance between the EMBH and the baryonic remnant, we see that this degeneracy is further extended: while the three PEM pulse eras are quite different, the common PEMB pulse era is largely insensitive to the location of the baryonic remnant, see Fig.~\ref{3gamma2b}. We have plotted the three gamma factors in the PEM pulse era corresponding to the different configurations of Fig.~\ref{3dens} and $B=10^{-2}$, in the two cases the baryonic remnant is positioned at different distances from the EMBH. 

If the PEM pulse has reached extreme relativistic regimes, the common value $\gamma_{coll}$ to which the three gamma factors drop in the collision with the baryonic matter of the remnant can be simply expressed by the large gamma limit of Eq.(\ref{gamma_circ})
\begin{equation}
\gamma_{coll}=\frac{B+1}{\sqrt{B^2+2B}}\, ,
\label{gammacollb}
\end{equation}
while the internal energy $E_{coll}$ developed in that collision is simply given by the corresponding limit of Eq.(\ref{E_int/E})
\begin{equation}
\frac{E_{coll}}{E_{dya}}=-B+\sqrt{B^2+2B}\, .
\label{eintcollb}
\end{equation}
This approximation applies when the final gamma factor at the end of the PEM pulse era is larger than $\gamma_{coll}$, upper panel in Fig.~\ref{3gamma2b}.

Turning from these general considerations to the GRB data, this degeneracy in the PEMB pulse eras and their dependence on only two parameters $E_{dya}$ and $B$ has far reaching astrophysical implications for the identification of the source of GRBs. As we will see in the conclusions all the information obtainable from GRBs with a large value of the parameter $B$  will lead to the determination of the above two parameters. An entire family of degenerate astrophysical solutions in the range of charges and masses given in Fig.~\ref{muxi} are possible. The direct knowledge of the mass and charge of the EMBH can only be gained from the PEM pulse or from GRBs with very small values of $B$ --- the so called ``short bursts'', see section~\ref{new} and the conclusions.

\section{The approach to transparency: the thermodynamical quantities}\label{at}

As the condition of transparency expressed by Eq.(\ref{thin_1}) is reached the {\em injector phase} terminates. The electromagnetic energy of the PEMB pulse is released in the form of free-streaming photons --- the proper GRB. The remaining energy of the PEMB pulse is released as an accelerated-baryonic-matter (ABM) pulse.

We now proceed to the analysis of the approach to the transparency condition. It is then necessary to turn from the pure dynamical description of the PEMB pulse described in the previous sections to the relevant thermodynamic parameters. Also such a description at the time of transparency needs the knowledge of the thermodynamical parameters in all previous eras of the GRB. 

As above we shall consider as  a typical case an EMBH of $E_{dya}=3.1\times 10^{54}$ erg and $B=10^{-2}$. The considerations will refer to a dyadosphere configuration described by the two limiting approximations shown in Fig.~\ref{dens}.

\begin{figure}
\includegraphics[width=8.5cm,clip]{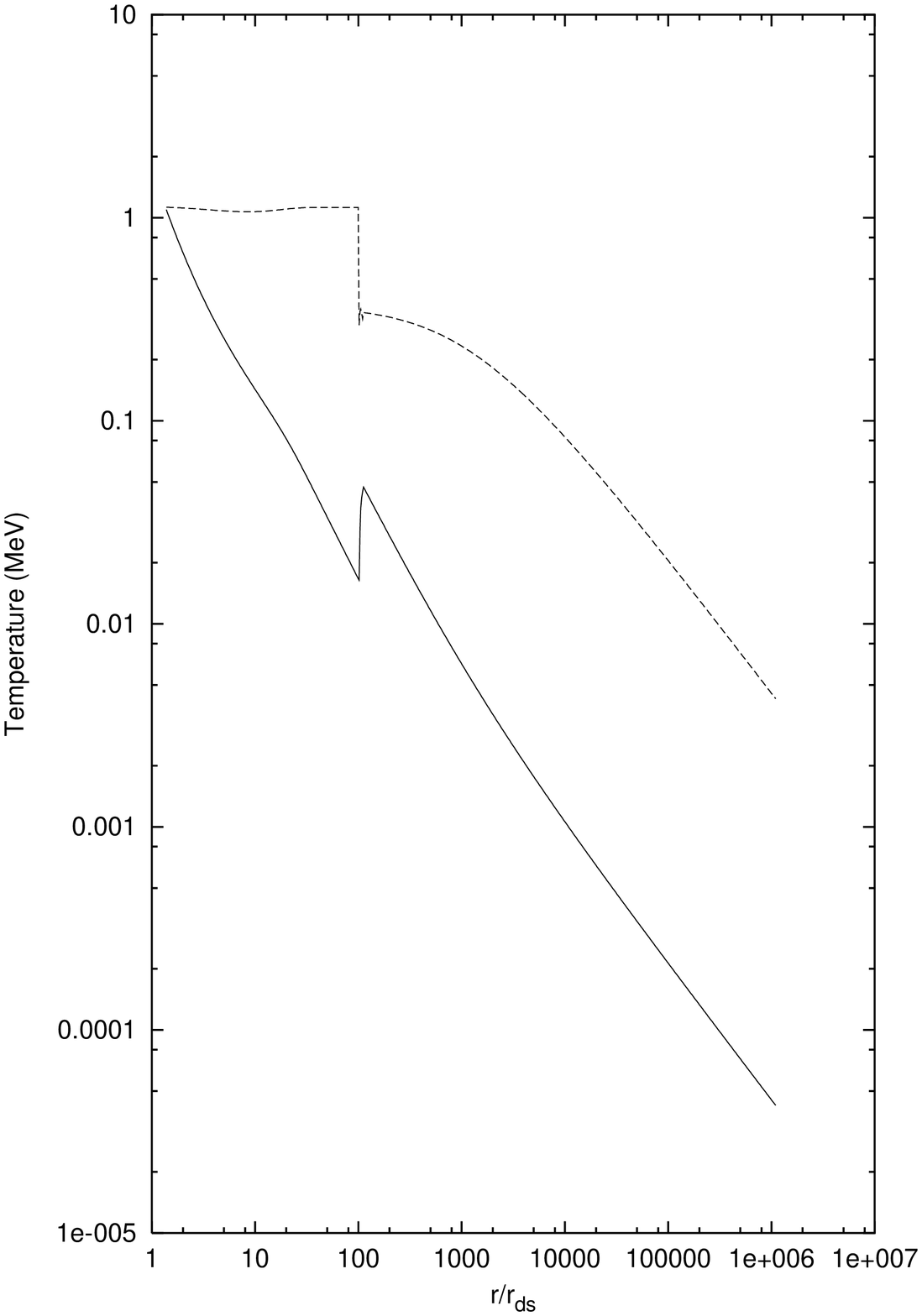}
\includegraphics[width=8.5cm,clip]{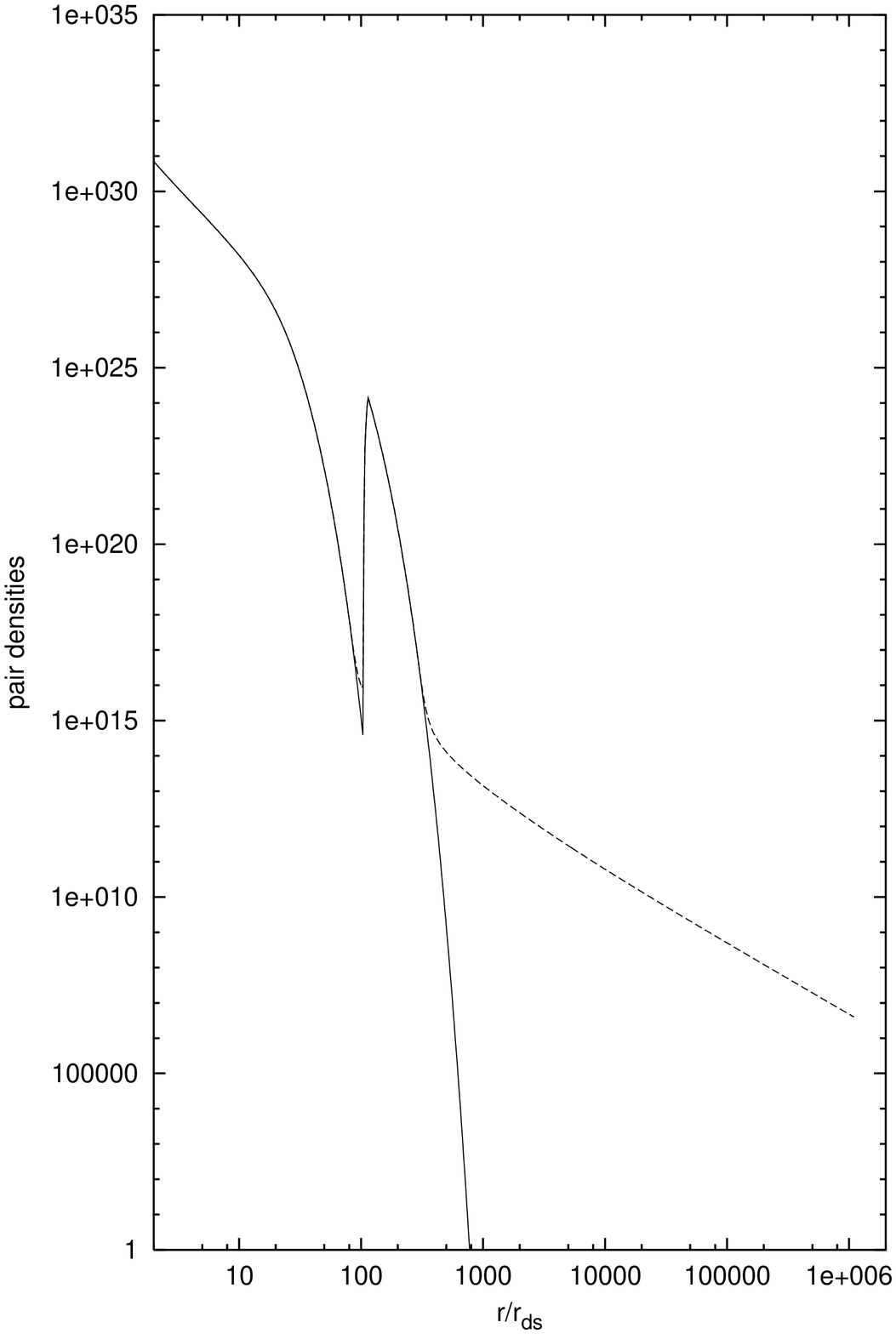}
\caption{{\bf Left)} The temperature of the plasma in the comoving frame $T'$(MeV) (the solid line) and in the laboratory frame $\bar\gamma T'$ (the dashed line) are plotted as functions of the radius in the unit of the dyadosphere radius $r_{\rm ds}$. {\bf Right)} The number densities $n_{e^+e^-}(T)$ (the solid line) computed by the Fermi integral and $n_{e^+e^-}$ (the dashed line) computed by the rate equation (see section~\ref{hydro_pem}) are plotted as functions of the radius. $T'\ll m_ec^2$, two curves strongly divergent due to $e^+e^-$-pairs frozen out of the thermal equilibrium. The peak at $r\simeq100r_{\rm ds}$ is due to the internal energy developed in the collision.}
\label{tem}
\label{pair}
\end{figure}

One of the key thermodynamical parameters is represented by the temperature of the PEM and PEMB pulses. It is given as a function of the radius both in the comoving and in the laboratory frames in Fig.~\ref{tem}. 
Before the collision the PEM pulse expands keeping its temperature in the laboratory frame constant while its temperature in the comoving frame falls (see \textcite{rswx99}). In fact Eqs.(\ref{res'},\ref{result'}) are equivalent to
\begin{equation}
{d(\epsilon\gamma^2{\cal V})\over dt}=0,
\label{tc}
\end{equation}
where the baryon mass-density is $\rho_B=0$ and the thermal energy-density of photons and $e^+e^-$-pairs is $\epsilon=\sigma_B T^4(1+f_{e^+e^-})$, $\sigma_B$ is the Boltzmann constant and $f_{e^+e^-}$ is the Fermi-integral for $e^+$ and $e^-$. This leads to
\begin{equation}
\epsilon\gamma^2{\cal V}=E_{\rm dya},\hskip0.3cm T^4\gamma^2{\cal V}={\rm const.}
\label{econ}
\end{equation}
Since $e^+$ and $e^-$ in the PEM pulse are extremely relativistic, we have the equation of state $p\simeq\epsilon/3$ and the thermal index (\ref{state}) $\Gamma\simeq 4/3$ in the evolution of PEM pulse. Eq.(\ref{econ}) is thus equivalent to
\begin{equation}
T^3\bar\gamma {\cal V}\simeq {\rm const.}
\label{encon}
\end{equation}
These two equations (\ref{tc}) and (\ref{encon}) result in the constancy of the laboratory temperature $T\bar\gamma$ in the evolution of the PEM pulse.

It is interesting to note that Eqs.(\ref{econ}) and (\ref{encon}) hold as well in
the cross-over region where $T\sim m_ec^2$ and $e^+e^-$ annihilation takes place.  
In fact from the conservation of entropy it follows that asymptotically we have
\begin{equation}
      \frac{(V T^3)_{T<m_ec^2}}{(V T^3)_{T>m_ec^2}}  =\frac{11}{4}\ ,
\label{reheat}
\end{equation}
exactly for the same reasons and physics scenario discussed in the cosmological framework by Weinberg, see e.g. Eq.~(15.6.37) of Weinberg (1972). 
The same considerations when
repeated for the conservation of the total energy 
$\epsilon\gamma V=\epsilon\gamma^2{\cal V}$
following from Eq.~(\ref{tc}) then lead to
\begin{equation}
      \frac{(V T^4 \gamma)_{T<m_ec^2}}{(V T^4 \gamma)_{T>m_ec^2}}  
             =\frac{11}{4}\ .
\end{equation}
The ratio of these last two quantities gives asymptotically 
\begin{equation}
      T_\circ= (T \gamma)_{T>m_ec^2}= (T \gamma)_{T<m_ec^2},
\label{rt}
\end{equation}
where $T_\circ$ is the initial average temperature of the dyadosphere at rest.

During the collision of the PEM pulse with the remnant we have an increase in the number density of $e^+e^-$ pairs (see Fig.~\ref{pair}). This transition corresponds to an {\em increase} of the temperature in the comoving frame and a {\em decrease} of the temperature in the laboratory frame as a direct effect of the dropping of the gamma factor (see Fig.~\ref{3gamma}). 

\begin{figure}
\includegraphics[width=8.5cm,clip]{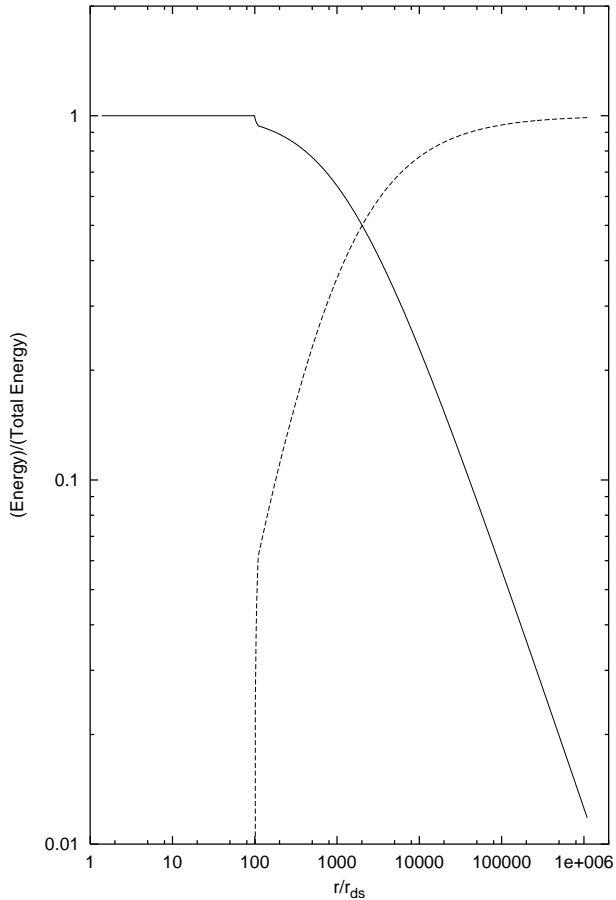}
\caption{The energy of the non baryonic components of the PEMB pulse (the solid line) and the kinetic energy of the baryonic matter (the dashed line) in unit of the total energy are plotted as functions of the radius in the unit of the dyadosphere radius $r_{\rm ds}$.}
\label{intkin}
\end{figure}

After the collision we have the further acceleration of the PEMB pulse (see Fig.~\ref{3gamma}). The temperature now decreases both in the laboratory and the comoving frame (see Fig.~\ref{tem}). Before the collision the total energy of the $e^+e^-$ pairs and the photons is constant and equal to $E_{\rm dya}$. After the collision
\begin{equation}
E_{\rm dya}=E_{\rm Baryons}+E_{e^+e^-}+E_{\rm photons},
\label{Etotal}
\end{equation}
which includes both the total energy $E_{e^+e^-}+E_{\rm photons}$ of the nonbaryonic components
and the kinetic energy $E_{\rm Baryons}$ of the baryonic matter
\begin{equation}
E_{\rm Baryons}=\bar\rho_B V(\bar\gamma -1).
\label{kinetic}
\end{equation}
In Fig.~\ref{intkin} we plot both the total energy $E_{e^+e^-}+E_{\rm photons}$ of the nonbaryonic components and the kinetic energy $E_{\rm Baryons}$ of the baryonic matter as functions of the radius for the typical case $E_{\rm dya}=3.1\times 10^{54}$ erg and $B=10^{-2}$. Further details are given in \textcite{rswx00}.

\section{The P-GRBs and the ``short bursts''. The end of the injector phase.}\label{new}

We now analyze the approach to the transparency condition given by Eq.(\ref{thin_1}). For selected values of $B$ we give the energy $E_{P\hbox{-}GRB}$ of the P-GRB, and $E_{\rm Baryons}$ of the ABM pulse. We clearly have
\begin{equation}
E_{dya}= E_{P\hbox{-}GRB} + E_{\rm Baryons}\, .
\label{esum}
\end{equation}

Taking into account the results shown in Figs. \ref{tem}--\ref{intkin}, we can  repeat all the considerations for selected values of $B$. We shall examine values of $B$ ranging from $B=10^{-8}$ only up to $B=10^{-2}$: for larger values of $B$ our constant slab approximation breaks down. We will see in the following that this range does indeed cover the most relevant observational features of the GRBs. 

As clearly shown in Fig.~\ref{3gamma} both the final value of the gamma factor and the radial coordinate at which the transparency condition is reached depend very strongly on $B$. Therefore a strong dependence on $B$ is also found in the relative values of $E_{P\hbox{-}GRB}$ and $E_{\rm Baryons}$.

We are now finally ready to give in Fig.~\ref{fintkin} the crucial diagram representing the values of $E_{P\hbox{-}GRB}$ and $E_{\rm Baryons}$ in units of the $E_{dya}$ as functions of $B$. This diagram, a universal one, is very important and is essential for the understanding of the GRB structure. 

We find that for small values of $B$ (around $10^{-8}$) almost all the $E_{dya}$ is emitted in the P-GRB (see also our previous paper \textcite{rswx99}) and very little energy is left in the baryons. While for $B\simeq10^{-2}$ roughly only $10^{-2}$ of the total initial energy of the dyadosphere is radiated away in the P-GRB and almost all energy is transferred to the baryons.

This behaviour is at the heart of the fundamental difference between the so called {\em short bursts} and {\em long bursts}. We have proposed in \textcite{lett2} that the {\em short bursts} must be identified with the P-GRBs in the case of very small $B$. There are a variety of reasons supporting this identification:

\begin{enumerate}
\item For small values of $B$, $E_{\rm Baryons}$ is negligible, see Fig.~\ref{fintkin}, and consequently the intensity of the afterglow is also negligible and the entire energy $E_{dya}$ is released into the P-GRB. This is clearly consistent with the absence of observed afterglows in the short bursts.
\item The temperature of the P-GRB in the laboratory frame $\bar\gamma T$ at the transparency point is a strongly decreasing function of $B$, see Fig.~\ref{energypeak}. $\bar\gamma T$ is related to the energy corresponding to the peak of the photon-number spectrum, as described in  \textcite{rswx99}. This is also in very good agreement with the observed decrease of the hardness ratio between the {\em short bursts} and the {\em long bursts} (\textcite{ka93}).
\item The time $T_{90}$, the duration of 90\% of the energy emission as used in the current literature and discussed in \textcite{rswx00} is plotted in Fig.~\ref{t90bs} for selected values of $E_{dya}$ and for different values of $B$.

\begin{figure}
\includegraphics[width=10cm,clip]{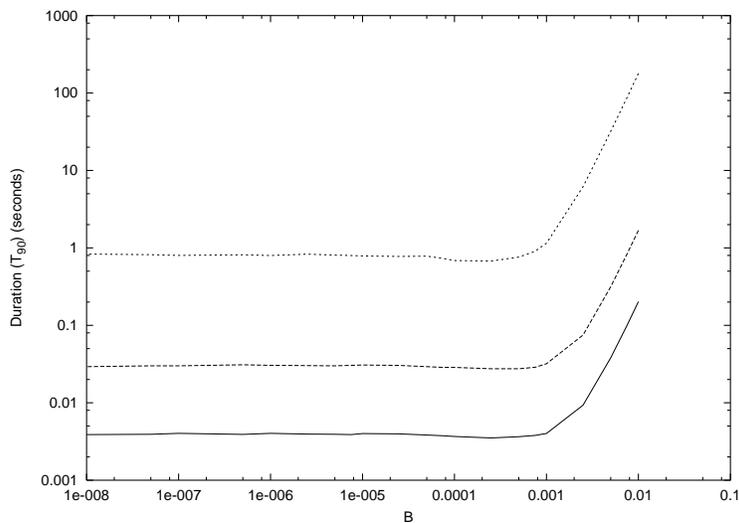}
\caption{The duration computed with the $T_{90}$ criterion is represented as a function of the $B$ parameter for three selected EMBH respectively with $E_{dya}=4.4\times 10^{52}$ erg, $E_{dya}=3.1\times 10^{54}$ erg, $E_{dya}=4.1\times 10^{58}$ erg going from the lower curve to the upper one.}
\label{t90bs}
\end{figure}
\end{enumerate}

Before concluding a word of caution is needed about how to use the above results: all these considerations are based on the drastic approximations in the description of the dyadosphere presented in section \ref{dyadosphere}, see also Fig.~\ref{3dens}. This treatment is very appropriate in estimating the general dependence of the energy of the P-GRB, the kinetic energy of the ABM pulse and consequently the intensity of the afterglow. Especially powerful is the establishment of the dependence of $E_{P\hbox{-}GRB}$ and $E_{\rm Baryons}$ on $B$ (see Fig.~\ref{fintkin}). As we will see in the next sections, this approximation is similarly powerful in determining the overall time structure of the GRB and especially the time of the release of the P-GRB with respect to the moment of gravitational collapse and the afterglow. 

If, however, we turn to the detailed temporal structure of the P-GRB and its detailed spectral distribution, it is clear that the approximations given in section \ref{dyadosphere} is no longer valid. The detailed description of the formation of the dyadosphere as qualitatively expressed in Fig.~\ref{dyaform} is now needed in all mathematical rigour with the full development of all its governing equations. Progress in this direction is being made at this moment (\textcite{crv02,rv02a,rv02b,rvx02}). This situation, however, provides a unique opportunity to follow in real time the general relativistic effects of the approach to the EMBH horizon as it occurs. In other words all direct general relativistic effects of the GRBs are encoded in the fine structure of the P-GRB. For the reasons given in section \ref{fp} the information on the EMBH mass and charge can only come from the short bursts.

This terminates the {\em injector phase}. We now turn to the {\em Beam-Target phase} in which the ABM pulse collides with the interstellar medium target and the afterglow is generated. We shall in the following sections review the basic theoretical treatment necessary for the description of these remaining eras and proceed then to the confrontation of the EMBH theory with the data. 

\section{The Era IV: the ultrarelativistic and relativistic regimes in the afterglow}\label{era4}

In the introduction we have already expressed the basic assumptions which we have adopted for the description of the collision of the ABM pulse with the ISM. In analogy and by extension of the results obtained for the PEM and PEMB pulse cases, we also assume that the expansion of the ABM pulse through the ISM occurs keeping its width constant in the laboratory frame, although the results are quite insensitive to this assumption. We assume then that this interaction can be represented by a sequence of inelastic collisions of the expanding ABM pulse with a large number of thin and cold ISM spherical shells at rest with respect to the central EMBH. Each of these swept up shells of thickness $\Delta r$ has a mass $\Delta M_{\rm ism}$ and is assumed to be located between two radial distances $r_1$ and $r_2$ (where $r_2-r_1 = \Delta r \ll r_1$) in the laboratory frame. These collisions create an internal energy $\Delta E_{\rm int}$.

We indicate by $\Delta\epsilon$ the increase in the proper internal energy density due to the collision with a single shell and by $\rho_B$ the proper energy density of the swept up baryonic matter. This includes the baryonic matter composing the remnant around the central EMBH, already swept up in the PEMB pulse formation, and the baryonic matter from the ISM swept up by the ABM pulse:

\begin{equation}
\rho_B=\frac{\left(M_B+M_{\rm ism}\right)c^2}{V}.
\label{rhob}
\end{equation}

Here $V$ is the ABM pulse volume in the comoving frame, $M_B$ is the mass of the baryonic remnant and $M_{\rm ism}$ is the ISM mass swept up from the transparency point through the $r$ in the laboratory frame:
\begin{equation}
M_{\rm ism}=m_pn_{\rm ism}{4\pi\over3}\left(r^3-{r_\circ}^3\right)\, ,
\label{dgm1}
\end{equation}
where $m_p$ the proton mass and $n_{\rm ism}$ the number density of the ISM in the laboratory frame.

The energy conservation law in the laboratory frame at a generic step of the collision process is given by

\begin{equation}
\rho_{B_1} {\gamma_1}^2{\cal V}_1 + \Delta M_{\rm ism} c^2 = \left(\rho_{B_1}\frac{V_1}{V_2} + {{\Delta M_{\rm ism} c^2}\over V_2} + \Delta\epsilon \right){\gamma_2}^2{\cal V}_2,
\label{ecc}
\end{equation}
where the quantities with the index ``$1$'' are calculated before the collision of the ABM pulse with an elementary shell of thickness $\Delta r$ and the quantities with ``$2$'' after the collision, $\gamma$ is the gamma factor and ${\cal V}$ the volume of the ABM pulse in the laboratory frame so that $V=\gamma {\cal V}$.
 
The momentum conservation law in the laboratory frame is given by
\begin{equation}
\rho_{B_1} \gamma_1 U_{r_1} {\cal V}_1 = \left(\rho_{B_1}\frac{V_1}{V_2} + {\Delta M_{\rm ism} c^2\over V_2} + 
\Delta\epsilon \right)\gamma_2 U_{r_2} {\cal V}_2,
\label{pcc}
\end{equation}
where $U_r=\sqrt{{\gamma}^2 - 1}$ is the radial covariant component of the four-velocity vector (see \textcite{rswx99,rswx00} and Eq.\ref{asww}).

We thus obtain
\begin{eqnarray}
\Delta\epsilon & = & \rho_{B_1} {\gamma_1 U_{r_1} {\cal V}_1 \over \gamma_2 U_{r_2} {\cal V}_2} - \left(\rho_{B_1}\frac{V_1}{V_2} + {\Delta M_{\rm ism} c^2\over V_2} \right) ,\label{heat}\\
\gamma_2 & = & {a\over\sqrt{a^2-1}},\hskip0.5cm a\equiv {\gamma_1  \over  
U_{r_1}}+ {\Delta M_{\rm ism} c^2\over \rho_{B_1} \gamma_1 U_{r_1} {\cal V}_1}.
\label{dgamma}
\end{eqnarray}

We can use for $\Delta \varepsilon$ the following expression
\begin{equation}
\Delta \varepsilon = \frac{E_{{\rm int}_2}}{V_2}-\frac{E_{{\rm int}_1}}{V_1} = \frac{E_{{\rm int}_1}+\Delta E_{\rm int}}{V_2}-\frac{E_{{\rm int}_1}}{V_1} = \frac{\Delta E_{\rm int}}{V_2}
\label{deltaexp}
\end{equation}
because we have assumed a ``fully radiative regime'' and so $E_{{\rm int}_1}=0$.
Substituting Eq.(\ref{dgamma}) in Eq.(\ref{heat}) and applying Eq.(\ref{deltaexp}), we obtain:
\begin{equation}
 \Delta E_{\rm int}  = \rho_{B_1} {V_1}\sqrt {1 + 2\gamma_1 \frac{{\Delta M_{\rm ism} c^2 }}{{\rho_{B_1} V_1 }} + \left( {\frac{{\Delta M_{\rm ism} c^2 }}{{\rho_{B_1} V_1 }}} \right)^2 }
  - \rho_{B_1}{V_1} \left( 1 + \frac{{\Delta M_{\rm ism} c^2 }}{\rho_{B_1} V_1} \right)\, , \label{heat2}
\end{equation}
\begin{equation}
 \gamma_2  = \frac{{\gamma_1  + \frac{{\Delta M_{\rm ism} c^2 }}{{\rho_{B_1} V_1 }}}}{{\sqrt {1 + 2\gamma_1 \frac{{\Delta M_{\rm ism} c^2 }}{{\rho_{B_1} V_1 }} + \left( {\frac{{\Delta M_{\rm ism} c^2 }}{{\rho_{B_1} V_1 }}} \right)^2 } }}\, . \label{dgamma2} 
\end{equation}
These relativistic hydrodynamic (RH) equations have to be numerically integrated.

These are the actual set of equations we have integrated in the EMBH theory. In order to compare and contrast our results with the ones in the current literature, in section \ref{approximation} we have introduced the continuous limit of our  equations and proceeded to have piecewise approximate power law solutions.
We examine as well in section \ref{substructures} still under the above assumptions, the effects of a possible departure from homogeneity in the interstellar medium, still keeping the average density $n_{ism}=const$. Although these inhomogeneities are not relevant for the overall behaviour of the afterglow which we address here, they are indeed important for the actual observed flux and its temporal structures (see \textcite{lett5}). Also these considerations are affected by the angular spreading (\textcite{rbcfx02a_sub}).

\section{The Era V: the approach to the nonrelativistic regimes in the afterglow}\label{era5}

The only reason for addressing this last era is that the issue of the approach to nonrelativistic behaviour has been extensively discussed in the literature. In our treatment these results do not show any particular problems and the relativistic equations of the previous section continue to hold. In the specific example of GRB~991216 we will present in section \ref{approximation} some analytic asymptotic expansions of these equations.

This concludes the exposition of the different eras of the EMBH theory. It goes without saying that for the description of each era, all the preceding eras must necessarily be known in order to determine the space-time grid in the laboratory frame and its relation to the arrival times as seen by a distant observer. This is the basic message expressed in the RSTT paradigm.

We can now turn to the comparison of the EMBH theory with the observational data.

\section{The best fit of the EMBH theory to the GRB~991216: the global features of the solution}\label{bf} 

For reasons already explained in the introduction, we use the GRB~991216 as a prototype. We will then later apply the EMBH theory to other GRBs. The relevant data of GRB~991216 are reproduced in Fig.~\ref{grb991216}: the data on the burst as recorded by \textcite{brbr99} and the data on the afterglow from the RXTE satellite (\textcite{cs00}) and the Chandra satellite (\textcite{p00}), see also \textcite{ha00}.

The data fitting procedure relies on three basic assumption:
\begin{enumerate}
\item In the E-APE region, the source luminosity is mainly in the energy band 50--300~KeV, so we consider the flux observed by BATSE a good approximation of the total flux.
\item In the decaying part of the afterglow, we assume that during the R-XTE and Chandra observations the source luminosity is mainly in the energy band 2--10~KeV, so we can again assume that the flux observed by these satellites is a good approximation of the total one.
\item We have neglected in this paper the optical and radio emissions, since they are always negligible with respect to the X and $\gamma$ ray fluxes. In fact, even in the latest afterglow phases up to where the X-ray data are available, they are one order of magnitude smaller then the X-ray flux.
\end{enumerate}
These assumptions were initially adopted for the sake of simplicity, but have now also been justified on the basis of the spectral description of the afterglow (\textcite{rbcfx02c_spectrum}).

As already emphasized in the previous sections, in the EMBH theory there are only two free parameters characterising the afterglow: the energy of the dyadosphere, $E_{dya}$, and the baryonic matter in the remnant of the progenitor star, parametrized by the dimensionless parameter $B$. The location of the remnant has been assumed $\sim 10^{10}$ cm. As discussed in \textcite{lett1} and section \ref{fp}, the results are rather insensitive to the actual density and location of the baryonic component but they are very sensitive to the value of $B$ (\textcite{rswx00}).

In Fig.~\ref{ii-fig2} we present the actual first results of fitting our EMBH theory to the data from the R-XTE and Chandra satellites, corresponding to selected values of $E_{dya}$ and $B$. There are three distinct features which are clearly evident as a function of the arrival time at the detector: an initial rising part in the afterglow luminosity which reaches a peak followed by a monotonically decreasing part. 

\begin{figure}
\includegraphics[width=10cm,clip]{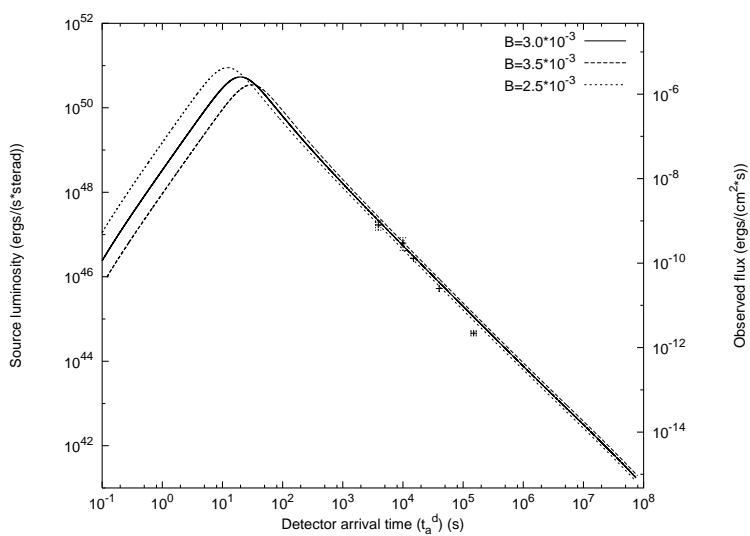}
\caption{Fine tuning of the best fit of the afterglow data of Chandra, RXTE as well as of the range of variability of the BATSE data on the major burst by a unique afterglow curve leading to the parameter values $E_{dya}=4.83\times 10^{53} erg, B=3\times 10^{-3}$.}
\label{fit_var}
\end{figure}

We have then proceeded to fine tune the two parameters in Fig.~\ref{fit_var}. The main conclusions from our model are the following:

1) The slope of the afterglow in the region where the experimental data are present is $n=-1.6$ and is in perfect agreement with the observational data. The index $n$ in this region is rather insensitive to the values of the parameters $E_{dya}$ and $B$. The physical reason for this universality of the slope is rather remarkable since it depends on a variety of factors including the ultrarelativistic energy of the baryons in the ABM pulse, the assumption of constant average density in the ISM, the ``fully radiative'' conditions leading predominantly to X-ray emission, as well as all the different relativistic effects described in the RSTT paradigm (see also section \ref{approximation}).

2) The afterglow fit does not depend directly on the parameters $\mu, \xi$ but only through their combination $E_{dya}$.  Thus there is a 1-parameter family of values of the pair $(\mu,\xi)$ allowed by a given viable value of $E_{dya}$ (see Fig~\ref{muxi} and section \ref{fp}).

3) By fine tuning the parameters of the best fit of the luminosity profile and time evolution of the afterglow the following parameters have been found:

\begin{equation}
E_{dya}=4.83\times 10^{53}erg,\; \; B=3\times 10^{-3} \ .
\label{values}
\end{equation}

After fixing in Eq.(\ref{values}) the two free parameters of the EMBH theory, modulo the mass-charge relationship which fixes $E_{dya}$,  we can derive all the space-time parameters of the GRB~991216 (see Tab.~\ref{tab1}) as well as the explicit dependence of the gamma factor as a function of the radial coordinate (see Fig.~\ref{gamma}).

Of special interest is the fundamental diagram of Fig.~\ref{tvsta}. Its role is essential in interpreting all quantities measured in arrival time (the time of an observer in an inertial frame at the detector) and their relations to the ones measured in the laboratory time by an observer in an inertial frame at the GRB source. The two times are clearly related by light signals (see Fig.~\ref{ttasch}) and expressed by the integral Eq.(\ref{tadef}) and are also affected by the cosmological expansion (see section \ref{arrival_time}).

\section{The explanation of the ``long bursts'' and the identification of the proper gamma ray burst (P-GRB)}\label{shortlongburst}

\begin{figure}
\includegraphics[width=10cm,clip]{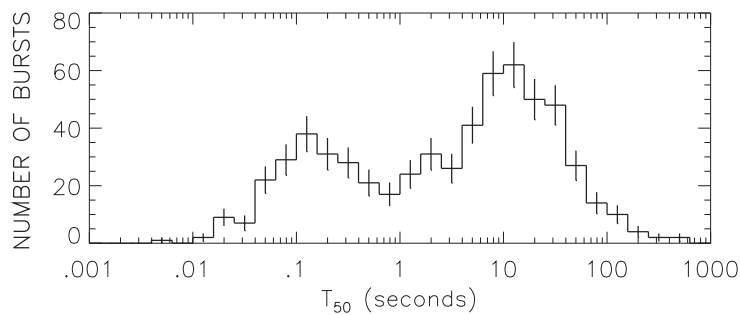}
\caption{The distribution of the burst durations clearly shows two different classes of events: the ``short bursts'' and the ``long bursts'' (reproduced from \textcite{batse4b}).}
\label{t50}
\end{figure}

Having determined the two free parameters of the EMBH theory, any other feature is a new prediction. An unexpected result soon became apparent, namely that the average luminosity of the main burst observed by BATSE can be fit by the afterglow curve (see Fig.~\ref{fit_1}). This led us to the identification of the long bursts observed by BATSE with the extended afterglow peak emission (E-APE). The peak of this E-APE occurs at $\sim 19.87\, s$ and its intensity and time scale are in excellent agreement with the BATSE observations (see also \textcite{lett5}). It is clear that this E-APE is {\em not} a burst, but is seen as such by BATSE due to its high noise threshold (see also \textcite{lett5}). Thus the outstanding unsolved problem of explaining the long GRBs (see e.g. \textcite{wmm96,swm00,p01}) is radically resolved: the so called ``long bursts'' do not exist, they are just E-APEs (see Fig.~\ref{t50}).

We now turn to the most cogent question to be asked: where does one find the burst which is emitted when the condition of transparency against Thomson scattering is reached? We have referred to this as the proper gamma ray burst (P-GRB) in order to distinguish it from the global GRB phenomena (see \textcite{lett1,brx00}). We are guided in this search by two fundamental diagrams (see Fig.~\ref{crossen} and Fig.~\ref{dtab}):
\begin{enumerate}
\item In \textcite{rswx00} it is shown that for a fixed value of $E_{dya}$ the value of $B$ uniquely determines the energy $E_{P\hbox{-}GRB}$ of the P-GRB and the kinetic energy $E_{Baryons}$ of the ABM pulse which gives origin to the afterglow  (see Fig.~\ref{crossen}). For the particular values of the parameters given in Eq.~(\ref{values}), we find
\begin{equation}
E_{P\hbox{-}GRB}=7.54\times 10^{51}erg\, , \quad E_{Baryons}= 9.43\times 10^{52}erg \label{fittedvalues}
\end{equation}
and then:
\begin{equation}
\frac{E_{P\hbox{-}GRB}}{E_{Baryons}}=1.58\times 10^{-2}\, .
\label{fittedvalues2}
\end{equation}
\item One important additional piece of information comes from the differences in arrival time between the P-GRB and the peak of the E-APE, see Fig.~\ref{dtab}. Using the results of this figure and the numerical values given in Tab.~\ref{tab1}, we can retrace the P-GRB by reading off the time parameters of point 4 in Fig.~\ref{gamma}. Transparency is reached at $21.57\, s$ in comoving time at a radial coordinate $r=1.94\times 10^{14}$ cm in the laboratory frame and at $8.41\times 10^{-2}\, s$ in arrival time at the detector.
\end{enumerate}

\begin{figure}
\includegraphics[width=10cm,clip]{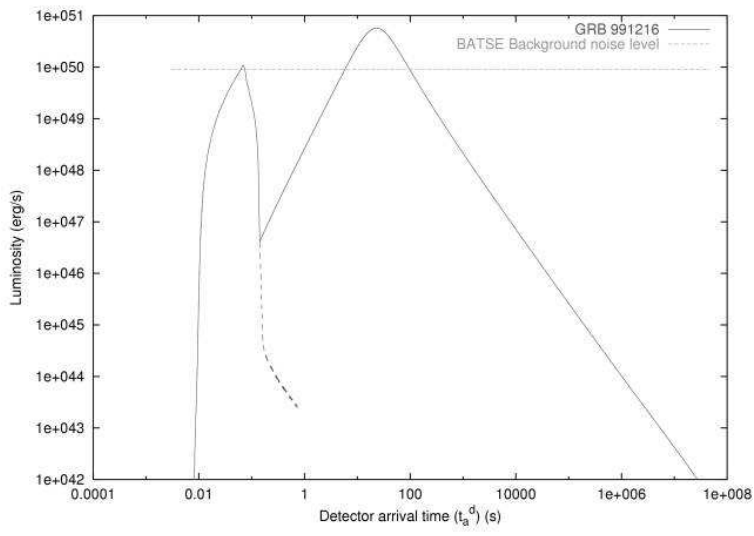}
\caption{A qualitative diagram showing the full picture of the model, with both P-GRB and E-APE.}
\label{final}
\end{figure}

All this, namely the energy predicted in Eq.(\ref{fittedvalues}) for the intensity of the burst and its time of arrival, leads to the unequivocal identification of the P-GRB with the apparently inconspicuous initial burst in the BATSE data. We have estimated from the BATSE data the ratio of the P-GRB to the E-APE over the noise threshold to be $\sim 10^{-2}$, in excellent agreement with the result in Eq.~(\ref{fittedvalues2}), see Fig.~\ref{final}.

\begin{figure}
\includegraphics[width=\hsize,clip]{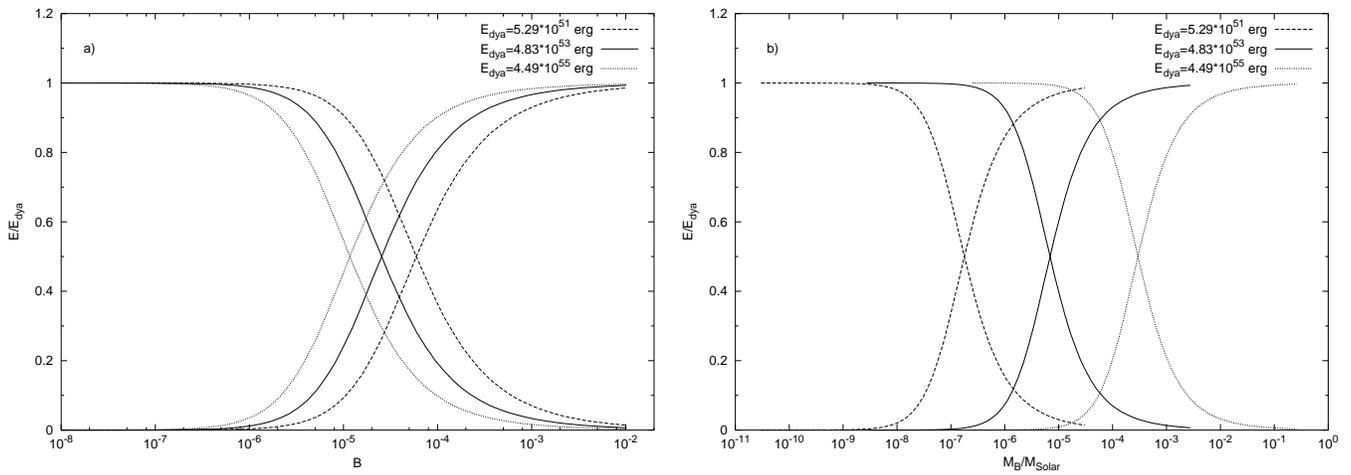}
\caption{{\bf a)} The same diagram of Fig.~\ref{fintkin} is plotted for three different $E_{dya}$ values: $E_{dya}=5.29\times 10^{51}$ erg (dashed lines), $E_{dya}=4.83\times 10^{53}$ erg (solid lines) and $E_{dya}=4.49\times 10^{55}$ erg (dotted lines). {\bf b)} Same as in a) but plotted as a function of the baryonic mass $M_B$ in units of solar masses instead of $B$.}
\label{b_multi2}
\end{figure}

It is important to emphasize that the diagrams in Fig.~\ref{fintkin} and Fig.~\ref{crossen} are not universal, but depend on the dyadosphere energy. The corresponding diagrams for three selected $E_{dya}$ values ($E_{dya}=5.29\times 10^{51}$ erg, $E_{dya}=4.83\times 10^{53}$ erg and $E_{dya}=4.49\times 10^{55}$ erg) are given in Fig.~\ref{b_multi2}a where we have plotted the energy of the P-GRB and of the E-APE as a function of $B$. The crossing of the intensity of P-GRB and E-APE occurs respectively at $B_1=6.0\times 10^{-5}$, $B_2=2.5\times 10^{-5}$ and $B_3=1.2\times 10^{-5}$ where $B_1>B_2>B_3$. In Fig.~\ref{b_multi2}b the same quantities are plotted as a function of the baryon mass $M_B$ in units of solar masses and the opposite dependence occurs: $M_1<M_2<M_3$.

\begin{figure}
\includegraphics[width=\hsize,clip]{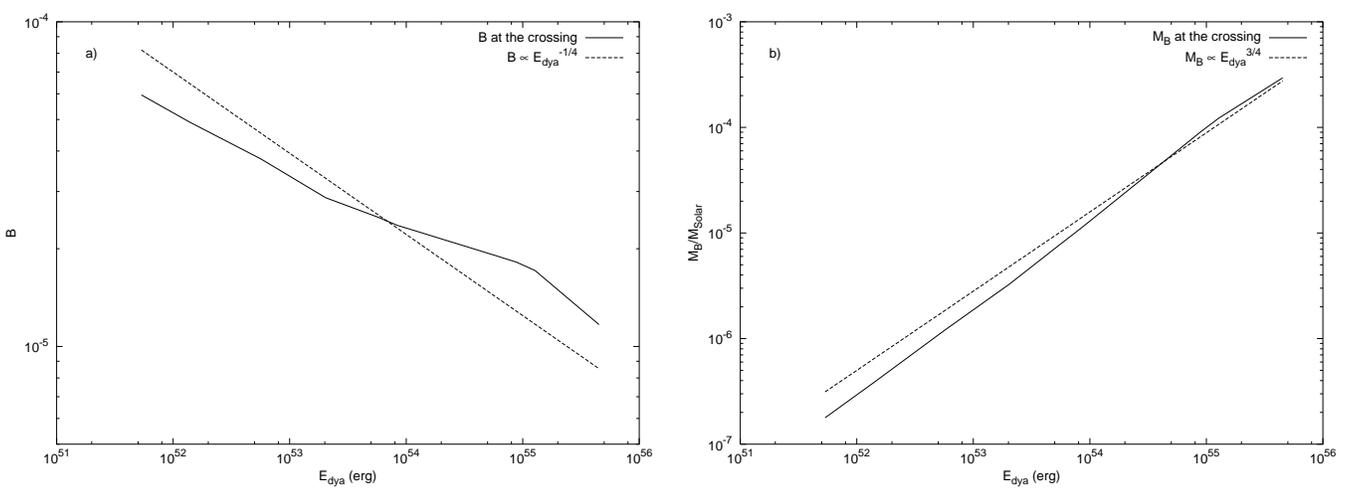}
\caption{{\bf a)} The $B$ values corresponding to the crossings in Fig.~\ref{b_multi2}a are plotted versus $E_{dya}$ (solid line). The function $B\propto E_{dya}^{-1/4}$ obtained from a qualitative theoretical estimate (see Eq.(\ref{bmbboh})) is also plotted (dashed line). {\bf b)} The $M_B$ values corresponding to the crossings in Fig.~\ref{b_multi2}b are plotted versus $E_{dya}$ (solid line). The function $M_B\propto E_{dya}^{3/4}$ obtained from a qualitative theoretical estimate (see Eq.(\ref{bmbboh})) is also plotted (dashed line).}
\label{b_cross2}
\end{figure}

The physical reasons beyond these results is the following. We recall that the kinetic energy $E_{Baryons}$ and mass $M_B$ of PEMB pulse are
\begin{equation}
E_{Baryons}=(\gamma -1 )M_B\quad M_B\equiv BE_{dya}
\end{equation}
at the crossing point defined by
\begin{equation}
E_{Baryons}=E_{P\hbox{-}GRB}={1\over2}E_{dya}.
\end{equation}
From these two equations, we obtain
\begin{equation}
B={1\over2(\gamma_\circ-1)}\simeq {1\over2\gamma_\circ},
\end{equation}
$\gamma_\circ$ is the Lorentz gamma factor of the PEMB pulse at the transparency point, where (see section~\ref{at})
\begin{equation}
(n_{pair}+n_B)\sigma_T\simeq n_B\sigma_T=1,\hskip0.5cm n_B={M_B\over 4\pi r_\circ^2\Delta\gamma_\circ},
\end{equation}
$\Delta_t$ is the PEMB pulse thickness and $r_\circ$ the radial position at the transparency point. In addition, from the total energy conservation, we have
\begin{equation}
(\epsilon+n_B)\gamma^2_\circ4\pi r_\circ^2\Delta =const.,
\end{equation}
where $\epsilon$ is the thermal energy of the PEMB pulse. In the regime $n_B\gg\epsilon$, we have 
\begin{equation}
\gamma_\circ\simeq {E_{dya}\over M_B},
\end{equation}
and in the regime $n_B\ll\epsilon$, we have 
\begin{equation}
\gamma_\circ\sim r_\circ.
\label{g0rtapp}
\end{equation}
Considering the crossing point to occur in the second regime, we obtain at the crossing point
\begin{equation}
B\sim (E_{dya})^{-{1\over4}}, \quad M_B\sim (E_{dya})^{3\over4}.
\label{bmbboh}
\end{equation}
These results are plotted in Figs.~\ref{b_cross2}a--b. The agreement with the computed results is quite satisfactory. The differences can be attributed to the approximation adopted in Eq.(\ref{g0rtapp}) which is modified for high $B$ values.

The conclusion is that for increasing $E_{dya}$ also the baryonic mass corresponding to the cross increases, but in percentage it increases less than $E_{dya}$.

\section{Considerations on the P-GRB spectrum and the hardness of the short bursts}

\begin{figure}
\includegraphics[width=10cm,clip]{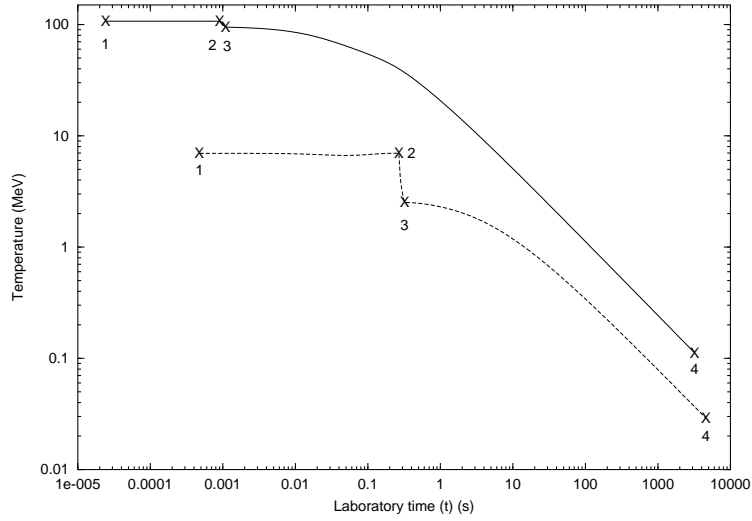}
\caption{The temperature of the pulse in the laboratory frame for the first three eras of Fig.~1 of \textcite{lett1} is given as a function of the laboratory time. The numbers 1, 2, 3, 4 represent the beginning and end of each era. The two curves refer to two extreme approximations adopted in the description of the dyadosphere. Details are given in \textcite{rswx00} and in section~\ref{fp}.}
\label{temp}
\end{figure}

Regarding the P-GRB spectrum, the initial energy of the electron-positron pairs and photons in the dyadosphere for given values of the parameters can be easily computed following the work of \textcite{prx98}. We obtain respectively $T=1.95$ MeV and $T=29.4$ MeV in the two approximations we have used for the average energy density of the dyadosphere (see section~\ref{fp}). It is then possible to follow in the laboratory frame the time evolution of the temperature of the electron-positron pairs and photons through the different eras, see Fig.~\ref{temp}. The condition of transparency is reached at temperatures in the range of $\sim 15-55$ KeV at the detector, in agreement with the BATSE results. We emphasize that in the limit of $B$ going to $10^{-8}$ in which the P-GRB coincides with the ``short bursts'' the spectrum of the P-GRB becomes harder in agreement with the observational data (see Fig.~\ref{energypeak} and \textcite{na86,ba93,dbc99,fa00}).

All the above are average values derived from the two approximations used in Fig.~\ref{dens}. If one wishes to compare the EMBH theoretical results with the fine temporal details of the observational data on the P-GRB, a departure from this average approach will be needed and the fully time varying relativistic analysis outlined in Fig.~\ref{dyaform} applies as will be further discussed in section~\ref{gc}.

\section{Approximations and power laws in the description of the afterglow}\label{approximation}

In addition to the BATSE data, there is also clearly perfect agreement with the decaying part of the afterglow data from the RXTE and Chandra satellites. 

We can also establish at this point a first set of conclusions on the luminosity power law index ``$n$'' which is a function depending strongly on the transformation $t \rightarrow  t_a \rightarrow  t_a^d$ (see Fig.~\ref{tvsta}). In the current literature such transformations and the corresponding $n$ values are incorrect. Our theoretical value $n_{theo} = -1.6$ obtained for spherical symmetry for fully radiative conditions and constant density of the ISM is in agreement with observed $n_{obs} = -1.616 \pm 0.067$. No evidence of beaming is found in GRB~991216. We shall return to this point in the conclusions.

An extremely large number of papers in the literature deal with the power law index in the afterglow era. This issue has been particularly debated in connection with the aim of decreasing the energy requirements of GRBs by the effect of beaming (see e.g. \textcite{my94,dbpt94}). It is currently very popular to infer the existence of beaming from the direct observations of breakings in the power-law index of the afterglow (see e.g. \textcite{mr97a,r97,mrw98,pmr98,dc98,sph99,pm99,r99,ha00,gdhl01}). Our aim here is to underline an often neglected point that the power law index of the afterglow is the result of a variety of factors including the very different regimes in the relation between the laboratory time $t$ and the detector arrival time $t_a^d$ presented in Fig.~\ref{tvsta}. No meaningful statements on the values of the power-law index of the afterglow can be made neglecting these necessary considerations expressed in the RSTT paradigm. This becomes particularly transparent from the power law expansion in the semianalytic treatments we present below. It is therefore not so surprising, as we will show in the next session, that the results obtained in the EMBH theory differ from the ones in the current literature.

\subsection{The approximate expression of the hydrodynamic equations}

We proceed to a first approximation and expand Eqs.(\ref{heat2}, \ref{dgamma2}) to second order in the quantity
\begin{equation}
\frac{\Delta M_{\rm ism} c^2}{\rho_{B_1} V_1} \ll 1\,.
\label{expansion1}
\end{equation}
We obtain the following expressions:
\begin{equation}
\Delta E_{{\rm int}}  = \left( {\gamma _1  - 1} \right)\Delta M_{{\rm ism}} c^2 - \frac{1}{2}\frac{{\gamma _1^2  - 1}}{{M_B  + M_{{\rm ism}} }}\left( {\Delta M_{{\rm ism}} } \right)^2 c^2\, , \label{Eint2}
\end{equation}
\begin{equation} 
\Delta \gamma  =  - \frac{{\gamma _1^2  - 1}}{{M_B  + M_{{\rm ism}} }}\Delta M_{{\rm ism}} + \frac{3}{2}\gamma _1 \frac{{\gamma _1^2  - 1}}{{\left( {M_B  + M_{{\rm ism}} } \right)^2 }}\left( {\Delta M_{{\rm ism}} } \right)^2\, , \label{gammadecel2}
\end{equation}
where we set $\Delta \gamma \equiv \gamma_2-\gamma_1$ and have used the fact that $\rho_{B_1}V_1\equiv \left(M_B+M_{\rm ism}\right)c^2$. In the limit $\Delta E_{\rm int} \rightarrow dE_{\rm int}$, $\Delta \gamma\rightarrow d\gamma$,  and $\Delta M_{\rm ism}\rightarrow dM_{\rm ism}$, neglecting also second order terms, where
\begin{equation}
dM_{\rm ism}=4\pi r^2m_p n_{\rm ism}dr=4\pi r^2m_pn_{\rm ism}v dt,\hskip0.3cm v={dr\over dt},
\label{dm}
\end{equation}
and where the ISM number density $n_{\rm ism}$ is assumed for simplicity to be $n_{\rm ism}=1\, {\rm cm}^{-3}$, we obtain:
\begin{equation}
dE_{\rm int} = \left(\gamma - 1\right) dM_{\rm ism} c^2\,,
\label{Eint}
\end{equation}
\begin{equation}
d\gamma = - \frac{{\gamma}^2 - 1}{M_B + M_{\rm ism}} dM_{\rm ism}\,.
\label{gammadecel}
\end{equation} 
Eqs.(\ref{Eint}, \ref{gammadecel}) are limiting cases of Taub's hydrodynamical equations (\textcite{taub,bor02,ll}). They have been at times referred into the GRB literature as the Blandford-McKee equations (see \textcite{bm76}). It is clear that the application of these equations holds if Eq.(\ref{expansion1}) applies. The behaviour of $\frac{\Delta M_{\rm ism} c^2}{\rho_{B_1} V_1}$ as a function of the radius when $M_{\rm ism} \ll M_B$ is:
\begin{equation}
\frac{\Delta M_{\rm ism} c^2}{\rho_{B_1} V_1} \sim \frac{r^2\Delta r}{M_B}.
\label{expansion1a}
\end{equation}
\begin{figure}
\includegraphics[width=10cm,clip]{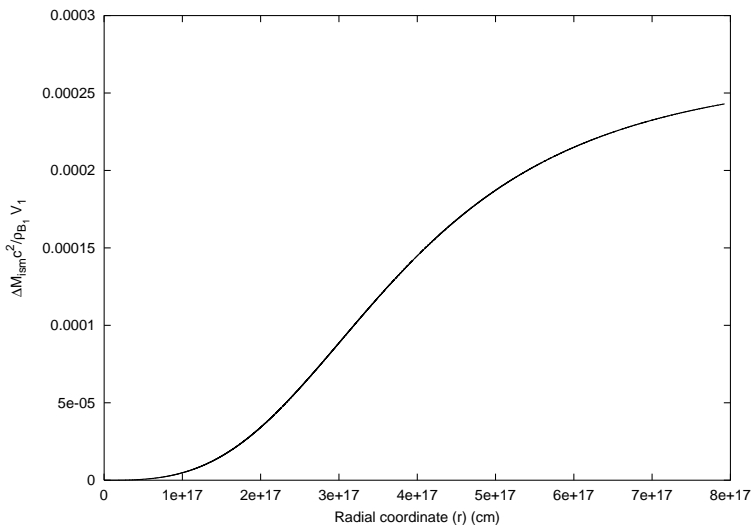}
\caption{The factor $\frac{\Delta M_{\rm ism} c^2}{\rho_{B_1} V_1}$ is represented as a function of the radial coordinate. It is manifestly an increasing function.}
\label{dm_m_fig}
\end{figure}
The condition $M_{\rm ism} \ll M_B$ holds for GRB~991216 during the entire evolution of the system and so Eq.(\ref{expansion1}) is valid (see Fig.~\ref{dm_m_fig}).

Eqs.(\ref{Eint},\ref{gammadecel}) can be simply solved analytically (see e.g. \textcite{bm76}). We then have:

\begin{equation}
\gamma={(M_B+M_{\rm ism})^2+C\over (M_B+M_{\rm ism})^2-C},
\label{dg}
\end{equation}
where

\begin{equation}
C={M_B}^2{\gamma_\circ-1\over\gamma_\circ +1},
\label{dgm2}
\end{equation}
where we recall that $r_\circ$ and 
$\gamma_\circ$ are the radial coordinate and the gamma factor at the transparency point and $M_B$ is the initial baryonic mass of the ABM pulse.

Eq.(\ref{dg}) is a differential equation for $r\left(t\right)$, namely

\begin{equation}
1 - \left( {\frac{{dr}}{{cdt}}} \right)^2  = \left[{(M_B+M_{\rm ism})^2+C\over (M_B+M_{\rm ism})^2-C}\right]^{-2}\,,
\label{dgeq}
\end{equation}
which can be integrated analytically with solution (see e.g. \textcite{intbook})

\begin{eqnarray}
2c\sqrt C \left( {t - t_\circ } \right) = \left( {M_B  - m_i^\circ } \right)\left( {r - r_\circ } \right)
+\frac{1}{4}m_i^\circ r_\circ \left[ {\left( {\frac{r}{{r_\circ }}} \right)^4  - 1} \right]
+\frac{{Cr_\circ }}{{6m_i^\circ B^2 }} \ln \left[ {\frac{{\left( {B + \frac{r}{{r_\circ }}} \right)^3 }}{{B^3  + \left( {\frac{r}{{r_\circ }}} \right)^3 }}\frac{{B^3  + 1}}{{\left( {B + 1} \right)^3 }}} \right] \label{analsol} \\ \nonumber
+\frac{{Cr_\circ }}{{3m_i^\circ B^2 }} \left[\sqrt 3 \arctan \frac{{2\frac{r}{{r_\circ }} - B}}{{B\sqrt 3 }} - \sqrt 3 \arctan \frac{{2 - B}}{{B\sqrt 3 }}\right],
\end{eqnarray}
where $m_i^\circ=\frac{4}{3}\pi m_p n_{\rm ism} r_\circ^3$, $B=\left(\frac{M_B-m_i^\circ}{m_i^\circ}\right)^{1/3}$ and we recall that $t_\circ$ is the laboratory time at the transparency point. Clearly the fulfilment of Eq.(\ref{expansion1}) has to be checked to ensure the validity of this solution.

\subsection{The approximate expression of the emitted flux}\label{app_expr_em_flux}

\begin{figure}
\includegraphics[width=\hsize,clip]{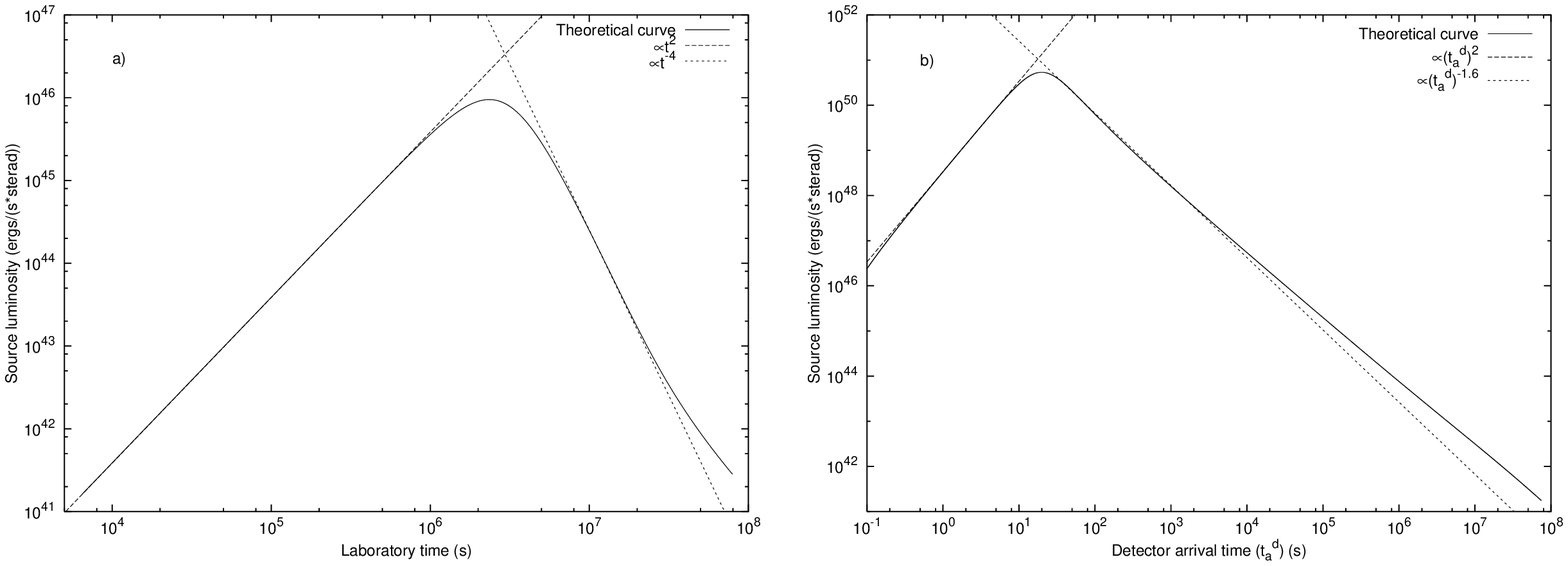}
\caption{{\bf a)} The GRB flux emitted in laboratory time. {\bf b)} the flux emitted in the arrival time, measured by an observer at rest with respect to the detector (see section~\ref{approximation}).}
\label{fluxes}
\end{figure}

From Eqs.(\ref{Eint},\ref{gammadecel}), it follows that the emitted flux in the laboratory frame is given by (see Fig.~\ref{fluxes}a)

\begin{equation}
\frac{{dE}}{{dt}} = 4\pi r^2 n_{{\rm ism}} m_p v\gamma \left( {\gamma  - 1} \right)c^2, 
\label{fluxgen}
\end{equation}
and the corresponding flux in detector arrival time (see Fig.~\ref{fluxes}b) by

\begin{equation}
\frac{{dE}}{{dt_a^d}} =\left[\frac{dt}{dt_a^d}\frac{dE}{dt}\right]_{t=t\left(t_a^d\right)} = 4\pi n_{{\rm ism}} m_p c^2 \left[vr^2 \gamma \left( {\gamma  - 1} \right)\frac{dt}{dt_a^d}\right]_{t=t\left(t_a^d\right)}.
\label{fluxgenarr}
\end{equation}

For the solution of these equations we distinguish four different phases 
(A--D). The first two correspond to era V.

\subsection*{Phase A}

Just after the transparency condition is reached, the ISM matter involved is so small that we can approximately neglect the $M_{\rm ism}$ term in Eq.(\ref{dg}) and we have:

\begin{equation}
\gamma\simeq\gamma_\circ .
\label{gA}
\end{equation}
In the specific case of GRB 991216 we have $\gamma_\circ = 310.1$, $r_\circ = 1.94\times 10^{14}\, cm$, $t_\circ = 6.48\times 10^3\, s$, $t_{a_\circ}\simeq 4.21\times 10^{-2}\, s$ and $t_{a_\circ}^d\simeq 8.41\times 10^{-2}\, s$, where the index ``$\circ$'' refers to the quantities at the transparency point. We can then establish the following equation describing the ABM pulse motion in this phase: $r\left(t\right)=vt$ with $v\simeq c$. We can than use the following relation between laboratory time and arrival time:

\begin{equation}
t = 2 {\gamma_\circ}^2 t_a = \frac{2 {\gamma_\circ}^2 }{1+z}t_a^d ,
\label{appA}
\end{equation}
which is in perfect agreement with the full numerical computation (see Fig.~\ref{tvsta}).

We can substitute these equations into Eqs.(\ref{fluxgen},\ref{fluxgenarr}), obtaining:

\begin{equation}
\frac{{dE}}{{dt}} \propto \gamma _\circ^2 n_{{\rm ism}} t^2 
\label{fluxEA}
\end{equation}
in laboratory time and

\begin{equation}
\frac{{dE}}{{dt_a^d }} \propto \frac {\gamma _\circ^8 n_{{\rm ism}}}{\left(1+z\right)^3} {\left({t_a^d}\right)}^2 
\label{fluxAA}
\end{equation}
in arrival time, assuming $\gamma\left(\gamma-1\right)\simeq \gamma^2$. The results of the numerical integration of Eqs.(\ref{heat},\ref{dgamma}) are in perfect agreement with these approximations (see Fig.~\ref{fluxes}).

\subsection*{Points P -- the two maxima of the energy flux}

Since the contribution of the ISM mass in Eqs.(\ref{dg}--\ref{dgm2}) can no longer be neglected, the value of $\gamma$ starts to significantly decrease (see Fig.~\ref{gamma}) and the flux reaches a maximum value. We integrate Eq.(\ref{fluxgen}) and Eq.(\ref{fluxgenarr}) using Eq.(\ref{dg}) for $\gamma$, assuming $r\left(t\right)=vt$ with $v\simeq c$ and Eq.(\ref{appA}) for the relation between the laboratory time and the arrival time (see Figs.~\ref{rvst}--\ref{tvsta}). We can now obtain the point where the emitted flux reaches its maximum. In general, the location of the maximum of the flux, point $P$ in \textcite{lett1}, will occur at different events, if considered in the arrival time $\left(P_A\right)$ or in the laboratory time $\left(P_L\right)$. In this second case, the point $P_L$ is determined by equating to zero the first derivative of Eq.(\ref{fluxgen}), and we have:

\begin{equation}
\gamma_{P_L}\simeq\frac{2}{3}\gamma_\circ, \quad 
\left. {\frac{M_B}{M_{\rm ism}}} \right|_{P_L}\simeq 2\gamma_\circ ,
\label{LabB}
\end{equation}
which in the case of GRB 991216 gives $\gamma_{P_L} = 206.7$ and 
$\left. {\frac{M_B}{M_{\rm ism}}} \right|_{P_L}\simeq 620.2$. 
The maximum of the observed flux is determined by equating to zero the first derivative of Eq.(\ref{fluxgenarr}). We obtain:

\begin{equation}
\gamma_{P_A}\simeq\frac{5}{6}\gamma_\circ, \quad 
\left. {\frac{M_B}{M_{\rm ism}}} \right|_{P_A}\simeq 5\gamma_\circ ,
\label{ArrB}
\end{equation}
which in the case of GRB 991216 gives $\gamma_{P_A}\simeq 258.4$ and $\left. {\frac{M_B}{M_{\rm ism}}} \right|_{P_A}\simeq 1550.5$.

The results of the numerical integration of Eqs.(\ref{heat},\ref{dgamma}) are in perfect agreement with these approximations (see Fig.~\ref{fluxes}).

\subsection*{Phase B -- the ``golden value'' $n=-1.6$}

In this phase $\gamma$ can no longer be considered constant and strongly decreases (see Fig.~\ref{gamma}). $M_{\rm ism}$ is increasing, but $v$ is still almost constant, equal to $c$. As a consequence, we can still say that $r\left(t\right)=vt$ with $v=c$, but the relation between laboratory time and arrival time given in Eq.(\ref{appA}) is no longer valid, and also Eq.(\ref{taexp3}) is no longer applicable in this phase (see Fig.~\ref{tvsta}). We can instead write the following ``effective'' relation:

\begin{equation}
t\propto {\left(t_a^d\right)}^{0.20} ,
\label{appC}
\end{equation}
which is a result of a best fit of the numerical data in this region. Expanding the squares in Eq.(\ref{dg}), neglecting $M_{\rm ism}^2$ with respect to $M_B^2$ but retaining the terms in $M_{\rm ism}$ and assuming $\gamma_\circ \gg 1$ we obtain:

\begin{equation}
\gamma\sim\frac{M_B}{M_{\rm ism}}\sim 
\gamma_{P_L}\frac{r_{P_L}^3}{r^3}=\gamma_{P_L}\frac{t_{P_L}^3}{t^3} ,
\label{gammaC}
\end{equation}
where $r_{P_L}$ and $t_{P_L}$ are the values of $r$ and $t$ at point $P_L$. Substituting this result into Eqs.(\ref{fluxgen}), we obtain the emitted flux in the laboratory frame, given by
\begin{equation}
\frac{{dE}}{{dt}} \propto \gamma _P^2 t_P^6 n_{{\rm ism}} t^{-4} \, ,
\label{fluxEC}
\end{equation}
and this is in good agreement with the full numerical computation (see Fig.~\ref{fluxes}).

To obtain an analytic formula for the observed flux on the detector, we can still try to use the approximate relation between $t$ and $t_a^d$ given by Eq.(\ref{taexp3}):
\begin{equation}
t = 2 {\gamma\left(t\right)}^2 t_a = \frac{2 {\gamma\left(t\right)}^2 }{1+z}t_a^d ,
\label{appC1}
\end{equation}
where $\gamma\left(t\right)$ is given by Eq.(\ref{gammaC}). We obtain:
\begin{equation}
t=\left(\frac{2\gamma_{P_L}^2t_{P_L}^6}{1+z}t_a^d\right)^{1/7}\, .
\label{appC2}
\end{equation}

Using this formula in Eq.(\ref{fluxgenarr}), we finally obtain:
\begin{equation}
\frac{{dE}}{{dt_a^d }} \propto \frac{ \gamma_P^{\frac{8}{7}} t_P^{\frac{24}{7}} n_{{\rm ism}}}{\left(1+z\right)^{-\frac{17}{7}}} {\left(t_a^d\right)}^{-\frac{10}{7}} 
\label{fluxACW}
\end{equation}
where we again assumed $\gamma\left(\gamma-1\right)\simeq \gamma^2$. This results are not in agreement with the observational data, because the power-law index for the observed flux is $-10/7\simeq -1.43$, instead of the observed value $-1.6$.

This is a confirmation that Eq.(\ref{appC1}) cannot be applied in this phase, as instead has been done by many authors in the current literature. We instead have to use Eq.(\ref{appC}). In fact, doing so we obtain the correct value:
\begin{equation}
\frac{{dE}}{{dt_a^d }} \propto n_{{\rm ism}} {\left(t_a^d\right)}^{-1.6} ,
\label{fluxAC}
\end{equation}
The results of the numerical integration of Eqs.(\ref{heat},\ref{dgamma}) are in perfect agreement with these approximations (see Fig.~\ref{fluxes}), which implies that the approximate Eq.(\ref{Eint},\ref{gammadecel}) can still be used in this regime, but not Eq.(\ref{taexp3}), which has to be replaced by an ``effective'' local power-law behaviour (see Eq.(\ref{appC})).

\subsection*{Phase C}

This new phase begins when $\gamma$ has decreased so much that the approximation $r=ct$ is no longer valid (see Fig.~\ref{rvst}). In the case of GRB~991216 this happens when $\gamma\simeq 3.0$, $t\simeq 1.5\times 10^7$ s, $t_a^d\simeq 2.9\times 10^5$ s and $r\simeq 4.4\times 10^{17}$ cm. In this entire phase, $r\left(t\right)$ manifests the following behaviour typical of damped motion:

\begin{figure}
\includegraphics[width=10cm,clip]{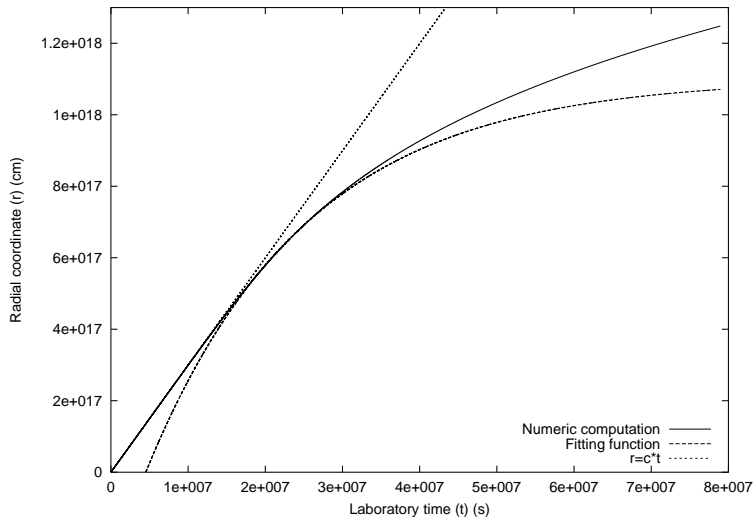}
\caption{The exact numerical solution for $r\left(t\right)$ (solid line), together with the line $r=ct$ (dotted line) and the fitting function given in Eq.(\ref{rdit}) (dashed line).}
\label{rvst}
\end{figure}

\begin{equation}
r\left( t \right) = \hat r \left( {1 - e^{ - \frac{{t - t^ \star }}{\tau }} } \right),
\label{rdit}
\end{equation}
where $\hat r$, $t^\star$ and $\tau$ are constants that can be determined by the best fit of the numerical solution. In the present case of GRB 991216 we obtain:

\begin{equation}
\hat r\simeq 1.101\times 10^{18} cm,\quad 
\tau\simeq 2.072\times 10^7 s,\quad 
t^\star\simeq 4.52\times 10^6 s.
\label{fitted}
\end{equation}

It is important to note that this interesting behaviour, typical of a damped motion, does not lead to any power-law relationship for the emitted flux as a function of the laboratory time (see Fig.~\ref{fluxes}). However, if we look at the observed flux as a function of the detector arrival time, we see that a power-law relationship still can be established, fitting the numerical solution. The result is:
\begin{equation}
\frac{{dE}}{{dt_a^d }} \propto {\left(t_a^d\right)}^{-1.36}.
\label{fluxCfit}
\end{equation}
This quite unexpected result can be explained because the relation between $t$ and $t_a^d$ depends on $r\left(t\right)$ in a nonpower-law behaviour. This fact balances the complex behaviour of the emitted flux as a function of the laboratory time, leading finally again to a power-law behaviour arrival time.

In this last phase, however, the flux decreases markedly, and from the point of view of the GRB observations, the most relevant regions are phases {\em A} and {\em B} described above, as well as the peak separating them.

\subsection*{Phase D}

This last phase starts when the system approaches a Newtonian regime. In the case of GRB~991216 this occurs when $\gamma\simeq 1.05$, $t\simeq 5.0\times 10^7$ s, $t_a^d\simeq 3.1\times 10^7$ s and $r\simeq 1.0\times 10^{18}$ cm. In this phase $r\left(t\right)$ is again approaching a linear behaviour, due to the velocity decreasing less steeply than in Phase C. The emitted flux as a function of the laboratory time still does not show a power-law behaviour, while the observed flux as a function of detector arrival time does, with an index $n=-1.45$ (see Fig.~\ref{fluxes}).

\section{The power-law index of the afterglow and inferences on beaming in GRBs}\label{power-law}

The results obtained in the previous sections have emphasized the relevance of the proper application of the RSTT paradigm to the determination of the power-law index of the afterglow. Particularly interesting is the subtle interplay between the different regimes in the relation between the laboratory time and the arrival time at the detector clearly expressed by Fig.~\ref{tvsta} and the corresponding different regimes encountered in the first order expansion of the relativistic hydrodynamic equations of \textcite{taub} (see section~\ref{approximation}). It is interesting to compare and contrast our treatment with selected results of the current literature, in order to illustrate some relevant points (see Tab.~\ref{tab2}). We will consider the results in the literature only with reference to the limiting case which we address in our work: the condition of fully radiative emission.

\begin{table}
\centering
\caption{We compare and contrast the results on the power-law index {\em n} of the afterglow in the EMBH theory with other treatments in the current literature, in the limit of high energy and fully radiative conditions. The differences between the values of $-10/7\sim -1.43$ (Dermer) and the results $-1.6$ in the EMBH theory can be retraced to the use of the two different approximation in the arrival time versus the laboratory time given in Fig.~\ref{tvsta}. See details in section~\ref{approximation}.\label{tab2}}
\begin{tabular}{c|c|c|c|c|c}
&& \textcite{cd99} & \textcite{p99} &&\\
& EMBH theory & \textcite{dcb99} & \textcite{sp99} & \textcite{v97} & \textcite{ha00}\\
&& \textcite{bd00} & \textcite{p01} &&\\
\hline \hline
&&&&&\\
Ultra-relativistic & $\displaystyle{\gamma=\gamma_\circ}$  & $\displaystyle{\gamma=\gamma_\circ}$ & $\displaystyle{\gamma=\gamma_\circ}$ & &\\
&&&&&\\
& $\gamma_\circ=310.1$ &&&& \\
&&&&&\\
 & $n=2$ & $n=2$ & $n\simeq2$ && \\
&&&&&\\
\hline
&&&&&\\
Relativistic & $\displaystyle{\gamma \simeq r^{-3}}$ & $\displaystyle{\gamma \sim r^{-3}}$ & $\gamma \sim r^{-3}$  && 
$n>-1.47$\\
&&&&&\\
& $3.0<\gamma<258.5$ &&&&  \\
&&&&&\\
 & $n=-1.6$ & $n=-\frac{10}{7}=-1.43$ & $n=-\frac{5.5}{4}=-1.375$ && \\
&&&&&\\
\hline
&&&&&\\
Non-relativistic & $n=-1.36$ &&& $n=-1.7$ & \\
&&&&&\\
& $1.05<\gamma<3.0$ &&&&\\
&&&&&\\
\hline
&&&&&\\
Newtonian & $n=-1.45$ &&&&\\
&&&&&\\
& $1<\gamma<1.05$ &&&&\\
&&&&&\\
\hline
\multicolumn{3}{c}{}\\
\end{tabular}
\end{table}

The first line of Tab.~\ref{tab2} describes the ultrarelativistic regime, corresponding to an increasing energy flux of the afterglow as a function of the arrival time (phase A in previous section). Our treatment and the results in the literature by Dermer et al. (see e.g. \textcite{cd99,dcb99,bd00}) coincide. They agree as well with the results by Piran et al. (see e.g. \textcite{p99,sp99,p01}).

The second line corresponds to the relativistic regime, in which the energy flux of the afterglow, after having reached the maximum (point P in previous section), monotonically decreases (phase B in previous section). The dependence we have found of the gamma factor on the radial coordinate of the expanding ABM pulse does coincide with the one given by Dermer et al. and Piran et al. Our power law index $n$ in this regime, which perfectly fits the data, however, is markedly different from the others. Particularly interesting is the difference between our results and those of Dermer et al: the two treatments coincide up to the last relation between the laboratory time and the arrival time at the detector. As explained in Eqs.(\ref{fluxACW}-\ref{fluxAC}), the two treatments differ in the approximation adopted in relating the laboratory time to the arrival time at the detector, illustrated in Fig.~\ref{tvsta}. Dermer et al. incorrectly adopted the approximation represented by the lower curve in Fig.~\ref{tvsta} and consequently they do not find agreement with the observational data. We have not been able to retrace in the treatment by Piran et al. the steps which have led to their different results. Special mention must be made of a result stated by \textcite{ha00}, the last entry in line 2, that an absolute lower limit for the power-law index $n-1.47$ can be established on theoretical grounds. Such a result, clearly not correct also on the basis of our analysis, has been erroneously used ti support the existence of beaming in GRBs, as we will see below.

The third line in Tab.~\ref{tab2} is also interesting, treating the nonrelativistic limit (Phase C in previous section). This regime has been analysed by \textcite{v97}, avoiding the exact integration of the equations and relying on simple qualitative arguments. These results are not confirmed by the integration of the equations we have performed. This is an interesting case to be examined for its pedagogical consequences. Having totally neglected the relation between the laboratory time and the time of arrival at the detector, which we have illustrated in Fig.~\ref{tvsta}, and identifying $t_a^d\equiv t$, Vietri reaches a very different power law from our. Moreover, his solution brings to an underestimation of the radial coordinate: he estimated a radial coordinate of $1.1\times 10^{15}\,cm$ at $t_a^d=3.5\times 10^4\,s$, while the exact computation shows a result greater than $3.0\times 10^{17}\,cm$ (see Tab.~\ref{tab1}). On the other hand if one assumes, from the above mentioned identity $t_a^d\equiv t$, $t=3.5\times 10^4\,s$, one obtains a gamma factor of $\sim 300$ (see Tab.~\ref{tab1}) in total disagreement with the nonrelativistic approximation adopted by Vietri. Quite apart from this pedagogical value, this nonrelativistic phase is of little interest from the observational point of view, due to the smallness of the flux emitted. 

For completeness, we have also shown our estimates of the index $n$ as the Newtonian phase approaches in the last line of Tab.~\ref{tab2}.

The perfect agreement between our theoretically predicted value for the power-law index, $n_{theo}$, and the observed one, $n_{obs}$,
\begin{equation}
n_{theo}=-1.6,\quad n_{obs}=-1.616\pm0.067,
\label{nembh}
\end{equation}
confirms the validity of our major assumptions:
\begin{enumerate}
\item The fully radiative regime.
\item The constant average density of the ISM ($n_{ism}=1\, proton/cm^3$).
\item The spherical symmetry of the emission and the absence of beaming in GRB~991216.
\end{enumerate}

After the work of \textcite{my94} pointing to the possibility of introducing beaming to reduce the energetics of GRBs and after the discovery of the afterglow, many articles have appeared trying to obtain theoretical and observational evidence for beamed emission in GRBs. The observations have ranged from radio (see e.g. \textcite{rh99,f00}) to optical (see e.g. \textcite{ga00,ha00,sa00,s00}) all the way to X-rays. Particular attention has been devoted to relating the existence of beaming to possible breaks in the light curve slope, generally expected at a value of the gamma factor
\begin{equation}
\gamma=\frac{1}{\vartheta_0},
\label{beam}
\end{equation}
where $\vartheta_0$ is the beam opening angle. There are many articles on this subject; to mention only the most popular ones, we recall \textcite{r97,r97b,r99,mrw98,pm99,sph99}. Far from having reached a standard formulation, these approaches differ from each other in the expected time at which the break should take place up to a factor of $20$ (see e.g. \textcite{sph99}). They differ as well for the opening angle of the beam, up to a factor of $3$ (see e.g. \textcite{sph99}). Disagreement still exists on the number of breaking points: two in the case of \textcite{pm99}, one in the case of \textcite{sph99}, one again in the case of \textcite{r97,r97b,r99} but differing in position from the one of \textcite{sph99}. It has also been noticed that other authors have shown through numerical simulations that such a transition, if visible at all, is not very sharp (see e.g. \textcite{ha00}).

Ample observational data have been obtained for the GRB~991216, in addition to the X-ray band, also in the optical and radio. For the reason mentioned at the beginning of section~\ref{bf}, we only address in this article the problem of the $\gamma$- and the X-ray emission. In that respect, the main article addressing the issue of beaming in the X-rays for GRB~991216 is the one of \textcite{ha00}. The key argument is based on the theoretical inequality claimed to exist for the power-law index $n>-1.47$ (see above). The fact that the observed X-ray decay rate is found to be $n_{obs}=1.616\pm 0.067$ is interpreted by the authors as evidence for beaming. Moreover, the fact that the decay rate $n=-1.6$ has been observed before a steepening in the optical decay occurred at approximately 1 day of arrival time authorized an even more extreme proposal of a narrower beam in the X-rays within the optical beam.

It is clear from the entire treatment which we have presented and the results of the EMBH theory given by $n_{theo}=-1.6$ that there is no evidence for such a beaming, as already stated above. The motivation by \textcite{ha00} stems from the incorrect theoretical assumption of the existence of a lower limit in the afterglow power-law index $n>-1.47$. From our theoretical analysis the existence of $n=-1.6$ is clear proof of isotropic emission in the GRB~991216 and a clear test of the complete relativistic treatment of the source. The fact that the break in the index should be ``achromatic'' and the absence of beaming in the X-rays imply an absence of beaming also in the optical and radio bands. The observed steepening in the optical decay has to find an alternative explanation. Although this is not the subject of our present work for the above mentioned reasons, we have found interesting the considerations by \textcite{pk01}, which find that ``there are some major difficulties to apply a jet model to GRB~991216''. They also state, still for GRB~991216, that ``the steepening of the optical decay of a few days is not due to a jet effect, as suggested by \textcite{ha00}, but to the passage of a spectral break''.

Concerning our own position on the possibility of beaming in GRBs, we would like just to remark that, from a preliminary analysis of beamed emission within the EMBH model, we have found some new features which are not encompassed by the results in the current literature, and they could become a distinctive signature for the discrimination of the existence or nonexistence of beaming (\textcite{rbcfx02b_beam}). The study of the steepening in the optical and radio decay is addressed within the EMBH theory in \textcite{rbcfx02c_spectrum}.

\section{Substructures in the E-APE due to inhomogeneities in the Interstellar medium}\label{substructures}

\begin{figure}
\begin{center}
\includegraphics[width=\hsize,clip]{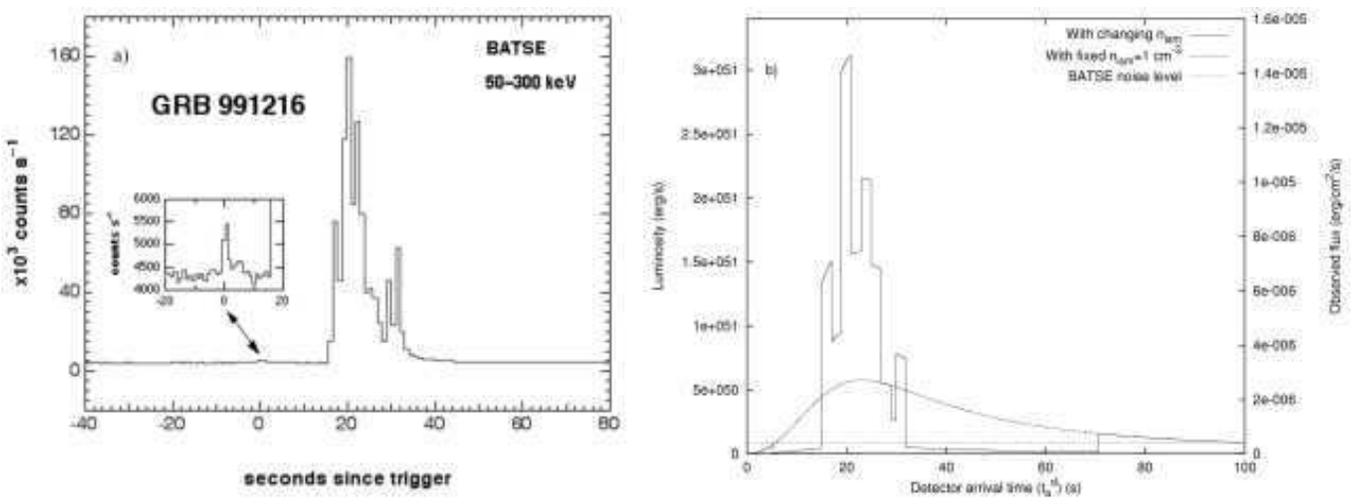}
\caption{{\bf a)} Flux of GRB 991216 observed by BATSE. The enlargement clearly shows the P-GRB (see \textcite{lett2}). {\bf b)} Flux computed in the collision of the ABM pulse with an ISM cloud with the density profile given in Fig.~\ref{denstity_prof}. The dashed line indicates the emission from an uniform ISM with $ n = 1 cm^{-3} $. The dotted line indicates the BATSE noise level.}
\label{fit_subs_1}
\end{center}
\end{figure}

The afterglow is emitted as the ABM pulse plows through the interstellar matter engulfing new baryonic material. In our previous articles we were interested in explaining the overall energetics of the GRB phenomena and in this sense, we have adopted the very simplified assumption that the interstellar medium is a constant density medium with $n_{ism}=1/cm^3$. Consequently, the afterglow emission obtained is very smooth in time. We are now interested in seeing if in this framework we can also explain most of the time variability observed by BATSE (see e.g. \textcite{fm95}), all of which except for the P-GRB should correspond to the beam-target phase in the IBS paradigm.

We pursue this treatment neglecting the angular spreading due to off-axis scattering in the radiation of the afterglow, which will be presented in sections~\ref{angle}--\ref{structure_angle}.

Our goal is to focus in this simplified model on the basic energetic parameters as well as on the drastic consequences of the space-time variables expressed in the RSTT paradigm.

Having obtained the two results presented in Fig.~\ref{gamma} and Fig.~\ref{fluxes}, we can proceed to attack the specific problem of the time variability observed by BATSE.

The fundamental point is that in both regimes {\em the flux observed in the arrival time is proportional to the interstellar matter density}: any inhomogeneity in the interstellar 
medium $\Delta n_{ism}/ \overline{n}_{ism}$ will lead correspondingly to a proportional variation in the intensity  $\Delta I/ \overline{I}$ of the afterglow. This result has been erroneously interpreted in the current literature as a burst originating in an unspecified ``inner engine''.
 
\begin{figure}
\begin{center}
\includegraphics[width=10cm,clip]{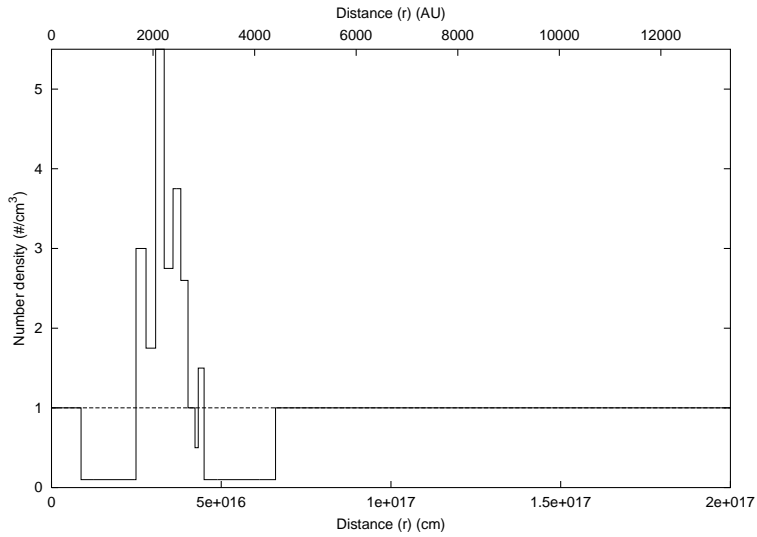}
\caption{The density contrast of the ISM cloud profile introduced in order to fit the observation of the burst of GRB991216. The dashed line indicates the average uniform density $ n = 1 cm^{-3} $.}
\label{denstity_prof}
\end{center}
\end{figure}

\begin{figure}
\begin{center}
\includegraphics[width=\hsize,clip]{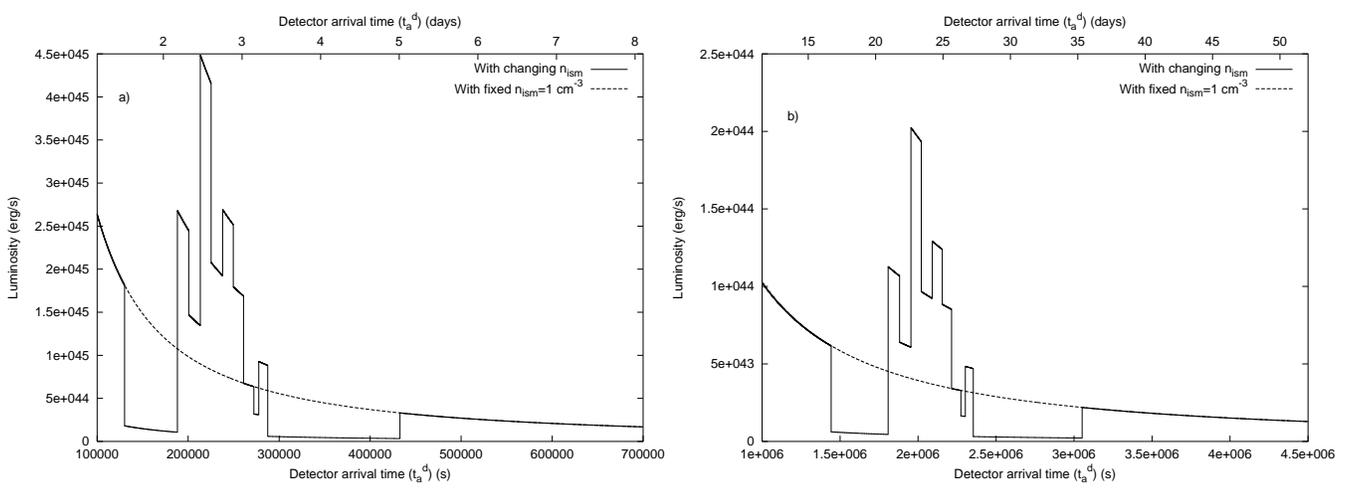}
\caption{{\bf a)} Same as Fig.~\ref{fit_subs_1}b with the ISM cloud located at a distance of $3.17\times 10^{17} cm$ from the EMBH, the time scale of the burst now extends to $\sim 1.58 \times 10^5 s $. {\bf b)} Same as a) with the ISM cloud at a distance of $4.71\times 10^{17} cm$ from the EMBH, the time scale of the burst now extends to $\sim 1.79 \times 10^6 s $.}
\label{fit_subs_2}
\end{center}
\end{figure}

In particular, for the main burst observed by BATSE (see Fig.~\ref{fit_subs_1}a) we have
\begin{equation}
 \left(\Delta 
I/ \overline{I} \right) =\left(\Delta n_{ism}/n_{ism}\right) \sim 5 .
\label{disui}
\end{equation}
There are still a variety of  physical circumstances which may lead to such density inhomogeneities.

The additional crucial parameter in understanding the physical nature of such inhomogeneities is 
the time scale of the burst observed by BATSE. Such a burst lasts $ \Delta t_a \simeq 20 s $ and shows substructures on a time scale of $ \sim 1s$ (see Fig.~\ref{fit_subs_1}a). In order to infer the nature of the structure emitting such a burst we must express these times scales in the laboratory time (see \textcite{lett1}). Since we are at the peak of the GRB we have $ \gamma_{P_A} \sim 258.5 $ (see Eq.(\ref{ArrB})) and $ \Delta t_a $ corresponds in the laboratory time to an interval
\begin{equation}
\Delta t \sim 1.0\times 10^6 s ,
\label{deltatl}
\end{equation}
which determines the characteristic size of the inhomogeneity creating the burst $\Delta L\sim 5.0 \times 10^{16} cm$ (see Tab.~\ref{tab1} and Fig.~\ref{tvsta}).

It is immediately clear from Eq.(\ref{disui}) and Eq.(\ref{deltatl}) that these are the typical 
dimensions and density contrasts corresponding to a small interstellar cloud. As an explicit example we have shown in Fig.~\ref{denstity_prof} the density contrasts and dimensions of an interstellar cloud with an {\em average density} $<n>=1/cm^3$. Such a cloud is located at a distance of $\sim 8.7\times 10^{15}cm$ from the EMBH, gives rise to a signal similar to the one observed by BATSE (see Fig.~\ref{fit_subs_1}b).

It is now interesting to see the burst that would be emitted, if our present approximation would still apply, by the interaction of the ABM pulse with the same ISM cloud encountered at later times during the evolution of the afterglow. Fig.~\ref{fit_subs_2}a shows the expected structure of the burst at a distance $4.1\times 10^{17}cm$, corresponding to an arrival time delay of $\sim 2$ days, where the gamma factor is now $\gamma_\star \sim 3.6$. It is interesting that the overall intensity would be smaller, the intensity ratio of the burst relative to the average emission would remains consistent with Eq.(\ref{disui}), but the time scales of the burst would be longer by a factor $ \left( \frac{\gamma_{P_A}}{\gamma_\star} \right)^2 \simeq  5 \times 10^3$. Fig.~\ref{fit_subs_2}b shows the corresponding quantities for the same ISM cloud located at a distance $6.4\times 10^{17}cm$ from the EMBH, corresponding to an arrival time delay of $\sim 1$ month, where the gamma factor is $\sim 1.5$.

We are going to analyze in the coming sections the modifications of this basic theory by the effect of the angular spreading: it will increase the accuracy of the fit obtained in Fig.~\ref{fit_subs_1} and will wash away all the features at late arrival time in the afterglow (see Fig.~\ref{fit_subs_2}).

\section{Considerations on the relativistic beaming angles and on the arrival time}\label{angle}

We now generalize the results obtained in section~\ref{arrival_time} to consider also the effects due to the size of the emitting surface and of its curvature. The frequency $\omega$ and wave-vector ${\bf k}$ of photons emitted from the ABM pulse (see Fig.~\ref{schema}) expressed in the laboratory frame are:
\begin{figure}
\begin{center}
\includegraphics[width=10cm,clip]{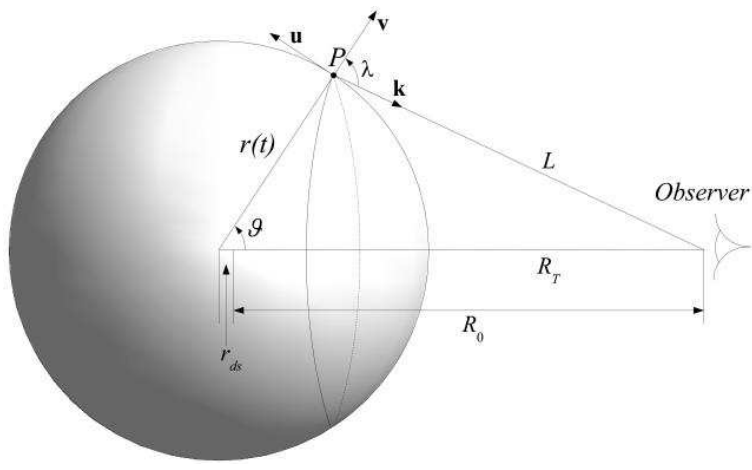}
\caption{Qualitative description of the kinematics of the system. The big sphere is the expanding ABM pulse interacting with the ISM (not shown in the picture). The radius of the ABM pulse at time $t$ is $r\left(t\right)$. The generic point $P$ on the ABM pulse, from which the photon is emitted, corresponds to a displacement angle $\vartheta$ from the line of sight. $L$ is the distance of $P$ from the observer. $R_{\rm T}$ is the distance of the EMBH from the observer. $r_{\rm ds}$ is the dyadosphere radius. $R_0$ is defined by $R_0 \equiv R_{\rm T} - r_{\rm ds}$. ${\bf v}$ is a unit vector along the radial expansion velocity. ${\bf u}$ is a unit vector orthogonal to ${\bf v}$ oriented toward rising $\vartheta$. ${\bf k}$ is the momentum of the photons emitted toward the observer. Note that we have assumed $\vartheta\equiv\lambda$, i.e. ${\bf k}\parallel R_{\rm T}$ (see text).}
\label{schema}
\end{center}
\end{figure} 
\begin{equation}
{\bf k} = \frac{\omega}{c} \left(-\sin\vartheta{\bf u} +\cos\vartheta{\bf v}\right)\, , \quad 
\left|{\bf k}\right|=\frac{\omega}{c}\, ,
\label{kdirection}
\end{equation}
where $\vartheta$ is the angle (in the laboratory frame) between the radial expansion velocity and the line of sight, ${\bf v}$ is a unit vector along the radial expansion velocity of the ABM pulse, and ${\bf u}$ is a unit vector orthogonal to ${\bf v}$ oriented toward rising $\vartheta$. We are assuming here that ${\bf k}$ and $R_{\rm T}$ are parallel, also for photons emitted with $\vartheta\neq 0$, so that $\lambda\equiv\vartheta$. This is clearly a good approximation, because the distance $R_{\rm T}$ corresponds to a redshift $z\sim 1$, while the radius of the emitting region is less than a light year in order of magnitude. Then the Lorentz boost along ${\bf v}$ to the comoving frame of the ABM pulse yields the corresponding comoving quantities:
\begin{equation}
\omega _ \circ   = \gamma \omega \left( {1 - \frac{v}{c}\cos \vartheta } \right),\quad \omega _ \circ   = \left| {{\bf k}_\circ } \right|c,
\label{entran}
\end{equation}
\begin{equation}
{\bf k}_\circ = -\left |{\bf k}\right|\sin\vartheta{\bf u}+\gamma \left|{\bf k}\right|\left(\cos\vartheta-\frac{v}{c}\right){\bf v},
\label{kntran}
\end{equation}
In the comoving frame photons radiating out of the ABM pulse must have (see Eq.(\ref{kntran})):
\begin{equation}
\cos\vartheta\ge {v\over c},
\label{cos}
\end{equation}
because the component of the photon momentum in the comoving frame along the radial expansion velocity direction must be positive in order to escape. There will then be a maximum allowed $\vartheta$ value $\vartheta_{max}$ defined by $\cos\vartheta_{max}=\left(v/c\right)$ 
(see Figs.~\ref{openang}--\ref{opening}).

\begin{figure}
\begin{center}
\includegraphics[width=8.5cm,clip]{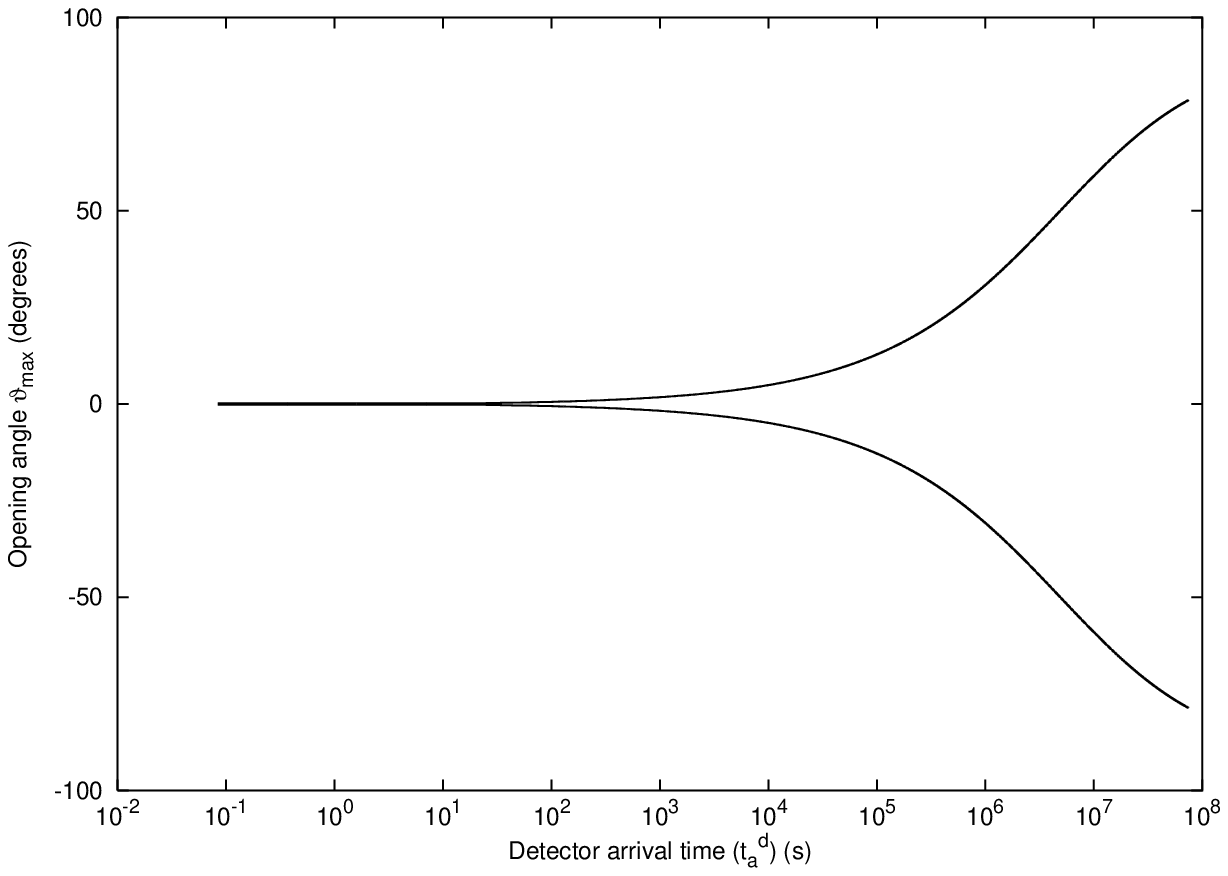}
\includegraphics[width=8.5cm,clip]{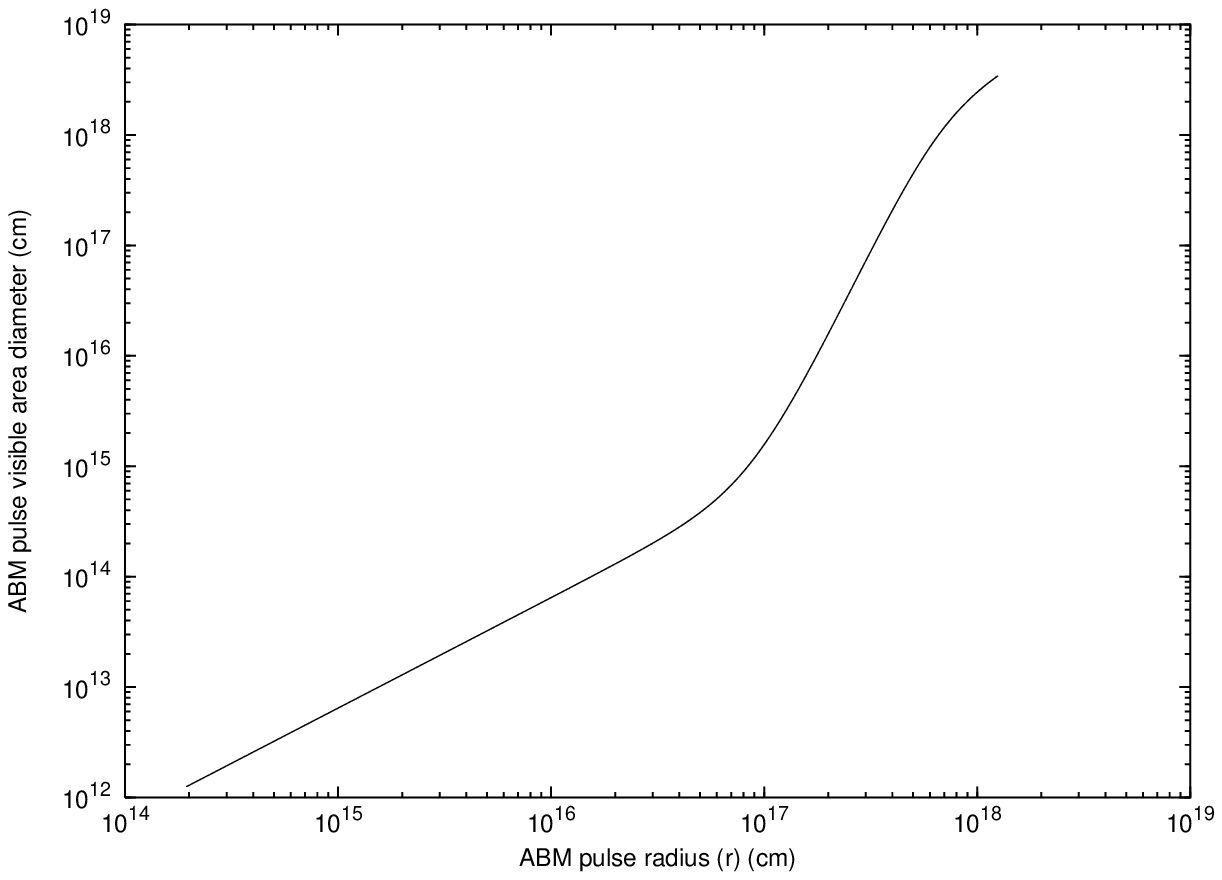}
\caption{{\bf Left)} Not all values of $\vartheta$ are allowed. Only photons emitted at an angle such that $\cos\vartheta \ge \left(v/c\right)$ can be viewed by the observer. Thus the maximum allowed $\vartheta$ value $\vartheta_{max}$ corresponds to $\cos\vartheta_{max} = (v/c)$. In this figure we represent $\vartheta_{max}$ (i.e. the angular amplitude of the visible area of the ABM pulse) in degrees as a function of the arrival time at the detector for the photons emitted along the line of sight (see text). In the earliest GRB phases $v\sim c$ and so $\vartheta_{max}\sim 0$. On the contrary, in the latest phases of the afterglow the ABM pulse velocity decreases and $\vartheta_{max}$ tends to the maximum possible value, i.e. $90^\circ$. {\bf Right)} The diameter of the visible area is represented as a function of the ABM pulse radius. In the earliest expansion phases ($\gamma\sim 310$) $\vartheta_{max}$ is very small (see left pane and Fig.~\ref{opening}), so the visible area is just a small fraction of the total ABM pulse surface. On the other hand, in the final expansion phases $\vartheta_{max} \to 90^\circ$ and almost all the ABM pulse surface becomes visible.}
\label{openang}
\label{opensrad}
\end{center}
\end{figure}

Due to the high value of the Lorentz gamma factor $\left(\sim 300\right)$ for the bulk motion of the expanding ABM pulse, the spherical waves emitted from its external surface appear extremely distorted to a distant observer. Let us indicate by $t_a$ the arrival time at a detector of a photon emitted at a laboratory time $t$ by the spherical surface of the relativistically expanding shell (see also section~\ref{arrival_time}). Photons arriving at the same time $t_a$ will be emitted at different $t$ as a function of the angle $\vartheta$ (see Fig.~\ref{schema}). The relation between $t$ and $t_{\rm a}$ in the case of a constant $\gamma\sim 5$ for expanding radio sources was found by Rees (see \textcite{r66}):
\begin{equation}
t_{\rm a}  = t\left( {1 - \frac{v}{c}\cos \vartheta } \right)\, .
\label{tar}
\end{equation}
For a constant expansion speed, the radius $r\left( t \right)$ of the source is given by:
\begin{equation}
r\left( t \right)=vt \, .
\label{rvt}
\end{equation}
From Eqs.(\ref{tar}--\ref{rvt}) we find the equation describing the ``surface'' emitting the photons detected at arrival time $t_{\rm a}$:
\begin{equation}
r= \frac{{v\;t_{\rm a}}}{{1 - \frac{v}{c}\cos \vartheta }}  ,
\label{ETSC}
\end{equation}
which describes an ellipsoid of eccentricity $\frac{v}{c}$ (see \textcite{r66}).

In our case the ABM pulse Lorentz gamma factor is not constant (see Fig.~\ref{gamma}), and so we must generalize Eqs.(\ref{tar},\ref{ETSC}) to nonconstant expansion velocity. This can be done using the geometry of Fig.~\ref{schema}. We set $t=0$ when the plasma starts to expand, so that $r\left( 0 \right)=r_{\rm ds}$, i.e. the dyadosphere radius. Let a photon be emitted at time $t$ from the point $P$. Its distance from the observer is $L$. The time it takes to arrive at the detector is of course $\frac{L}{c}$. Thus its arrival time, measured from the arrival of the first photon a time $\frac{R_0}{c}$ after its emission at $t=0$, is:
\begin{equation}
t_{\rm a}=t+\frac{L}{c}-\frac{R_0}{c} \, ,
\label{ta}
\end{equation}
where we have defined $t_{\rm a}=0$ when a photon emitted at $t=0$ and $\vartheta=0$ reaches the observer. $L$ is clearly given by:
\begin{equation}
L = \sqrt {R_{\rm T}^2  + r\left( t \right)^2  - 2\,R_{\rm T}\,r\left( t \right)\cos \vartheta } \, ,
\label{Lex}
\end{equation}
where at any given value of emission time $t$, $\cos \vartheta$ can assume any value between $\left(\frac{v\left( t \right)}{c}\right)$ and $1$ as noted above, where $v\left( t \right)$ is the expansion speed of the ABM pulse at time $t$ (see Eq.(\ref{cos})). Now $r\left( t \right)$ is less than one light year in order of magnitude while $R_{\rm T}$ corresponds to a redshift $z\sim 1$. Thus we can expand the right hand side of equation (\ref{Lex}) in powers of $\frac{r\left( t \right)}{R_{\rm T}}$ to first order:
\begin{equation}
L \simeq R_{\rm T}\left( {1 - \frac{{r\left( t \right)}}{R_{\rm T}}\,\cos \vartheta } \right)\, ,
\label{Lapp}
\end{equation}
which corresponds to assuming $L$ to be equal to its projection on the line of sight (see Fig.~\ref{schema}). Substituting (\ref{Lapp}) into (\ref{ta}) yields:
\begin{equation}
t_{\rm a}  = t +  \frac{{r_{\rm ds} }}{c} - \frac{{r\left( t \right)}}{c}\cos \vartheta\, ,
\label{taapp_2}
\end{equation}
where we have used the fact that $R_{\rm T}=R_0+r_{\rm ds}$ (see Fig.~\ref{schema}). For $r\left( t \right)$ we can use the following expression:
\begin{equation}
r\left( t \right) = \int_0^t {v\left( {t'} \right)dt'}  + r_{\rm ds} ,
\label{rdiv}
\end{equation}
so that equation (\ref{taapp_2}) can be written in the form:
\begin{equation}
t_{\rm a}  = t - \frac{{\int_0^t {v\left( {t'} \right)dt'}  + r_{\rm ds} }}{c}\cos \vartheta  + \frac{{r_{\rm ds} }}{c},
\label{ta_fin}
\end{equation}
which reduces to Eq.(\ref{tar}) only if $v$ is constant and $r_{\rm ds}$ is negligible with respect to $r\left( t \right)$.

\begin{figure}
\begin{center}
\includegraphics[width=6.2cm,clip]{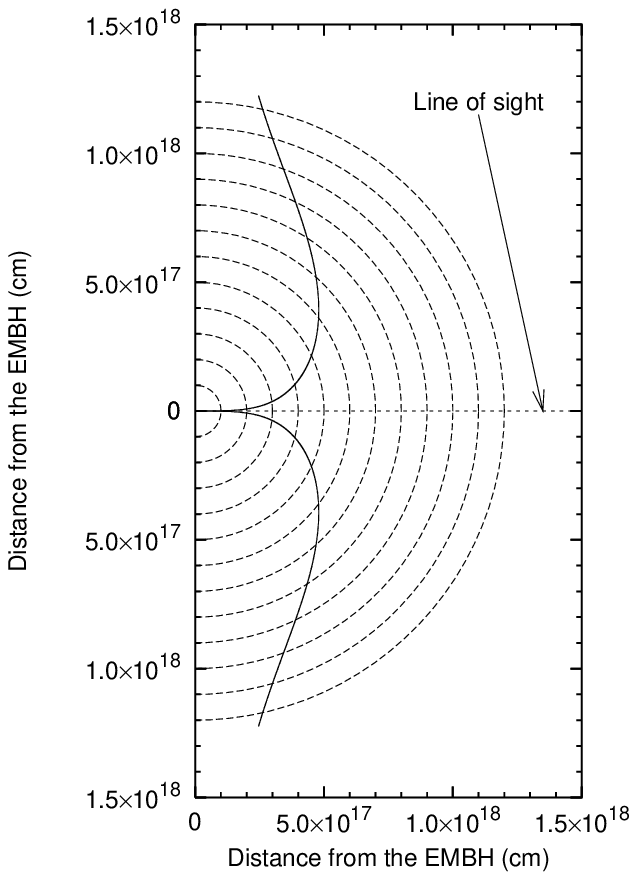}
\includegraphics[width=11cm,clip]{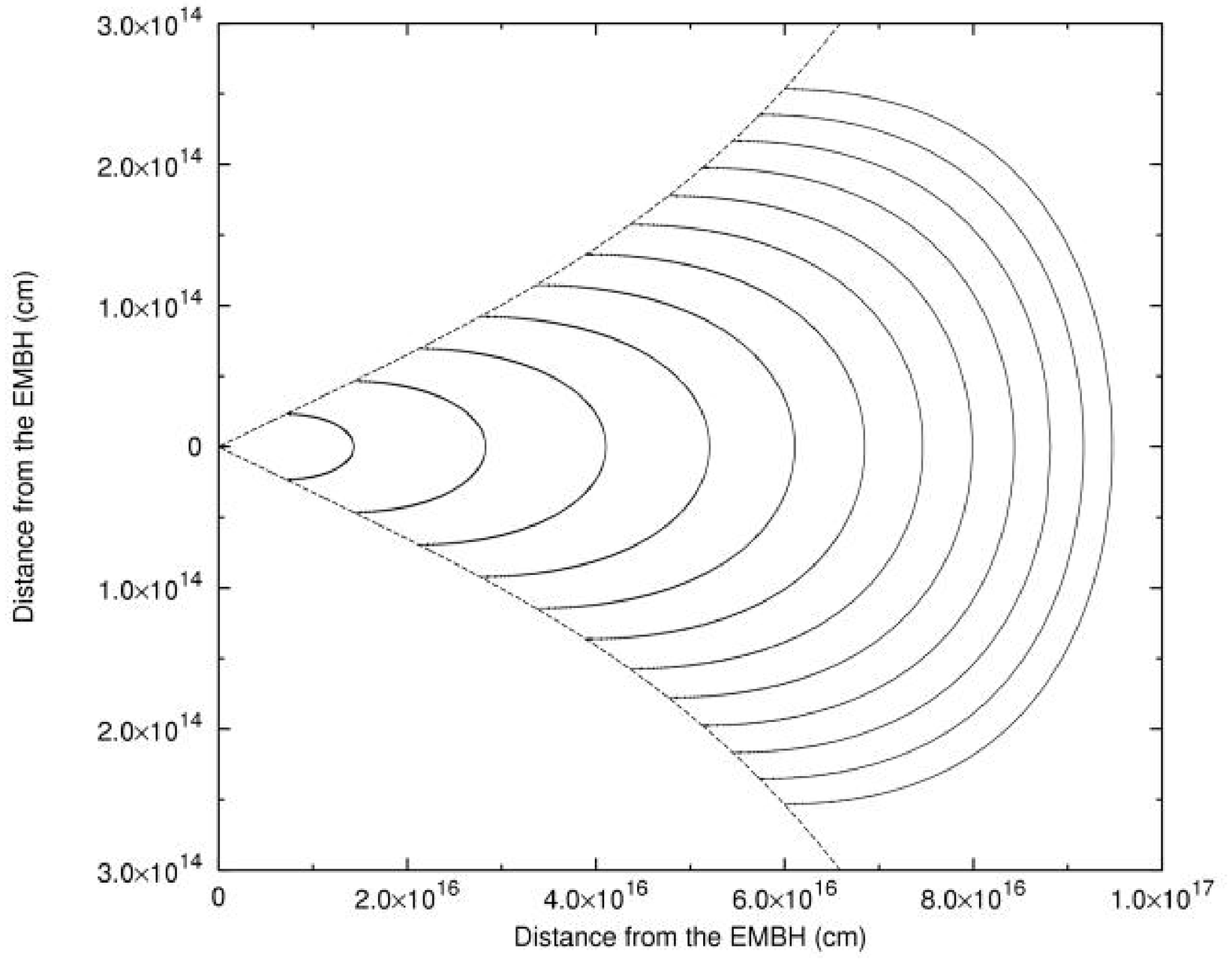}
\caption{{\bf Left)} This figure shows the temporal evolution of visible area of the ABM pulse. The dashed half-circles are the expanding ABM pulse at radii corresponding to different laboratory times. The black curve marks the boundary of the visible region. The EMBH is located at position (0,0) in this plot. Again, in the earliest GRB phases the visible region is squeezed along the line of sight, while in the final part of the afterglow phase almost all the emitted photons reach the observer. This time evolution of the visible area is crucial to the explanation of the GRB temporal structure. {\bf Right)} Due to the extremely high and extremely varying Lorentz gamma factor, photons reaching the detector on the Earth at the same arrival time are actually emitted at very different times and positions. We represent here the surfaces of photon emission corresponding to selected values of the photon arrival time at the detector: the {\em equitemporal surfaces} (EQTS). Such surfaces differ from the ellipsoids described by Rees in the context of the expanding radio sources with typical Lorentz factor $\gamma\sim 4$ and constant. In fact, in GRB~991216 the Lorentz gamma factor ranges from $310$ to $1$. The EQTSes represented here (solid lines) correspond respectively to values of the arrival time ranging from $5\, s$ (the smallest surface on the left of the plot) to $60\, s$ (the largest one on the right). Each surface differs from the previous one by $5\, s$. To each EQTS contributes emission processes occurring at different values of the Lorentz gamma factor. The dashed lines are the boundaries of the  visible area of the ABM pulse and the EMBH is located at position $(0,0)$ in this plot. Note the different scale on the two axes, indicating the very high EQTS ``effective eccentricity''. The time interval from $5\, {\rm s}$ to $60\, {\rm s}$ has been chosen to encompass the E-APE emission, ranging from $\gamma=308.8$ to $\gamma=56.84$.}
\label{opening}
\label{ETSNCF}
\end{center}
\end{figure}

Also from Eq.(\ref{ta_fin}) we can obtain the equation describing the surface that emits the photons detected at an arrival time $t_{\rm a}$. In this case, we no longer have ellipsoids of constant eccentricity $\frac{v}{c}$. Since the velocity is strongly varying from point to point, we have more complicated surfaces like the profiles reported in Fig.~\ref{ETSNCF} where at every point there will be a tangent ellipsoid of a given eccentricity, but such an ellipsoid varies in eccentricity from point to point.

For a fixed time $t$ of emission in Eq.(\ref{ta_fin}), the allowed angular interval $\frac{v}{c} \le \cos\vartheta \le 1$ leads to a corresponding smearing of the arrival time $t_a$ over the interval
\begin{equation}
\Delta t_a = \frac{r}{\gamma^2c\left(1+\frac{v}{c}\right)}\, .
\label{angular}
\end{equation}

We need now to correct Eq.(\ref{ta_fin}) for the cosmological expansion effects to get the wanted relation between $t$ and $t_a^d$. We recall that (see section~\ref{arrival_time})
\begin{equation}
t_a^d=\left(1+z\right)t_a\, ,
\label{taddef_2}
\end{equation}
where $z$ is the cosmological redshift. Our final relation is therefore:
\begin{equation}
t_a^d  =\left(1+z\right)\left(t - \frac{{\int_0^t {v\left( {t'} \right)dt'}  + r_{\rm ds} }}{c}\cos \vartheta  + \frac{{r_{\rm ds} }}{c}\right)\, .
\label{tad_fin}
\end{equation}

\section{The emission process taking off-axis contributions into account}\label{off-axis}

We now take into consideration the contributions of the off-axis emission to the afterglow to see if the previous positive results still hold and if some of the problems just stated can be overcome by a more detailed and relativistic treatment. The corresponding computation for the P-GRB structure will be presented elsewhere, where the time evolutions of the dyadosphere formation and its consequences on the P-GRB structures are presented following the work of \textcite{crv02,rv02a,rv02b,rvx02}. The effects on the P-GRB structure of the dyadosphere formation dominate those due to the angular spreading.

Following Eqs.(\ref{heat2}--\ref{dgamma2}), we recall that in the comoving frame of the expanding ABM pulse we suppose that the internal energy due to kinetic collision is instantly radiated away and that the corresponding  emission is isotropic. As in section~\ref{int}, let $\Delta \varepsilon$ be the internal energy density developed in the collision. In the comoving frame the energy per unit of volume and per solid angle is simply
\begin{equation}
\left(\frac{dE}{dV d\Omega}\right)_{\circ}  =  \frac{\Delta \varepsilon}{4 \pi}
\label{dEo}
\end{equation}
due to the fact that the emission is isotropic in this frame. The total number of photons emitted is an invariant quantity independent of the frame used. Thus we can compute this quantity as seen by an observer in the comoving frame (which we denote with the subscript ``$\circ$'') and by an observer in the laboratory frame (which we denote with no subscripts). Doing this we find
\begin{equation}
\frac{dN_\gamma}{dt d \Omega d \Sigma}= \int_{shell}  \left(\frac{dN_\gamma}{dt d \Omega d \Sigma} \right)_{\circ} \Lambda^{-3}
\cos \vartheta
\, , 
\end{equation}
where $\cos\vartheta$ comes from the projection of the elementary surface of the shell on the direction of propagation and $\Lambda = \gamma ( 1 - \beta \cos \vartheta )$ is the Doppler factor introduced in the two following differential transformation
\begin{equation}
d \Omega_{\circ} = d \Omega \times \Lambda^{-2}
\end{equation}
for the solid angle transformation and
\begin{equation}
d t_{\circ} = d t \times \Lambda^{-1}
\end{equation}
for the time transformation. The integration in $d \Sigma$ is performed over the visible area of the ABM pulse at laboratory time $t$, namely with $0\le\vartheta\le\vartheta_{max}$ and $\vartheta_{max}$ defined in section~\ref{angle} (see Eq.(\ref{cos}) and Figs.~\ref{openang}--\ref{opening}). An extra $\Lambda$ factor  comes from the energy transformation:
\begin{equation}
E_{\circ} = E \times \Lambda\, .
\end{equation}
See also \textcite{cd99}. Thus finally we obtain:
\begin{equation}
\frac{dE}{dt d \Omega d \Sigma} = \int_{shell}  \left(\frac{dE}{dt d \Omega d \Sigma} \right)_{\circ} \Lambda^{-4} \cos \vartheta \, .
\end{equation}
Doing this we clearly identify  $  \left(\frac{dE}{dt d \Omega d \Sigma} \right)_{\circ} $ 
as the energy density in comoving frame up to a factor $\frac{v}{4\pi}$ (see Eq.(\ref{dEo})). Then we have:
\begin{equation}
\frac{dE}{dt d \Omega } = \int_{shell} \frac{\Delta \varepsilon}{4 \pi} \; v \; \cos \vartheta \; \Lambda^{-4} \; d \Sigma\, ,
\label{fluxlab}
\end{equation}
where the integration in $d \Sigma$ is performed over the ABM pulse visible area at laboratory time $t$, namely with $0\le\vartheta\le\vartheta_{max}$ and $\vartheta_{max}$ defined in section~\ref{angle}.

Eq.(\ref{fluxlab}) gives us the energy emitted toward the observer per unit solid angle and per unit laboratory time $t$ in the laboratory frame. But what we really need is the energy emitted per unit solid angle and per unit detector arrival time $t_a^d$, so we must use the complete relation between $t_a^d$ and $t$ given in Eq.(\ref{tad_fin}). First we have to multiply the integrand in Eq.(\ref{fluxlab}) by the factor $\left(dt/dt_a^d\right)$ to transform the energy density generated per unit of laboratory time $t$ into the energy density generated per unit arrival time $t_a^d$. Then we have to integrate with respect to $d \Sigma$ over the {\em equitemporal surface} (EQTS, see section~\ref{angle}) of constant arrival time $t_a^d$ instead of the ABM pulse visible area at laboratory time $t$. The analog of Eq.(\ref{fluxlab}) for the source luminosity in detector arrival time is then:
\begin{equation}
\frac{dE_\gamma}{dt_a^d d \Omega } = \int_{EQTS} \frac{\Delta \varepsilon}{4 \pi} \; v \; \cos \vartheta \; \Lambda^{-4} \; \frac{dt}{dt_a^d} d \Sigma\, .
\label{fluxarr}
\end{equation}
It is important to note that, in the present case of GRB~991216, the Doppler factor $\Lambda^{-4}$ in Eq.(\ref{fluxarr}) enhances the apparent luminosity of the burst, as compared to the intrinsic luminosity, by a factor which at the E-APE is in the range between $10^{10}$ and $10^{12}$!

To perform the numerical integration of Eq.(\ref{fluxarr}) we have implemented the following procedure for each fixed value of the laboratory time $t$:
\begin{enumerate}
\item We fix the laboratory time $t$.
\item We divide the interval of the allowed values $\left(v\left(t\right)/c\right)\le\cos\vartheta\le 1$ into $N$ small steps, each one of amplitude
\begin{equation}
\Delta_N \left(\cos\vartheta\right)=\frac{1-\left(v\left(t\right)/c\right)}{N}\, .
\label{dct_def}
\end{equation}
\item We select $n$ directions defined by:
\begin{equation}
\cos\vartheta_n=1-n\Delta \left(\cos\vartheta\right)\, ,
\label{ct_def}
\end{equation}
where $n$ is an integer, $0\le n\le N$ and so $\vartheta_0=0$ and $\vartheta_n=\vartheta_{max}$.
\item For each $\vartheta_n$ we compute with Eq.(\ref{fluxarr}) the contribution to the afterglow luminosity arising from an angular aperture corresponding to $\Delta_N\left(\cos\vartheta\right)$ around such a direction.
\item We compute for each value of $n$ the corresponding values of the arrival time $t_a^d$ using Eq.(\ref{tad_fin}).
\end{enumerate}
To obtain the total luminosity at arrival time $t_a^d$ we sum together all the above contributions corresponding to the same $t_a^d$.

We first apply this treatment to the analysis of the afterglow using assumptions 1 and 2 of section.~\ref{int}, namely that the ISM density is constant $n_{ism}=<n_{ism}>=1\, {\rm particle}/{\rm cm}^3$ and that the ABM is spherically symmetric.

\begin{figure}
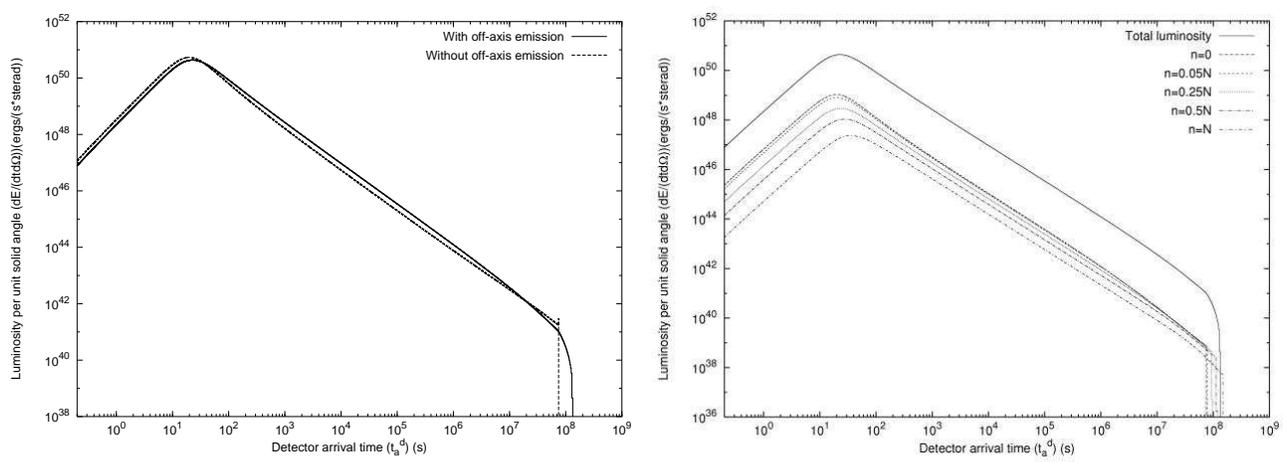

\begin{center}
\includegraphics[width=8.5cm,clip]{afterang}
\includegraphics[width=8.5cm,clip]{angspre}
\caption{{\bf Left)} The predicted afterglow curve for GRB~991216 assuming a constant ISM density equal to $1\, {\rm particle}/{\rm cm}^3$ and taking into account all the effects due to off-axis emission (solid line). For comparison we plot also  the corresponding curve obtained in the simple radial approximation (dashed line). We see that this last curve falls sharply to zero when the ABM pulse reaches $\gamma=1$, while the first one has a much smoother behavior due to the time delay in the arrival of the photons emitted at large $\vartheta$. Recall that when $\gamma$ tends to $1$, the maximum allowed values of $\vartheta$ tend to $90^\circ$. {\bf Right)} This figure shows how the radiation emitted from different angles contributes to the afterglow luminosity. The solid line on the top of the picture is the total luminosity as in the previous plots. The other dashed and dotted curves represent the radiation components corresponding to selected values of $n$ in Eq.(\ref{ct_def}). From the upper to the lower one they corresponds respectively to $n=0$, $n=0.05N$, $n=0.25N$, $n=0.5N$, $n=N$, where in this plot $N=200$. We can easily see that the radiation emitted at large angles ($n=N$) is time shifted with respect to that emitted near the line of sight ($n=0$).}
\label{afterang}
\label{angspre}
\end{center}
\end{figure}

Fig.~\ref{afterang} compares the new result for the afterglow luminosity as a function of the detector arrival time with the previous one obtained in section~\ref{bf} by neglecting off-axis emission. The main conclusions are:
\begin{enumerate}
\item The total energy emitted both in the radial approximation and in the full computation with the off-axis emission is conserved. This is a necessary condition for checking the consistency of the model.
\item The slope of the decreasing part of the afterglow is unchanged. We emphasize once more the great advantage of the radial approximation which has allowed to obtain an analytic expression for this slope.
\item The final phase of the afterglow ($\gamma<2$) is largely affected by the late arrival of the radiation emitted at large angles. In fact in the radial approximation the luminosity goes abruptly to zero when $\gamma$ reaches $1$ while in the new complete treatment the behavior is much smoother due to the delayed arrival of the radiation emitted at large angles. Consequently, enforcing the energy conservation, in the rising part of the afterglow the luminosity in the new treatment is shown to be slightly smaller than in the radial case.
\end{enumerate}

In order to acquire a better understanding of the effects of angular spreading, we have found it helpful to analyze the radiation emitted from selected angles $\vartheta$ between $0$ and $\vartheta_{max}$. This is in addition to the integration results presented in Fig.~\ref{afterang}. In Fig.~\ref{angspre} we show the results of such an analysis plotting the contributions to the total luminosity corresponding to selected values of $n$ in Eq.(\ref{ct_def}). We easily see that radiation emitted at large angles is time shifted with respect to that emitted near the line of sight. In fact the afterglow peak occurs later going to higher $n$ values (see Fig.~\ref{angspre}).

\section{The E-APE temporal substructures taking into account the off-axis emission}\label{structure_angle}

\begin{figure}
\includegraphics[width=10cm,clip]{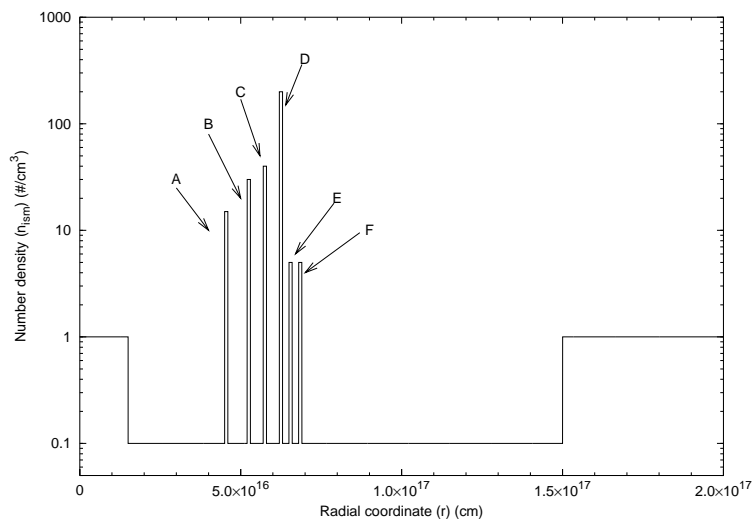}
\caption{The density profile (``mask") of an ISM cloud used to reproduce the GRB~991216 temporal structure. As before, the radial coordinate is measured from the black hole. In this cloud we have six ``spikes" with overdensity separated by low density regions. Each spike has the same spatial extension of $10^{15}\, {\rm cm}$. The cloud average density is $<n_{ism}>=1\, {\rm particle}/{\rm cm}^3$.}
\label{maschera}
\end{figure}

We are now ready to reconsider the problem of the ISM inhomogeneity generating the temporal substructures in the E-APE by integrating on the EQTS surfaces and improving on the considerations based on the purely radial approximation. We have created (see details in \textcite{rbcfx02e_paperII}) an ISM inhomogeneity ``mask" (see Fig.~\ref{maschera} and Tab.~\ref{tab3}) with the main criteria that the density inhomogeneities and their spatial distribution still fulfill $<n_{ism}>=1\, {\rm particle}/{\rm cm}^3$.

\begin{table}
\centering
\caption{For each ISM density peak represented in Fig.~\ref{maschera} we give the initial radius $r$, the corresponding comoving time $\tau$, laboratory time $t$, arrival time at the detector $t_a^d$, diameter of the ABM pulse visible area $d_{v}$, Lorentz factor $\gamma$ and observed duration $\Delta t_a^d$ of the afterglow luminosity peaks generated by each density peak. In the last column, the apparent motion in the radial coordinate, evaluated in the arrival time at the detector, leads to an enormous ``superluminal" behavior, up to $9.5\times 10^4\,c$. \label{tab3}}
\begin{ruledtabular}
\begin{tabular}{c|e{8}|e{8}|e{8}|e{1}|e{8}|e{3}|e{2}|e{9}}
Peak & r (cm) & \tau (s) & t (s) & t_a^d (s) & d_v (cm) & \Delta t_a^d (s) & \gamma &  \begin{array}{c} {\rm ``Superluminal"} \\ v\equiv\frac{r}{t_a^d} \\ \\ \end{array}\\
\hline
A & 4.50\times10^{16} & 4.88\times10^3 & 1.50\times10^6 & 15.8 & 2.95\times10^{14} & 0.400 & 303.8 & 9.5\times10^4c\\
B & 5.20\times10^{16} & 5.74\times10^3 & 1.73\times10^6 & 19.0 & 3.89\times10^{14} & 0.622 & 265.4 & 9.1\times10^4c\\
C & 5.70\times10^{16} & 6.54\times10^3 & 1.90\times10^6 & 22.9 & 5.83\times10^{14} & 1.13  & 200.5 & 8.3\times10^4c\\
D & 6.20\times10^{16} & 7.64\times10^3 & 2.07\times10^6 & 30.1 & 9.03\times10^{14} & 5.16  & 139.9 & 6.9\times10^4c\\
E & 6.50\times10^{16} & 9.22\times10^3 & 2.17\times10^6 & 55.9 & 2.27\times10^{15} & 10.2  & 57.23 & 3.9\times10^4c\\
F & 6.80\times10^{16} & 1.10\times10^4 & 2.27\times10^6 & 87.4 & 2.42\times10^{15} & 10.6  & 56.24 & 2.6\times10^4c\\
\end{tabular}
\end{ruledtabular}
\end{table}

\begin{figure}
\includegraphics[width=8.5cm,clip]{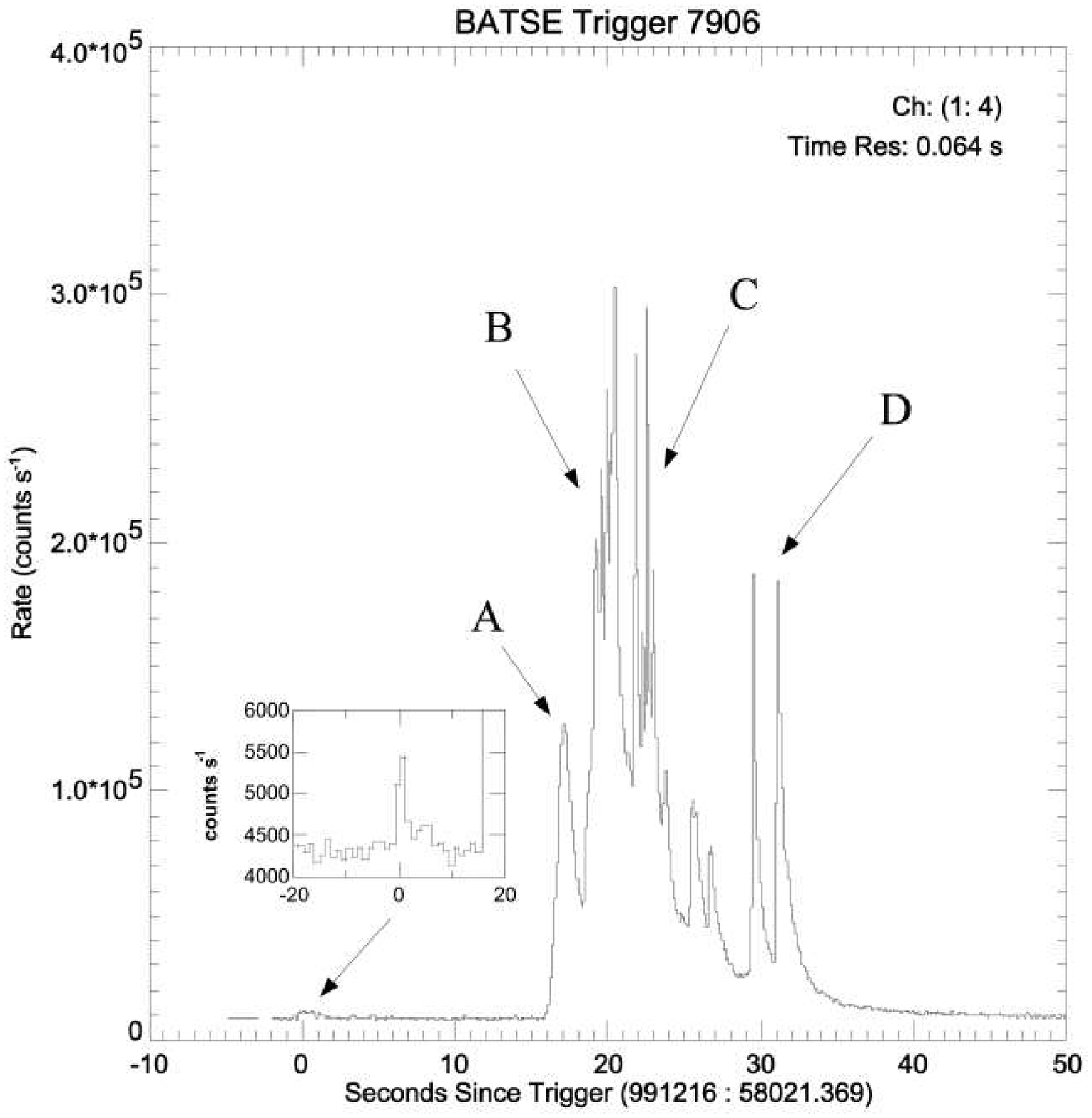}
\includegraphics[width=8.5cm,clip]{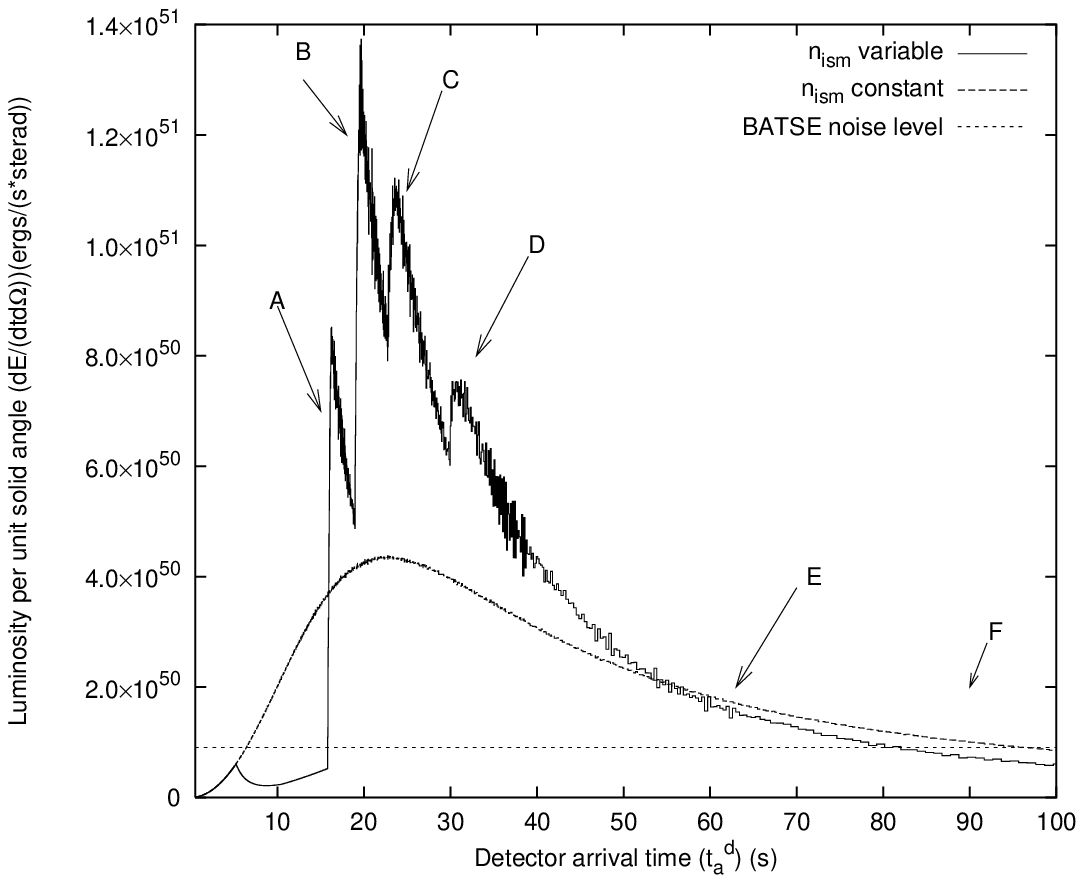}
\caption{{\bf Left)} The BATSE data on the E-APE of GRB~991216 (source: \textcite{grblc99}) together with an enlargement of the P-GRB data (source: \textcite{brbr99}). For convenience each E-APE peak has been labeled by a different uppercase Latin letter. {\bf Right)} The source luminosity connected to the mask in Fig.~\ref{maschera} is given as a function of the detector arrival time (solid ``spiky" line) with the corresponding curve for the case of constant $n_{ism}=1\, {\rm particle}/{\rm cm}^3$ (dashed smooth line) and the BATSE noise level (dotted horizontal line). The ``noise" observed in the theoretical curves is due to the discretization process adopted, described in \textcite{rbcfx02e_paperII}, for the description of the angular spreading of the scattered radiation. For each fixed value of the laboratory time we have summed $500$ different contributions from different angles. The integration of the equation of motion of this system is performed in $22,314,500$ contributions to be considered. An increase in the number of steps and in the precision of the numerical computation would lead to a smoother curve.}
\label{substr_peak}
\end{figure}

The results are given in Fig.~\ref{substr_peak}. We obtain, in perfect agreement with the observations:
\begin{enumerate}
\item the theoretically computed intensity of the A, B, C peaks as a function of the ISM inhomogneities;
\item the fast rise and exponential decay shape for each peak;
\item a continuous and smooth emission between the peaks.
\end{enumerate}

Interestingly, the signals from shells E and F, which have a density inhomogeneity comparable to A, are undetectable. The reason is due to a variety of relativistic effects and partly to the spreading in the arrival time, which for A, corresponding to $\gamma=303.8$ is $0.4 s$ while for E (F) corresponding to $\gamma=57.23$ $(56.24)$ is of $10.2\, {\rm s}$ $(10.6\, {\rm s})$ (see Tab.~\ref{tab3} and \textcite{rbcfx02a_sub,rbcfx02e_paperII}).

In the case of D, the agreement with the arrival time is reached, but we do not obtain the double peaked structure. The ABM pulse visible area diameter at the moment of interaction with the D shell is $\sim 1.0\times 10^{15}\, {\rm cm}$, equal to the extension of the ISM shell (see Tab.~\ref{tab3} and \textcite{rbcfx02a_sub,rbcfx02e_paperII}). Under these conditions, the concentric shell approximation does not hold anymore: the disagreement with the observations simply makes manifest the need for a more detailed description of the three dimensional nature of the ISM cloud.

The physical reasons for these results can be simply summarized: we can distinguish two different regimes corresponding in the afterglow of GRB~991216 respectively to $\gamma > 150$ and to $\gamma < 150$. For different sources this value may be slightly different. In the E-APE region ($\gamma > 150$) the GRB substructure intensities indeed correlate with the ISM inhomogeneities. In this limited region (see peaks A, B, C) the Lorentz gamma factor of the ABM pulse ranges from $\gamma\sim 304$ to $\gamma\sim 200$. The boundary of the visible region is smaller than the thickness $\Delta R$ of the inhomogeneities (see Fig.~\ref{opening} and Tab.~\ref{tab3}). Under this condition the adopted spherical approximation is not only mathematically simpler but also fully justified. The angular spreading is not strong enough to wipe out the signal from the inhomogeneity spike.

As we descend in the afterglow ($\gamma < 150$), the Lorentz gamma factor decreases markedly and in the border line case of peak D $\gamma\sim 140$. For the peaks E and F we have $\gamma\sim 50$ and, under these circumstances, the boundary of the visible region becomes much larger than the thickness $\Delta R$ of the inhomogeneities (see Fig.~\ref{ETSNCF} and Tab.~\ref{tab3}). A three dimensional description would be necessary, breaking the spherical symmetry and making the computation more difficult. However we do not need to perform this more complex analysis for peaks E and F: any three dimensional description would {\em a fortiori} augment the smoothing of the observed flux. The spherically symmetric description of the inhomogeneities is already enough to prove the overwhelming effect of the angular spreading (\textcite{rbcfx02e_paperII}).

On this general issue of the possible explanation of the observed substructures with the ISM inhomogeneities, there exists in the literature two extreme points of view: the one by Fenimore and collaborators (see e.g. \textcite{fmn96,fcrsyn99,f99}) and Piran and collaborators (see e.g. \textcite{sp97,p99,p00,p01}) on one side and the one by Dermer and collaborators (\textcite{d98,dbc99,dm99}) on the other.

Fenimore and collaborators have emphasized the relevance of a specific signature to be expected in the collision of a relativistic expanding shell with the ISM, what they call a fast rise and exponential decay (FRED) shape. This feature is confirmed by our analysis (see peaks A, B, C in Fig.~\ref{substr_peak}). However they also conclude, sharing the opinion by Piran and collaborators, that the variability observed in GRBs is inconsistent with causally connected variations in a single, symmetric, relativistic shell interacting with the ambient material (``external shocks") (\textcite{fcrsyn99}). In their opinion the solution of the short time variability has to be envisioned within the protracted activity of an unspecified ``inner engine'' (\textcite{sp97}); see as well \textcite{rm94,pm98,mr00,mr01,m01}.

On the other hand, Dermer and collaborators, by considering an idealized process occurring at a fixed $\gamma=300$, have reached the opposite conclusions and they purport that GRB light curves are tomographic images of the density distributions of the medium surrounding the sources of GRBs (\textcite{dm99}).

From our analysis we can conclude that Dermer's conclusions are correct for $\gamma\sim 300$ and do indeed hold for $\gamma > 150$. However, as the gamma factor drops from $\gamma\sim 150$ to $\gamma\sim 1$ (see Fig~\ref{gamma}), the intensity due to the inhomogeneities markedly decreases also due to the angular spreading (events E and F). The initial Lorentz factor of the ABM pulse $\gamma\sim 310$ decreases very rapidly to $\gamma\sim 150$ as soon as a fraction of a typical ISM cloud is engulfed (see Tab.~\ref{tab3}). We conclude that the ``tomography" is indeed effective, but uniquely in the first ISM region close to the source and for GRBs with $\gamma > 150$.

One of the most striking feature in our analysis is clearly represented by the fact that the inhomogeneities of a mask of radial dimension of the order of $10^{17}\, {\rm cm}$ give rise to arrival time signals of the order of $20\, {\rm s}$. This outstanding result implies an apparent ``superluminal velocity'' of $\sim 10^5c$ (see Tab.~\ref{tab3}). The ``superluminal velocity'' here considered, first introduced in \textcite{lett1}, refers to the motion along the line of sight. This effect is proportional to $\gamma^2$. It is much larger than the one usually considered in the literature, within the context of radio sources and microquasars (see e.g. \textcite{mr95}), referring to the component of the velocity at right angles to the line of sight (see details in \textcite{rbcfx02e_paperII}). This second effect is in fact proportional to $\gamma$ (see \textcite{r66}). We recall that this ``superluminal velocty'' was the starting point for the enunciation of the RSTT paradigm (\textcite{lett1}), emphasizing the need of the knowledge of the {\em entire} past worldlines of the source. This need has been further clarified here in the determination of the EQTS surfaces (see Fig.~\ref{opening} which indeed depend on an integral of the Lorentz gamma factor extended over the {\em entire} past worldlines of the source. In turn, therefore, the agreement between the observed structures and the theoretical predicted ones (see Figs.~\ref{grb991216}--\ref{substr_peak}) is also an extremely stringent additional test on the values of the Lorentz gamma factor determined as a function of the radial coordinate within the EMBH theory (see Fig.~\ref{gamma}).

\section{On the istantaneous spectrum of GRBs}\label{spectrum}

Variability on the shortest time scale ever observed in nature is the main message we have acquired from the theoretical understanding of GRB astrophysical phenomena (see sections~\ref{arrival_time},\ref{era1}--\ref{at}). This situation is made even more extreme by the fact that astronomical and astrophysical observations are carried out in the ``pathological" time coordinate of the photon arrival time at the detector (see section~\ref{arrival_time}), whereby the first $10^4$ seconds of the GRB phenomena are further compressed in $\sim 0.1$ seconds (see Tab.~\ref{tab1}) and further enhanced. The understanding that in these first $10^4$ seconds four different physical eras of the GRB phenomena occur has led us to a sentiment of natural skepticism toward any global or average description of the GRB phenomenon. We start to realize that such average descriptions mediate on totally different physical processes and lead to very questionable results. Such skepticism was even strengthened as soon as we realized that the characteristic quantities usually adopted for the description of the bursts, e.g. $T_{50}$ and $T_{90}$, which so many tried for years to explain within the context of the internal shock model (see e.g. \textcite{rm94,px94,sp97,f99,fcrsyn99,p99} and references therein) were actually referring not at all to the bursts but to the extended emission from the peak of the afterglow: the E-APE! In this sense they were quite irrelevant for understanding the nature of the GRB source and were at most of interest for inquiring the structure of the ISM a few light months away from the source! It has been then with this sentiment of marked skepticism toward a global approach that we have started to consider the problem of the spectrum of GRBs and the validity of the band relation (\textcite{b93}). To attempt an integral description of the spectra of the GRBs extending over $10^6$ seconds in arrival time is clearly meaningless. It mediates on two conceptually physically different phases of GRBs: the injector phase and the beam-target phase (\textcite{lett2}). In addition, in each of these phases many specific eras are present and each one of these eras needs due attention and can lead in principle to a different instantaneous spectrum. The fact that the spectral distribution observed by Band was a non-thermal one has been a very strong objection to consider any thermal spectrum. The situation became so extreme in the recent years that the sole appearance of a thermal spectrum in any part of a theoretical paper was considered a good reason for rejecting the paper by a refereed journal and to discard the validity of that work.

Having developed the very powerful theoretical tool of the EQTS surfaces (see section~\ref{angle} and \textcite{rbcfx02a_sub,rbcfx02e_paperII}) and having been successful in having established the substructure of the E-APE, in addition to the features of the afterglow, we have decided to approach the instantaneous spectra of the GRBs in \textcite{rbcfx02c_spectrum}. In the abstract of that paper, we summarize as follows the results: {\em ``A theoretical attempt to identify the physical process at the basis of the afterglow emission of GRBs is presented, assuming GRB~991216 as a prototype. Such a physical process is identified in a mechanism leading to a thermal emission occurring in the comoving frame of the shock wave originating the GRBs. For the determination of the actually observed GRB luminosities and spectra at a given arrival time, the concept of equitemporal surfaces (EQTS, see \textcite{rbcfx02a_sub}) has to be implemented: the final results comprehend an integration over an infinite number of planckian spectra, weighted by appropriate relativistic transformations, each one corresponding to a different viewing angle in the past light cone of the observer. The relativistic transformations have been computed on the ground of the knowledge of the already determined equations of motion of GRBs within the EMBH theory (\textcite{lett1,lett2,rbcfx02a_sub}). The only free parameter of the present theory is then the dimension of the ``effective cavity'' where the thermalization process occurs. A precise fit $\left(\chi^2\simeq1.08\right)$ of the observed luminosity in the $2$--$10$ keV band of GRB~991216 is presented as well as a detailed estimate of the observed luminosity in the $50$--$300$ keV band and of the expected one in the $10$--$50$ keV band. The long awaited explanation of the observed hard-to-soft transition in GRBs is also presented''} (\textcite{rbcfx02c_spectrum}). It is interesting that this theoretical result, which up to few years ago were hardly testable due to the paucity of photons collected by the detectors, have now become a necessity in order to interpret the splendid observational results of the new families of space observatory like Chandra and XMM (see e.g. \textcite{bt03,wa02a,wa02b}).

Prior to our work, the possibility that the non-thermal looking spectrum of GRBs can be found as a superposition of a set of thermal blackbody spectrum was forcefully expressed in a simple paper by \textcite{bkp99}. These three authors have expressed in an analytic treatment that indeed the time integration of the black body planckian spectrum with a temperature varying with time following a simple power-law and expanding with another power-law can lead to a non-thermal spectrum in agreement with the observed Band relations. To obtain this result, they use two indexes for their qualitative analysis to be fitted by the observational data. Toward the end of their paper they finally quoted {\em ``In reality, not only time, but also space integration takes place. As shown by \textcite{r66}, (see also \textcite{dp97,s98}) in the case of an expanding emitting shell an observer simultaneously detects radiation produced in different moments of time (thus, with different temperatures) on the ellipsoidal or egg-like surface. The integration over this surface can give the same effect as the integration over time done in this paper, but we do not perform this here because the result strongly depends on the unknown geometry of the emitting surface''} (\textcite{bkp99}). This treatment which they outline but they discard due to the difficulty of defining the geometry of the EQTS is exactly what we have done. Our treatment has only one free parameter and can fit the data of GRBs in a range between a few seconds all the way up to $10^6$ seconds. There is a basic observational feature between our treatment and the one by \textcite{bkp99}: their instantaneous spectral distribution has necessarily to be a blackbody one, while in our case is represented by an integration over an infinite number of planckian spectra, weighted by appropriate relativistic transformations, each one corresponding to a different viewing angle in the past light cone of the observer. The difference between such unique spectra should be simply discernible using the observations of XMM, Chandra, and of future space observatories.

\section{The observation of the iron lines in GRB~991216: on a possible GRB-Supernova time sequence}\label{gsts}

We have seen in the previous sections how the time structure of the E-APE gives information on the composition of the interstellar matter at distances of the order of $5\times 10^{16}\, {\rm cm}$ from the source. We would like now to point out that the data on the iron lines from the Chandra satellite on the GRB~991216 (\textcite{p00}) and similar observations from other sources (\textcite{p99b,a00,p00}) make it possible to extend this analysis to a larger distance scale, possibly all the way out to a few light years, and consequently probe the distribution of stars in the surroundings of the newly formed EMBH.

Most importantly, these considerations lead to a new paradigm for the interpretation of the supernova-GRB correlation (see \textcite{lett3}). Indeed a correlation between the occurrence of GRBs and supernova events exists and has been established by the works of \textcite{b99,g98b,g98c,g00,k98,p98a,p99,r99,vp00}.

Such an association has been assumed to indicate that GRBs are generated by supernova explosions (see e.g. \textcite{k98}). In turn, such a point of view has implied further consequences: the optical and radio data of the supernova have been attributed to the GRB afterglow, and many theorists have tried to encompass these data and explain them as a genuine component of the GRB scenario.

We propose instead an alternative point of view implying a very clear distinction between the GRB phenomenon and the supernova: if relativistic effects presented in the RSTT paradigm are properly taken into account, then a kinematically viable explanation can be given of the supernova-GRB association. We still use GRB~991216 as a prototypical case.

The GRB-Supernova Time Sequence paradigm, which we have indicated for short as GSTS paradigm (see \textcite{lett3}), states that: {\em A massive GRB-progenitor star $P_1$ of mass $M_1$ undergoes gravitational collapse to an EMBH. During this process a dyadosphere is formed and subsequently the P-GRB and the E-APE are generated in sequence. They propagate and impact, with their photon and neutrino components, on a second supernova-progenitor star $P_2$ of mass $M_2$. Assuming that both stars were generated approximately at the same time, we expect to have $M_2 < M_1$. Under some special conditions of the thermonuclear evolution of the supernova-progenitor star $P_2$, the collision of the P-GRB and the E-APE with the star $P_2$ can induce its supernova explosion}.

Especially relevant to our paradigm are the following data from the Chandra satellite (see \textcite{p00}):
\begin{enumerate}
\item At the arrival time of 37 hr after the initial burst there is evidence of
iron emission lines for GRB~991216.
\item The emission lines are present during the entire observation period of $10^4$ s. The iron lines could also have been produced earlier, before Chandra was observing. Thus the times used in these calculations are not unique: they do serve to provide an example of the scenario.
\item The emission lines appear to have a peak at an energy of $3.49 \pm 0.06$ keV which, at a redshift $z=1.00 \pm 0.02$ corresponds to an hydrogen-like iron line at 6.97 keV at rest. This source does not appear to have any significant motion departing from the cosmological flow. The iron lines have a width of 0.23 keV consistent with a radial velocity field of $0.1c$. The iron lines are only a small fraction of the observed flux.
\end{enumerate}

On the basis of the explicit computations of the different eras presented in the above sections, we make three key points:
\begin{enumerate}
\item An arrival time of $37$ hr in the detector frame corresponds to a radial distance from the EMBH travelled by the ABM pulse of $3.94\times10^{17}$ cm in the laboratory frame (see Tab.~\ref{tab1}).
\item It is likely that a few stars are present within that radius as members of a cluster. It has 
become evident from observations of dense clusters of star-forming regions that a stellar average density of typically $ 10^2 \mbox{pc}^{-3} $ (\textcite{btk00}) should be expected. There is also the distinct possibility for this case and other systems that the stars $P_1$ and  $P_2$ are members of a binary system.
\item The possible observations at different wavelengths of the supernova crucially depend on the relative intensities between the GRB and the supernova as well as on the value of the distance and the redshift of the source. In the present case of GRB~991216, the expected optical and radio emission from the supernova are many orders of magnitude smaller than the GRB intensity. The opposite situation will be encountered in GRB~980425 (\textcite{rbcfx02d_supernova}).
\end{enumerate}

\begin{figure}
\begin{center}
\includegraphics[width=10cm,clip]{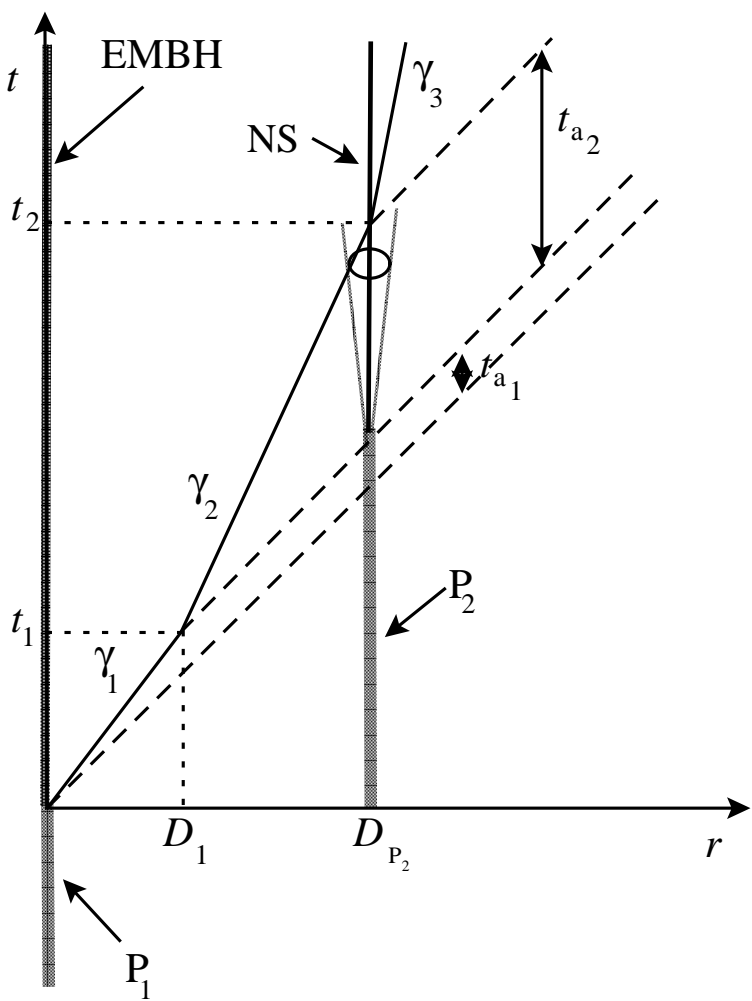}
\caption{A qualitative simplified space-time diagram (in arbitrary units) illustrating the GSTS paradigm. The EMBH, originating from the gravitational collapse of a massive GRB-progenitor star $P_1$, and the massive supernova-progenitor star $P_2$-neutron star ($P_2$-NS) system, separated by a radial distance $D_{P_2}$, are assumed to be at rest in in the laboratory frame. Their worldlines are represented by two parallel vertical lines. The supernova shell moving at $0.1c$ generated by the $P_2$-NS transition is represented by the dotted line cone. The solid line represents the motion of the pulse, as if it would move with an ``effective'' constant gamma factor $\gamma_1$ during the eras reaching the condition of transparency. Similarly, another ``effective'' constant gamma factor $\gamma_2<\gamma_1$ applies during era IV up to the collision with the $P_2$-NS system. A third ``effective'' constant gamma factor $\gamma_3 < \gamma_2$ occurs during era V after the collision as the nonrelativistic regime of expansion is reached. The dashed lines at 45 degrees represent signals propagating at speed of light.}
\label{iii-fig1}
\end{center}
\end{figure}

In order to reach an intuitive understanding of these complex computations we present a schematic very simplified diagram (not to scale) in Fig.~\ref{iii-fig1}.

We now describe the sequence of events and the specific data corresponding to the GSTS paradigm:
\begin{enumerate}
\item The two stars $P_1$ and $P_2$ are separated by a distance $D_{P_2} = 3.94\times 10^{17}$ cm in the laboratory frame, see Fig.~\ref{iii-fig1}. Both stars are at rest in the inertial laboratory frame. At laboratory time $t=0$ and at comoving time $\tau=0$, the gravitational collapse of the GRB-progenitor star $P_1$ occurs, and the initial emission of gravitational radiation or a neutrino burst from the event then synchronizes this event with the arrival times $t_a=0$ at the supernova-progenitor star $P_2$ and $t_a^d=0$ for the distant observer at rest with the detector. The electromagnetic radiation emitted by the gravitational collapse process is instead practically zero, due to the optical thickness of the material at this stage (\textcite{brx00}, see Tab.~\ref{tab1}).

\item From Tab.~\ref{tab1}, at laboratory time $t_1 = 6.48 \times 10^3\,s$ and at a distance from the EMBH of $D_1=1.94\times 10^{14}\,cm$, the condition of transparency for the PEMB pulse is reached and the P-GRB is emitted (see section~\ref{era3}). This time is recorded in arrival time at the detector $t_{a_1}^d = 8.41\times 10^{-2}\,s$, and, at $P_2$, at $t_{a_1}=4.20\times 10^{-2}\, s$. The fact that the PEMB pulse in an arrival time of $8.41\times 10^{-2}\,s$ covers a distance of $1.94\times 10^{14}\,cm$ gives rise to an apparent ``superluminal'' effect. This apparent paradox can be straightforwardly explained by introducing an ``effective'' gamma factor, see \textcite{lett3}.

\item At laboratory time $t = 1.73\times 10^6\,s$ and at a distance from the EMBH of $5.18\times 10^{16}\,cm$ in the laboratory frame, the peak of the E-APE is reached which is recorded at the arrival time $t_a=9.93\,s$ at $P_2$ and $t_a^d=19.87\,s$ at the detector. This also gives rise to an apparent ``superluminal'' effect.

\item At a distance $D_{P_2} = 3.94 \times 10^{17}\,cm $, the two bursts described in the above points 2) and 3) collide with the supernova-progenitor star $P_2$ at arrival times $t_{a_1}=4.20\times 10^{-2}\,s$ and $ t_a  = 9.93\,s $ respectively. They can then induce the supernova explosion of the massive star $P_2$.

\item The associated supernova shell expands with velocity $0.1c$.

\item The expanding supernova shell is reached by the ABM pulse generating the afterglow with a delay of $t_{a_2}=18.5\,hr$ in arrival time following the arrival of the P-GRB and the E-APE. This time delay coincides with the interval of laboratory time separating the two events, since the $P_2$ is at rest in the inertial laboratory frame (see \textcite{lett3}). The ABM pulse has travelled in the laboratory frame a distance $D_{P_2}-D_1\simeq D_{P_2}=3.94\times 10^{17}\,cm$ in a laboratory time $t_2-t_1 \simeq t_2=1.32\times 10^7\,s$ (neglecting the supernova expansion).
\end{enumerate}

The collision of the pulse with the supernova shell occurs at $\gamma\simeq 4.0$. By this time the supernova shell has reached a dimension of $1.997\times 10^{14}\,cm$, which is consistent with the observations from the Chandra satellite.

In these considerations on GRB~991216 the supernova remnant has been assumed to be close to but not exactly along the line of sight extending from the EMBH to the distant observer. If such an alignment should exist for other GRBs, it would lead to an observation of iron absorption lines as well as to an increase in the radiation observed in the afterglow corresponding to the crossing of the supernova shell by the ABM pulse. In fact, as the ABM pulse engulfs the baryonic matter of the remnant, above and beyond the normal interstellar medium baryonic matter, the conservation of energy and momentum implies that a larger amount of internal energy is available and radiated in the process (see section~\ref{era4}). This increased energy-momentum loss will generally affect the slope of the afterglow decay, approaching more rapidly a nonrelativistic expansion phase (details are given in section~\ref{approximation}).

It is quite clear that as soon as the relativistic transformations of the RSTT paradigm are duly taken into account, the sequence of events between the supernova and the GRB occurrences are exactly the opposite of the one postulated in the so-called ``supranova'' scenario (\textcite{vs98,vs99,va99}). This can be considered a very appropriate pedagogical example of how classical nonrelativistic applied to ultrarelativistic regimes can indeed subvert the very causal relation between events.

If we now turn to the possibility of dynamically implementing the scenario, there are at least three different possibilities:
\begin{enumerate}
\item Particularly attractive is the possibility that a massive star $P_2$ has rapidly evolved during its thermonuclear evolution to a white dwarf (see e.g. \textcite{c78}). It it then sufficient that the P-GRB and the E-APE implode the star sufficiently as to reach a central density above the critical density for the ignition of thermonuclear burning. Consequently, the explosion of the star $P_2$ occurs, and a significant fraction of a solar mass of iron is generated. These configurations are currently generally considered precursors of some type I supernovae (see e.g. \textcite{f97} and references therein).
\item Alternatively, the massive star $P_2$ can have evolved to the condition of being close to the point of gravitational collapse, having developed the formation of an iron-silicon core, type II supernovae. The above transfer of energy momentum from the P-GRB and the E-APE may enhance the capture of the electrons on the iron nuclei and consequently decrease the Fermi energy of the core, leading to the onset of gravitational instability (see e.g. \textcite{b91} p. 270 and followings). Since the time for the final evolution of a massive star with an iron-silicon core is short, this event requires a well tuned coincidence.
\item The pressure wave may trigger massive and instantaneous nuclear burning process, with corresponding changes in the chemical composition of the star, leading to the collapse.
\end{enumerate}

The GSTS paradigm has been applied to the case of the GRB 980425 - SN1998bw which, with a red shift of 0.0083, is one of the closest and weaker GRBs observed. In this case, the radio and the optical emission of the supernova is distinctively observed. For this particular case, the EMBH appears to have a significantly lower value of the parameter $\xi$ and the validity of the GSTS paradigm presented here is confirmed (see \textcite{rbcfx02d_supernova}).

\section{General considerations on the EMBH formation}\label{gc}

Before concluding let us consider the problem of the EMBH formation. Such a problem has been debated for many years since the earliest discussions in 1970 in Princeton and has been finally clarified and addressed in general terms to justify the plausibility of the hypothesis in \textcite{r01mg9}. There has been a basic change of paradigm. All the considerations on the electric charge of stars were traditionally directed, following the classical work by \textcite{s70} all the way to the fundamental book by \textcite{punsly_book}, to the presence of a net charge on the star surface in a steady state condition. The star can be endowed of rotation and magnetic field and surrounded by plasma, like in the case of \textcite{gj69}, or, in the case of absence of both magnetic field and rotation, the electrostatic processes can be related to the depth of the gravitational well, like in the treatment of \textcite{s70}. However, in neither cases it is possible to reach the condition of the overcritical field needed for pair creation nor have the condition of no baryonic contamination discussed in sections~\ref{dyadosphere},~\ref{era1} and essential for the dyadosphere formation. The basic conceptual point is that GRBs are maybe the most violent transient phenomenon occurring in the universe and so the condition for the dyadosphere creation have to be searched in a transient phenomenon. The solution is related to the most transient phenomenon occurring in the life of a star: the process of gravitational collapse.

Having acquired such a fundamental understanding, the next step is to estimate the amount of polarization needed in order to reach the fully relativistic condition
\begin{equation}
\frac{Q}{M\sqrt{G}}=1\; .
\label{gc_eq1}
\end{equation}
Recalling that the charge to mass ratio of a proton is $q_p/\left(m_p\sqrt{G}\right)=1.1\times 10^{18}$, it is enough to have an excess of one quantum of charge every $10^{18}$ nucleons in the core of the collapsing star to obtain an extreme EMBH after the occurrence of the gravitational collapse. Physically this means that we are dealing with a process of charge segregation between the core and the outer part of the star which has the opposite sign of net charge in order to enforce the overall charge neutrality condition. We here emphasize the name ``charge segregation'' instead of the name ``charge separation'' in order to contrast a very mild charge surplus created in different part of the star, keeping the overall charge neutrality, from the much more extreme condition of charge separation in which all the charges of the atomic component of the star are separated. It is indeed reassuring that such a core, endowed with charge segregation, is indeed stable with respect to the Fermi-Chandrasekhar criteria for the stability of self-gravitating stars duly extended from the magnetic to the electric case: the electric energy of such a core is consistently smaller than its gravitational energy (see \textcite{bor02}).

Such a condition of charge segregation between the core and the oppositely charged star surface layer can be reached under a very large number of physical conditions. We consider, for simplicity, one of the oldest example: the one of a star endowed with both a magnetic field and rotation. It is proved that a typical magnetic field expected for the ISM is $B_\circ\sim 10^{-5}\, G$ (\textcite{fe01}). We further assume, consistently with the data which we have acquired and verified in the present article (see sections~\ref{era4},~\ref{power-law}), that also in the galaxy where GRB~991216 occurred the ISM has an average density of $n_{ism}=1\, proton/cm^3$. From this value of density we have that an ISM cloud with mass $M\sim 10M_\odot$ occupies a sphere of radius $R_\circ\sim 1.4\times 10^{19}\, cm$. If this sphere collapse to a star with radius $R=R_\odot$, from the flux conservation we obtain that it is enough for this star to rotate with the most reasonable angular speed
\begin{equation}
\Omega\sim\frac{\xi Mc\sqrt{G}}{R_\odot R_\circ^2 B_\circ}
\label{omega}
\end{equation}
to conclude that the progenitor star core is endowed of a charge to mass ratio equal to $\xi$. In the extreme case of Eq.(\ref{gc_eq1}) we have $\xi=1$ and so the angular speed is $\Omega\sim 1.1\times 10^{-3}\, rad/s$ --- i.e. one round in $1.5\, hr$ --- and correspondingly we have smaller $\Omega$ values for $\xi < 1$ (see \textcite{bor02}). Clearly the overall neutrality is guaranteed by the oppositely charged baryonic matter which is the one measured by the $B$ parameter in the EMBH model (see sections~\ref{era2}--\ref{era3}). The smallness of the $B$ value clearly points to the absence of an extended envelope of the progenitor star.

The formation process of such an electromagnetised progenitor star will be clearly affected by the presence of differential rotation, the consequent amplification of the magnetic field and a variety of magnetohydrodynamical problems which will affect somewhat the simplicity of the heuristic Eq.(\ref{omega}). Similarly the process of gravitational collapse of such a progenitor star endowed with rotation will lead to complex phenomena of ``gravitationally induced electromagnetic radiation'' (\textcite{ja73}) and of ``electromagnetically induced gravitational radiation'' (\textcite{ja74}) which will tend to reduce both the eccentricity and the angular velocity of the collapsing core. The general outcome of gravitational collapse will be a Kerr-Newmann spacetime. It is interesting that such a general case will break the degeneracy in $\left(\mu,\xi\right)$ described in section~\ref{fp} (see \textcite{rbcfx02e_paperII}). In this article we have addressed the much simpler case of a solution in which $\left(cL\right)/\left(GM^2\right) \ll 1$ and the treatment can be well approximated by a collapse described by a Reissner-Nordstr\"{o}m geometry.

\begin{figure}
\includegraphics[width=\hsize,clip]{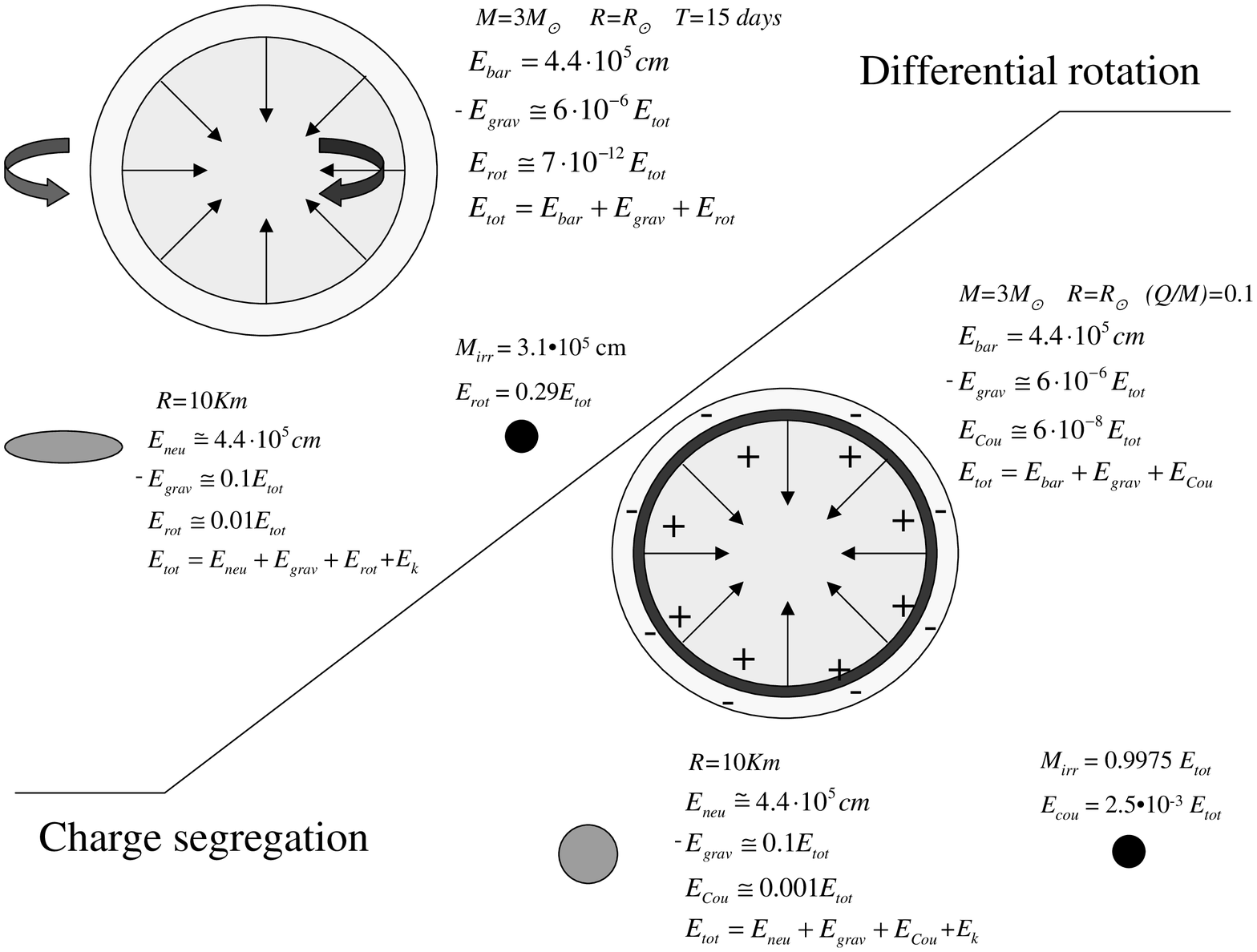}
\caption{Quantitative description of the gravitational collapse to a neutron star and to a black hole of the core of a rotating progenitor rotating. The core is estimated to have a mass equal to $3M_{\odot}$, to have an initial radius $r=r_{\odot}$ and a rotation period of $15$ days. Although the initial rotational energy is of the order of $10^{-11}$ of the total energy, the total rotational energy, in principle extractable, of the rotating black hole can be as high as of the order of $29\%$. On the lower-right side the same considerations are applied to the case of a neutral star formed by a core oppositely charged from its outermost envelope. The core is expected to have a mass of $3M_{\odot}$, a radius equal to $r_{\odot}$ and electromagnetic energy $Q/M=0.1$. Although the initial Coulomb energy is only $\sim 10^{-7}$ of the total energy, which is in turn hundred times smaller than the gravitational energy, the final Coulomb energy can be as high as $2.5\times 10^{-3}$ of the total energy. In both cases, the amplification of the rotational energy and of the Coulomb energy, which indeed are the only two extractable forms of energy from a black hole, is due to the process of gravitational collapse.}
\label{diapmg9}
\end{figure}

In addition to this scenario, based on the role of magnetic field and rotation, we are as well pursuing the possible generation of the charge segregation by quantum effects at the surface of the Fermi semi-degenerate core. In this framework, it is particularly interesting to consider the purely electric analog of the \textcite{cf53} paper on the gravitational stability of self-gravitating magnetized stars. The stability condition, based on the virial theorem, is simply that the Coulomb energy of the inner core of a charged star should be smaller or equal than the gravitational energy of the star (\textcite{bor02}). Previous to the collapse, the gravitational energy can be much smaller than the rest energy of the star and be amplified during the process of gravitational collapse reaching overcritical intensity of the electric field (see Fig.~\ref{diapmg9} and \textcite{r01mg9}). It is interesting that the Chandrasekhar-Fermi inequality just leads to an extreme Reissner-Nordstr\"{o}m solution.

In both these cases the Reissner-Nordstr\"{o}m geometry appears indeed to be the relevant model for GRB~991216 as discussed in the previous sections. We shall return to non spherical configuration in forthcoming publications and/or when requested by observational evidence (see \textcite{rbcfx02e_paperII}).

\section{Some propaedeutic analysis for the dynamical formation of the EMBH}\label{luca}

While the formation in time of the dyadosphere is the fundamental phenomena we are interested in, we can get an insight on the issue of gravitational collapse of an electrically charged star core studying in details a simplified model, namely a thin shell of charged dust. In \textcite{i66,idlc67} it is shown that the problem of a collapsing charged shell in general relativity can be reduced to a set of ordinary differential equations. We reconsider here the following relativistic system: a spherical shell of electrically charged dust which is moving radially in the Reissner-Nordstr\"{o}m background of an already formed nonrotating EMBH of mass $M_{1}$ and charge $Q_{1}$, with $Q_{1}\leq M_{1}$.

The world surface spanned by the shell divides the space-time into two regions: an internal one $\mathcal{M}_{-}$ and an external one $\mathcal{M}_{+}$. The line element in Schwarzschild like coordinate is (\textcite{crv02})
\begin{equation}
ds^{2}=\left\{
\begin{array}[c]{l}
-f_{+}dt_{+}^{2}+f_{+}^{-1}dr^{2}+r^{2}d\Omega^{2}\qquad\text{in } \mathcal{M}_{+}\\
-f_{-}dt_{-}^{2}+f_{-}^{-1}dr^{2}+r^{2}d\Omega^{2}\qquad\text{in } \mathcal{M}_{-}
\end{array}
\right.  , \label{E0}
\end{equation}
where $f_{+}=1-\tfrac{2M}{r}+\tfrac{Q^{2}}{r^{2}}$, $f_{-}=1-\tfrac{2M_{1}} {r}+\tfrac{Q_{1}^{2}}{r^{2}}$ and $t_{-}$ and $t_{+}$ are the Schwarzschild-like time coordinates in $\mathcal{M}_{-}$ and $\mathcal{M}_{+}$ respectively. $M$ is the total mass-energy of the system formed by the shell and the EMBH, measured by an observer at rest at infinity and $Q=Q_{0}+Q_{1}$ is the total charge: sum of the charge $Q_{0}$ of the shell and the charge $Q_{1}$ of the internal EMBH.

Indicating by $R$ the radius of the shell and by $T_{\pm}$ its time coordinate, the equations of motion of the shell become (\textcite{rv02a})
\begin{align}
\left(  \tfrac{dR}{d\tau}\right)  ^{2}  &  =\tfrac{1}{M_{0}^{2}}\left(M-M_{1}+\tfrac{M_{0}^{2}}{2R}-\tfrac{Q_{0}^{2}}{2R}-\tfrac{Q_{1}Q_{0}}{R}\right)  ^{2}-f_{-}\left(  R\right) \nonumber\\
&  =\tfrac{1}{M_{0}^{2}}\left(  M-M_{1}-\tfrac{M_{0}^{2}}{2R}-\tfrac{Q_{0}^{2}}{2R}-\tfrac{Q_{1}Q_{0}}{R}\right)  ^{2}-f_{+}\left(  R\right),\label{EQUYa}\\
\tfrac{dT_{\pm}}{d\tau}  &  =\tfrac{1}{M_{0}f_{\pm}\left(  R\right)  }\left(M-M_{1}\mp\tfrac{M_{0}^{2}}{2R}-\tfrac{Q_{0}^{2}}{2R}-\tfrac{Q_{1}Q_{0}}{R}\right)  , \label{EQUYb}
\end{align}
where $M_{0}$ is the rest mass of the shell and $\tau$ is its proper time. Eqs.(\ref{EQUYa},\ref{EQUYb}) (together with Eq.(\ref{E0})) completely describe a 5-parameter ($M$, $Q$, $M_{1}$, $Q_{1}$, $M_{0}$) family of solutions of the Einstein-Maxwell equations. Note that Eqs.(\ref{EQUYa},\ref{EQUYb}) imply that
\begin{equation}
M-M_{1}-\tfrac{Q_{0}^{2}}{2R}-\tfrac{Q_{1}Q_{0}}{R}>0
\label{Constraint}
\end{equation}
holds for $R>M+\sqrt{M^{2}-Q^{2}}$ if $Q<M$ and for $R>M_{1}+\sqrt{M_{1}^{2}-Q_{1}^{2}}$ if $Q>M$.

For astrophysical applications (\textcite{rvx02}) the trajectory of the shell $R=R\left(  T_{+}\right)  $ is obtained as a function of the time coordinate $T_{+}$ relative to the space-time region $\mathcal{M}_{+}$. In the following we drop the $+$ index from $T_{+}$. From Eqs.(\ref{EQUYa},\ref{EQUYb}) we have
\begin{equation}
\tfrac{dR}{dT}=\tfrac{dR}{d\tau}\tfrac{d\tau}{dT}=\pm\tfrac{F}{\Omega} \sqrt{\Omega^{2}-F},
\label{EQUAISRDLC}
\end{equation}
where
\begin{align}
F  & \equiv f_{+}\left(  R\right)  =1-\tfrac{2M}{R}+\tfrac{Q^{2}}{R^{2}},\\
\Omega & \equiv\Gamma-\tfrac{M_{0}^{2}+Q^{2}-Q_{1}^{2}}{2M_{0}R},\\
\Gamma & \equiv\tfrac{M-M_{1}}{M_{0}}.
\end{align}
Since we are interested in an imploding shell, only the minus sign case in (\ref{EQUAISRDLC}) will be studied. We can give the following physical interpretation of $\Gamma$. If $M-M_{1}\geq M_{0}$, $\Gamma$ coincides with the Lorentz $\gamma$ factor of the imploding shell at infinity; from Eq.(\ref{EQUAISRDLC}) it satisfies
\begin{equation}
\Gamma=\tfrac{1}{\sqrt{1-\left(  \frac{dR}{dT}\right)  _{R=\infty}^{2}}}\geq1.
\end{equation}
When $M-M_{1}<M_{0}$ then there is a \emph{turning point} $R^{\ast}$, defined by $\left.  \tfrac{dR}{dT}\right|  _{R=R^{\ast}}=0$. In this case $\Gamma$ coincides with the ``effective potential'' at $R^{\ast}$ :
\begin{equation}
\Gamma=\sqrt{f_{-}\left(  R^{\ast}\right)  }+M_{0}^{-1}\left(  -\tfrac{M_{0}^{2}}{2R^{\ast}}+\tfrac{Q_{0}^{2}}{2R^{\ast}}+\tfrac{Q_{1}Q_{0}}{R^{\ast}}\right)
\leq1.
\end{equation}
The solution of the differential equation (\ref{EQUAISRDLC}) is given by:
\begin{equation}
\int dT=-\int\tfrac{\Omega}{F\sqrt{\Omega^{2}-F}}dR.
\label{GRYD}
\end{equation}
The functional form of the integral (\ref{GRYD}) crucially depends on the degree of the polynomial $P\left(  R\right)  =R^{2}\left(  \Omega^{2}-F\right)  $, which is generically two, but in special cases has lower values. We therefore distinguish the following cases:

\begin{enumerate}
\item {\boldmath$M=M_{0}+M_{1}$}; {\boldmath$Q_{1}=M_{1}$}; {\boldmath $Q=M$}: $P\left(  R\right)  $ is equal to $0$, we simply have
\begin{equation}
R(T)=\mathrm{{const}.}
\end{equation}
\item {\boldmath$M=M_{0}+M_{1}$}; {\boldmath$M^{2}-Q^{2}=M_{1}^{2}-Q_{1}^{2}$}; {\boldmath$Q\neq M$}: $P\left(  R\right)  $ is a constant, we have
\begin{align}
T  & =\mathrm{const}+\tfrac{1}{2\sqrt{M^{2}-Q^{2}}}\left[  \left(
R+2M\right)  R\right.  \nonumber\\
& \left.  +r_{+}^{2}\log\left(  \tfrac{R-r_{+}}{M}\right)  +r_{-}^{2}%
\log\left(  \tfrac{R-r_{-}}{M}\right)  \right]  .\label{CASO1}%
\end{align}
\item {\boldmath$M=M_{0}+M_{1}$}; {\boldmath$M^{2}-Q^{2}\neq M_{1}^{2}-Q_{1}^{2}$}: $P\left(  R\right)  $ is a first order polynomial and
\begin{align}
T &  =\mathrm{const}+2R\sqrt{\Omega^{2}-F}\left[  \tfrac{M_{0}R}{3\left(
M^{2}-Q^{2}-M_{1}^{2}+Q_{1}^{2}\right)  }\right.  \nonumber\\
&  \left.  +\tfrac{\left(  M_{0}^{2}+Q^{2}-Q_{1}^{2}\right)  ^{2}%
-9MM_{0}\left(  M_{0}^{2}+Q^{2}-Q_{1}^{2}\right)  +12M^{2}M_{0}^{2}%
+2Q^{2}M_{0}^{2}}{3\left(  M^{2}-Q^{2}-M_{1}^{2}+Q_{1}^{2}\right)  ^{2}%
}\right]  \nonumber\\
&  -\tfrac{1}{\sqrt{M^{2}-Q^{2}}}\left[  r_{+}^{2}\mathrm{arctanh}\left(
\tfrac{R}{r_{+}}\tfrac{\sqrt{\Omega^{2}-F}}{\Omega_{+}}\right)  \right.
\nonumber\\
&  \left.  -r_{-}^{2}\mathrm{arctanh}\left(  \tfrac{R}{r_{-}}\tfrac
{\sqrt{\Omega^{2}-F}}{\Omega_{-}}\right)  \right]  ,\label{CASO2}%
\end{align}

where $\Omega_{\pm}\equiv\Omega\left(  r_{\pm}\right)  $.

\item {\boldmath$M\neq M_{0}+M_{1}$}: $P\left(  R\right)  $ is a second order polynomial and
\begin{align}
T &  =\mathrm{const}-\tfrac{1}{2\sqrt{M^{2}-Q^{2}}}\left\{  \tfrac
{2\Gamma\sqrt{M^{2}-Q^{2}}}{\Gamma^{2}-1}R\sqrt{\Omega^{2}-F}\right.
\nonumber\\
&  +r_{+}^{2}\log\left[  \tfrac{R\sqrt{\Omega^{2}-F}}{R-r_{+}}+\tfrac
{R^{2}\left(  \Omega^{2}-F\right)  +r_{+}^{2}\Omega_{+}^{2}-\left(  \Gamma
^{2}-1\right)  \left(  R-r_{+}\right)  ^{2}}{2\left(  R-r_{+}\right)
R\sqrt{\Omega^{2}-F}}\right]  \nonumber\\
&  -r_{-}^{2}\log\left[  \tfrac{R\sqrt{\Omega^{2}-F}}{R-r_{-}}+\tfrac
{R^{2}\left(  \Omega^{2}-F\right)  +r_{-}^{2}\Omega_{-}^{2}-\left(  \Gamma
^{2}-1\right)  \left(  R-r_{-}\right)  ^{2}}{2\left(  R-r_{-}\right)
R\sqrt{\Omega^{2}-F}}\right]  \nonumber\\
&  -\tfrac{\left[  2MM_{0}\left(  2\Gamma^{3}-3\Gamma\right)  +M_{0}^{2}%
+Q^{2}-Q_{1}^{2}\right]  \sqrt{M^{2}-Q^{2}}}{M_{0}\left(  \Gamma^{2}-1\right)
^{3/2}}\log\left[  \tfrac{R\sqrt{\Omega^{2}-F}}{M}\right.  \nonumber\\
&  \left.  \left.  +\tfrac{2M_{0}\left(  \Gamma^{2}-1\right)  R-\left(
M_{0}^{2}+Q^{2}-Q_{1}^{2}\right)  \Gamma+2M_{0}M}{2M_{0}M\sqrt{\Gamma^{2}-1}%
}\right]  \right\}  .\label{CASO3}%
\end{align}
\end{enumerate}

\begin{figure}
\includegraphics[width=8.5cm,clip]{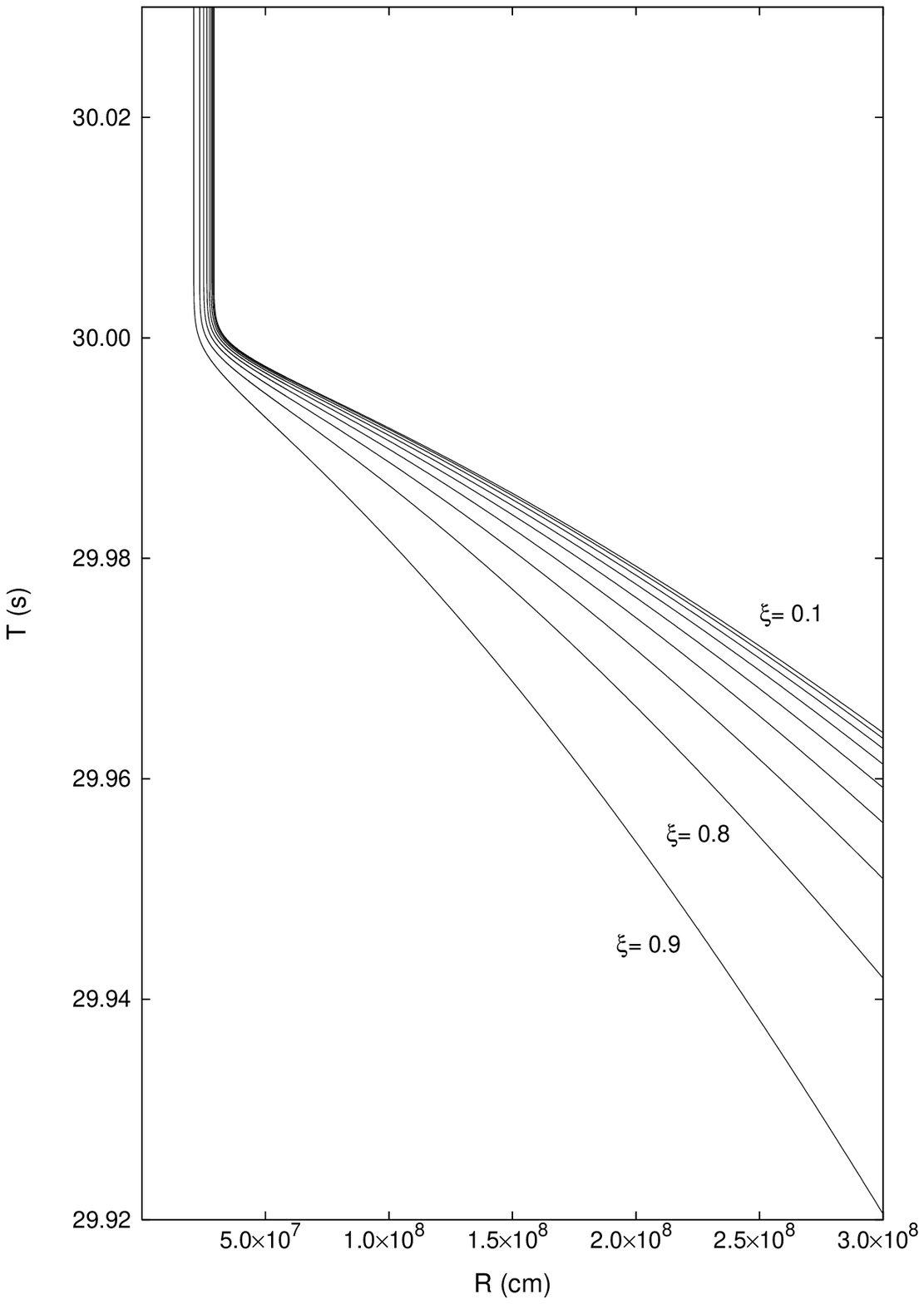}
\includegraphics[width=8.5cm,clip]{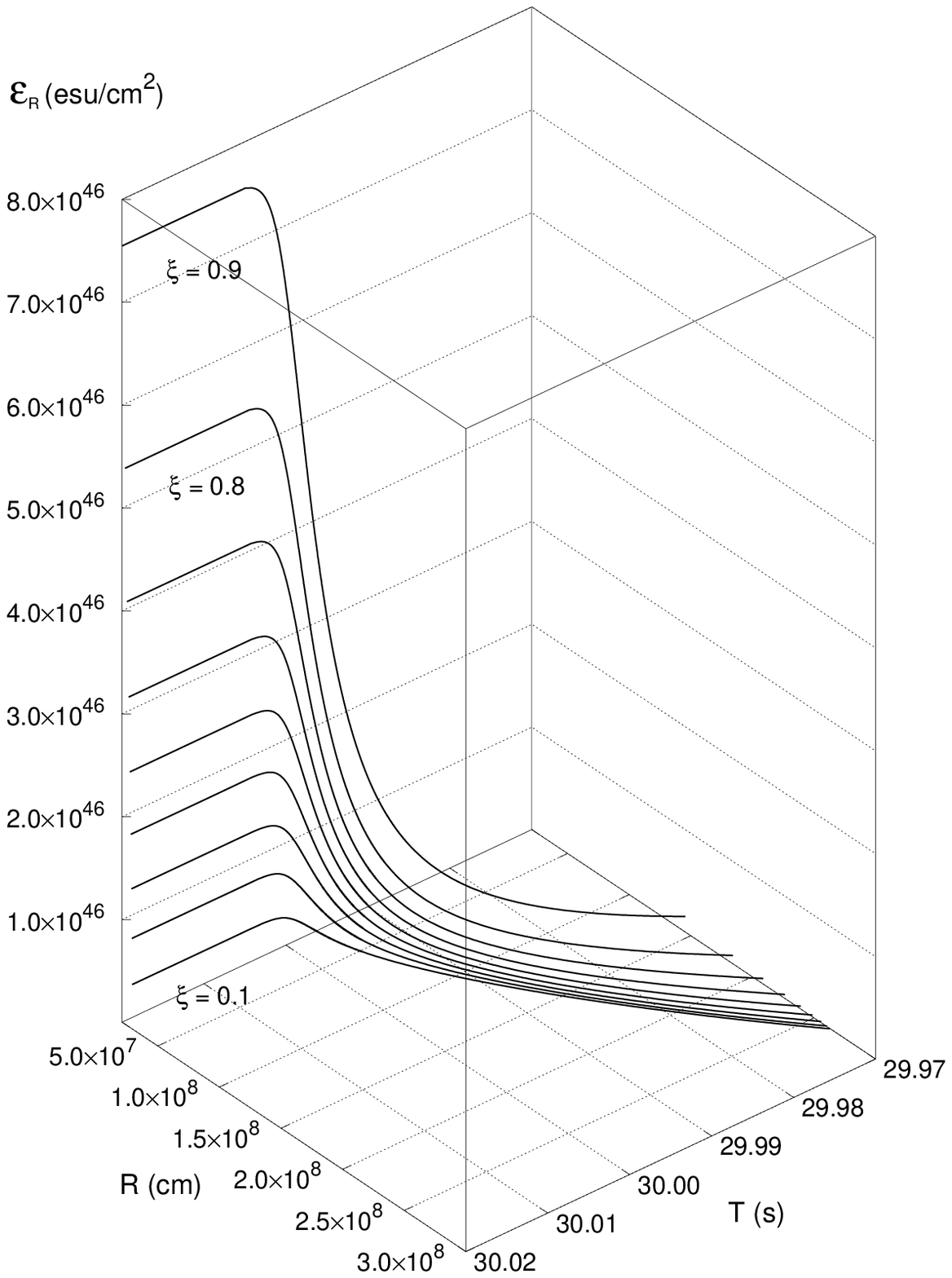}
\caption{{\bf Left)} Collapse curves in the plane $(T,R)$ for $M=20M_{\odot}$ and for different values of the parameter $\xi$. The asymptotic behavior is the clear manifestation of general relativistic effects as the horizon of the EMBH is approached. {\bf Right)} Electric field behaviour at the surface of the shell for $M=20M_{\odot}$ and for different values of the parameter $\xi$. The asymptotic behavior is the clear manifestation of general relativistic effects as the horizon of the EMBH is approached.}
\label{elec}
\label{elec3d}
\end{figure}

Of particular interest is the time varying electric field $\mathcal{E}%
_{R}=\tfrac{Q}{R^{2}}$ on the external surface of the shell. In order to study
the variability of $\mathcal{E}_{R}$ with time it is useful to consider in the
tridimensional space of parameters $(R,T,\mathcal{E}_{R})$ the parametric
curve $\mathcal{C}:\left(  R=\lambda,\quad T=T(\lambda),\quad\mathcal{E}%
_{R}=\tfrac{Q}{\lambda^{2}}\right)  $. In astrophysical applications
(\textcite{rvx02}) we are specially interested in the family of solutions such that
$\frac{dR}{dT}$ is 0 when $R=\infty$ which implies that $\Gamma=1$. In Fig.~
\ref{elec} we plot the collapse curves in the plane $(T,R)$ for different
values of the parameter $\xi\equiv\frac{Q}{M}$, $0<\xi<1$. The initial data
$\left(  T_{0},R_{0}\right)  $ are chosen so that the integration constant in
equation (\ref{CASO2}) is equal to 0. In all the cases we can follow the
details of the approach to the horizon which is reached in an infinite
Schwarzschild time coordinate. In Fig.~\ref{elec3d} we plot the parametric
curves $\mathcal{C}$ in the space $(R,T,\mathcal{E}_{R})$ for different values
of $\xi$. Again we can follow the exact asymptotic behavior of the curves
$\mathcal{C}$, $\mathcal{E}_{R}$ reaching the asymptotic value $\frac{Q}%
{r_{+}^{2}}$. The detailed knowledge of this asymptotic behavior is of great
relevance for the observational properties of the EMBH formation (see e.g. \textcite{rv02a}).

In the case of a shell falling in a flat background ($M_{1}=Q_{1}=0$) Eq.(\ref{EQUYa}) reduces to
\begin{equation}
\left(  \tfrac{dR}{d\tau}\right)  ^{2}=\tfrac{1}{M_{0}^{2}}\left(
M+\tfrac{M_{0}^{2}}{2R}-\tfrac{Q^{2}}{2R}\right)  ^{2}-1. \label{EQUY2}%
\end{equation}
Introducing the total radial momentum $P\equiv M_{0}u^{r}=M_{0}\tfrac
{dR}{d\tau}$ of the shell, we can express the kinetic energy of the shell as
measured by static observers in $\mathcal{M}_{-}$ as $T\equiv-M_{0}u_{\mu}%
\xi_{-}^{\mu}-M_{0}=\sqrt{P^{2}+M_{0}^{2}}-M_{0}$. Then from equation
(\ref{EQUY2}) we have
\begin{equation}
M=-\tfrac{M_{0}^{2}}{2R}+\tfrac{Q^{2}}{2R}+\sqrt{P^{2}+M_{0}^{2}}%
=M_{0}+T-\tfrac{M_{0}^{2}}{2R}+\tfrac{Q^{2}}{2R}. \label{EQC}%
\end{equation}
where we choose the positive root solution due to the constraint
(\ref{Constraint}). Eq.(\ref{EQC}) is the \emph{mass formula} of the shell,
which depends on the time-dependent radial coordinate $R$ and kinetic energy
$T$. If $M\geq Q$, an EMBH is formed and we have
\begin{equation}
M=M_{0}+T_{+}-\tfrac{M_{0}^{2}}{2r_{+}}+\tfrac{Q^{2}}{2r_{+}}\,, \label{EQL}%
\end{equation}
where $T_{+}\equiv T\left(  r_{+}\right)  $ and $r_{+}=M+\sqrt{M^{2}-Q^{2}}$
is the radius of external horizon of the EMBH. We know from the
Christodoulou-Ruffini EMBH mass formula that
\begin{equation}
M=M_{\mathrm{irr}}+\tfrac{Q^{2}}{2r_{+}}, \label{irrmass}%
\end{equation}
so it follows that {}
\begin{equation}
M_{\mathrm{irr}}=M_{0}-\tfrac{M_{0}^{2}}{2r_{+}}+T_{+}, \label{EQM}
\end{equation}
namely that $M_{\mathrm{irr}}$ is the sum of only three contributions: the rest
mass $M_{0}$, the gravitational potential energy and the kinetic energy of the
rest mass evaluated at the horizon. $M_{\mathrm{irr}}$ is independent of the
electromagnetic energy, a fact noticed by Bekenstein (\textcite{b71}). We have taken
one further step here by identifying the independent physical contributions to
$M_{\mathrm{irr}}$.

Next we consider the physical interpretation of the electromagnetic term
$\tfrac{Q^{2}}{2R}$, which can be obtained by evaluating the conserved Killing
integral
\begin{align}
\int_{\Sigma_{t}^{+}}\xi_{+}^{\mu}T_{\mu\nu}^{\mathrm{(em)}}d\Sigma^{\nu}  &
=\int_{R}^{\infty}r^{2}dr\int_{0}^{1}d\cos\theta\int_{0}^{2\pi}d\phi
\ T^{\mathrm{(em)}}{}{}_{0}{}^{0}\nonumber\\
& =\tfrac{Q^{2}}{2R}\,,\label{EQR}%
\end{align}
where $\Sigma_{t}^{+}$ is the space-like hypersurface in $\mathcal{M}_{+}$
described by the equation $t_{+}=t=\mathrm{const}$, with $d\Sigma^{\nu}$ as
its surface element vector and where $T_{\mu\nu}^{\mathrm{(em)}}=-\tfrac
{1}{4\pi}\left(  F_{\mu}{}^{\rho}F_{\rho\nu}+\tfrac{1}{4}g_{\mu\nu}%
F^{\rho\sigma}F_{\rho\sigma}\right)  $ is the energy-momentum tensor of the
electromagnetic field. The quantity in Eq.(\ref{EQR}) differs from the purely
electromagnetic energy
\[
\int_{\Sigma_{t}^{+}}n_{+}^{\mu}T_{\mu\nu}^{\mathrm{(em)}}d\Sigma^{\nu}%
=\tfrac{1}{2}\int_{R}^{\infty}dr\sqrt{g_{rr}}\tfrac{Q^{2}}{r^{2}},
\]
where $n_{+}^{\mu}=f_{+}^{-1/2}\xi_{+}^{\mu}$ is the unit normal to the
integration hypersurface and $g_{rr}=f_{+}$. This is similar to the analogous
situation for the total energy of a static spherical star of energy density
$\epsilon$ within a radius $R$, $m\left(  R\right)  =4\pi\int_{0}^{R}%
dr\ r^{2}\epsilon$, which differs from the pure matter energy $m_{\mathrm{p}%
}\left(  R\right)  $ $=4\pi\int_{0}^{R}dr\sqrt{g_{rr}}r^{2}\epsilon$ by the
gravitational energy (see \textcite{mtw73}). Therefore the term $\tfrac{Q^{2}}%
{2R}$ in the mass formula (\ref{EQC}) is the \emph{total} energy of the
electromagnetic field and includes its own gravitational binding energy. This
energy is stored throughout the region $\Sigma_{t}^{+}$, extending from $R$ to infinity.

We now turn to the problem of extracting the electromagnetic energy from an
EMBH see (see \textcite{cr71}). We can distinguish between two conceptually physically
different processes, depending on whether the electric field strength
$\mathcal{E}=\frac{Q}{r^{2}}$ is smaller or greater than the critical value
$\mathcal{E}_{\mathrm{c}}=\tfrac{m_{e}^{2}c^{3}}{e\hbar}$. Here $m_{e}$ and
$e$ are the mass and the charge of the electron. As already mentioned in this
paper an electric field $\mathcal{E}>\mathcal{E}_{\mathrm{c}}$ polarizes the
vacuum creating electron-positron pairs (see \textcite{he35}). The maximum value
$\mathcal{E}_{+}=\tfrac{Q}{r_{+}^{2}}$ of the electric field around an EMBH is
reached at the horizon. We then have the following:

\begin{enumerate}
\item  For $\mathcal{E}_{+}<\mathcal{E}_{\mathrm{c}}$ the leading energy
extraction mechanism consists of a sequence of discrete elementary decay
processes of a particle into two oppositely charged particles. The condition
$\mathcal{E}_{+}<\mathcal{E}_{\mathrm{c}}$ implies
\begin{align}
\xi & \equiv\tfrac{Q}{\sqrt{G}M}\nonumber\\
& \lesssim\left\{
\begin{array}
[c]{r}%
\tfrac{GM/c^{2}}{\lambda_{\mathrm{C}}}\tfrac{\sqrt{G}m_{e}}{e}\sim
10^{-6}\tfrac{M}{M_{\odot}}\quad\text{if }\tfrac{M}{M_{\odot}}\leq10^{6}\\
1\quad\quad\quad\quad\quad\quad\quad\quad\text{if }\tfrac{M}{M_{\odot}}>10^{6}%
\end{array}
\right.  ,\label{critical3}%
\end{align}
where $\lambda_{\mathrm{C}}$ is the Compton wavelength of the electron.
\textcite{dr73} and \textcite{dhr74} have defined as the \emph{effective ergosphere} the region around an EMBH
where the energy extraction processes occur. This region extends from the
horizon $r_{+}$ up to a radius
\begin{align}
r_{\mathrm{Eerg}}  & =\tfrac{GM}{c^{2}}\left[  1+\sqrt{1-\xi^{2}\left(
1-\tfrac{e^{2}}{G{m_{e}^{2}}}\right)  }\right]  \nonumber\\
& \simeq\tfrac{e}{m_{e}}\tfrac{Q}{c^{2}}\,.\label{EffErg}%
\end{align}
The energy extraction occurs in a finite number $N_{\mathrm{PD}}$ of such
discrete elementary processes, each one corresponding to a decrease of the EMBH
charge. We have
\begin{equation}
N_{\mathrm{PD}}\simeq\tfrac{Q}{e}\,.
\end{equation}
Since the total extracted energy is (see Eq.~(\ref{irrmass})) $E^{\mathrm{tot}%
}=\tfrac{Q^{2}}{2r_{+}}$, we obtain for the mean energy per accelerated
particle $\left\langle E\right\rangle _{\mathrm{PD}}=\tfrac{E^{\mathrm{tot}}%
}{N_{\mathrm{PD}}}$
\begin{equation}
\left\langle E\right\rangle _{\mathrm{PD}}=\tfrac{Qe}{2r_{+}}=\tfrac{1}%
{2}\tfrac{\xi}{1+\sqrt{1-\xi^{2}}}\tfrac{e}{\sqrt{G}m_{e}}\ m_{e}c^{2}%
\simeq\tfrac{1}{2}\xi\tfrac{e}{\sqrt{G}m_{e}}\ m_{e}c^{2},
\end{equation}
which gives
\begin{equation}
\left\langle E\right\rangle _{\mathrm{PD}}\lesssim\left\{
\begin{array}
[c]{r}%
\left(  \tfrac{M}{M_{\odot}}\right)  \times10^{21}eV\quad\text{if }\tfrac
{M}{M_{\odot}}\leq10^{6}\\
10^{27}eV\quad\quad\text{if }\tfrac{M}{M_{\odot}}>10^{6}%
\end{array}
\right.  . \label{UHECR}%
\end{equation}
One of the crucial aspects of the energy extraction process from an EMBH is
its back reaction on the irreducible mass expressed in \textcite{cr71}. Although
the energy extraction processes can occur in the entire effective ergosphere
defined by Eq. (\ref{EffErg}), only the limiting processes occurring on the
horizon with zero kinetic energy can reach the maximum efficiency while
approaching the condition of total reversibility (see Fig.~2 in \textcite{cr71} for details). 
The farther from the horizon that a decay occurs, the more it
increases the irreducible mass and loses efficiency. Only in the complete
reversibility limit (\textcite{cr71}) can the energy extraction process from an
extreme EMBH reach the upper value of $50\%$ of the total EMBH energy.

\item  For $\mathcal{E}_{+}\geq\mathcal{E}_{\mathrm{c}}$ the leading
extraction process is a \emph{collective} process based on an
electron-positron plasma generated by the vacuum polarization, (see Fig.~\ref{dyaon}) 
as discussed in section \ref{dyadosphere} The condition $\mathcal{E}_{+}%
\geq\mathcal{E}_{\mathrm{c}}$ implies
\begin{equation}
\tfrac{GM/c^{2}}{\lambda_{\mathrm{C}}}\left(  \tfrac{e}{\sqrt{G}m_{e}}\right)
^{-1}\simeq2\cdot10^{-6}\tfrac{M}{M_{\odot}}\leq\xi\leq1\,.
\end{equation}
This vacuum polarization process can occur only for an EMBH with mass smaller
than $2\cdot10^{6}M_{\odot}$. The electron-positron pairs are now produced in
the dyadosphere of the EMBH, (note that the dyadosphere is a subregion of the
effective ergosphere) whose radius $r_{ds}$ is given in Eq.(\ref{rc}).
We have $r_{ds}\ll r_{\mathrm{Eerg}}$. The number of particles
created and the total energy stored in dyadosphere are given in Eqs.(\ref{tn},\ref{tee})
respectively and we have approximately
\begin{align}
N^\circ_{e^+e^-}  & \simeq\left(  \tfrac{r_{ds}}%
{\lambda_{\mathrm{C}}}\right)  \tfrac{Q}{e}\,,\label{numdya}\\
E_{dya}  & \simeq\tfrac{Q^{2}}{2r_{+}}\,
\end{align}
The mean energy per particle produced in the dyadosphere $\left\langle
E\right\rangle _{\mathrm{ds}}=\tfrac{E_{dya}}{N^\circ_{e^+e^-}}$ is
then
\begin{equation}
\left\langle E\right\rangle _{\mathrm{ds}}\simeq\tfrac{3}{8}\left(
\tfrac{\lambda_{\mathrm{C}}}{r_{ds}}\right)  \tfrac{Qe}{r_{+}%
}\,,\label{meanenedya}%
\end{equation}
which can be also rewritten as
\begin{equation}
\left\langle E\right\rangle _{\mathrm{ds}}\simeq\tfrac{1}{2}\left(
\tfrac{r_{\mathrm{ds}}}{r_{+}}\right)  \ m_{e}c^{2}\sim\sqrt{\tfrac{\xi
}{M/M_{\odot}}}10^{5}keV\,.\label{GRB}%
\end{equation}
Such a process of vacuum polarization, occurring not at the horizon but in the
extended dyadosphere region ($r_{+}\leq r\leq r_{\mathrm{ds}}$) around an
EMBH, has been observed to reach the maximum efficiency limit of $50\%$ of the
total mass-energy of an extreme EMBH (see e.g. \textcite{prx98}). The conceptual
justification of this result follows from the present work: the $e^{+}e^{-}$
creation process occurs at the expence of the Coulomb energy given by Eq.
(\ref{EQR}) and does not affect the irreducible mass given by Eq. (\ref{EQM}),
which indeed, as we have proved, does not depend of the electromagnetic
energy. In this sense, $\delta M_{\mathrm{irr}}=0$ and the transformation is
fully reversible. This result will be further validated by the study of the
dynamical formation of the dyadosphere, which we have obtained using the
present work and \textcite{crv02} (see \textcite{rvx02}).
\end{enumerate}

Let us now compare and contrast these two processes. We have
\begin{align}
r_{\mathrm{Eerg}}  & \simeq\left(  \tfrac{r_{ds}}{\lambda
_{\mathrm{C}}}\right)  r\\
N_{\mathrm{dya}}  & \simeq\left(  \tfrac{r_{ds}}{\lambda
_{\mathrm{C}}}\right)  N_{\mathrm{PD}},\\
\left\langle E\right\rangle _{\mathrm{dya}}  & \simeq\left(  \tfrac
{\lambda_{\mathrm{C}}}{r_{ds}}\right)  \left\langle E\right\rangle
_{\mathrm{PD}}.
\end{align}
Moreover we see (Eqs. (\ref{UHECR}), (\ref{GRB})) that $\left\langle
E\right\rangle _{\mathrm{PD}}$ is in the range of energies of UHECR, while for
$\xi\sim0.1$ and $M\sim10M_{\odot}$, $\left\langle E\right\rangle
_{\mathrm{ds}}$ is in the gamma ray range. In other words, the discrete
particle decay process involves a small number of particles with ultra high
energies ($\sim10^{21}eV$), while vacuum polarization involves a much larger
number of particles with lower mean energies ($\sim10MeV$).

\begin{figure}
\includegraphics[width=10cm,clip]{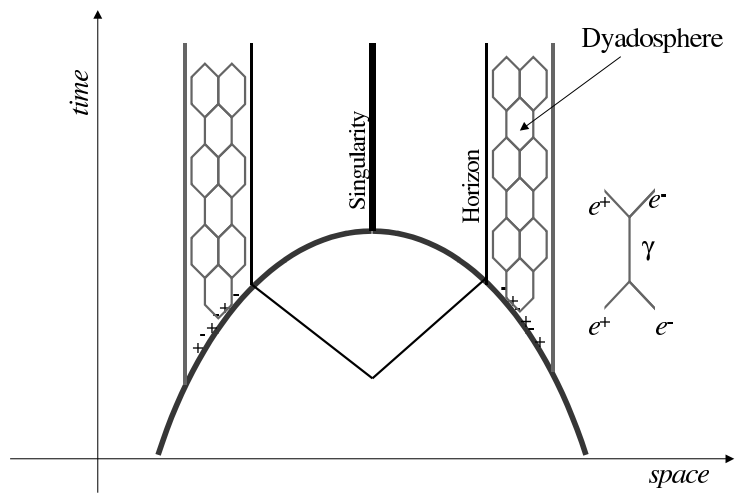}
\caption{Space-time diagram of the collapse process leading to the formation of the dyadosphere. As the collapsing core crosses the dyadosphere radius the pair creation process starts, and the pairs thermalize in a neutral plasma configuration. Then also the horizon is crossed and the singularity is formed.}
\label{dyaform}
\end{figure}

Having so established and clarified the basic conceptual processes of the energetic of the EMBH, we are now ready to approach, using the new analytic solution obtained, the dynamical process of vacuum polarization occurring during the formation of an EMBH as qualitatively represented in Fig.~\ref{dyaform}. The study of the dyadosphere dynamical formation as well as of the electron-positron plasma dynamical evolution will lead to the first possibility of directly observing the general relativistic effects approaching the EMBH horizon.

Before closing we would like to emphasize once more a basic point: all the considerations presented in the description of the preceding eras are based on the approximations in the description of the dyadosphere presented in section \ref{dyadosphere}. This treatment is very appropriate in estimating the general dependence of the energy of the P-GRB, the kinetic energy of the ABM pulse and consequently the intensity of the afterglow, as well as the overall time structure of the GRB and especially the time of the release of the P-GRB in respect to the moment of gravitational collapse and its relative intensity with respect to the afterglow. If, however, is addressed the issue of the detailed temporal structure of the P-GRB and its detailed spectral distribution, the above dynamical considerations on the dyadosphere formation are needed (see also \textcite{rvx02}). In turn, this detailed analysis is needed if the general relativistic effects close to the horizon formation have to be followed. As expressed already in section.~\ref{new}, all general relativistic quantum field theory effects are encoded in the fine structure of the P-GRB. As emphasized in section~\ref{fp}, the only way to differentiate between solutions with same $E_{dya}$ but different EMBH mass and charge is to observe the P-GRBs in the limit $B\to0$, namely, to observe the short GRBs.

\section{Contribution of the EMBH model to the black hole theory}\label{luca2}

The aim of this section is to point out how the knowledge obtained from the EMBH model is of relevance also for the basic theory of black holes and further how very high precision verification of general relativistic effects in the very strong field near the formation of the horizon should be expected in the near future.

We shall first see how Eq.(\ref{EQM}) for $M_{\mathrm{irr}}$,
\begin{equation}
M_{\mathrm{irr}}=M_{0}-\tfrac{M_{0}^{2}}{2r_{+}}+T_{+}\, ,
\end{equation}
leads to a deeper physical understanding of the role of the gravitational interaction in the maximum energy extraction process of an EMBH. This formula can also be of assistance in clarifying some long lasting epistemological issue on the role of general relativity, quantum theory and thermodynamics.

It is well known that if a spherically symmetric mass distribution without any electromagnetic structure undergoes free gravitational collapse, its total mass-energy $M$ is conserved according to the Birkhoff theorem: the increase in the kinetic energy of implosion is balanced by the increase in the gravitational energy of the system. If one considers the possibility that part of the kinetic energy of implosion is extracted then the situation is very different: configurations of smaller mass-energy and greater density can be attained without violating Birkhoff theorem.

We illustrate our considerations with two examples: one has found confirmation from astrophysical observations, the other promises to be of relevance for gamma ray bursts (GRBs) (see \textcite{rv02a}). Concerning the first example, it is well known from the work of \textcite{l32} that at the endpoint of thermonuclear evolution, the gravitational collapse of a spherically symmetric star can be stopped by the Fermi pressure of the degenerate electron gas (white dwarf). A configuration of equilibrium can be found all the way up to the critical number of particles
\begin{equation}
N_{\mathrm{crit}}=0.775\tfrac{m_{Pl}^{3}}{m_{0}^{3}},
\end{equation}
where the factor $0.775$ comes from the coefficient $\tfrac{3.098}{\mu^{2}}$ of the solution of the Lane-Emden equation with polytropic index $n=3$, and $m_{Pl}=\sqrt{\tfrac{\hbar c}{G}}$ is the Planck mass, $m_{0}$ is the nucleon mass and $\mu$ the average number of electrons per nucleon. As the kinetic energy of implosion is carried away by radiation the star settles down to a configuration of mass
\begin{equation}
M=N_{\mathrm{crit}}m_{0}-U, \label{BE}%
\end{equation}
where the gravitational binding energy $U$ can be as high as $5.72\times 10^{-4}N_{\mathrm{crit}}m_{0}$.

Similarly Gamov (see \textcite{g51}) has shown that a gravitational collapse process to still higher densities can be stopped by the Fermi pressure of the neutrons (neutron star) and Oppenheimer (\textcite{ov39}) has shown that, if the effects of strong interactions are neglected, a configuration of equilibrium exists also in this case all the way up to a critical number of particles
\begin{equation}
N_{\mathrm{crit}}=0.398\tfrac{m_{Pl}^{3}}{m_{0}^{3}},
\end{equation}
where the factor $0.398$ comes now from the integration of the
Tolman-Oppenheimer-Volkoff equation (see e.g. \textcite{htww65}). If the kinetic energy of implosion is again carried away by radiation of photons or neutrinos and antineutrinos the final configuration is characterized by the formula (\ref{BE}) with $U\lesssim2.48\times10^{-2}N_{\mathrm{crit}}m_{0}$. These considerations and the existence of such large values of the gravitational binding energy have been at the heart of the explanation of astrophysical phenomena such as red-giant stars and supernovae: the corresponding measurements of the masses of neutron stars and white dwarfs have been carried out with unprecedented accuracy in binary systems (\textcite{gr75}).

From a theoretical physics point of view it is still an open question how far such a sequence can go: using causality nonviolating interactions, can one find a sequence of braking and energy extraction processes by which the density and the gravitational binding energy can increase indefinitely and the mass-energy of the collapsed object be reduced at will? This question can also be formulated in the mass-formula language of a black hole given in \textcite{cr71} (see also \textcite{rv02a}): given a collapsing core of nucleons with a given rest mass-energy $M_{0}$, what is the minimum irreducible mass of the black hole which is formed?

Following \textcite{crv02} and \textcite{rv02a}, consider a spherical shell of rest mass $M_{0}$ collapsing in a flat space-time. In the neutral case the irreducible mass of the final black hole satisfies the equation (see \textcite{rv02a})
\begin{equation}
M_{\mathrm{irr}}=M=M_{0}-\tfrac{M_{0}^{2}}{2r_{+}}+T_{+}, \label{Mirr2}%
\end{equation}
where $M$ is the total energy of the collapsing shell and $T_{+}$ the kinetic energy at the horizon $r_{+}$. Recall that the area $S$ of the horizon is (\textcite{cr71})
\begin{equation}
S=4\pi r_{+}^{2}=16\pi M_{\mathrm{irr}}^{2} \label{S}%
\end{equation}
where $r_{+}=2M_{\mathrm{irr}}$ is the horizon radius. The minimum irreducible mass $M_{\mathrm{irr}}^{\left(  {\mathrm{min}}\right)  }$ is obtained when the kinetic energy at the horizon $T_{+}$ is $0$, that is when the entire kinetic energy $T_{+}$ has been extracted. We then obtain the simple result
\begin{equation}
M_{\mathrm{irr}}^{\left(  \mathrm{min}\right)  }=\tfrac{M_{0}}{2}.
\label{Mirrmin}%
\end{equation}
We conclude that in the gravitational collapse of a spherical shell of rest mass $M_{0}$ at rest at infinity (initial energy $M_{\mathrm{i}}=M_{0}$), an energy up to $50\%$ of $M_{0}c^{2}$ can in principle be extracted, by braking processes of the kinetic energy. In this limiting case the shell crosses the horizon with $T_{+}=0$. The limit $\tfrac{M_{0}}{2}$ in the extractable kinetic energy can further increase if the collapsing shell is endowed with kinetic energy at infinity, since all that kinetic energy is in principle extractable.

\begin{figure}
\begin{center}
\includegraphics[width=10cm,clip]{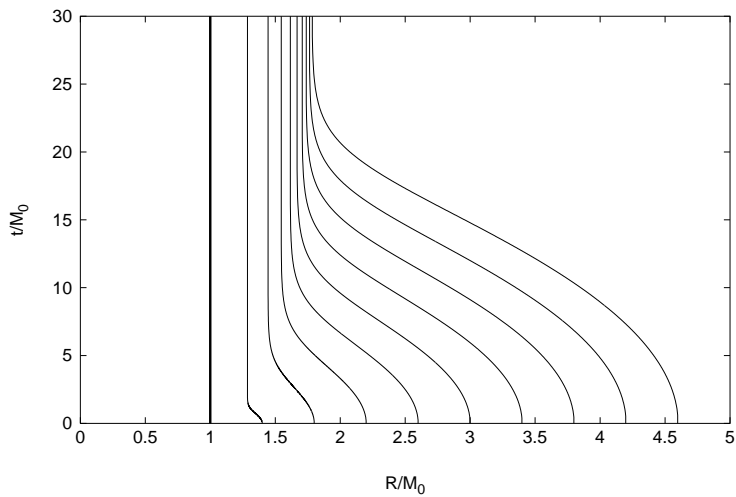}
\caption{Collapse curves for neutral shells with rest mass $M_{0}$ starting at rest at selected radii $R^{\ast}$ computed by using the exact solutions given in \textcite{crv02}. A different value of $M_{\mathrm{irr}}$ (and therefore of $r_{+}$) corresponds to each curve. The time parameter is the Schwarzschild time coordinate $t$ and the asymptotic behaviour at the respective horizons is evident. The limiting configuration $M_{\mathrm{irr}}=\tfrac{M_{0}}{2}$ (solid line) corresponds to the case in which the shell is trapped, at the very beginning of its motion, by the formation of the horizon.}
\label{fig1l2}
\end{center}
\end{figure}

\begin{figure}
\begin{center}
\includegraphics[width=10cm,clip]{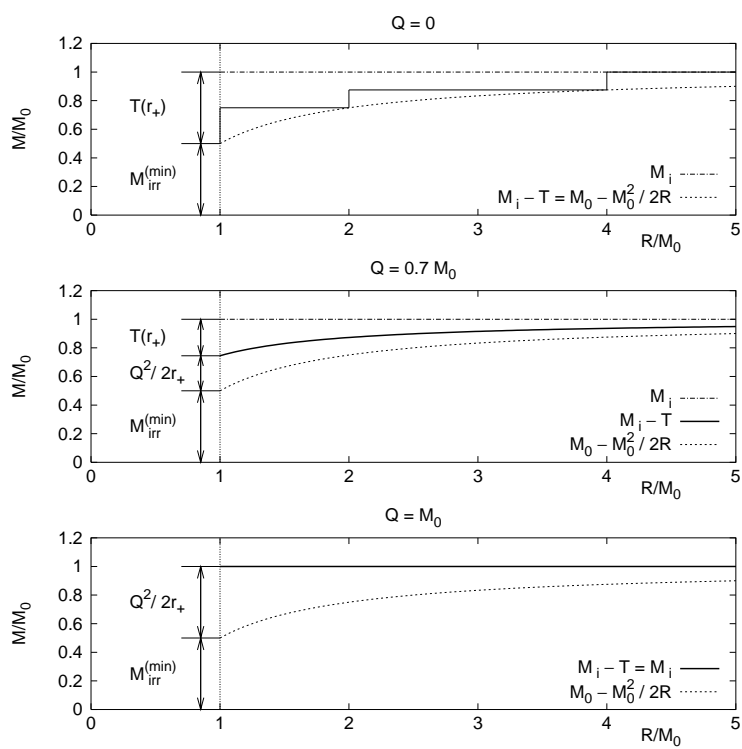}
\caption{Energetics of a shell such that $M_{\mathrm{i}}=M_{0}$, for selected values of the charge. In the first diagram $Q=0$; the dashed line represents the total energy for a gravitational collapse without any braking process as a function of the radius $R$ of the shell; the solid, stepwise line represents a collapse with suitable braking of the kinetc energy of implosion at selected radii; the dotted line represents the rest mass energy plus the gravitational binding energy. In the second and third diagram $Q/M_{0}=0.7$, $Q/M_{0}=1$ respectively; the dashed and the dotted lines have the same meaning as above; the solid lines represent the total energy minus the kinetic energy. The region between the solid line and the dotted line corresponds to the stored electromagnetic energy. The region between the dashed line and the solid line corresponds to the kinetic energy of collapse. In all the cases the sum of the kinetic energy and the electromagnetic energy at the horizon is 50\% of $M_{0}$. Both the electromagnetic and the kinetic energy are extractable. It is most remarkable that the same underlying process occurs in the three cases: the role of the electromagnetic interaction is twofold: a) to reduce the kinetic energy of implosion by the Coulomb repulsion of the shell; b) to store such an energy in the region around the EMBH. The stored electromagnetic energy is extractable as shown in \textcite{rv02a}.}
\label{fig3l2}
\end{center}
\end{figure}

In order to illustrate the physical reasons for this result, using the formulas of \textcite{crv02}, we have represented in Fig.~\ref{fig1l2} the world lines of spherical shells of the same rest mass $M_{0}$, starting their gravitational collapse at rest at selected radii $R^{\ast}$. These initial conditions can be implemented by performing suitable braking of the collapsing shell and concurrent kinetic energy extraction processes at progressively smaller radii (see also Fig.~\ref{fig3l2}). The reason for the existence of the minimum (\ref{Mirrmin}) in the black hole mass is the ``self closure'' occurring by the formation of a horizon in the initial configuration (thick line in Fig.~\ref{fig1l2}).

Is the limit $M_{\mathrm{irr}}\rightarrow\tfrac{M_{0}}{2}$ actually attainable without violating causality? Let us consider a collapsing shell with charge $Q$. If $M\geq Q$ an EMBH is formed. As pointed out in \textcite{rv02a} the irreducible mass of the final EMBH does not depend on the charge $Q$. Therefore Eqs.~(\ref{Mirr2}) and (\ref{Mirrmin}) still hold in the charged case with $r_{+}=M+\sqrt{M^{2}-Q^{2}}$. In Fig.~\ref{fig3l2} we consider the special case in which the shell is initially at rest at infinity, i.e. has initial energy $M_{\mathrm{i}}=M_{0}$, for three different values of the charge $Q$. We plot the initial energy $M_{i}$, the energy of the system when all the kinetic energy of implosion has been extracted as well as the sum of the rest mass energy and the gravitational binding energy $-\tfrac{M_{0}^{2}}{2R}$ of the system (here $R$ is the radius of the shell). In the extreme case $Q=M_{0}$, the shell is in equilibrium at all radii (see \textcite{crv02}) and the kinetic energy is identically zero. In all three cases, the sum of the extractable kinetic energy $T$ and the electromagnetic energy $\tfrac{Q^{2}}{2R}$ reaches $50\%$ of the rest mass energy at the horizon, according to Eq.~(\ref{Mirrmin}).

What is the role of the electromagnetic field here? If we consider the case of a charged shell with $Q\simeq M_{0}$, the electromagnetic repulsion implements the braking process and the extractable energy is entirely stored in the electromagnetic field surrounding the EMBH (see \textcite{rv02a}). In \textcite{rv02a} we have outlined two different processes of electromagnetic energy extraction. We emphasize here that the extraction of $50\%$ of the mass-energy of an EMBH is not specifically linked to the electromagnetic field but depends on three factors: a) the increase of the gravitational energy during the collapse, b) the formation of a horizon, c) the reduction of the kinetic energy of implosion. Such conditions are naturally met during the formation of an extreme EMBH but are more general and can indeed occur in a variety of different situations, e.g. during the formation of a Schwarzschild black hole by a suitable extraction of the kinetic energy of implosion (see Fig.~\ref{fig1l2} and Fig.~\ref{fig3l2}).

Now consider a test particle of mass $m$ in the gravitational field of an already formed Schwarzschild black hole of mass $M$ and go through such a sequence of braking and energy extraction processes. Kaplan (\textcite{k49}) found for the energy $E$ of the particle as a function of the radius $r$
\begin{equation}
E=m\sqrt{1-\tfrac{2M}{r}}.\label{pointtest}
\end{equation}
It would appear from this formula that the entire energy of a particle could be extracted in the limit $r\rightarrow2M$. Such $100\%$ efficiency of energy extraction has often been quoted as evidence for incompatibility between General Relativity and the second principle of Thermodynamics (see \textcite{b73} and references therein). J. Bekenstein and S. Hawking have gone as far as to consider General Relativity not to be a complete theory and to conclude that in order to avoid inconsistencies with thermodynamics, the theory should be implemented through a quantum description (\textcite{b73,h74}). Einstein himself often expressed the opposite point of view (see e.g. \textcite{d02}).

The analytic treatment presented in \textcite{crv02} can clarify this fundamental issue. It allows to express the energy increase $E$ of a black hole of mass $M_{1}$ through the accretion of a shell of mass $M_{0}$ starting its motion at rest at a radius $R$ in the following formula which generalizes Eq.~(\ref{pointtest}):
\begin{equation}
E\equiv M-M_{1}=-\tfrac{M_{0}^{2}}{2R}+M_{0}\sqrt{1-\tfrac{2M_{1}}{R}},
\end{equation}
where $M=M_{1}+E$ is clearly the mass-energy of the final black hole. This formula differs from the Kaplan formula (\ref{pointtest}) in three respects: a) it takes into account the increase of the horizon area due to the accretion of the shell; b) it shows the role of the gravitational self energy of the imploding shell; c) it expresses the combined effects of a) and b) in an exact closed formula.

The minimum value $E_{\mathrm{\min}}$ of $E$ is attained for the minimum value of the radius $R=2M$: the horizon of the final black hole. This corresponds to the maximum efficiency of the energy extraction. We have
\begin{equation}
E_{\min}=-\tfrac{M_{0}^{2}}{4M}+M_{0}\sqrt{1-\tfrac{M_{1}}{M}}=-\tfrac
{M_{0}^{2}}{4(M_{1}+E_{\min})}+M_{0}\sqrt{1-\tfrac{M_{1}}{M_{1}+E_{\min}}},
\end{equation}
or solving the quadratic equation and choosing the positive solution for physical reasons
\begin{equation}
E_{\min}=\tfrac{1}{2}\left(  \sqrt{M_{1}^{2}+M_{0}^{2}}-M_{1}\right)  .
\end{equation}
The corresponding efficiency of energy extraction is
\begin{equation}
\eta_{\max}=\tfrac{M_{0}-E_{\min}}{M_{0}}=1-\tfrac{1}{2}\tfrac{M_{1}}{M_{0}%
}\left(  \sqrt{1+\tfrac{M_{0}^{2}}{M_{1}^{2}}}-1\right)  , \label{efficiency}%
\end{equation}
which is strictly \emph{smaller than} 100\% for \emph{any} given $M_{0}\neq0$. It is interesting that this analytic formula, in the limit $M_{1}\ll M_{0}$, properly reproduces the result of equation (\ref{Mirrmin}), corresponding to an efficiency of $50\%$. In the opposite limit $M_{1}\gg M_{0}$ we have
\begin{equation}
\eta_{\max}\simeq1-\tfrac{1}{4}\tfrac{M_{0}}{M_{1}}.
\end{equation}
Only for $M_{0}\rightarrow0$, Eq.~(\ref{efficiency}) corresponds to an efficiency of 100\% and correctly represents the limiting reversible transformations introduced in \textcite{cr71}. It seems that the difficulties of reconciling General Relativity and Thermodynamics are ascribable not to an incompleteness of General Relativity but to the use of the Kaplan formula in a regime in which it is not valid. The generalization of the above results to stationary black holes is being considered.

\section{Conclusions}\label{conclusions}

The EMBH theory has been here applied for the first time to fit the experimental data of GRB~991216. This process has given us the opportunity to rethink the entire GRB process in an unitary description starting from the moment of gravitational collapse all the way up to the latest phases of the afterglow. We have identified the three fundamental actors of the GRB phenomenon in:
\begin{enumerate}
\item $E_{dya}$. Having reanalyzed in section~\ref{dyadosphere} the physics of the dyadosphere we have pointed out in Fig.~\ref{muxi} that the same value of $E_{dya}$ can be obtained from an entire family of $\left(\mu,\xi\right)$ parameters (i.e. $E_{dya}$ is degenerate in $\left(\mu,\xi\right)$). We have then shown in the reexamination of all the GRB eras that all the results depend only on the value of $E_{dya}$ and not on the particular value of $\left(\mu,\xi\right)$ (see sections~\ref{era2},\ref{era3},\ref{era4},\ref{era5}). The only exception to this occurs in the era I (see section~\ref{era1}) which is the only one relevant for short GRBs.
\item $B$. The crucial role played by the baryonic remnant of the progenitor star in determining the relative intensity ratio and the time delay between the P-GRB and the E-APE has been summarized already in the two Figs.~\ref{crossen}--\ref{dtab} in the introduction.
\item ISM. The density $n_{ism}$ of the interstellar medium and its inhomogeneities appears to have a fundamental role in the intensity and the temporal substructures of the E-APE and the afterglow.
\end{enumerate}

The observational data agree with the predictions of the model on:
\begin{enumerate}
\item the intensity ratio, $1.58\times 10^{-2}$,  between the P-GRB and the E-APE, which strongly depends on the parameter $B$;
\item the absolute intensities for both the P-GRB and the E-APE, respectively $7.54\times 10^{51}$ erg  and  $4.75\times 10^{53}$;
\item the arrival time of the P-GRB and the peak of the E-APE, respectively $8.41 \times 10^{-2}$ s and $19.87$ s;
\item the power-law index $n$ of the afterglow, predicted $n_{theo} = -1.6$ and observed $n_{obs} = -1.616 \pm 0.067$ (see sections~\ref{approximation},~\ref{power-law});
\item the temporal structure of the E-APE and its correlation with the inhomogeneity in the ISM;
\item the spectral distribution of the X-ray and $\gamma$-ray emission.
\end{enumerate}

Concerning the total energy of GRB~991216, $E_{dya}=4.83\times 10^{53}$ erg is found in the EMBH theory. This value is systematically larger than the ones quoted in the current literature by \textcite{pk01} and by \textcite{ha00} due to the fact that they respectively consider beaming angles of $3^\circ-4^\circ$ and $6^\circ$. These considerations have been shown to be untenable in section~\ref{power-law}. There is still a difference of $\sim 28\%$ between the total energy implied by the EMBH theory ($4.83\times 10^{53}$ erg) and the value quoted by Halpern ($E_{dya}=6.7\times 10^{53}$ erg) in the case of spherical emission. We trust that this is a consequence of the underlying assumption of the spectral distribution of the radiation assumed by \textcite{ha00} (see e.g. \textcite{f01}), which should be reassessed on the ground of our theoretical results (see also \textcite{rbcfx02e_paperII}).

These results can certainly be considered the success of the EMBH theory.

Before closing, we like to stress how GRBs, if duly theoretically interpreted, can open a main avenue of inquiring on totally new physical and astrophysical regimes. This program is very likely one of the greatest computational efforts in physics and astrophysics and cannot be actuated using shortcuts.

From the point of view of fundamental physics new regimes are explored:
\begin{enumerate}
\item The process of energy extraction from black holes. It is interesting that the analysis of GRBs has promoted a new effort in developing new theoretical tools for approaching the dynamical phase of collapse as expressed in section.~\ref{luca}. These results have further clarified some basic issue related to the energy extraction process from black hole (see e.g. \textcite{r01mg9}). It was already known from the definition of the ergosphere (see \textcite{rw71b}) that the rotational energy extraction process do occur in an extended region around the horizon of a black hole. The fortunate situation that the energy extraction process in GRBs occurs in a condition of almost perfect spherical symmetry have allowed us to focus on the second fundamental parameter of black holes, namely the electric energy. The spherical symmetry has allowed as well to develop some powerful theoretical tools (see section~\ref{luca}) which have allowed to reach a better understanding of the role of kinetic energy of implosion in the process of gravitational collapse, in the storage of electromagnetic energy in the region around black holes and to establish as well a new upper limit in the energy extraction process in the gravitational collapse up to $50\%$ of the initial rest mass of the system (see section~\ref{luca2}). These results are of general validity and do transcend the work on the EMBH theory, although they are motivated by these researches. Interestingly this work, by giving a new expression for the efficiency of transforming gravitational energy into mechanical work (see section~\ref{luca2}), has opened up a new opportunity of debating the relation between general relativity, thermodynamics and quantum theory, which is certainly one of the most profound and important topic of research in the entire realm of fundamental physics.
\item The quantum and general relativistic effects of matter-antimatter creation near the black hole horizon. It is well known that one of the most important topics pursued in the last seventy years in physics has been the possibility, postulated by Sauter, Heisemberg, Euler, Schwinger to create matter-antimatter from the vacuum. In order to have the first experimental and observational evidence for this phenomenon, three major approaches are being followed:\\
a) In central collisions of heavy ions near the Coulomb barrier, as first proposed in \textcite{gz69,gz70} (see also \textcite{pr71,p72,zp72}). Efforts in experimentally implementing this idea at GSI were made since early 80's. Despite some apparently encouraging result (\textcite{s83}), such efforts have failed so far due to the small contact time of the colliding ions (see e.g. \textcite{aa95,ga96,la97,ba95,ha98}). Typically the electromagnetic energy involved in the collisions of heavy ions with impact parameter $l_{1}\sim10^{-12}$cm is $E_{1}\sim10^{-6}$erg.\\
b) At the focus of an X-ray free electron laser (XFEL) (see \textcite{rsv02,rw01} and references therein). This idea will be possibly testable at DESY, where the XFEL is part of the design of the collider TESLA, as well as at SLAC, where the so-called Linac Coherent Light Source (LCLS) has been proposed. The electromagnetic energy at the focus of an XFEL is $E_{2}\sim10^{6}$erg concentrated in a region of linear extension $l_{2}\sim10^{-8}$ cm (\textcite{rw01}).\\
c) Around an electromagnetic black hole (EMBH) (\textcite{dr75,prx98,prxprl}), giving rise to the observed phenomenon of GRBs (see e.g. \textcite{lett1,lett2,lett3,rbcfx02a_sub}). The electromagnetic energy of an EMBH of mass $M\sim 10M_\odot$ and charge $Q\sim 0.1M/\sqrt{G}$ is $E_3\sim 10^{54}\, {\rm ergs}$ and it is deposited in a region of linear extension $l_3\sim 10^8\, {\rm m}$ (\textcite{prx98,rv02a}).\\
There is the very distinct possibility that in this race the success will be reached by the observations in relativistic astrophysics more than from the high energy experiments on the Earth. This will be certainly a splendid success which will be only second to the discovery of Helium first in the stars and then on the Earth! Quite apart from the discovery in itself, the detection of vacuum polarization in the astrophysical settings presents distinctively new physical phenomena as \textcite{rvx03}. The very important topic to be covered in the forthcoming months is the study of the dynamical phase of gravitational collapse and to follow the effects of such process of vacuum polarization in the dynamical phase. It will be also important to follow the development of this process all the way to the emission of the P-GRB (\textcite{rv03}).
\item The physics of ultrarelativisitc shock waves with Lorentz gamma factor $\gamma > 100$. We are expecting much progress in this topic from the understanding of the instantaneous spectrum of GRBs. Some preliminary results along this line are presented in \textcite{rbcfx02c_spectrum}. See also section~\ref{spectrum}.
\end{enumerate}

From the point of view of astronomy and astrophysics also new regimes are explored:
\begin{enumerate}
\item The occurrence of gravitational collapse to a black hole from a critical mass core of mass $M\agt 10M_\odot$, which clearly differs from the values of the critical mass encountered in the study of stars ``catalyzed at the endpoint of thermonuclear evolution" (white dwarfs and neutron stars).
\item The extremely high efficiency of the spherical collapse to a black hole, where almost $99.99\%$ of the core mass collapses leaving negligible remnant. The EMBH theory offers an unprecedented tool in order to map with great accuracy all the matter distribution around the newly formed EMBH from the horizon all the way to the ISM. This concept was pioneered by \textcite{dm99} who proposed to use GRB sources as ``tomographic images of the density distributions of the medium surrounding the sources of GRBs''. It is important to emphasize that the very precise reading of the matter distribution encoded in the data of the P-GRB, the E-APE and the afterglow in GRB~991216 is in marked disagreement with the matter distribution postulated by the ``collapsar'' scenario (see \textcite{p98,w93,mw99}). This conclusion is evidenced not only by the absence of beaming already mentioned above, but also for the paucity of the baryonic matter encountered by the PEM pulse in its way out from the EMBH. There is no evidence for the presence either of a baryonic disk component nor of a conspicuous baryonic remnant. We actually have $B=3.0\times 10^{-3}$. Unlike the case of formation of a neutron star, the mass of the remnant of the progenitor star is very small indeed. This mass, determined by $B$, is very accurately inferable from the relative intensity and temporal distance between the P-GRB and the E-APE (see above). In the present case we have $M_B \sim 8.1\times 10^{-4} M_{\odot}$. The presence of the remnant is also important for guaranteeing the overall charge neutrality of the system formed by the oppositely charged collapsing core and the remnant. It has been pointed out in section~\ref{gc} that this condition of charge separation between the collapsing core and the remnant occurs only during the relevant part of the gravitational collapse process which, we recall, for a $10M_{\odot}$ is of the order of 30 seconds.
\item The necessity of developing a fine tuning in the final phases of thermonuclear evolution of the stars, both for the star collapsing to the black hole and the surrounding ones, in order to explain the possible occurrence of the ``induced gravitational collapse".
\end{enumerate}

New regimes are as well encountered from the point of view of nature of GRBs:
\begin{enumerate}
\item The basic structure of GRBs is uniquely composed by a proper-GRB (P-GRB) and the afterglow. The most general GRB contains three different components: the P-GRB, the E-APE and the rest of the afterglow. The ratio between the P-GRB and the E-APE intensity and their temporal separation is a function of the $B$ parameter (see Figs.~\ref{crossen}--\ref{dtab}). The best fit is obtained for $B=3.0\times 10^{-3}$ (see section~\ref{bf}). We recall that in the present case for $B < 2.5\times10^{-5}$ the energy of the P-GRB would be larger than the one of the E-APE and the energy of the dyadosphere would be mainly emitted in what have been called the ``short bursts'', while for $B > 2.5\times10^{-5} $ the energy of the E-APE would predominate and the energy of the dyadosphere would be mainly carried by the ABM pulse and emitted in the afterglow.
\item The long bursts are then simply explained as the peak of the afterglow (the E-APE) and their observed time variability is explained in terms of inhomogeneities in the interstellar medium (ISM). The difficulties encountered by {\em all} theoretical models, through the years, in order to explain the so called ``long bursts'' are resolved in a drastic way (see section~\ref{shortlongburst}). The so called ``long bursts'' are {\em not} bursts at all. They represent just the E-APE which was interpreted as a burst only due to the noise threshold in the BATSE observations (see Fig.~\ref{fit_1}). The E-APE is emitted at distances from the EMBH in the range $1.0\times 10^{16}\sim 1.0\times 10^{17}$ cm, see Tab.~\ref{tab1}, namely well outside the size of the progenitor star and already deep in interstellar space. The fact that the crossing of such distance, which is a typical dimension of an interstellar cloud, appears to occur in arrival time in only $\sim 100$ seconds is perfectly explained by the relativistic transformations encoded in the RSTT paradigm corresponding to a gamma factor between $100$ and $300$ (see section~\ref{arrival_time} and Tab.~\ref{tab1}). This effect would be interpreted within a classical and incorrect astronomical picture by a ``superluminal'' behaviour propagating at $\sim 3.6\times 10^4c$ (see Tab.~\ref{tab1}).
\item The short bursts are identified with the P-GRBs and the crucial information on general relativistic and vacuum polarization effects are encoded in their spectra and intensity time variability. In the limit $B \to 0$ the entire dyadosphere energy is emitted in the P-GRB. These events represents the ``short bursts'' class, for which the afterglow intensity is smaller than the P-GRB emission and below the actual observational limits (see section~\ref{new}). It is interesting that the proposed differentiation between the ``short bursts'' and ``long bursts'' within the EMBH theory is merely due to the amount of baryonic matter in the remnant, described by the $B$ parameter, and totally independent from the process of gravitational collapse which is clearly identical in both cases. This explains at once the recently found conclusion that the distribution of short and long GRBs have essentially the same characteristic peak luminosity (\textcite{s01}). Also the result expressed in Fig.~\ref{energypeak} that the average temperature corresponding to the P-GRB emission does increase for decreasing values of the $B$ parameter can explain the observed fact that the ``short bursts'', which are obtained in the limit $B\to 0$, are systematically harder than ``long bursts'' (\textcite{ka93}).
\end{enumerate}

A new class of space missions to acquire information on such extreme new regimes are urgently needed. The detailed observations of the yet unexplored region in the range up to $10$ seconds in Fig.~\ref{final} and the corresponding observations of the ``short bursts'' by a new class of space missions with higher sensitivity than the BATSE instrument appear to be of great importance. Such observations should allow to directly observe for the first time the general relativistic and extreme quantum field theory effects connected to the process of formation of the EMBH. It can be of some interest to explore the possibility of observing in these regimes the ``gravitationally induced electromagnetic radiation'' (\textcite{ja73}) and the ``electromagnetically induced gravitational radiation'' (\textcite{ja74}) phenomena as well as to explore the possibility of developing neutrino detectors. This will need further developments of the predictions of the EMBH theory in these general relativistic and ultra-high-energy particle phenomena.

\end{document}